\documentclass{article}

\usepackage[T1]{fontenc}
\usepackage{amsmath,amssymb,amsfonts}
\usepackage{graphics, subfigure,color} 
\usepackage{latexsym}
\usepackage[dvips]{graphicx}
\usepackage{epsfig}
\usepackage{cite,hyperref}
\usepackage{chngcntr}

\usepackage[strings]{underscore}

\numberwithin{equation}{section}
\counterwithout{figure}{section}

\newtheorem{definition}{Definition}[section]
\newtheorem{prop}{Proposition}[section]
\newtheorem{lemma}[prop]{Lemma}
\newtheorem{thm}[prop]{Theorem}
\newtheorem{theorem}[prop]{Theorem}
\newtheorem{cor}[prop]{Corollary}

\newcommand{\bbbone}{\bf{ 1}}
\newcommand{\bbone}{{\bf 1}}

\newcommand{\cId}{{\rm{Id}}}

\newcommand{\R}{\mathbb{R}}
\newcommand{\N}{\mathbb{N}}
\newcommand{\Z}{\mathbb{Z}}

\newcommand{\cB}{{\cal{B}}}
\newcommand{\cL}{{\cal{L}}}
\newcommand{\cJ}{{\cal{J}}}
\newcommand{\cM}{{\cal{M}}}
\newcommand{\cV}{{\cal{V}}}
\newcommand{\cU}{{\cal{U}}}
\newcommand{\Ws}{{\cal W}^s}
\newcommand{\Wu}{{\cal W}^u}
\newcommand{\cE}{{\cal{E}}}
\newcommand{\cF}{{\cal{F}}}
\newcommand{\cT}{{\cal{T}}}
\newcommand{\cS}{{\mathcal S}}
\newcommand{\cW}{{\mathcal W}}

\newcommand{\ga}{\gamma}
\newcommand{\de}{\delta}
\newcommand{\al}{\alpha}

\newcommand{\Ren}{\mathfrak{R}}
\newcommand{\ren}{\mathsf{R}}
\newcommand{\eps}{\varepsilon}
\newcommand{\D}{{\bf{D}}}

\newcommand{\bee}{\begin{equation}}
\newcommand{\be}{\begin{equation}}
\newcommand{\ee}{\end{equation}}
\newcommand{\beq}{\begin{eqnarray}}
\newcommand{\eeq}{\end{eqnarray}}
\newcommand{\bqa}{\begin{eqnarray}}
\newcommand{\eqa}{\end{eqnarray}}
\newcommand{\bea}{\begin{eqnarray}}
\newcommand{\eea}{\end{eqnarray}}

\newcommand{\prf}{{\noindent \bf Proof:\;}}
\newcommand{\qed}{{\hfill $\Box$}}

\begin{document}

\thispagestyle{empty}

\begin{flushright}\footnotesize\ttfamily
ITP-UU-14/04\\
SPIN-14/04\\
HU-Mathematik-2014-01\\
HU-EP-14/01
\end{flushright}
\vspace{3em}

\begin{center}
\textbf{\Large\mathversion{bold} Renormalization: an advanced overview}

\vspace{2em}

\textrm{\large Razvan Gurau${}^{1,2}$, Vincent Rivasseau${}^{3,2}$\\ and Alessandro Sfondrini${}^{4,5}$}

\vspace{2em}

\textit{\small
1.  CPHT - UMR 7644, CNRS, \'Ecole Polytechnique,\\ 91128 Palaiseau cedex, France\\[0.2cm]
2.  Perimeter Institute for Theoretical Physics,\\
 31 Caroline St. N, N2L 2Y5, Waterloo, ON, Canada\\[0.2cm]
3. LPT - UMR 8627, CNRS, Universit\'e Paris 11,\\ 91405 Orsay Cedex, France\\[0.2cm]
4. Institute for Theoretical Physics and Spinoza Institute,\\ Utrecht University, 3508 TD Utrecht, The Netherlands\\[0.2cm]
5. Inst. f\"ur Mathematik \& Inst. f\"ur Physik, Humboldt-Universit\"at zu Berlin\\
IRIS Geb\"aude, Zum Grossen Windkanal 6, 12489 Berlin, Germany}

\vspace{1em}

\texttt{\small rgurau@cpht.polytechnique.fr, vincent.rivasseau@th.u-psud.fr, Alessandro.Sfondrini@physik.hu-berlin.de}

%%%%%%%%

\end{center}

\vspace{4em}

\begin{abstract}\noindent
We present several approaches to renormalization in QFT: the multi-scale analysis in perturbative renormalization, the functional methods \textit{\`a la} Wetterich equation, and the loop-vertex expansion in non-perturbative renormalization.
While each of these is quite well-established, they go beyond standard QFT textbook material, and may be little-known to specialists of each other approach. This review is aimed at bridging this gap.
\end{abstract}

%%%%%%%%%%%%%%%%%%%%%%%%%%%%%%%%%%%%%%%%%%%%%%%%%%%%%%%%%%%%%%%%%%%%%%%%%%%
\newpage

\tableofcontents

\newpage

\section{Introduction}
Quantum field theory (QFT) emerged as a framework to reconcile quantum physics with special
relativity, and has now gained a central role in theoretical physics. Since its origin, QFT has been plagued by the
problem of divergences, which led to the formulation of the theory of \emph{renormalization}. This procedure, that initially might have 
appeared as a computational trick, is now understood to be the heart of QFT. In fact, the so-called renormalization group approach 
explains why we are able to efficiently describe complicated systems, from ferromagnetism to the standard model, in terms of simple
theories that depend only on a small number of parameters.

In time,  many different approaches to renormalization have been developed and gained popularity in specific fields of application where
they are most useful. There exists several excellent texts that describe how to use QFT and renormalization theory in those fields. We feel 
however that it is sometimes hard to pinpoint the relative strengths and  limitations of each approach. This is the motivation for this review.

QFT and renormalization are enormous topics. Here we will focus on highlighting a few different and complementary approaches to 
renormalization, and deal only with a very simple theory---the interacting scalar field with quartic potential in $d$-dimensional 
Euclidean space time, denoted by $\phi^4_d$. Furthermore, in what follows we will restrict to the path integral approach.

In this approach $S_{\text{M}}(\phi)$ is a classical action on Minkowskian space time, and $S(\phi)$ its analytic continuation to
Euclidean space time. The quantization of the 
Euclidean field theory defined by $S(\phi)$ is given by the (potentially ill-defined) functional probability measure
\begin{equation}
\label{eq:formalmeasure}
\text{d}\nu(\phi)=\frac{1}{\mathcal{N}}\,\mathcal{D}\phi\,e^{-S(\phi)}\,,
\end{equation}
where $\mathcal{N}$ is an appropriate normalization and $\mathcal{D}\phi$ is formally defined to be the 
product of uncountably infinitely many Lebesgue measures at every point of the space,
\bee
\mathcal{D}\phi=\prod\limits_{x \in {\R}^{d} }\text{d}\phi(x)\,.
\ee
The moments of such a measure are the Schwinger functions
\begin{equation}
\label{eq:schwingerS}
\mathsf{S}_n(z_1,\dots z_n)=\Big{\langle}\phi(z_1)\cdots\phi(z_n)  \Big{\rangle} =\int  \phi(z_1)\cdots\phi(z_n) \; \; \text{d}\nu(\phi)\,,
\end{equation}
out of which any observable can in principle be computed. Our task in what follows will be to give a precise meaning to those formal expressions.

\subsection{Axioms for an Euclidean quantum field theory}
If we let $\text{d}\nu(\phi)$ be any Euclidean probability measure such as  \eqref{eq:formalmeasure} with action $S(\phi)$, there is no
guarantee that we can use it to construct a reasonable Minkowsian QFT
with action $S_{\text{M}}(\phi)$.  The Osterwalder-Schrader axioms \cite{Osterwalder:1973dx,Osterwalder:1974tc} are a set of properties for the Schwinger functions  of 
an Euclidean field theory which 
allows the analytic continuation of these functions to a set of  distributions in Minkowski space which form a sensible relativistic QFT, see also~\cite{Wightman1967,streater}. 
In particular, such distributions will satisfy all Wightman axioms.

The O.S. axioms consist of five properties:
\begin{itemize}
\item{(OS1)} {A regularity property}
\item{(OS2)} {Covariance under transformations of the Euclidean group}
\item{(OS3)} {A positivity requirement}
\item{(OS4)} {Symmetry under permutation of the evaluation points $z_i$}
\item{(OS5)} {Cluster decomposition property}
\end{itemize}

OS1 is technical; for the analytic continuation to Minkowski space to 
be feasible one must check that the sequence of moments $\mathsf{S}_{n}$ does not grow too fast with~$n$.

OS2  states that the Schwinger functions are covariant under a global 
Euclidean transformation. In the case of a scalar Bosonic field, this simply means that they are invariant. After analytic continuation,
this property ensures the proper covariance of the Wightman functions of the Minkovski theory under the Poincar\'e group.

OS3 is the most interesting Euclidean axiom. It guarantees the existence of the physical Hilbert space for the corresponding Minkowskian theory
and of unitary time evolution in this Hilbert space. It is quite non-trivial, and we will comment more on it later.

OS4 requires the full symmetry of the Schwinger functions under permutations of the coordinates related to Bosonic fields, and full antisymmetry 
for Fermionic fields. This ensures compatibility with statistics.

OS5 states that the Schwinger functions asymptotically
factorize when two sets of arguments are taken far apart. This ensures the 
unicity of the vacuum in the Wightman axioms. Physically, this means that experiments can be performed locally, without influences from 
arbitrarily far away. In theories where all particles are massive, the decay of correlations is exponential with the separation distance. 
For the two-point function, the rate of decay is called the mass gap.

The O.S. positivity axiom has stringent physical implications and it is worth exploring in more detail.
Any hyperplane $H$ of $\R^d$ separates $\R^d$ into two half-spaces $H_+$ and $H_-$.
Let $f$ be a sequence of test functions for an arbitrary number of points: $f = (f_0 , \dots f_n, \dots )$,
where $f_n$ is a function of $n$ variables. Let the product $f \times g$ be defined by 
\bee (f \times g)_n =\sum_{k=0}^n  f_{n-k}  \times g_k
\ee
We will say that $f\in\cS_+$
if the support of $f$ is included in $H_+$.

\begin{definition}[Osterwalder-Schrader Positivity]
Let $H$ be any hyperplane in $\R^d$ and $\Theta$ be the reflection operator about $H$. Consider
the Schwinger functions $\mathcal{S}_n (f_1, .... f_n)$. They are said to satisfy OS-positivity if the sum
\bee  \sum_{n,m} \mathsf{S}_{n+m} ( \Theta f_n \times f_m)  \; 
\ee
is positive $\forall \; f \in \cS_+$.
\end{definition}

In simpler terms, if we forget for a moment the distributional aspect and consider a quadratic action of the form
\bee
\label{eq:freetheory}
S(\phi)=\frac{1}{2}\int \text{d}^dx\,\text{d}^dy\,\;  \phi(x)\,C^{-1}(x,y)\,\phi(y)\,,
\ee
where $C$ is a positive quadratic form, ordinary positivity means
that the matrix
\bee
\Big(C(z_i, z_j)\Big)_{i,j=1,\dots n}\,,
\ee
is positive. In particular if the covariance $C$ is invariant under translation it is equivalent to the Fourier transform $\hat{C}(p)$ 
of $C(x-y)$ being positive. Instead OS positivity for the free field means that the different matrix
\bee
\Big(C(\Theta z_i, z_j)\Big)_{i,j=1,\dots n}\,,
\ee
is positive for any finite set of points $z_1, \cdots z_n$.
In fact for the Euclidean free field we have the following result, which is an Euclidean counterpart to the Minkovski 
K\"allen-Lehmann representation:

\begin{theorem}
\label{th:freecovariance}
The Euclidean free field measure of covariance $C$ with
\bee
\hat{C}(p) = \frac{1}{\vert p \vert^\alpha  + m^2}\,,
\ee
is O.S.-positive if and only if $0\leq\alpha\leq2$. 
\end{theorem}
This theorem shows that existence of a physical Hilbert space and unitary evolution require a propagator which is not too
convergent in the ultraviolet (large $p$) regime. The inverse of the Laplacian is the most convergent propagator allowed by OS positivity. 
The associated ultraviolet divergences have therefore a deep origin and cannot be ignored or suppressed through a cheap cutoff. Their 
renormalization constrains in a beautiful way the set of consistent quantum field theories.

As it is well know, the free case~\eqref{eq:freetheory} is the only one where the treatment of the formal measure~\eqref{eq:formalmeasure} is 
straightforward, since~$\text{d}\nu(\phi)$ is Gaussian. Assuming that the hypotheses of Theorem~\ref{th:freecovariance} hold, it is then possible 
to construct a consistent QFT from the Schwinger functions. Of course, the resulting theory is then free and of limited interest. Let us consider 
then what is probably the next simplest theory after the free one.

\subsection{The \texorpdfstring{$\phi^{4}_{d}$}{phi**4(d)} field theory}
The simplest stable Euclidean interacting field theory is the theory of a one component scalar 
bosonic field $\phi$ with quartic interaction $\lambda\phi^{4}$ in $d$-dimensions. The simpler cubic interaction would in fact lead to instabilities. 
For $d=2,3$ this model is 
superrenormalizable and has been built non perturbatively by constructive 
field theory techniques~\cite{Glimm:1987ng,Rivasseau:2011ri}. In these dimensions the model 
is unambiguously related to its perturbation
series \cite{eckmann1974,magnen1977} through Borel summability~\cite{Sokal:1980ey}. 
For $d=4$ the model is just renormalizable, and provides the simplest 
pedagogical introduction to perturbative renormalization theory. But 
because of the Landau ghost or triviality problem that we will briefly mention later, the model presumably does not exist as a true interacting
theory at the non perturbative level\footnote{%
For a discussion of this subtle issue, see refs.~\cite{Aizenman:1982ze,Frohlich:1982tw,Rivasseau:2011ri}
}.

Formally the Schwinger functions of $\phi^{4}_{d}$ are the moments of the 
measure $\text{d}\nu(\phi)$ of~\eqref{eq:formalmeasure} with action
\bee  S(\phi)  = \int\text{d}^dx \left(\frac{1}{2}\,a\,\partial_{i} \phi \partial ^{i}\phi+\frac{1}{2}\,m^2\,\phi^2+ \frac{1}{4!}\,\lambda\,\phi^{4}\right),
\label{eq:phi4daction}
\ee
where 
\begin{itemize}
\item $\lambda$ is the coupling constant, usually assumed positive or complex 
with positive real part; remark the arbitrary but convenient 1/4! factor to take into account
the symmetry of permutation of all fields at a local vertex.

\item $m$ is the mass, which fixes an energy scale for the theory;

\item $a$ is the wave function constant. It can be set to 1 by a rescaling of the field~$\phi$.
\end{itemize}

When $\lambda=0$, one recovers the free scalar theory of the previous subsection. However, the simple quartic
interaction is enough to describe a wealth of physical phenomena. One of the historically most relevant aspects 
from the point of view of renormalization theory is that  the quartic interaction describes the universality class 
of the Lenz-Ising model of ferromagnetism, and we will see how  in~$d=3$ the critical exponent of the Ising model can 
be extracted by renormalization techniques. It should also be mentioned that a scalar quartic interaction appears in the
theory of fundamental particles as the self-interaction of the Higgs field, which however we will not explore in what follows.

\subsection{Contents and plan of the review}

In order for this review to be self-contained, we have included some useful and somewhat more technical prerequisites in 
Section~\ref{sec:tools}. These are some elements of graph theory, an introduction to flows and dynamical systems with emphasis 
on renormalization techniques in that setting, and a brief discussion of analyticity and Borel summability. We also recall 
some standard notions on quantum field theory and renormalization in Section~\ref{sec:renorm-essetials}.

Then, focusing on $\phi^4_d$, with typically $d=2,3$ or~$4$, we will consider three distinct approaches to renormalization. In 
Section~\ref{sec:multiscale} we will investigate what is perhaps the more customary approach to renormalization, i.e. perturbative
renormalization to all-loops. Our emphasis will be on multiscale analysis and the problem of finite terms---\emph{renormalons}. 
Next, in Section~\ref{sec:functional}, we will present functional methods and in particular Wetterich's equation. This approach 
does not rely on any small-coupling expansion but adopts approximated truncation schemes, and as we will see can be useful when 
studying large-coupling issues such as finding the critical exponents of second order phase transitions. Finally, in 
Section~\ref{sec:nonperturbative} we will conclude with an invitation to the more rigorous theory of constructive renormalization. 
To avoid excessive technicalities, we limit ourselves in this final section to a presentation of the forest formula and loop vertex 
expansion.

\newpage

\section{Useful tools}
\label{sec:tools}

We present here an overview of some of the more important concepts required for the study of renormalization in QFT. The topics detailed here require some familiarity with QFTs and the specific issues they raise. The reader that is unfamiliar with these is advised to skip this section in a first reading and come back to it when needed.

\subsection{Graphs and combinatorial maps}

In this section we provide a brief overview of graph theory which is the combinatorial backbone of renormalization.

\label{graphsub}
\subsubsection{Generalities}

Graphs are truly ubiquitous structures appearing everywhere in science. Here we limit ourselves to some aspects of graph theory of particular interest 
in theoretical physics. We first give an overview of graphs and some of their most interesting immediate applications. 
We then explain how to move on graphs by defining random paths. Finally in quantum field theory Feynman graphs themselves become 
structures to be summed. We will detail the subtle interplay between graphs and combinatorial maps (sometimes called embedded graphs)
relevant in the QFT context. 

The mathematics and physics literature often use different words for the same objects. We shall mostly use in this review the graph theory language. 
We include a very short bibliography: the first two items \cite{bondy1976graph,bang2008digraphs} are general references on the subject.

\begin{definition}
A graph $G= (V, E)$ is a set of vertices $V$ and of edges $E$ which are lists of two (not necessarily distinct) elements in $V$.
That is $e\in E$ is
\begin{itemize}
 \item either a list $\bigl(a(e), b (e) \bigr)$ with $a(e), b(e) \in V$ and $a(e)\neq b(e)$. 
 \item or a list $\bigl( a(e) , a(e) \bigr)$ with $a(e)\in V$.
\end{itemize}
The vertices $a(e)$ and $b(e)$ are called the ends of the edge $e$.
\end{definition}
In the physics literature edges are often called lines or propagators.
The number of vertices and edges in a graph will be denoted $\vert V\vert $ and $\vert E\vert $. 
For the purposes of QFT, we explicitly allow graphs having multiple edges, that is two elements $e$ and $e'$ of $E$ having the same end vertices, 
and self loops (or tadpoles in the physics literature), that is an edge $e$ in $E$ with $b(e)=a(e)$. 

A very important (although often ignored) notion in QFT related to graphs is that of combinatorial map.

\begin{definition}
 A combinatorial map is given by three items
 \begin{itemize}
  \item A finite set $D$ of half edges (or darts),
  \item a permutation $\sigma$ on $D$,
  \item and an involution $\alpha$ on $D$ with no fixed points (a.k.a. a ``pairing'' of half edges).
 \end{itemize}
\end{definition}
The permutation $\sigma$ encodes the ``next half edge'' when turning clockwise around a vertex. The vertices are thus the cycles of $\sigma$. The 
involution $\alpha$ encodes the pairs of half edges which must be connected into an edge.

\begin{figure}[htb]
\begin{center}
\includegraphics[width=10cm]{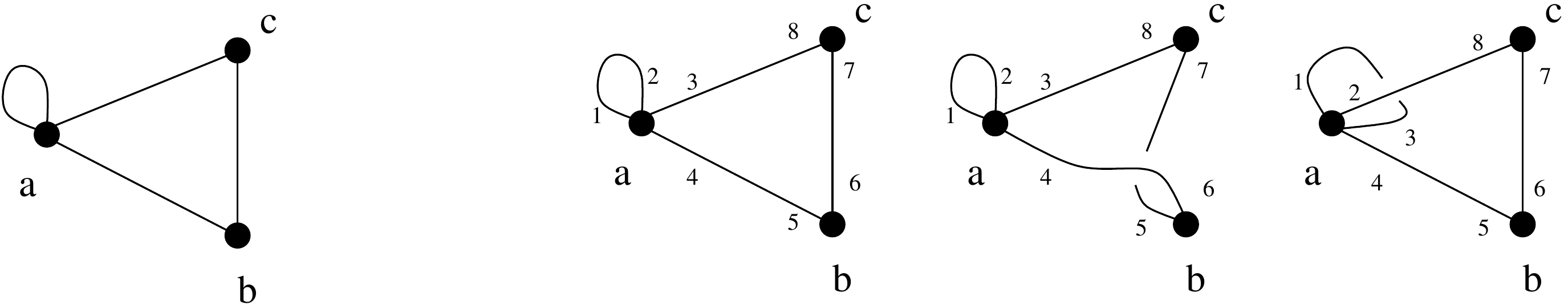}
\caption{A graph and several associated combinatorial maps.}
\label{fig:bas}
\end{center}
\end{figure}

A combinatorial map is an embedding of a graph in the plane. Naturally, there are several combinatorial maps (embeddings) associated to the same graph. 
Take the example in figure \ref{fig:bas}. The graph is $V=\{a,b,c\}$ , $E=\{ (a,a), (a,b), (a,c), (b,c) \}$ while the three combinatorial 
maps we represented are $D = \{1,2,3,4,5,6,7,8 \}$, $\sigma = (1234)(56)(78)$ and
\begin{itemize}
 \item $\alpha=\{ 1\leftrightarrow 2, 3\leftrightarrow 8, 4\leftrightarrow 5, 6\leftrightarrow 7 \}$
 \item $\alpha=\{ 1\leftrightarrow 2, 3\leftrightarrow 8, 4\leftrightarrow 6, 5\leftrightarrow 7 \} $
 \item $\alpha=\{ 1\leftrightarrow 3, 2\leftrightarrow 8, 4\leftrightarrow 5, 6\leftrightarrow 7 \} $
\end{itemize}

Graphs as well as combinatorial maps are labelled. That is all the vertices in a graph have a label ($a$, $b$ or $c$ in figure \ref{fig:bas})
and all the half edges in a map are labelled ($1$ to $8$ in figure \ref{fig:bas}). The first two combinatorial maps above can be transformed into 
one another by flipping the half edges $5$ and $6$ on the vertex $b$ (which comes to draw the anticlockwise). However, 
the third one is topologically distinct: it can not be drawn 
on a plane without an over/under crossing of two of its edges. While this distinction is not very important for the $\phi^4_d$ scalar QFT, it 
becomes crucial in matrix models \cite{DiFrancesco:1993nw}.
 
\begin{definition}
A {\emph{proper graph}} (also sometimes called a regular graph) is a graph $G$ without any self-loop. An orientation of a proper
graph is the choice of an arrow or direction for each edge. 
Hence a proper graph has $2^{\vert E \vert }$ orientations.
\end{definition}

\begin{definition}
The {\emph{complete graph}} $K_{|V|}$ is the proper graph in which every distinct pair of vertices is joined by an edge.
It has $\vert V \vert (\vert V\vert -1)/2$ edges.
\end{definition}

\begin{figure}[ht]
\centerline{   \includegraphics[width=10cm]{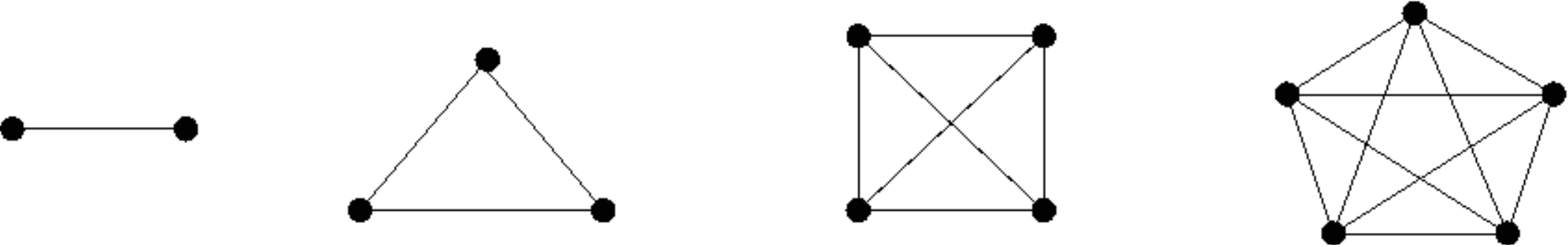}   }
\caption{The complete Graphs $K_2$, $K_3$, $K_4$ and $K_5$.}
\label{graph1x}
\end{figure}

A graph is said to be {\emph{connected}} if one can join any of its vertices to any other one through a chain of edges.

\begin{definition}[Connected Components]
A maximal subset of edges and vertices of a graph which is connected is called a {\emph{connected component}}.
Isolated vertices count as connected components.
Any graph is naturally partitioned as the disjoint union of its connected components. 
We define $k(A)$ as the number of 
connected components of $A$ (including isolated vertices). 
\end{definition}

An edge whose removal increases (by one) the number of connected components
of the graph is called a {\emph{bridge}} (in physics it may be called a one-particle-reducible edge).
An edge which is neither a bridge nor a self-loop is called regular. We shall call  \emph{semi-regular}
an edge which is not a self-loop, hence which joins two distinct vertices. Proper graphs have only semi-regular edges.

A cycle is a set of distinct edges $e_1=(v_1,v_2), e_2=(v_2, v_3) \cdots e_n=(v_n ,v_1)$ with $v_1, \cdots v_n$ 
all distinct.

An edge-subgraph of a graph $G$, also called for short a subgraph of $G$  is a set $(V, A)$ with $A \subset E$; hence a graph has
$2^{\vert E \vert }$ subgraphs. We always keep all the vertices of $G$ in the subgraph, even if they are isolated.

\subsubsection{Forests, trees and plane trees}

\begin{definition}
A \emph{forest} over $n$ vertices is an acyclic subgraph (a subgraph without any cycle) of $K_n$. 
A \emph{tree} is a forest of $K_n$ which is connected. 
\end{definition}

A tree over $n$ vertices has $n-1$ edges. There are $n^{n-2}$ distinct trees over $n$ {\it labelled} vertices. This statement 
can be proved in many ways. One of the  
most elegant is through the bijection between trees and Pr\"ufer sequences~\cite{prufer1918neuer},  which are words of length $n-2$ over the alphabet 
$1,\dots n$. A tree is called {\it rooted} if one vertex has been designated as the root. 
In any such rooted tree the edges can be given one of two canonical orientations, towards or away from the root. 

A rooted tree yields a canonically associated partial ordering on the vertices, which represents the filiation (or descent) if the 
tree is interpreted as a genealogy.
In that interpretation each vertex is a descendant of the root.
To each vertex $v$ is associated a unique path to the root and the number of edges
in this unique path is the order of descent. 
Vertices at distance 1 from the root (hence related to the root by an edge) are the first order descendants of the root (its children) 
and so on. We say that $u \le v$, or that $v$ is a descendant of $u$ if and only if the unique path from the root 
to $v$ passes through $u$.  

\begin{definition}
 A plane tree is a combinatorial map whose associated graph is a tree. A rooted (or planted) plane tree 
 is a plane tree supplemented by a mark on one of its vertices. The mark breaks the cyclic order of half edges 
 at the root vertex.
\end{definition}

A rooted plane tree can be seen as a tree for which an ordering is specified for the children of each vertex (first born, second born, etc.). 
Rooted plane trees with {\it unlabeled} vertices can be counted by a recursion relation.
We denote $C_n$ the number of rooted plane trees with $n$ edges. 
Take a plane tree with $n+1$ edges and cut the leftmost edge hooked to its root. This divides the tree into a plane tree with $k$ edges
and another one with $n-k$ edges. Both trees are rooted: one at the root of the original tree, and the other one at the half edge obtained by cutting 
the edge (the ``scar'' of the edge), hence:
\bee \label{reccata} C_{n+1} =   \sum_{i=0}^{n} C_i C_{n-i}\, , \quad   \quad C_0 =1 \; .
\ee
The solution of this recursion relation are the well known \emph{Catalan numbers}\footnote{These numbers were used before Catalan by 
the Mongolian mathematician Minggantu by 1730.} which arise in many enumeration problems related to trees:
\bea C_n &=& \frac{1}{n+1} \binom{2n}{n} \; .
\eea

Trees and forests can be defined with respect to any graph $G$, and not only the complete graph $K_n$. 

\begin{definition}
A \emph{forest} of $G$ is an acyclic subgraph of $G$, hence a subgraph without any cycle. 
A spanning forest of $G$ is a forest of $G$ which has as many connected components as $G$ itself. 
A \emph{(spanning) tree} is a forest of $G$ which is connected. 
\end{definition}

Only connected graphs can have spanning trees. They 
then coincide with the spanning forests of the graph, as they must connect all vertices of $G$. 
We may forget the word "spanning" where there is no ambiguity. 

\begin{definition}[Complexity]
The {\emph{complexity}} of a connected graph, denoted $\chi (G)$ 
is defined as the number of (spanning) trees of $G$.
The complexity of a non-connected graph is defined as the
product of the complexities of its connected components, hence as the number of its spanning forests.
\end{definition}

\begin{definition}[Rank, Nullity]
We define
$r(A) =  \vert V \vert  - k(A) $ as the {\emph{rank}} of the subgraph $A$ and 
$n(A) =\vert A \vert + k(A) - \vert V \vert  $ as its {\emph{nullity}} or cyclomatic number.
In physicists' language $n(A)$ is the number of independent loops (or cycles) in $A$.
\end{definition}

Remark that $r(A)$ is the number of edges in any spanning forest of $A$,
and $n(A)$ is the number of remaining edges in $A$ when a spanning forest
is suppressed, that is the number of \emph{independent cycles} in $A$.
While the number of independent cycles is a canonical notion, the cycles themselves are {\it not}. 
Once a spanning forest is chosen in a connected graph the remaining set of edges can be considered as generators 
for the independent cycles (e.g. they can be used for a flow attribution). But this flow attribution depends on the chosen tree.

In quantum field theory we shall need a second type of trees and forests which correspond to a higher category of objects
in set theory. Ordinary trees and forests are defined as subset of edges of a graph. We need a more subtle notion encoding
the inclusion relations between non overlapping subgraphs. To distinguish these higher category objects (first introduced in 
quantum field theory by Zimmermann~\cite{Zimmermann:1969jj,Zimmermann:1972te})  we shall call them \emph{inclusion} trees and forests, or in short i-trees and i-forests.

\begin{definition}\label{iforest}
A set $\cF$ of non-empty \emph{connected} subgraphs of $G$ is called an \emph{i-forest} (inclusion-forest) if 
for any pair of elements $g$ and $g'$ of $\cF$
either $g$ and $g'$ are disjoint (i.e. have no common edge) or
are included one into the other. An \emph{i-tree} is an i-forest which is connected for the inclusion relation.
\end{definition}
An i-tree is an i-forest with a maximal element (the root). In particular any i-forest containing $G$ itself is an i-tree (as any subgraph is included in $G$).

Let us better describe the relation between these i-forests and the ordinary forests.
For each set  $\cF$ of subgraphs of $G$ we consider the graph 
whose vertices are the elements
of $\cF$ and whose edges represent \emph{elementary inclusion relations}. 
They are defined as follows: for each couple $(g, g')$ of elements of $\cF$ we say
that there is an \emph{elementary inclusion relation} between $g$ and $g'$  if $g\subset g'$
and there is no other $h\in \cF, h \ne g, h \ne g'$ such that $g \subset h \subset g'$.
In this way we obtain a graph $(\cF, \cE)$  whose set of edges $\cE$ is the set of elementary
inclusion relations of elements of $\cF$. If $\cF$ is an i-forest, the graph $(\cF, \cE)$  is a forest; if $\cF$
is an i-tree, $(\cF, \cE)$ is an ordinary tree.

Furthermore we have the following important lemma:

\begin{lemma}\label{compat}
Let $\cF$ be an i-forest. There exists at least one ordinary forest $F$ of edges of $G$
compatible with $\cF$, that is such that $F \cap g$ is a spanning tree of $g$ for any $g \in \cF$.
\end{lemma}
\prf Choose $F$ by induction from the leaves (or ``smallest elements'') of $(\cF,\cE)$ towards the ``larger'' elements. That is, 
start from a $g\in \cF$ such that any $g'\in \cF, g'\neq g$, either $g' \cap g =\emptyset $ or $g'\supset g$, and chose a tree of ordinary edges in $g$. 
Now consider the graph obtained from $G$ by contracting $g$ to a single vertex and iterate.

\qed  

\subsubsection{Incidence, degree, adjacency and Laplacian matrices}
\label{sec:LaplacianMat} 
 
Any oriented proper graph, that is having no self loops (tadpoles) is fully characterized by its incidence
matrix ${\epsilon_G(v,e)}$. 

\begin{definition}[Incidence Matrix]
The  {\emph{incidence matrix}} of $G$ is the rectangular $\vert V\vert $ by  $\vert E\vert $ matrix 
with indices running over vertices and edges respectively, such that 
\begin{itemize}
\item
${\epsilon_G(v,e)}$ is +1 
if $e$  ends at $v$, 
\item
${\epsilon_G(v,e)}$ is -1 if $e$ starts at $v$,
\item
${\epsilon_G(v,e)}$ is  0 otherwise.
\end{itemize}
\end{definition}

For every edge $e$, only the two elements corresponding to its end vertices on the column $\epsilon_G(v,e)$ are non zero, and they equal $+1$ and $-1$,
hence the sum of coefficients in any column is 0. It is sometimes useful to introduce the positive incidence 
matrix $\eta_G(v,e) = \vert \epsilon_G(v,e) \vert$. This matrix can be then generalized to unoriented graphs including self-loops by defining
$\eta_G(v,e) =2$ for a self-loop attached at vertex $v$.

\begin{definition}[Degree, Adjacency] 
The {\emph{degree of a vertex}} (coordination in the physics literature) $v$ is the number of half-edges which touch  $v$. It is an integer 
noted $d_G(v)$. A self-loop counts for 2 in the degree of its vertex, so that
$d_G(v) = \sum_{e\in E} \eta_G(v,e)$. 

The diagonal $\vert V \vert\times\vert V \vert$ square matrix 
with entries  $d_G(v)$ on the diagonal, $D_G(v,w) = d_G(v) \delta_{vw} $ is called the {\emph{degree matrix}} of $G$.

The {\emph{adjacency matrix}} of a proper graph $G$ is the $\vert V \vert\times\vert V \vert$ matrix which has zero on the diagonal, $A_G(v,v) = 0 \;\; \forall v\in V$, and 
for $v \ne w$, $A_G(v,w)$  is the number of edges connecting $v$ and $w$.
\end{definition}

\begin{definition}[Laplacian Matrix] 
The {\emph{Laplacian matrix}} of a proper graph $G$ is $L_G = D_G - A_G$.
\end{definition}

For example, the Laplacian matrix of the graph $K_n$ of figure~\ref{graph1x} is
\bee
L_{K_n} = \begin{pmatrix} 
 n-1 & -1 & \dots & -1  \cr
-1 & n-1 &  \dots & -1  \cr
   \vdots \cr
-1 & -1 & \dots  & n-1 
\end{pmatrix} .
\ee

\begin{lemma}\label{incid}
Let $G = (V, E)$ be a proper graph, orient $G$ arbitrarily, and let $\epsilon_{G}$ be the incidence matrix of $G$. 
Then $\epsilon_G \cdot (\epsilon_G)^{T}= L_G$. 
\end{lemma}
\prf We have 
\bea
 \sum_{e\in E} \epsilon_G(v,e) \epsilon_G(w,e) = \begin{cases}
                                                   \sum_{e\in E} |\epsilon_G(v,e)| = d_G(v) \quad & \text{ if } v=w \cr \\
                                                   \sum_{e\in E, e=(v,w)} (-1) = -A_G(v,w) \quad & \text{ if } v\neq w 
                                                 \end{cases} \; .
\eea 

\qed

Although the incidence matrix depends on the orientations of the edges, the Laplacian matrix does not. 

\subsubsection{The symmetry factor}

As we will see later on, graphs arise in QFT as ``Feynman graphs'', having an amplitude and a symmetry factor.

In fact the terminology is somewhat confusing: as Feynman ``graphs'' are obtained by evaluating a Gaussian integral 
via Wick contractions, the perturbative series of QFT is {\it  not} indexed by graphs, but by combinatorial maps. 
Take for example the partition function $\langle 1 \rangle$.  When performing the perturbative expansion of 
the $\phi^4_d$ model at order $n$ one obtains:
\bea
 \langle 1 \rangle = \sum_{n=0}^{\infty} \frac{1}{n!} \frac{(-\lambda)^n}{(4!)^n}  \int d\mu_C 
   \int  \bigl( \prod_{i} d^dx_i  \bigr) \prod_i \phi^4(x_i)
\; ,
\eea
where $d\mu_C$ denotes the normalized Gaussian measure of covariance $C$. Before performing the Gaussian integral one has $n$ four 
valent vertices. The fields $\phi$ in the equation above are distinguished, hence one can assign labels to each field. Say we assign the 
labels $1,2,3,4$ to the half edges of a vertex, $5,6,7,8$ to the half edges of another one and so on. We thus obtain a set half 
edges $D= \{1,\dots ,4n\} $ and a permutation 
\bea
 \sigma= (1234)(5678)\dots (4n-3,4n-2,4n-1,4n) \; .
\eea 

The Gaussian integral is evaluated by the Wick theorem as a sum over contractions (pairings). Each Wick contraction scheme is a choice of an 
involution $\alpha$ over the labelled $4n$ half edges (and there are $(4n)!!$ such involutions). It follows that each term in the Feynman 
expansion is a combinatorial map. 
\bea
 \langle 1 \rangle =  \sum_{n=0}^{\infty} \frac{1}{n!} \frac{(-\lambda)^n}{(4!)^n}  \sum_{\text{labelled combinatorial maps with n vertices}} A(M) \; ,
\eea 
where $A(M)$ is the amplitude of the map $M$. As many maps have the same amplitude, one usually groups together 
the maps corresponding to the same unlabeled graph in the above sum. This leads to the somewhat involved combinatorial problem of counting how many labelled 
maps correspond to the same unlabeled graph.

The situation however simplifies greatly for the connected two-point function. Indeed, the sum 
over labelled combinatorial maps with two external half edges can be organized very easily in terms of
unlabeled combinatorial maps with 2 (labelled) external half edges. 

Indeed, consider an unlabeled combinatorial map with two external half edges (the external half edges are labelled 1 and 2).
Chose a plane tree in this map, and root it at the external half edge $1$. Starting from the root and going clockwise around the plane tree
we encounter a first vertex at a particular half edge, hence there are $4 \cdot n$ choices for connecting this vertex. For the next vertex we 
encounter we have $4\cdot(n-1)$ choices and so on. It follows that 
\bea
  \langle \phi(y_1) \phi(y_2) \rangle_{\text{connected}} 
  = \sum_{n=0}^{\infty} \frac{(-\lambda)^n}{(3!)^n}   
 \sum_{\genfrac{}{}{0pt}{}{\text{unlabeled combinatorial maps with n vertices} }{\text{and two (labelled) external half edges}}} A(M) \; .
\eea 

\subsection{Graph polynomials}

Much of the topological information about a graph can be captured by somewhat more manageable mathematical objects, such as matrix and polynomials canonically associated to it. 
As we will see later, these, and polynomials in particular, play a very important role in QFT.

\subsubsection{The matrix-tree theorem}

Let $M$ be a matrix and let $[M]_{\bar \imath , \bar   \jmath}$ denote the sub matrix of $M$ obtained by deleting row $i$ and column $j$ from $M$.
More generally
for subsets $S$ of the line indices and $T$  of the column indices we denote
$M_{\bar S, \bar T}$ the matrix $M$ where we have deleted the lines in $S$ and  the columns in $T$, 
$M_{\bar S, T}$ the matrix $M$ where we have deleted the lines in $S$ and kept the columns in $T$ and so on.

The Matrix-Tree Theorem computes the complexity of a connected proper graph in terms of its Laplacian matrix. The complexity
of more general graphs follows easily by erasing the self-loops and working connected component by connected component.

\begin{thm}[Matrix-Tree Theorem] Let $G = (V, E)$ be a proper connected graph, and let $L_G$ be the Laplacian matrix of $G$. 
Then for any $v \in V$,
\bee \chi (G) = \det \Bigl( [L_{G}]_{\bar v , \bar v}  \Bigr) \; .
\ee
\end{thm}

We can evaluate $\det \Bigl( [L_{G}]_{\bar v , \bar v} \Bigr) $, first identifying the non-zero
$\vert V\vert -1$ by $\vert V\vert -1$ sub determinants of the incidence matrix 
$\epsilon_G$, and then the using Binet-Cauchy formula.

\begin{prop}\label{proptree} 
Let $G = (V,E)$ be a connected proper oriented graph and let $\epsilon_G$ be its incidence matrix. Let $v \in V$ and $S \subset E$ 
be such that $|S| = |V|-1$. Then $[\epsilon_{G}]_{\bar v , S}$ is a $\vert V\vert -1$ by $\vert V\vert -1$ square matrix, and 
\begin{itemize}

\item $\det \Bigl(  [ \epsilon_{G} ]_{ \bar v , S} \Bigr) = \pm 1$ if $S$ is a tree,

\item $\det \Bigl( [ \epsilon_{G}]_{\bar v , S} \Bigr) = 0$ otherwise.
\end{itemize}  
\end{prop}

\prf If $S$ is not a tree it has  to contain a cycle $C$. Orient the edges of $C$ consistently, and for 
each $e \in C $ let  $\eta_C (e) = +1$ if the orientations of $e$ in $C$ and $G$ agree and let $\eta_C (e) = -1$ if they differ.
Then 
\bee
\sum_{e \in C}  \eta_C (e)   \epsilon_G( v, e)=0 \;, \quad \forall v \; ,
\ee
and the determinant is zero. 

Now suppose $S$ is a tree. We can prove that  $\det\Bigl(  [\epsilon_{G}]_{\bar v , S} \Bigr) = \pm 1$ by induction. 
It is obvious for $\vert V \vert =2$. 
If $S$ is a tree with $n$ edges on $n+1$ vertices, it has at least two leaves, hence one leaf $v'$ not equal to $v$. The line with 
index $v'$ in $ [\epsilon_{G}]_{\bar v , S}$ 
has a single non zero element which is $\pm 1$ (corresponding to the unique edge $e$ touching the leaf $v'$). 
Expanding $\det \Bigl( [\epsilon_{G}]_{\bar v , S} \Bigr) $ along that line
we obtain the determinant of the incidence matrix of $ S-e$, and we conclude.

\qed

\begin{prop}[The Binet-Cauchy Formula] 
Let $M$ be an $r\times m$ matrix, and let $P$ be an $m\times r$ matrix with $r\le m$. Then
\bea
\det MP =  \sum_{S,\; \vert S \vert  = r} \det M_{\cdot, S }  \det P_{S, \cdot} \;, 
\eea 
where the point $\cdot$ means we delete nothing.
\end{prop}

Returning to the matrix-tree theorem let $G = (V, E)$ be a proper oriented connected graph, and $L_G$ its Laplacian matrix. 
Let $v \in V$ be any vertex. Using Lemma \ref{incid}, proposition \ref{proptree} and the Binet-Cauchy formula we have:
\bea
&&\det \Bigl( [L_{G}]_{\bar v , \bar v}\Bigr)
= \sum_{S \subset E, \vert S\vert = \vert V \vert -1} \det \Bigl( [\epsilon_{G}]_{\bar v , S }\Bigr)  
     \det \Bigl( [\epsilon_{G}]_{\bar v , S }\Bigr)    \crcr
&&\qquad  =  \sum_{S \;\; {\rm tree\;\;  of \;\; } G} 1 = \chi (G) .
\eea

Let us assign a variable $y_e$ to each edge $e$ of $G$, and define the  $\vert V \vert $ by  $\vert V \vert $
{\emph {weighted Laplacian matrix}}  $L_{G, y } = \epsilon_G Y (\epsilon_G)^T $, that is 
\bee 
L_{G, y }  (v,v) =  \sum_{e \; {\rm incident\; to\; }  v}  y_e \; , 
\quad   L_{G, y }   (v,w) = -   \sum_{e \; {\rm incident\; to\; }  v \;  {\rm and}\;  w}  y_e  \; .
\ee
The matrix $L_{G, y } $ does not depend on the choice of orientation used to define $\epsilon_G$, and its rows and columns sum to zero.
The Matrix-Tree Theorem generalizes immediately to

\begin{thm}[Weighted Matrix-Tree Theorem]  \label{wmtt}
\bee\label{eq:weigtedmatrixtree}
\det \Bigl(  [L_{G, y }]_{\bar v , \bar v} \Bigr)  =  \sum_{T \; {\rm tree\; of   \; } G} \;\;  \prod_{e \in T}  y_e    \; .
\ee
\end{thm}
\begin{thm}[Principal Minors Weighted Matrix-Tree Theorem]  \label{pwmtt}
For any subset $R \subset V$ of vertices, we have
\bee
\det \Bigl( [L_{G, y}]_{ \bar R , \bar R} \Bigr)  =  \sum_{\cF \; {\rm R-forest\; of \; } G}  \;\; \prod_{e \in \cF}  y_e   \label{Rforests}
\ee
in which the sum over $R$-forests means a sum over all maximal 
forests $\cF$ of $ G$ for which each component of $\cF$ contains exactly one vertex of $R$.
\end{thm}

\subsubsection{Deletion, contraction}

\begin{figure}[ht]
\begin{center}
\includegraphics[scale=0.8,angle=-90]{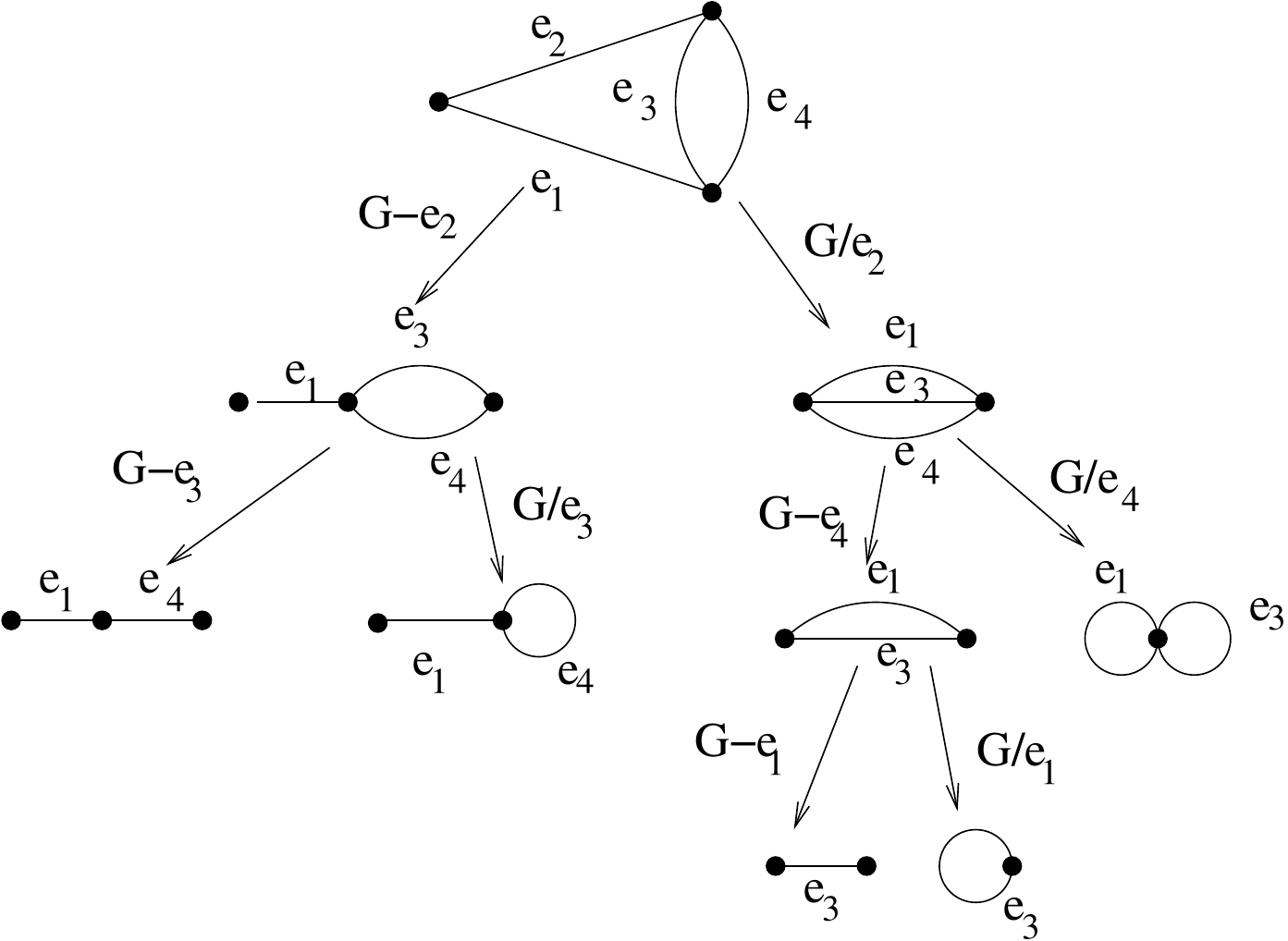}
\caption{An example of the contraction-deletion of a graph}
\label{figcond}
\end{center}
\end{figure}

There are two natural operations associated to an edge $e$
of a graph $G$, pictured in \ref{figcond}:

\begin{itemize}
\item the deletion of the edge, which leads to a graph denoted $G-e$,

\item the contraction of the edge, which leads to a graph denoted $G/e$. If $e$ is not a self-loop, it identifies the two vertices $v_1$and $v_2$ at the ends of $e$ 
into a new vertex $v_{12}$, attributing all the 
flags (half-edges) attached to $v_1$ and $v_2$ to $v_{12}$, and then it removes $e$.
If $e$ is a self-loop, $G/e$ is by definition the same as $G-e$.
\end{itemize}

A terminal form for the deletion-contraction process is a connected graph made solely
of bridges (one-particle-reducibility edges) and self-loops, hence an end point in the process pictured in  Figure \ref{figcond}.

\begin{definition}
A function $F_G(a,b, \cdots) $ obeys a linear deletion-contraction relation with coefficients $a$ and $b$ 
if for any connected graph $G$ and regular edge\footnote{That is neither a self-loop nor a bridge.}~$e$
\bee F_G(a,b,\cdots)  = aF_{G-e}(a,b,\cdots) + b F_{G/e}(a,b,\cdots) . \label{delcont}
\ee
\end{definition}

\begin{lemma}
The complexity obeys a linear deletion contraction relation with coefficients 1 and 1:
\bee \chi_G  = \chi_{G-e} +  \chi_{G/e} \label{delcomp1}
\ee
\end{lemma}
\prf  This is because the trees in $G$ can be classified into trees not containing $e$ (hence trees of $G-e$)
and trees containing $e$, which are in bijection with those of $G/e$ by contracting $e$.

\qed

Since the complexity of any terminal form is 1, the complexity of a (connected) graph counts the number of terminal forms
under any full deletion contraction such as those obtained in the process of  Figure \ref{figcond}. This proves that the {\emph{number}} of such terminal
forms is independent of the order in which regular edges are deleted or contracted: it is a function of $G$ alone.

Developing this idea in more generality leads to a two-variables generalization of complexity, called the Tutte 
polynomial~\cite{tutte1962}. It is one of the most interesting polynomials associated to a graph.

\subsubsection{The Tutte polynomial}

\begin{definition}[Sum overs subsets]
 If $G=(V,E)$ is a graph, the Tutte polynomial of
$G$, $T_G(x,y)$ is the polynomial in two variables $x$ and $y$ defined by
\bee
T_G (x,y)=\sum_{A\subset E}      (x-1)^{r(E)-r(A)} (y-1)^{n(A)},  \label{tutdef}
\ee
where $r(A) =  \vert V \vert  - k(A) $ is the rank of the subgraph $A$ and 
$n(A) =\vert A \vert + k(A) - \vert V \vert  $ is its nullity.
\end{definition}

\begin{thm} The Tutte polynomial has the following properties:
\begin{itemize}
 \item it obeys a linear deletion-contraction relation with coefficients 1 and 1:
\begin{equation}
T_G (x,y)=T_{G/e} (x,y)+T_{G-e} (x,y) \;. \label{tutdel}
\end{equation}
\item its value on terminal forms with $m$ bridges and $n$ self-loops is:
\begin{equation}
T_G(x,y)=x^m y^n \;.  \label{brsl}
\end{equation}
\item it factorizes over connected components:
\begin{equation}
T_G(x,y)= \prod_{i=1}^{k(G)} T_{G_i}(x,y) \; .
\end{equation}
\end{itemize}
Furthermore, it is the {\emph{unique polynomial}} which obeys these properties.
\end{thm}
\prf First consider a connected graph $G$ and choose $e$ a regular edge of $G$. We organize 
the sum over $A$ in \eqref{tutdef} according to whether $e$ belongs to $A$ or not:
\bea
 T_G (x,y)&=& \sum_{A\subset E, e\in A}      (x-1)^{r(E)-r(A)} (y-1)^{n(A)}  \crcr
    &&+ \sum_{A\subset E, e\notin A}      (x-1)^{r(E)-r(A)} (y-1)^{n(A)} \; .
\eea 

In the first term, $(V,A)/e$ is a subgraph of $G/e$, and any subgraph of $G/e$ can be written as $(V,A)/e$ for some $A$.
Furthermore, the number of connected components of $(V,A)/e$ (resp. $G/e$) is the same one as the number of connected 
components of $(V,A)$ (resp. $G$), and the number of edges (resp. vertices) of $(V,A)/e$ is the number 
of edges  (resp. vertices) of $(V,A)$ minus one. Hence $r(A/e) = r(A)-1$, $r(E/e) = r(E)-1$ and 
$n(A/e) = n(A)$. Thus the first term is $T_{G/e}(x,y)$.

In the second term  $(V,A)$ is a subgraph of $G-e$, and any subgraph of $G-e$ can be written like this. 
The number of connected components of $G-e$ is the same one as the number of connected
components of $G$ (as $e$ is not a bridge), hence $r(E-e) = r(E)$, and the second term is $T_{G-e}(x,y)$.

Second, for a terminal form, we organize the sum over $A$ in \eqref{tutdef} according to the number of bridges and self-loops in $A$:
\bea  
&& \sum_{A\subset E}      (x-1)^{r(E)-r(A)} (y-1)^{n(A)}  = \sum_{b=0}^m \sum_{s=0}^n \binom{m}{b} \binom{n}{s} (x-1)^{m-b} (y-1)^{s} \crcr
 \qquad && = \bigl(1 + (x-1) \bigr)^m \bigl(1 + (y-1) \bigr)^n= x^m y^n.
\eea

Third, eq. \eqref{tutdef} trivially factorizes over connected components.

Finally unicity is trivial, as any polynomial respecting the deletion-contraction relation can be written uniquely in terms of terminal forms 
by choosing a complete set of regular edges.

\qed

\begin{thm}[Universality]
Suppose a function $F_G(a,b,x,y)$ obeys:
\begin{itemize}
\item the linear deletion contraction relation with coefficients $a$ and $b$
\bee F_G(a,b,x,y)  = aF_{G-e}(a,b,x,y) + b F_{G/e}(a,b,x,y), \label{del}
\ee
\item the factorization properties 
\bee
F_{G_1 \cup G_2}(a,b,x,y)  = F_{G_1}(a,b,x,y)F_{G_2}(a,b,x,y) \label{facto1}
\ee for disjoint unions  and 
\bee F_{G_1 \star G_2}(a,b,x,y)  = F_{G_1}(a,b,x,y)F_{G_2}(a,b,x,y) \label{facto2}
\ee for vertex-joined graphs,

\item the terminal forms  $F_{bridge} = x$,  $F_{self-loop} = y$.
\end{itemize}
Then the function $F$ is:
\bee  F_G(a,b,x,y)  =  a^{n(G)} b^{r(G)} T_G (x/b , y/a) \; . \label{tuttun}
\ee
where $T_G$ is the Tutte polynomial, $r$ is the rank and $n$ the nullity of $G$.
\end{thm}

\prf  By the disjoint union property
it is sufficient to check the theorem for connected graphs. By the vertex-factorization property,
\eqref{tuttun} holds for terminal forms, hence connected graphs without regular edges. 

Finally \eqref{tuttun} is proved in the general case by induction on the number of regular edges 
using the linear deletion-contraction rule \eqref{delcont} and remarking that $r(G-e)=r(G)$, $n(G-e) = n(G)-1$,
$r(G/e) = r(G) - 1$ and $n(G/e) = n(G)$.

\qed

Evaluations of the Tutte polynomial at special values yield some interesting combinatorial countings.
\begin{lemma}
 In a graph G
\begin{itemize}\item $T_G(1,1)= \chi (G)$ counts the number of spanning forests in $G$,
and in particular the number of (spanning) trees in $G$ if $G$ is connected.
\item $T_G(2,1)$ counts the number of forests, i.e., the number of acyclic edge subsets.
\item $T_G(1,2)$ counts the number of connected spanning subgraphs.
\item $T_G(2,0)$ counts the number of acyclic orientations of $G$, i.e. orientations which do not allow any consistent
oriented cycle.
\item the ``chromatic polynomial'' $P_G(q):=  (-1)^{\vert V\vert - k(G)}  q^{k(G)}  T_G (1-q, 0) $
counts for integer $q$, the number of proper vertex colorings of $G$ using a set of $q$ colors
(a proper vertex coloring means that any vertices joined by an edge have a different color). 
\end{itemize}
\end{lemma}
\prf   We have already seen the first statement. One can check the other formulas on terminal forms,
then establish the deletion/contraction rule with coefficients 1 and 1. 

For the last point, the proof can be done for connected graphs only. One can check the factorization properties 
for $Q_G (q) = P_G(q) / q^{k(G)}$ and the linear deletion-contraction relation 
\bee Q_G (q) = P_{G-e} (q)  - P_{G/e}  (q)
\ee
because if $e$ joins $v_1$ to $v_2$, the colorings of $G$ are the same as the colorings of
$G-e$ where $v_1$ and $v_2$ have distinct colors, and the colorings of $G/e$ are the same as the colorings of
$G-e$ where $v_1$ and $v_2$ have the same color.
The terminal values of $Q_G (q)$ are $q-1$ for a bridge and 0 for a self-loop and one concludes through the universality theorem.  

\qed
 
\subsubsection{The Sokal polynomial}

This section follows closely the original work of Sokal~\cite{Sokal2005}. Interesting multivariate polynomials can be defined like the Tutte polynomial through 
a global "sum over subsets" formula. They
are also the unique ones to satisfy certain deletion-contraction rules with some specified terminal values. 
They occur in many physics problems, such as statistical models on graphs or the parametric representation of
Feynman amplitudes in quantum field theory.

The simplest multivariate such polynomial is the Sokal polynomial $Z_G( q,\{ y \} )$. It
has a different variable $y_e$ for each edge $e$,
plus another variable $q$ to count vertices.  It is defined
as a sum over subsets of edges:
\begin{definition}[Sum over subsets]
\begin{equation} \label{sok}
Z_G(q,y)=\sum_{A\subset  E}q^{k(A)}\prod_{e\in A}y_e ,
\end{equation}
where we recall that $k(A)$ is the number of connected components in the subgraph $(V,A)$.
\end{definition}

It obeys also a completely general  linear deletion-contraction relation. Separating the sum over $A$ above into a sum over 
$A, e\in A$ and another one over $A,e\notin A$ we obtain 
\begin{lemma}
For any edge $e$ (not necessarily regular)
\bee \label{multivartut}
Z_G( q,\{ y \} )= y_e Z_{G/e} (q, \{y \} \setminus \{y_e\}  ) + Z_{G-e} (q, \{ y \} \setminus   \{ y_e   \} ) \; .
\ee 
\end{lemma}

The terminal forms are graphs without edges, and with $v$ vertices; for such graphs $Z_G(q,\{y\})= q^v$.
The deletion/contraction relation together with the evaluation on terminal forms define  $Z_G(q,\{y\} )$
uniquely, since the result is again independent of the order of suppression of edges.
The Tutte polynomial can be obtained from the Sokal polynomial as
\bea
\big[ q^{- V }   Z_G(q,\{y\}) \big] {\Big\vert}_{ y_e = y-1, q = (x-1)(y-1)} =
(x-1)^{k(E)  -|V|}  T_G(x,y).
\eea
  
\subsubsection{Spanning polynomials, trees and forests polynomials}
We consider now rescalings of the Sokal polynomial
\bee
q^{- k(G) }Z_G (q,\{y\}).
\ee 
Taking the limit $q \to 0$,
that is retaining only the constant term in $q$
we obtain a sum over maximally spanning subgraphs $A$,
that is subgraphs with $k(A)=k(G)$:
\bee S_{G} (\{y\})=\sum_{A \subset E
\mathrm{ \; \; maximally  \; \; spanning  \; \; in \;\; }  G  } \quad
\prod_{e\in A} y_e .
\ee

If we now retain only the lowest degree of homogeneity
in $y$ we obtain a sum over maximally spanning graphs
with lowest number of edges, i.e. maximally spanning acyclic graphs or 
\emph{spanning forests} of $G$.
\bee F_{G} (y)=\sum_{\cF
\mathrm{ \; \; spanning  \; \;  forest    \; \;   of  \; \; }  G  } \quad
\prod_{e\in \cF} y_e .
\ee
This polynomial satisfies the factorization properties \eqref{facto1}-\eqref{facto2} on disjoint unions and on vertex-unions and
evaluated for $y_e=1$ yields the complexity $\chi (G)$. It plays a crucial role in the parametric representation of 
Feynman amplitudes.

Recall that we say that an edge is {\emph {semi-regular}} if it is either regular, or a bridge.
Proper graphs have only semi-regular edges.
The polynomial $F$ satisfies the deletion contraction-recursion
\begin{equation}\label{delcontrsymf1}
F_G(y)=F_{G- e}(y)+y_e F_{G/e}(y)
\end{equation}
for any semi-regular edge $e$, together with the terminal form evaluation
\bee \label{delcontrsymf2}
F_G (y) = 1
\ee 
on graphs made solely of self-loops.

Similarly we can generalize the terms appearing in the right hand side of
\eqref{Rforests}. We define, for a graph $G$ and a subset $R$ of vertices of $G$ 
containing {\emph {at least one vertex per connected component of $G$}},
a generalization of $F_G$ called the $R$-forest polynomial:
\bee  \label{secondsysy1} F^R_{G} (y)=\sum_{\cF
\mathrm{\; \;  R-forest    \; \;   of  \; \; }  G  } \quad
\prod_{e\in \cF} y_e .
\ee
where we recall that an $R$-forest is a maximal forest of $G$ 
containing exactly one vertex of $R$ per connected component of the forest.

For a graph $G = (V, E)$ and vertices $a, b \in  V$, let us define $G/ab$ as the graph 
obtained by merging the two vertices $a$ and $b$ together into a single vertex.
Another way to define this graph is to add between $a$ and $b$ an extra edge $e_{ab}$,
then contract this edge: $G/ab = G/e_{ab}$.
Then one can check that for a connected graph $G$ with two distinguished vertices $a$ and $b$
\bee  \label{secondsysy2} F^{ab}_{G} (y)=\sum_{\cF
\mathrm{ \; \; \;  ab-forest    \;   of   \; }  G  } \;\;
\prod_{e\in \cF} y_e   =  F_{G/ab}  (y)  .
\ee

\subsection{Flows as dynamical systems}

In this section we introduce some basic notions of the theory of dynamical systems, see also \cite{strogatz, devaney1992first, abraham1978foundations, ruelle}.
Dynamical systems are ubiquitous in physics, and in QFT the notion of renormalization group and fixed points thereof is of paramount importance.
Here we will consider simple examples which we divide in two categories depending on which notion of time we adopt that is, whether the 
system will have \emph{continuous time} $t\in\R$ or \emph{discrete time} $t\in\mathbb{Z}$. The former include the familiar ordinary differential equations (ODEs), and the latter feature dynamics that are close to a renormalization group setup. In particular we will briefly describe the renormalization form maps on the interval in relation to Feigenbaum's universality.

\subsubsection{Dynamical systems with continuous time}
\textbf{Generalities}\\
Let us consider autonomous, first order ordinary differential equations (ODEs) on an open set $\cM\in\R^n$, of the form
\be
\label{eq:ODE}
\dot{x}=f(x)\,,\quad\quad x\in\cM \,.
\ee
Let us further assume everything to be suitably regular, so that Cauchy's theorem guarantees the existence of the unique solution with 
initial condition $x\in\cM$ at time $t=0$. We will call this solution $\Phi^t(x)$. Furthermore we will assume that this solution exists
for all $t\in\R$ (or at least for all $t>0$). In fact, in what follows, we will not be interested in solving one specific Cauchy problem,
but on understanding the generic motion of a generic point $x\in\cM$, and in particular in what happens asymptotically, i.e. when $t\to\infty$. 
To this end, we can study the map
\be
\Phi^t:\quad \cM \to \cM\,,
\ee
which will be smooth under our assumptions. In the case where the solution exists for all $t\in\R$, one immediately notices the following properties
\be
\Phi^0=\mathrm{Id}\,,\quad
(\Phi^t)^{-1}=\Phi^{-t}\,,\quad
\Phi^t\circ\Phi^s=\Phi^{t+s}\,,
\ee
that imply that we can define a one-parameter Abelian group of diffeomorphisms
\be
\Phi=\{\Phi^t,\quad t\in\R\}\,.
\ee
One could also distinguish the case where $\Phi(x)$ is not invertible, and consequently $\Phi$ is only a \emph{semigroup}, but this is not important now.
We will say that $\Phi$ is the \emph{flow} of the differential equation (\ref{eq:ODE}).

We are now in a position to give a more formal definition. A \emph{continuous time regular dynamical system} is a couple $(\cM,\Phi)$ where 
$\cM$ is\footnote{More generally, $\cM$ can be a $n$-dimensional smooth manifold.} a regular open subset of $\R^n$ and $\Phi$ is a one 
parameter group of diffeomorphisms.

Our goal here will be to understand the properties of $\Phi(x)$ for a \emph{generic} $x\in\cM$, and ask asymptotic and often qualitative
questions, such as whether the motion remains bounded, whether it will tend to same particular point in~$\cM$, and so on.

\bigskip\noindent
\textbf{Asymptotic behavior: fixed points}\\
The simplest asymptotic behaviour for an autonomous ODE is the case when there exists one point $x^*\in\cM$ such that
\be
\Phi^t(x^*)=x^*\,,\quad \forall\ t>T\,,
\ee
for some $T$. It is clear that this is equivalent to requiring the above condition to hold for all $t$ and, in terms of (\ref{eq:ODE}), to requiring that
\be
f(x^*)=0\,.
\ee
We will call such an $x^*$ a \emph{fixed point} or \emph{critical point} for the dynamical system.

Of course it is not typical that the initial condition for a Cauchy problem is precisely $x(0)=x^*$. Fixed points are interesting because 
they influence the flow for any initial condition close to them. This is obvious in the case where $\cM=\R$ and there exists a unique
critical point $x^*$ such that $f'(x^*)\neq0$.\footnote{We regard the case $f'(x^*)=0$ as non-generic.} Then, two things may happen:
\begin{itemize}
\item $f'(x^*)>0$: then $f(x)<0$ to the left of $x^*$, and the flow pushes these points to smaller values, away from $x^*$;
similarly $f(x)>0$ to the right of $x^*$ and the flow pushes them to the right, again away from $x^*$. 
\item $f'(x^*)<0$: then $f(x)>0$ to the left of $x^*$, and the flow pushes these points to larger values, towards $x^*$; 
similarly $f(x)<0$ to the right of $x^*$ and the flow pushes them to the left, again towards $x^*$. As it is easy to prove, 
the flow cannot cross $x^*$, so that the motion tends to $x^*$.  
\end{itemize}

This simple example motivates the need to classify fixed points depending on their property to attract or repel the points 
in their neighbourhood under the flow $\Phi$. This classification, together with many useful criteria, was first put forward by Lyapunov.
Let $x^*$ be a critical point for $\Phi$, and let $\Phi$ exists for all $t\in\R$. Then
\begin{enumerate}
\item $x^*$ is \emph{attractive} (or \emph{asymptotically stable}) if there exists a neighbourhood $\cV$ of $x^*$ such that
\[x\in \cV\quad\quad\Rightarrow\quad\quad \lim_{t\to+\infty}\Phi^t(x)=x^*\,.\]
\item $x^*$ is \emph{stable} for all times\footnote{Stability only in the future or past amounts to restricting
to $t>t_0$ or $t<t_0$ respectively.} if for any neighbourhood $\cU$ of $x^*$ there exists a neighbourhood $\cV_0$ of $x^*$ such that
\[x\in \cV_0\quad\quad\Rightarrow\quad\quad\Phi^t(x)\in\cU\quad\quad\forall\ t\in\R\,.\]
\item $x^*$ is \emph{unstable} if it is not stable. 
\item $x^*$ is \emph{repulsive} if there exists a neighborhood $\cV$ of $x^*$ such that
\[x\in \cV\quad\quad\Rightarrow\quad\quad \lim_{t\to-\infty}\Phi^t(x)=x^*\,.\] 
\end{enumerate}

With this terminology, it is easy to classify the dynamical systems given by a \emph{linear} ODEs. As it is well known, we have in that case
\be
\dot{x}=A\,x\,,\quad\quad \Phi^t(x)=e^{t\,A}\,x\,,
\ee
where $e^{t\,A}$ is the exponential of a matrix, defined by the convergent series $e^{t\,A}=\sum_{k=0}^\infty t^k\,A^k/k!\,$.
Furthermore, if $A$ is normal\footnote{More generally, similar considerations can be made using the Jordan form 
of $A$, but they will not be important for us.} then it can be diagonalized, and it is enough to consider its eigenvalues $\lambda_1,\dots,\lambda_n$.
Clearly $x^*=0$ is a fixed point, and it is not hard to verify the following statements:
\begin{itemize}
\item If Re$(\lambda_i)<0$ for all eigenvalues, then $x^*$ is attractive.
\item If Re$(\lambda_i)>0$ for all eigenvalues, then $x^*$ is repulsive.
\item If Re$(\lambda_i)=0$ for all eigenvalues, then $x^*$ is stable.
\item If Re$(\lambda_i)<0$ for some eigenvalues and Re$(\lambda_j)>0$ for others, then $x^*$ is unstable.
\end{itemize}
We will also say that $x^*$ is \emph{hyperbolic} if Re$(\lambda_i)\neq0$ for all $i$.

Of course linear equations are not very interesting \textit{per se}. However, consider any ODE with (at least) one fixed point $x^*$. Then we can write
\be
\dot{x}=f(x)=A\,(x-x^*)+O\big(\|x-x^*\|^2\big)\,,\quad\quad{\rm with}\quad A=\frac{\partial f}{\partial x} \Big{|}_{x=x^*}\,,
\ee
simply by expanding around $x^*$. Our intuition suggests then that, at least if $x^*$ is hyperbolic, then the linearized analysis
should be enough to ``understand'' the flow, at least close to $x^*$. This is actually the case, as it follows from an important
theorem that we will state without proof (see e.g. \cite{ruelle}).

\begin{theorem}[Grobman-Hartman] If $\dot{x}=f(x)$ has a hyperbolic fixed point $x^*\in\cM=\R^n$, then there is a neighborhood 
of $x^*$ such that the flow there is homeomorphic to the flow of the linear system $\dot{x}=A\,x$. In other words, locally the 
nonlinear flow is conjugated by a continuous invertible map to the linear one. 
\end{theorem}

There are extensions of this theorem that guarantee (under additional hypotheses)  that the two flows are diffeomorphic, but
they are more subtle and we will not discuss them here.

As a corollary of this theorem, it follows that the stability property of an hyperbolic fixed point can be found from the 
corresponding linearized ODE, i.e. from the eigenvalues of the Jacobian at the fixed point, which is also known as Lyapunov's spectral method.

Before proceeding further, it is worth illustrating all this on a simple example. Consider the following dynamical system, which we spell out in coordinates:
\be
\dot{x}_1=x_1-x_1\,x_2\,,\quad\quad\dot{x}_2=-x_2+x_1^2\,.
\ee 
One easily finds three fixed points
\be
x^*_{(1)}=(0,0)\,,\quad\quad
x^*_{(2)}=(1,1)\,,\quad\quad
x^*_{(3)}=(-1,1)\,.
\ee
Consider for instance $x^*_{(1)}$. It is clearly hyperbolic and unstable, because the Jacobian there is the matrix $A={\rm diag}(1,-1)$. 
Therefore, by Grobman-Hartman theorem, it follows that the flow in the vicinity of $x^*_{(1)}$ is conjugated to the one of the associated 
linear system. The latter is very simple: in the linear system, the axis $x_1$ supports an exponentially repulsive motion, whereas the 
axis $x_2$ supports an exponentially attractive one. Generic initial conditions yield hyperbolae that asymptote to the coordinate axes.

\begin{figure}[!t]
  \begin{center}
    \subfigure[Full nonlinear flow.]{\label{fig:nonlinearflow}\includegraphics[width=0.4\textwidth]{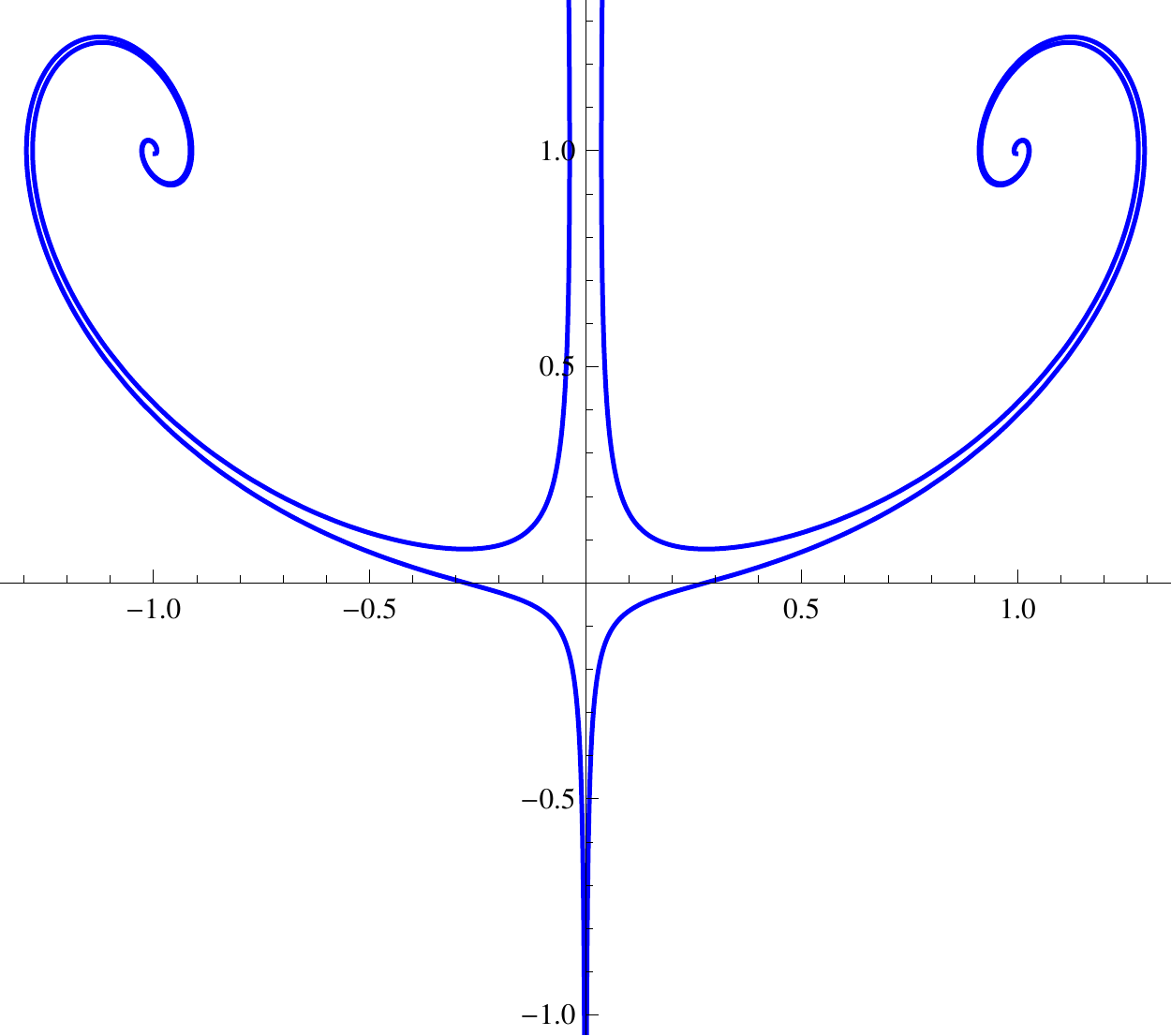}}
    $\quad\quad$
    \subfigure[Linearization around $(0,0)$.]{\label{fig:linearflow}\includegraphics[width=0.4\textwidth]{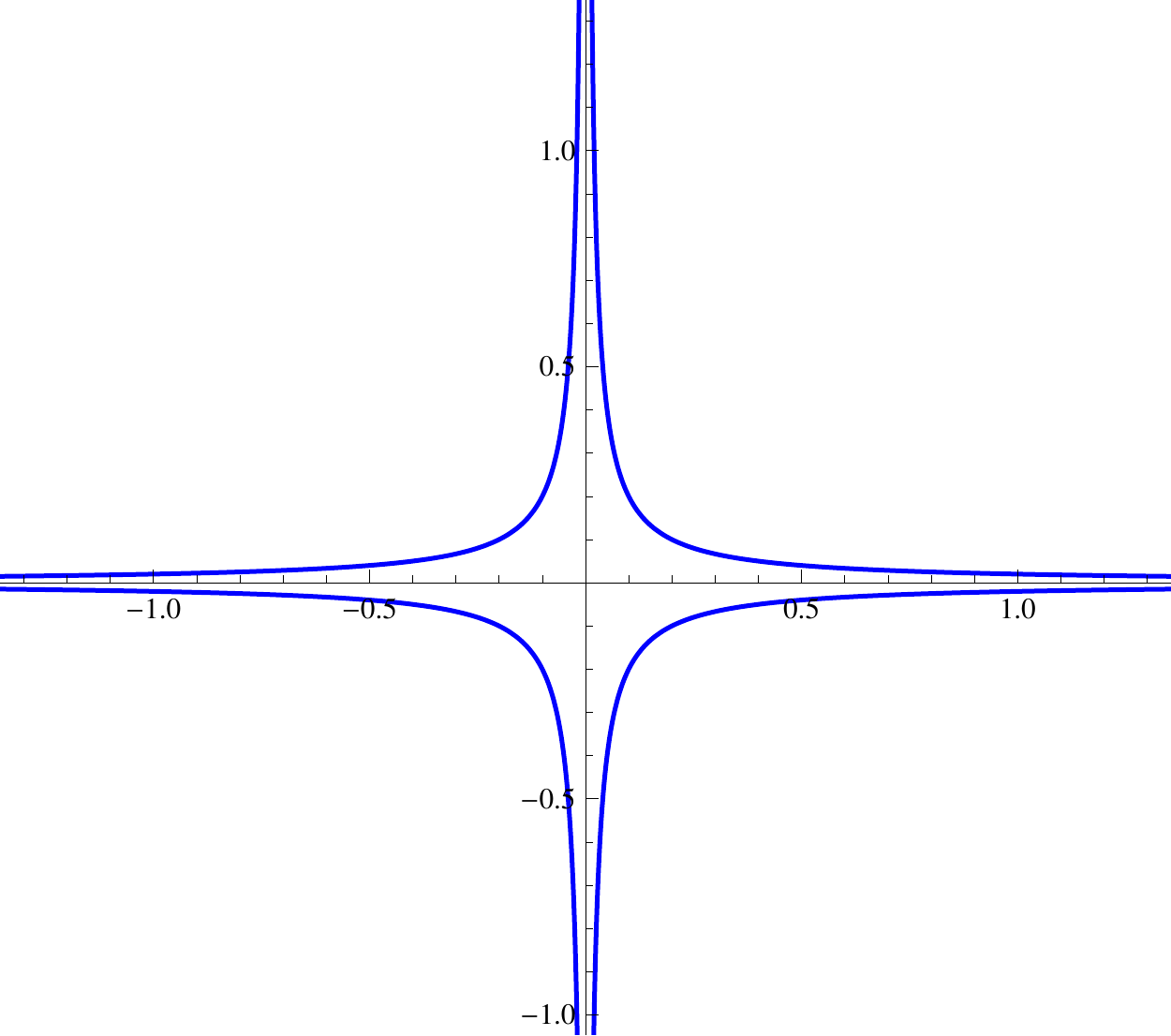}}
  \end{center}
  \caption{Flow for a nonlinear and linearized system.}
  \label{fig:flow}
\end{figure}

Looking at Figure \ref{fig:flow} it is clear that the nonlinear flow around  $x_{(1)}^*$ is similar to the linear one. It is interesting 
to look for some curves that play a role similar to the coordinate axes in the linear system, which helped us to understand the motion of 
a generic initial condition. A very important and useful result guarantees their existence.

\begin{theorem}[Stable manifold theorem (Perron-Hadamard)] 
If $x^*$ is a hyperbolic fixed point, then the two sets
\bea
\Ws&=&\left\{x\in\cM:\ \lim_{t\to+\infty}\Phi^t(x)=x^*\right\}\,,\\
\Wu&=&\left\{x\in\cM:\ \lim_{t\to-\infty}\Phi^t(x)=x^*\right\}\,,
\nonumber
\eea 
are regular manifolds called stable and unstable manifold, and are tangent to the hyperplanes generated by the eigenvectors of the
linearized system corresponding resp. to negative and positive eigenvalues (see e.g. \cite{abraham1978foundations,hirsch1970}).
\end{theorem}

We shall not prove this theorem, but let us mention that the difficulty is in proving that a regular stable (unstable) manifold exists
locally. Once one has constructed such $\Ws_{\rm loc.}$ ($\Wu_{\rm loc.}$) it is easy to obtain the whole manifolds by defining $
\Ws=\bigcup_{t\leq0} \Phi^t(\Ws_{\rm loc.})$ and $\Wu=\bigcup_{t\geq0} \Phi^t(\Wu_{\rm loc.})$.

A word of warning: we discussed only the dynamics around fixed points, but these are not the only objects that influence the asymptotic
behaviour. From dimension $n\geq2$, dynamical systems may present \emph{limit cycles}, and for $n\geq3$ they may have \emph{chaotic behavior}. 
The interested reader is invited to consult,~e.g.~\cite{strogatz}.

\bigskip\noindent
\textbf{Bifurcations}\\
We introduce here the notion of \emph{family of dynamical systems}, that we will analyze in more detail for discrete-time systems. Here we
simply allow for a (regular) dependence of (\ref{eq:ODE}) on one or more real parameters $\mu_1,\dots,\mu_n$, so that we have
\be
\dot{x}=f(x;\,\mu_1,\dots,\mu_n)\,,\quad\quad x\in\cM \,.
\ee
This generalization is quite natural, as it allows to study the evolution of a system depending on some external condition, such as the 
demographics as a function of resources, etc.

\begin{figure}[!t]
  \begin{center}
    \subfigure[Tangent bifurcation.]{\label{fig:tangentbif}\includegraphics[width=0.48\textwidth]{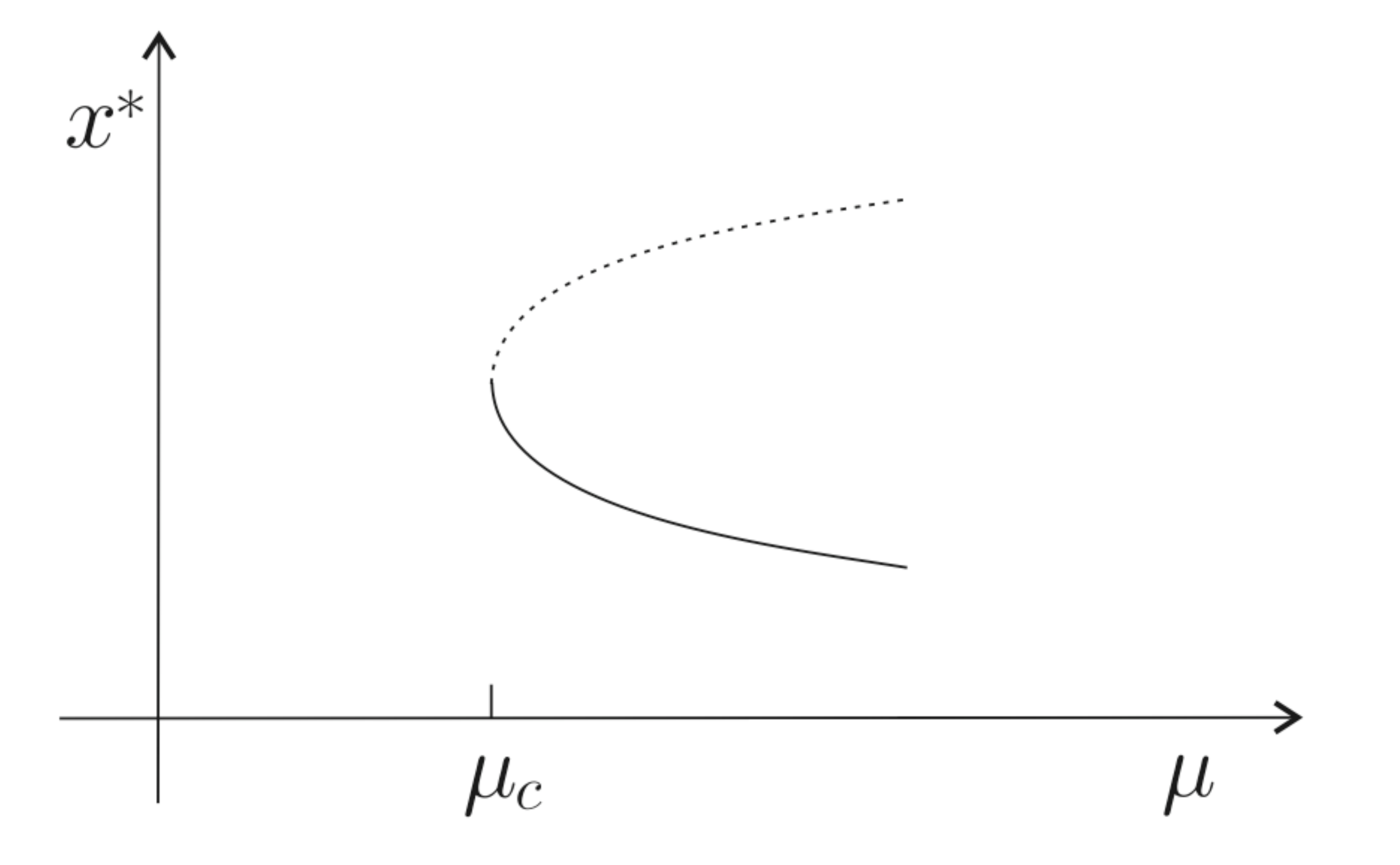}}
    \subfigure[Pitchfork bifurcation.]{\label{fig:pitchforkbif}\includegraphics[width=0.48\textwidth]{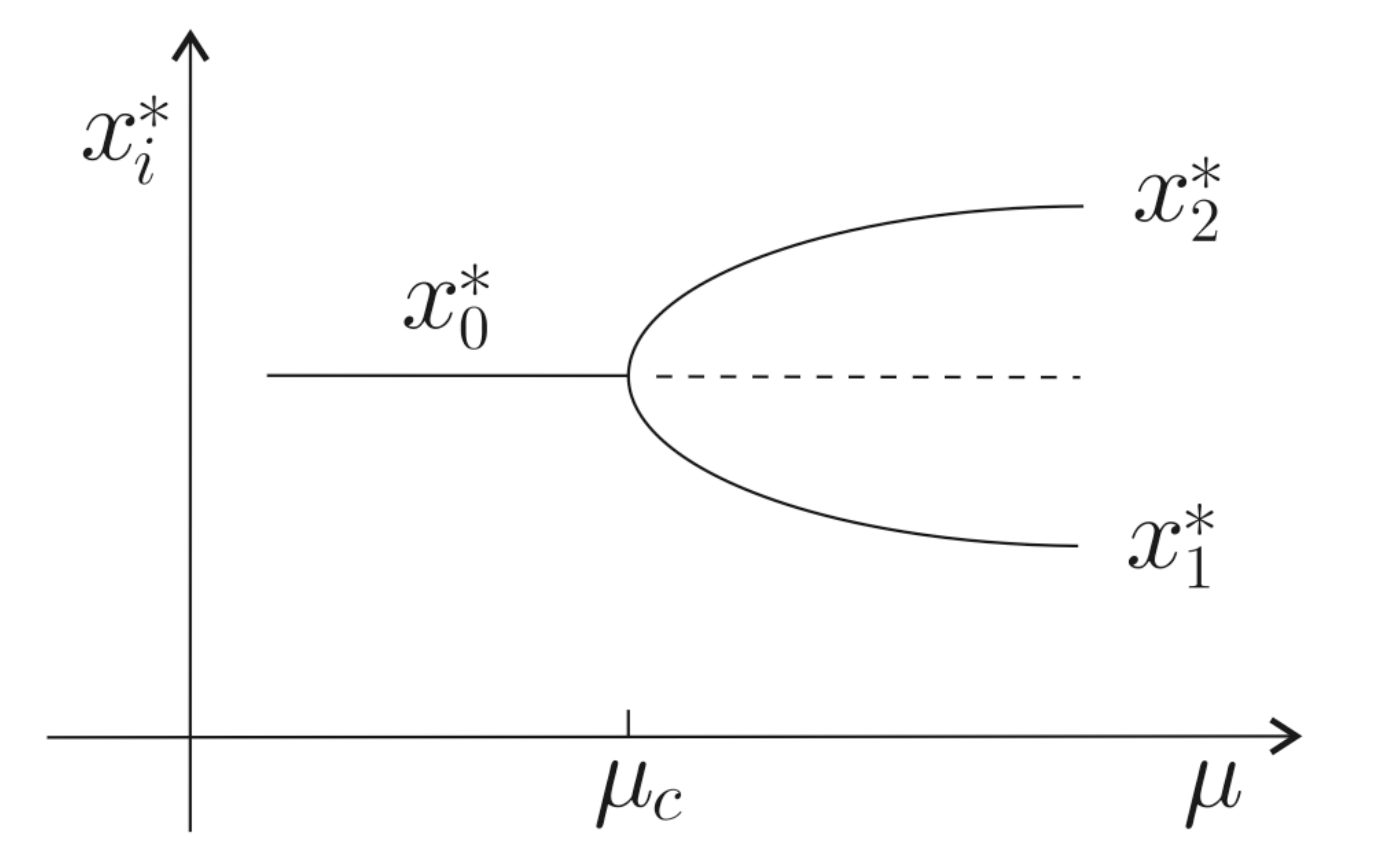}}
  \end{center}
  \caption{Bifurcations diagrams showing the fixed points as functions of $\mu$.}
  \label{fig:bifurcs}
\end{figure}

The simplest example is the linear equation
\be
\dot{x}=\mu\,x\,,\quad\quad x\in\cM=\R\,,\quad \mu\in\R\,,
\ee
and it is clear that the asymptotic properties of the system depend on the sign of $\mu$.
A similar, but somewhat less trivial example is
\be
\dot{x}=(x-\mu_1)^2-\mu_2\,,\quad\quad x\in\cM=\R\,,\quad \mu_i\in\R\,.
\ee
Here the parameter $\mu_1$ is largely inessential, as it can be reabsorbed by a translation. However, again the sign of $\mu_2$ is 
important: when it is negative there are no critical points, whereas when it turns positive a couple of critical points (one stable,
one unstable) is created. This is called the \emph{tangent bifurcation}. It is useful to plot the \emph{bifurcation diagram}
(Figure \ref{fig:tangentbif}), where one draws the position of the critical points as a function of the relevant parameter
(in this case, $\mu_2$). This gives an overview of the asymptotic properties of the system as a function of the external parameters. 

Another example, to which we will return later, is the so-called \emph{pitchfork bifurcation}, given for instance by
\be
\dot{x}=-x^3+\mu\,x\,,\quad\quad x\in\cM=\R\,,\quad \mu\in\R\,.
\ee
Here we have a stable critical point for $\mu<0$ which splits up into two stable and one unstable critical points when $\mu$ 
becomes positive. The bifurcation diagram of Figure \ref{fig:pitchforkbif} suggests the name.

\subsubsection{Dynamical systems with discrete time}
Some phenomena where observables can be measured only at some moment in time are more naturally described by using a discrete 
time variable. Examples are the abundance of a certain species after each reproductive cycle, the amount of crops collected every year, etc.

To discuss dynamical systems with discrete time we just have to rephrase what we said in the previous section.

\bigskip\noindent
\textbf{Generalities, fixed points, Lyapunov exponents}\\
Rather than being defined by an ODE such as (\ref{eq:ODE}), the typical definition of a discrete-time dynamical system is a recursion law
\be
x_{n+1}=f(x_n)\,,\quad\quad f:\ \cM\to\cM\,.
\ee
We immediately obtain that the (discrete-time) flow $\Phi$ satisfies, for $n\in\N$,
\be
\label{eq:discreteflow}
\Phi^0={\rm Id}\,,\quad\quad \Phi^1(x)=f(x)\,,\quad\quad \Phi^n=\left(\Phi^1\right)^n=f\circ\dots\circ f\,,
\ee
where the last equation indicates the $n$-fold composition of functions. Depending on whether $f(x)$ is invertible, one can add the additional property
\be
 \Phi^{-1}=\left(\Phi^1\right)^{-1}=f^{-1}\,,
\ee
and extend (\ref{eq:discreteflow}) to $n\in\Z$, in which case the flow will be a group, rather than just a semigroup.

As seen in the previous section, it is interesting to look at \emph{fixed points} $x^*\in\cM$ that satisfy
\be
\Phi^n(x^*)=x^*\quad\quad\Longleftrightarrow\quad\quad x^*=f(x^*)\,.
\ee
Again, it will be important to understand whether $x^*$ attracts or repels the nearby points. In what follows, let us restrict to the case 
where $\cM=I\subset\R$ is a (possibly unbounded) interval on the real line, and study the flow of $x_0=x^*+\eps$. Then
\be
\label{eq:slope}
x_1=f(x_0)\approx x^*+\eps\,f'(x^*)\,,\quad\quad x_n=f^n(x_0)\approx x^*+\eps\,\left(f'(x^*)\right)^n\,.
\ee
Clearly, the asymptotic behaviour around $x^*$ depends on whether the modulus of the slope $|f'(x^*)|$ is larger than one 
(repulsive fixed point) or smaller than one (attractive fixed point).

It is useful to introduce the \emph{Lyapunov exponent}, that is a tool to understand the behaviour of two neighbouring ($\eps$-close)
generic points in $\cM$: will they remain close together, eventually get squeezed to the same attractive fixed point, or will they 
become more and more separated\footnote{ The latter is a typical feature of \emph{chaotic systems} that show unpredictable behaviour
for a generic initial condition.}? The natural quantity to consider is the limit
\be
\de(x,n)=\lim_{\eps\to0}\frac{\left|\Phi^n(x+\eps)-\Phi^n(x)\right|}{|\eps|}=\left|f'\left(\Phi^{n-1}(x)\right)\right|\,.
\ee
To remove the dependence on $n$ one can take the average of $\de(x,n)$ along the orbit. Finally, (\ref{eq:slope}) suggests that the 
separation grows geometrically, so that we write
\be
\label{eq:lyapunovexp}
\ga(x)=\lim_{N\to\infty} \frac{1}{N}\sum_{n=0}^{N-1}\log\left|f'\left(\Phi^{n}(x)\right)\right|\,.
\ee
We will say (if the above limit exists) that $\ga(x)$ is \emph{the Lyapunov exponent of $x$}. A theorem by Oseledec \cite{oseledec} 
guarantees that indeed the limit exists for almost every $x\in\cM$, and it is immediate to see that $\ga(x)$ will be the same for any $x\in\cM$ with
the same asymptotic behaviour. In the cases of our interest, in fact, there will be only one Lyapunov exponent, so that we will from now on drop the dependence on~$x$.

We conclude this section with two simple examples of one-dimensional discrete time-systems, that are also called \emph{iterated maps of the interval}. 
As we have already seen it is interesting to allow for dependence on one or more parameters $\mu_i$.  The simplest example is the \emph{Malthusian growth}, 
a simple model for population expansion with unlimited resources. The law is linear
\be
x_{n+1}=f_\mu(x_n)\,,\quad\quad f_\mu(x)=\mu\,x\,,\quad\quad x\in[0,+\infty)\,,\quad\mu>0\,,
\ee
and the recursion can be solved immediately to give
\be
\Phi^n(x)=\mu^n\,x\,,\quad\quad n\in\Z\,,
\ee
with the Lyapunov exponent $\gamma=\log\mu$.
As indicated, the flow can be extended to negative $n$, and dictates a simple geometric behavior, similar to the one of the 
continuous-time system $\dot{x}=\log\mu\ x$.

One could have the impression that discrete time systems are just a trivial modification of ODEs, but this is not the case. In fact,
discrete-time systems given by a simple law can yield an extremely complex behavior. The model immediately more complicated than
Malthusian growth is the \emph{logistic growth}, which is a modification of the former to include finite resources. It can be written as
\be
\label{eq:logisitc}
x_{n+1}=f_\mu(x_n)\,,\quad\quad f_\mu(x)=\mu\,x\,(1-x)\,,\quad\quad x\in[0,1]\,,\quad0<\mu\leq4\,,
\ee
and a partial study of it, in its simplest regime, will occupy the rest of this section.

This surprising complexity comes from the fact that discrete-time systems can be seen as arising from continuous-time ones in higher
dimensions, by defining $\Phi^n(x)$ as the intersection of $\Phi^t(x)$ with some submanifold embedded in $\cM$, a procedure called
Poincar\'e section. It is then clear that discrete-time systems do not suffer the same topological limitation of continuous time 
ones\footnote{Consider, for instance, the Poincar\'e-Bendixon theorem \cite{strogatz}.}, and therefore can exhibit a richer behavior also in low dimension.

\bigskip\noindent
\textbf{The logistic map}\\
Let us start the study of the logistic map. As we said this is defined by the one-parameter family of functions on $\cM=[0,1]$
\be
f_\mu\,:\ [0,1] \to [0,1]\,,\quad\quad f_\mu\,:\ x\mapsto \mu\,x\,(1-x)\,.
\ee 
In order to have that $f_\mu([0,1])\subset[0,1]$, it must be that $0<\mu\leq4$. Let us start our analysis from small $\mu$.

\begin{figure}[!t]
  \begin{center}
    \subfigure[$\mu=0.9$, equilibrium at $x^*=0$.]{\label{fig:09}\includegraphics[width=0.4\textwidth]{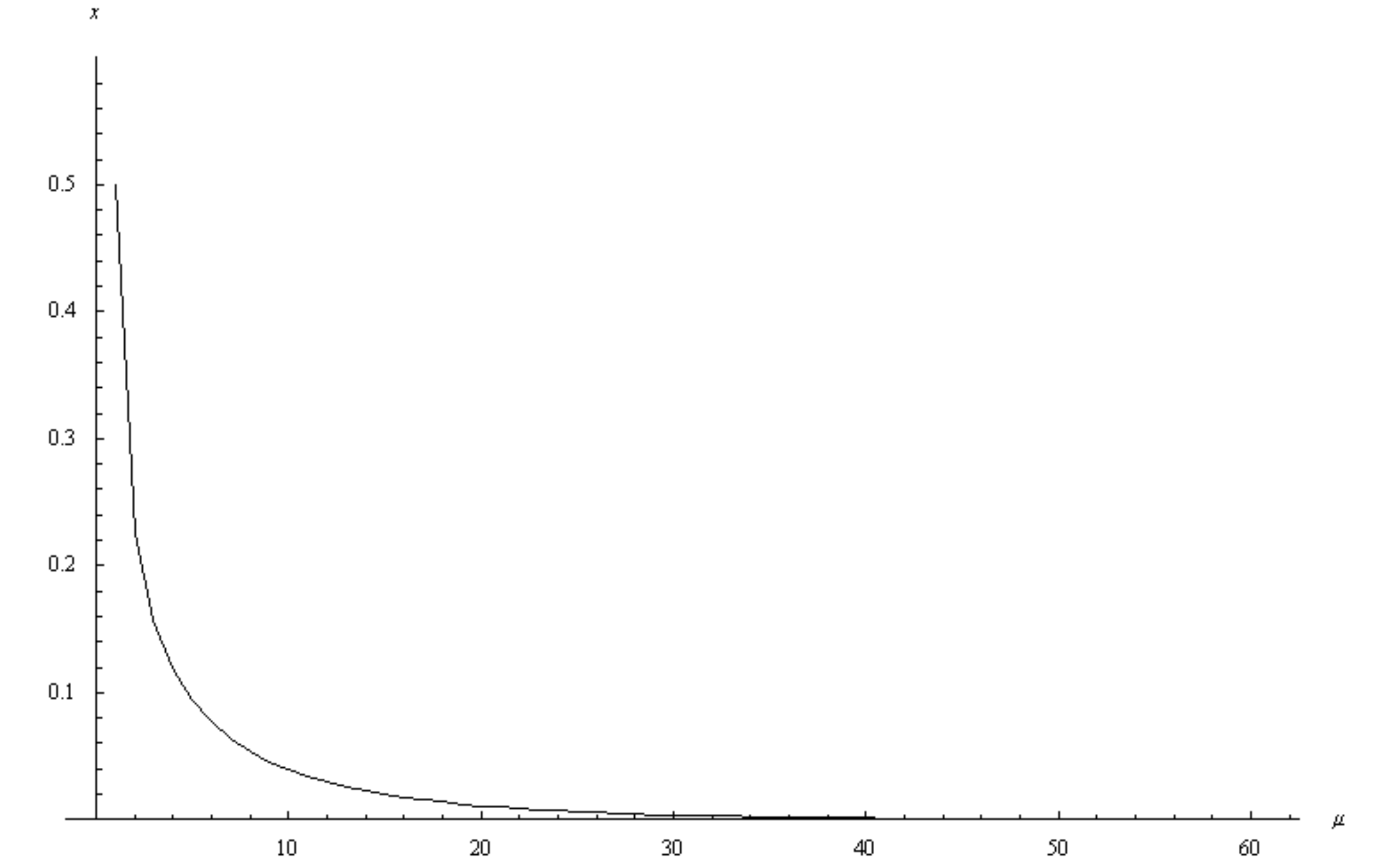}}
    \subfigure[$\mu=2.85$, equilibrium at $x^*\neq0$.]{\label{fig:285}\includegraphics[width=0.4\textwidth]{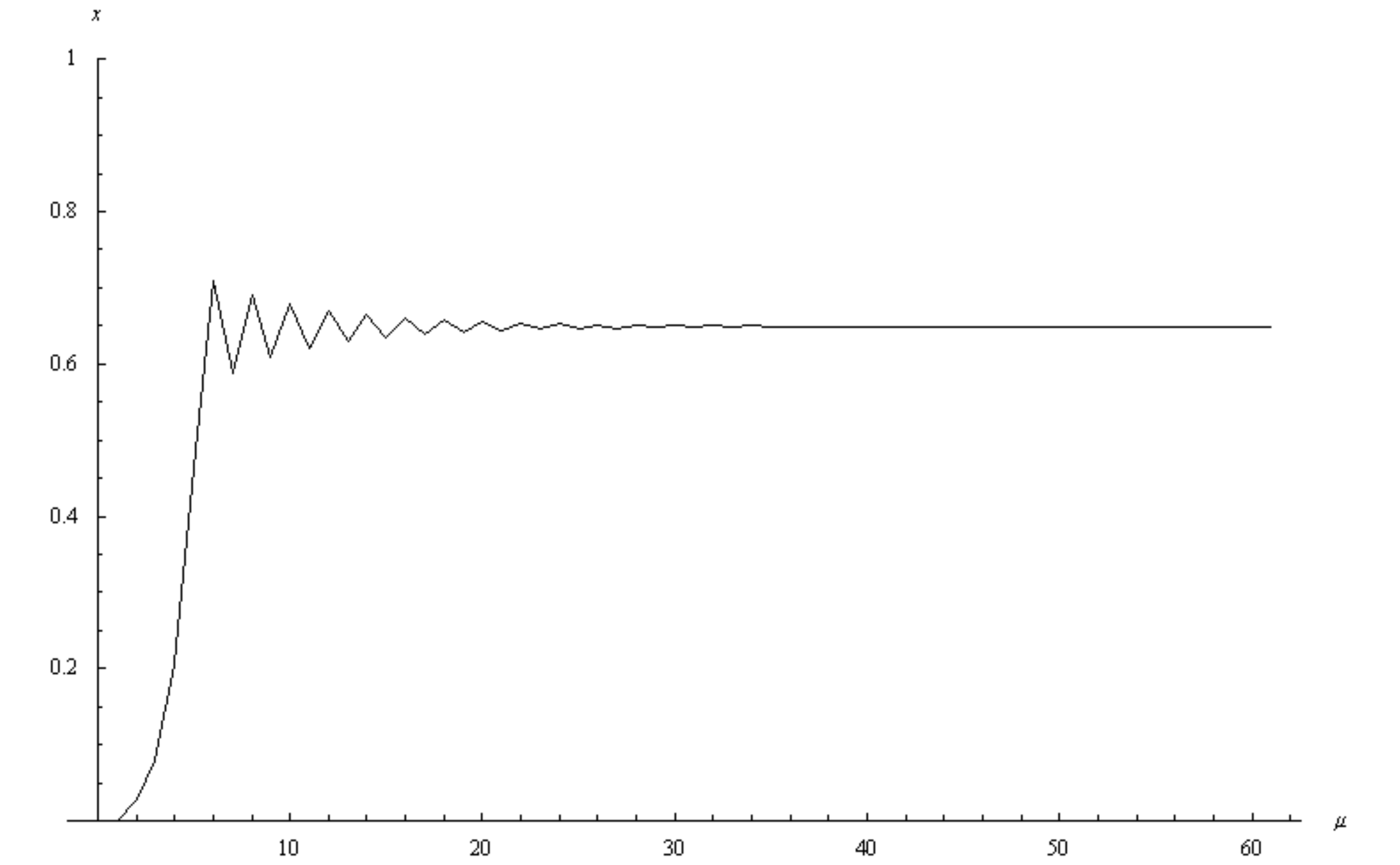}}\\
    \subfigure[$\mu=3.2$, two-cycle.]{\label{fig:32}\includegraphics[width=0.4\textwidth]{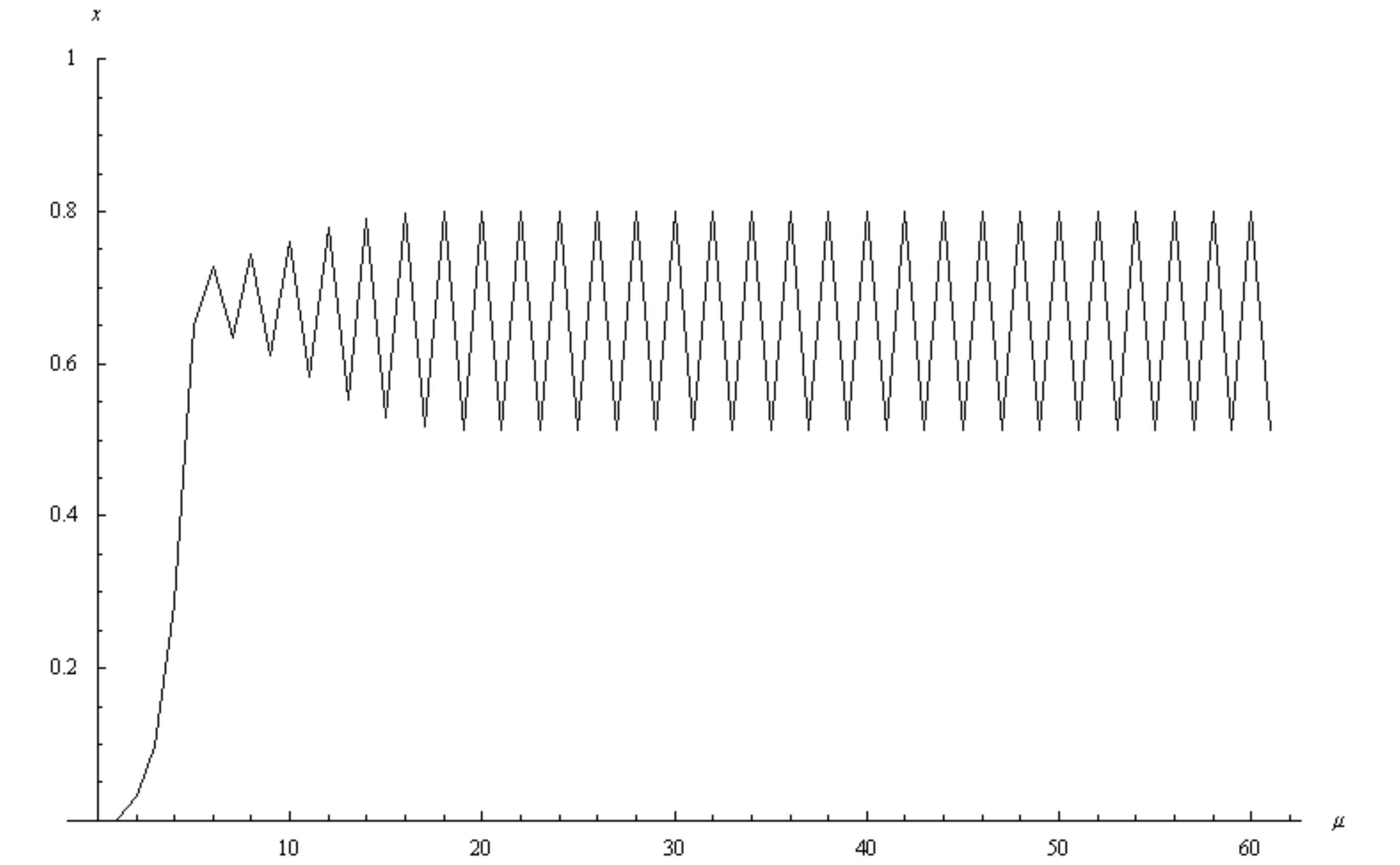}}
    \subfigure[$\mu=3.5$, four-cycle.]{\label{fig:35}\includegraphics[width=0.4\textwidth]{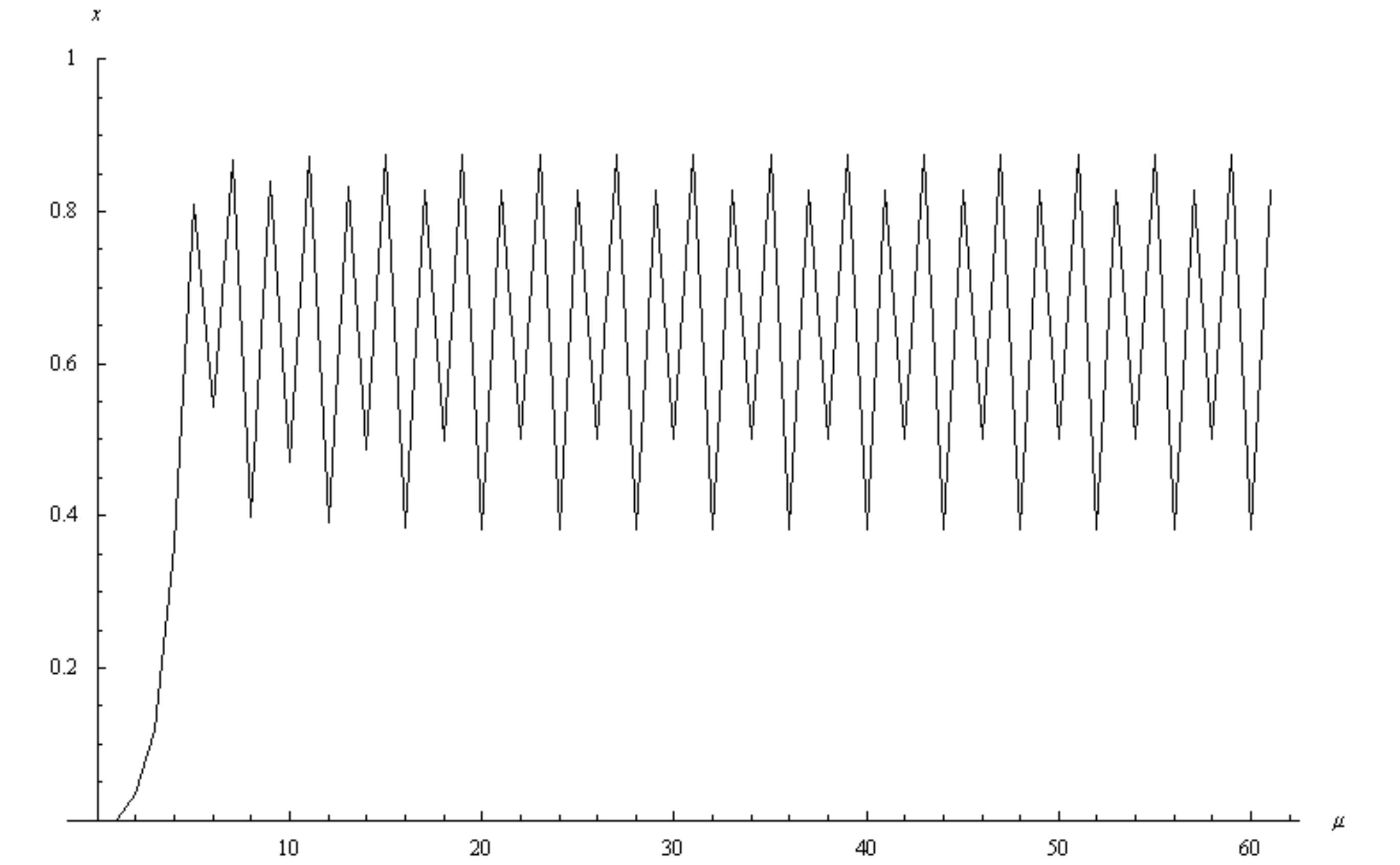}}\\
    \subfigure[$\mu=3.9$, chaotic behavior.]{\label{fig:39}\includegraphics[width=0.7\textwidth]{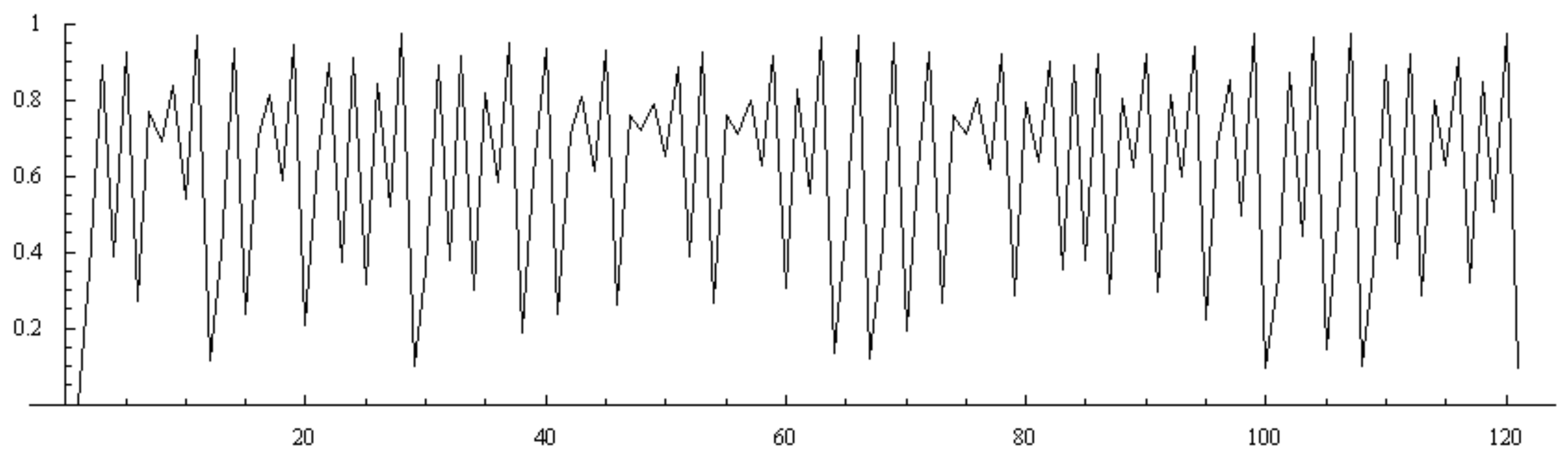}}
  \end{center}
  \caption{Asymptotic behavior of the logistic map for some $\mu$'s.}
  \label{fig:bifurcations}
\end{figure}

When $0<\mu<1$ the fixed point equation $f_\mu(x)=x$ has only one solution in $x^*=0$, which is a stable fixed point. For larger 
values of $\mu$, $x^*=0$ becomes unstable and a new stable fixed point is generated at $x^*=\frac{\mu-1}{\mu}$. Its stability can be checked by looking at
\be
|f'(x^*)|=|2-\mu|\,,
\ee
so that this fixed point is stable for $1< \mu<3$. To understand what happens when $\mu$ gets larger than $3$ we can simulate the
behaviour of this system with a computer and plot the resulting orbits.

In Figure \ref{fig:bifurcations} some orbits are plotted. As expected, for $\mu=0.9$ one has that $x_n\to x^*=0$, whereas for
$\mu=2.85$ the attractive fixed point is at $x^*\approx0.65$. It is interesting to notice that at $\mu=3.2$ there are no fixed 
points, but $x_n$ oscillates between two points. We will say in this case that there is an \emph{attractive 2-cycle}. When $\mu$ is
further increased to $\mu=3.5$, the motion oscillates between four points--an attractive 4-cycle. Finally, for very large values of
$\mu$ such as $\mu=3.9$, there is no apparent pattern for $x_n$; indeed it will turn out that there the motion is \emph{chaotic}.

\begin{figure}[t!]
  \begin{center}
  \includegraphics[width=0.85\textwidth]{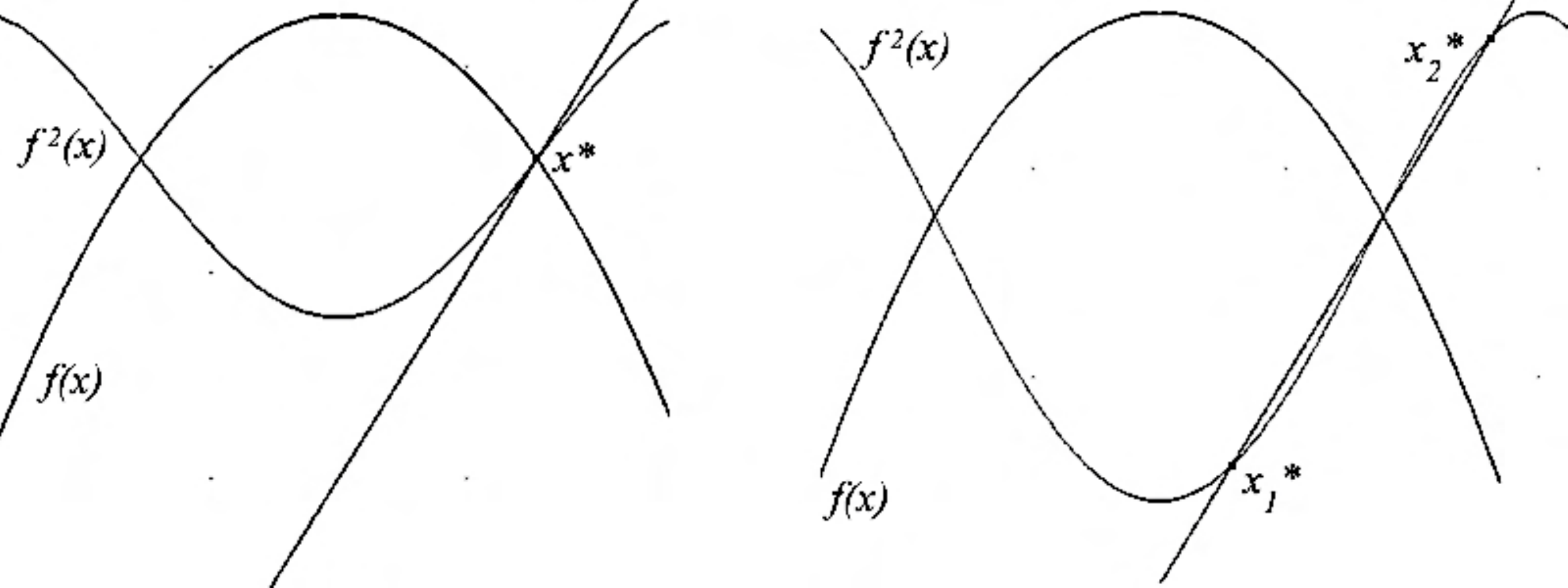}
  \end{center}
  \caption{A period-doubling bifurcation.}
  \label{fig:perioddoubling}
\end{figure}

It is worth investigating in more detail what happens when passing $\mu=3$. First, let us remark that when $f_\mu$ has a two-cycle,
$(f_\mu)^2=f_\mu\circ f_\mu$ must have two fixed points:
\be
f_\mu(x_1)=x_2\,,\quad f_\mu(x_2)=x_1 \quad\quad
\Longrightarrow\quad\quad (f_\mu)^2(x_1)=x_1\,,\quad(f_\mu)^2(x_2)=x_2\,.
\ee
Let us now set $\mu=3-\eps$. Then $x^*=\frac{\mu-1}{\mu}$ is a stable fixed point with slope $f_\mu'(x^*)=\eps-1$. Correspondingly,
the composition $f_\mu\circ f_\mu$ has also a fixed point there, with slope $(f_\mu\circ f_\mu)'(x^*)=(\eps-1)^2<1$, again stable.
Let now $\mu=3+\eps$. Then $x^*=\frac{\mu-1}{\mu}$ is unstable, with slope $f_\mu'(x^*)=-\eps-1<-1$. The same fixed point for the 
composition has then slope $(f_\mu\circ f_\mu)'(x^*)=(\eps+1)^2>1$, and it is also unstable. Furthermore, as depicted in Figure \ref{fig:perioddoubling},
by continuity a couple of fixed points are created to the left and to the right of $x^*$, and it is easy to see that they are stable.

\begin{figure}[t!]
  \begin{center}
  \includegraphics[width=0.9\textwidth]{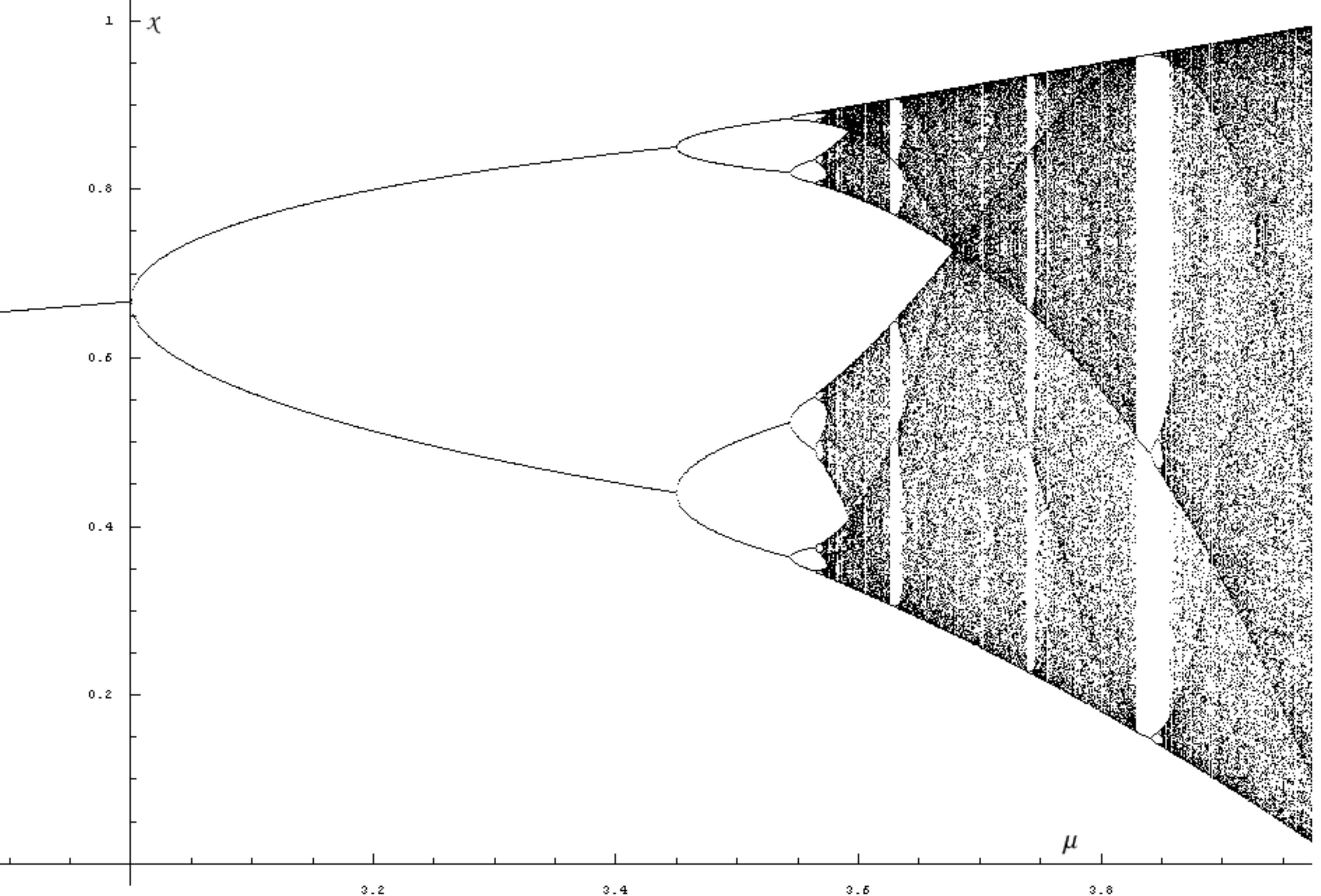}
  \end{center}
  \caption{Bifurcation diagram for the logistic map for $\mu\geq3$, where the first bifurcation occurs.}
  \label{fig:bifurcationdiag}
\end{figure}

The above reasoning seems to be applicable not only when going from period-one to period-two, but every time we double the period of the attractive
cycle. We have already seen that for larger $\mu$ there exists an attractive four-cycle. It is worth plotting the \emph{bifurcation diagram} for 
the logistic map, that indicates for any $\mu$ the set to which the motion is attracted. Looking at Figure \ref{fig:bifurcationdiag} we see that
at several points $\mu_0,\mu_1,\dots,\mu_n,\dots$ a bifurcation occurs, where the number of attractive points doubles (i.e. one goes from 
a $2^n$-cycle to a $2^{n+1}$ one). As we discussed, $\mu_0=3$, and one can see from the plot that $\mu_1\approx3.45$, $\mu_2\approx3.55$, etc.

\begin{figure}[ht!]
  \begin{center}
  \includegraphics[width=0.75\textwidth]{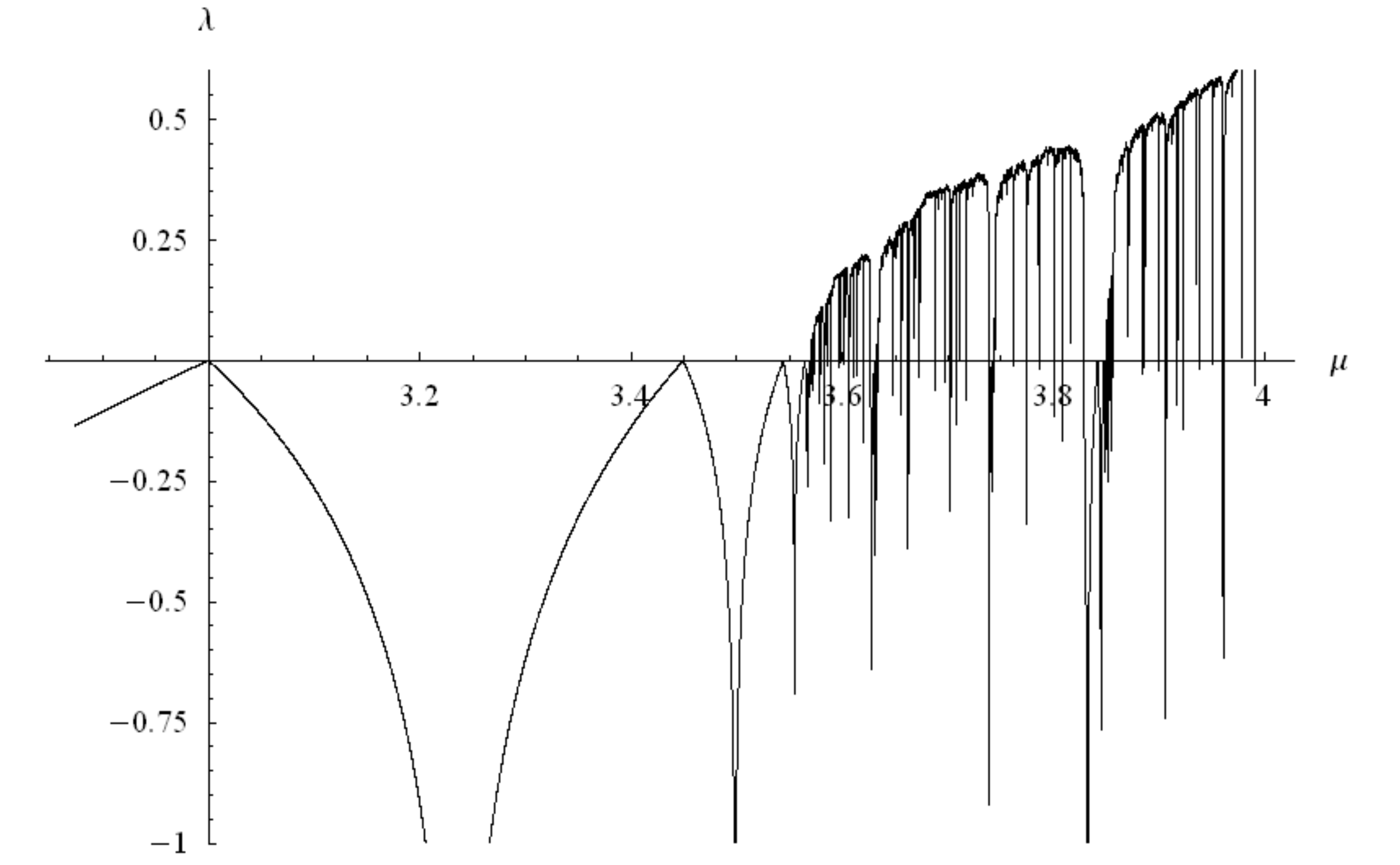}
  \end{center}
  \caption{Lyapunov exponent for the logistic map for $\mu\geq3$.}
  \label{fig:lyapunov}
\end{figure}

Furthermore, these points seem to accumulate to some $\mu_\infty\approx3.6$ after which the trajectory jumps wildly between many points. 
To better understand this, let us look at Figure \ref{fig:lyapunov}. One sees that the Lyapunov exponent $\ga(\mu)$ is smaller than zero
in presence of a $2^n$-cycle, and vanishing at the bifurcation point as it should. However, for $\mu>\mu_\infty$ one sees that $\ga(\mu)>0$
which signals the beginning of chaotic behavior\footnote{For $\mu$ even larger there exists some interval where the behavior is back to 
periodic, with $\ga(\mu)<0$. These are the so-called periodic-windows, which we will not discuss here.}. It is also interesting to notice 
that between any couple of bifurcation points there is  (by continuity) a point where $\ga(\tilde{\mu}_n)=-\infty$, with
$\mu_{n-1}<\tilde{\mu}_n<\mu_{n}$. This happens when one of the points in the cycle is the maximum $x=1/2$, and we will then say 
that there the system has $2^n$-\emph{superstable cycle}.

We are not interested in what happens when $\mu>\mu_\infty$. What is important to us, and we have established numerically, is that there 
is a sequence of period-doubling bifurcations at $\mu_n$ and, correspondingly, a sequence of superstable $2^n$-cycles at $\tilde{\mu}_n$, 
that converge to some $\mu_\infty$.

\begin{table}
\begin{minipage}[b]{0.4\linewidth}
\centering
\begin{tabular}{ l | l | l }
$j$&$\tilde{\mu}_j$&$\delta_j$\\ \hline
0 & 3.00000000$\dots$   &                 \\
1 & 3.44948974$\dots$   &  4.751        \\
2 & 3.54409035$\dots$   &  4.656        \\
3 & 3.56440726$\dots$   &  4.668        \\
4 & 3.56875941$\dots$   &  4.668        \\
5 & 3.56969160$\dots$   &  4.669        \\
6 & 3.56989125$\dots$   & 4.669 \\
7 & 3.56993401$\dots$   & 4.668  \\
8 & 3.56994317$\dots$   & 4.667  \\
9 & 3.56994514$\dots$   & 4.671  \\
10& 3.5699455573883578  & 4.673  \\
11& 3.5699456473525193  &  4.66955 \\
12& 3.5699456666186404  &  4.66966\\
13& 3.5699456707444445  &  4.66935\\
14& 3.5699456716280371  &  4.66917\\
15& 3.5699456718175778  &                 \\
\end{tabular}
\caption{Values of ${\mu}_j$ and $\delta_j$.}
\label{tab:deltas}
\end{minipage}
\hspace{0.7cm}
\begin{minipage}[b]{0.5\linewidth}
\centering
\begin{tabular}{ l | r | r | r }
$n$&$\tilde{\mu}_n$&$d_n$&$\alpha_n$ \\ \hline
1 & 3.236067977 & -0.190983 & -2.68555\\
2 & 3.498561698 & 0.0711151 &  -2.52528\\
3 & 3.554640862 & -0.028161 & -2.50880\\
4 & 3.566667379 & 0.0112250 &  -2.50400\\
5 & 3.569243531 & -0.004482 & -2.50316\\
6 & 3.569795293 & 0.0017908 &  -2.50296\\
7 & 3.569913465 & -0.000715 & -2.50295\\
8 & 3.569938774 & 0.0017908 &  -2.50293\\
9 & 3.569944194 & -0.0007155 & -2.50293\\
\end{tabular}
\vspace{2.7cm}
\caption{Values of $\tilde{\mu}_n$, $d_n$ and $\alpha_n$.}
\label{tab:alphas}
\end{minipage}
\end{table}

It is possible to compute numerically (e.g. by Newton's method) the values of the first few $\mu_n$'s to a good precision. These are
written in Table \ref{tab:deltas}, and it is not hard to see that the sequence $\mu_n\to\mu_\infty\approx3.5699$ converges, at least
approximately, geometrically. We will call the number
\be
\delta=\lim_{n\to\infty}\frac{\mu_{n}-\mu_{n-1}}{\mu_{n+1}-\mu_{n}}\approx4.669201609\,,
\ee
\emph{Feigenbaum's $\delta$}.\footnote{It is interesting how Mitchell Feigenbaum found this rate of convergence; he was studying the 
sequence $\mu_n$ on a pocket calculator, and needed to guess the next bifurcation point as well as he could in order not to waste 
computer time. In doing so, he realized that the convergence was geometric.} Notice that $\frac{\mu_{n}-\mu_{n-1}}{\mu_{n+1}-\mu_{n}}$ 
is not exactly equal to $\delta$ when $n$ is finite. Clearly the same rate dictates the convergence of $\tilde{\mu}_n$ as well.

\begin{figure}[t]
\centering
\includegraphics[width=9cm]{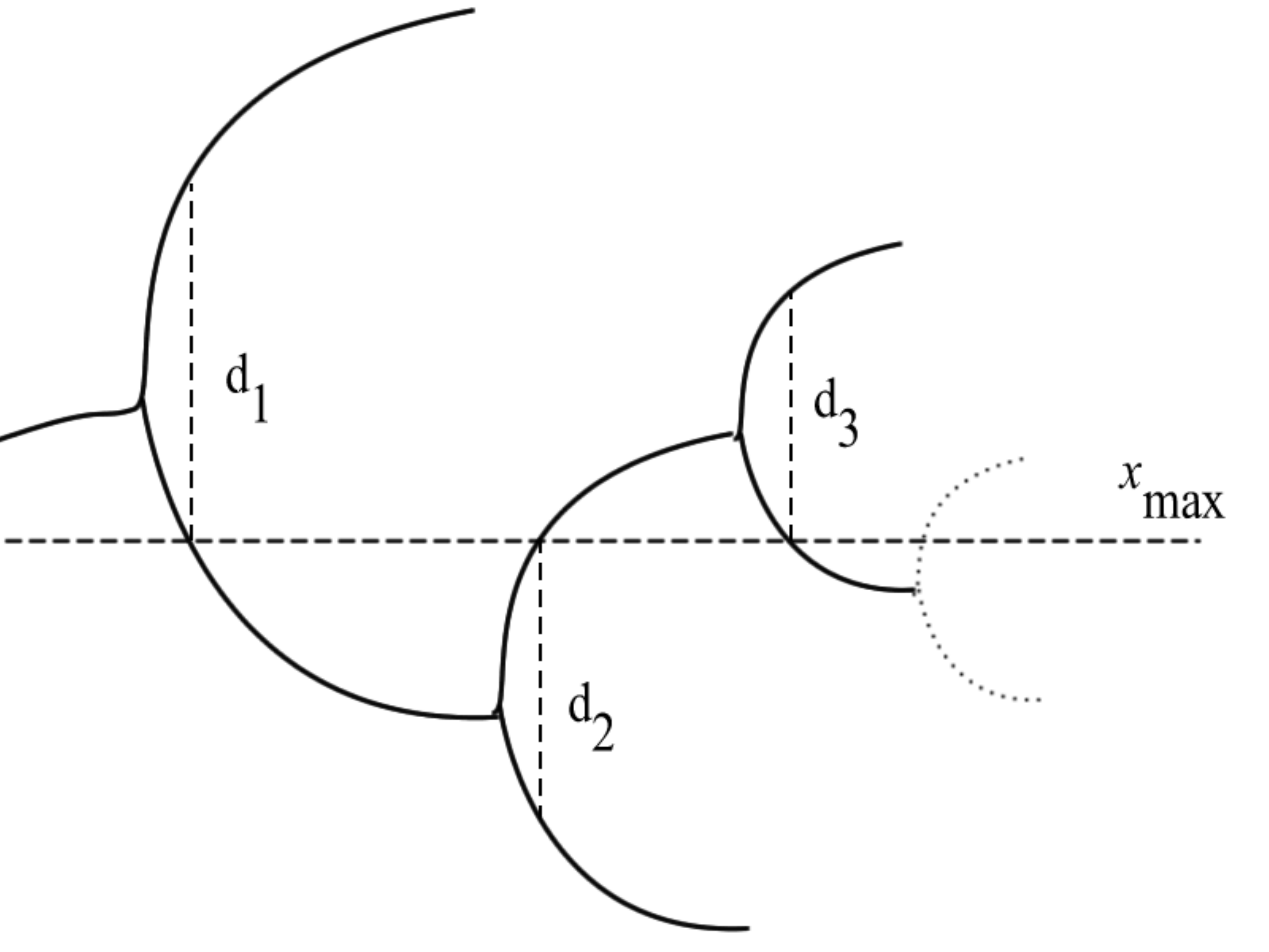}
\caption{The $d_n$ sequence used to define $\alpha$.}
\label{fig:Usequence}
\end{figure}

Looking back at Figure \ref{fig:bifurcationdiag} we see that the constant  $\delta$ dictates the horizontal scale in the sequence of 
bifurcations. Clearly enough, there is also a vertical scale: in fact, after each bifurcation, the couple of new attractive points 
that are generated spread out in a $\sf C$-shaped figure\footnote{Or $\sf U$-shaped: in fact historically this goes under the name 
of $\sf U$-sequence.} as $\mu$ increases. The size of this $\sf C$ shrinks bifurcation after bifurcation, and it makes a lot of 
sense to suspect that it does so geometrically. The ``size'' $d_n$ can be defined as the distance between two neighboring points
in a superstable cycle. This can be found by  looking at the intersection of the bifurcation sequence with the line $x=x_{max}=1/2$,
as depicted in Figure \ref{fig:Usequence}.

In Table \ref{tab:alphas} the first few values of $d_n$, as found numerically, are written. Indeed they converge geometrically,
and one can define \emph{Feigenbaum's $\alpha$} as
\be
\alpha=\lim_{n\to\infty}\frac{d_{n}}{d_{n+1}}\approx-2.502907875\,,
\ee
which is negative because we keep into account the sign of $d_n$, see Figure  \ref{fig:Usequence}.

\bigskip\noindent
\textbf{Universality}\\
So far we have established that a certain family of maps of the interval into itself exhibits a sequence of period-doubling bifurcations,
with geometric rate tending to $\delta\approx4.669$, leading to a chaotic behavior. What makes this story more interesting is that this 
behavior is \emph{common to many maps as well as real-world systems}, i.e. it is, at least to some extent, \emph{universal}.

In fact, one can check numerically that the same sequence of period-doubling bifurcations occurs for dynamical systems defined by the maps on $I=[0,1]$
\bea
\nonumber
f_\mu(x)&=&\mu\,(1-x^2)(2x-x^2)\,,\quad\quad 0\leq\mu\leq\frac{9}{16}\,,\\
f_\mu(x)&=&\mu\,\sin(\pi x)\,,\quad\quad\quad\quad\quad\quad\quad 0\leq\mu\leq1\,,
\label{eq:othermaps}
\eea
as well as many others. What is more remarkable, the bifurcations points $\mu_n$ converge geometrically to some $\mu_\infty$ with a
rate that tends to $\delta\approx4.669$. This happens also for dynamical systems of different kind, such as Mandelbrot's set, or R\"ossler's 
system of ODEs, see e.g. \cite{strogatz}.

Furthermore, and what is more important from a physicist's perspective, the period-doubling cascade towards chaos occurs also in real-world 
systems. For instance, let us consider a fluid-dynamics experiment of Rayleigh-B\'enard convection, following Libchaber and Maurer \cite{libchaberA}.

Consider a box containing a fluid. The bottom of the box is kept at temperature $T$, whereas the top is kept at $T+\Delta T$. The temperature
difference (or rather a related dimensionless quantity called Rayleigh's number) is the external parameter that the experimenter may vary. 
When the temperature difference is small, heat is conduced to the colder upper surface. However, increasing the gradient, the familiar
convective motions are generated. These consists of several counter-rotating cylinders that drive steadily the hotter fluid
upwards.\footnote{In practice, in order to obtain a stable enough convective motion, great care has to be taken in setting the experiment,
such as picking appropriate shape and dimension of the box, and of course an appropriate fluid.} 
Further increasing $\Delta T$ leads to a more complicated dynamics of the fluid:  the heat flow is not steady any more, but fluctuates, 
as it can be seen by measuring the time-evolution of the local temperature at a given point in the upper surface.

This is a discrete-time dynamical system\footnote{In fact, it makes sense to make measurements with time scales that are large with respect 
to the ones of the microscopic degrees of freedom.}: $\Delta T$ plays the role of $\mu$, and the oscillations of the local temperature the
role of $\Phi^n(x)$. What was found then is that, as one increases $\Delta T\sim \mu$, one goes from the steady temperature (fixed point) 
to a two-cycle, then to a four-cycle, and so on. Even more strikingly, the period-doubling bifurcations occur geometrically with rate
$4.4\pm0.1$ compatible with $\delta$. This has been shown to occur in a number of experiments in hydrodynamics \cite{libchaberA,libchaber83,berge1986order,PhysRevLett.47.243,smith},
electronics \cite{linsay,testa,PhysRevLett.49.94,yeh}, charged gases \cite{PhysRevLett.59.613} and chemistry \cite{simoyi}.

 It is worth pointing out that obtaining these experimental results is quite hard. On top of difficulties such as suppressing the noise and avoid 
 generating chaotic behaviour due to other kind of turbulence, a key obstacle is that  (due to the geometric progression) it is in practice possible 
 to measure only the first few bifurcations. On the other hand, $\frac{\mu_{n}-\mu_{n-1}}{\mu_{n+1}-\mu_{n}}$ will approach $\de$ only asymptotically. 
 This makes these results even more remarkable.
 
 Of course, the word ``universal'' should be taken with a pinch of salt. The maps (\ref{eq:othermaps}) are not terribly general, and indeed they 
 share some features with the logistic map:
 \begin{enumerate}
\item $f_\mu(x)$ is regular\footnote{All our examples are analytic functions, which is a very strong requirement. We will not go into the details
of \emph{how regular} we need $f_\mu(x)$ to be.}.
\item $f_\mu(x)$ is unimodal, i.e. it has one maximum $x_{max}$ and satisfies
\be
f'_\mu(x)>0\,,\quad x<x_{max}\,\quad\quad\&\quad\quad 
f'_\mu(x)<0\,,\quad x>x_{max}\,.
\ee
\item $f_\mu(x)$ has a quadratic maximum
\be
f''_\mu(x_{max})<0\,.
\ee
\end{enumerate}

\begin{figure}[t]
\begin{minipage}[t]{0.54\linewidth}
\centering
\includegraphics[width=\linewidth]{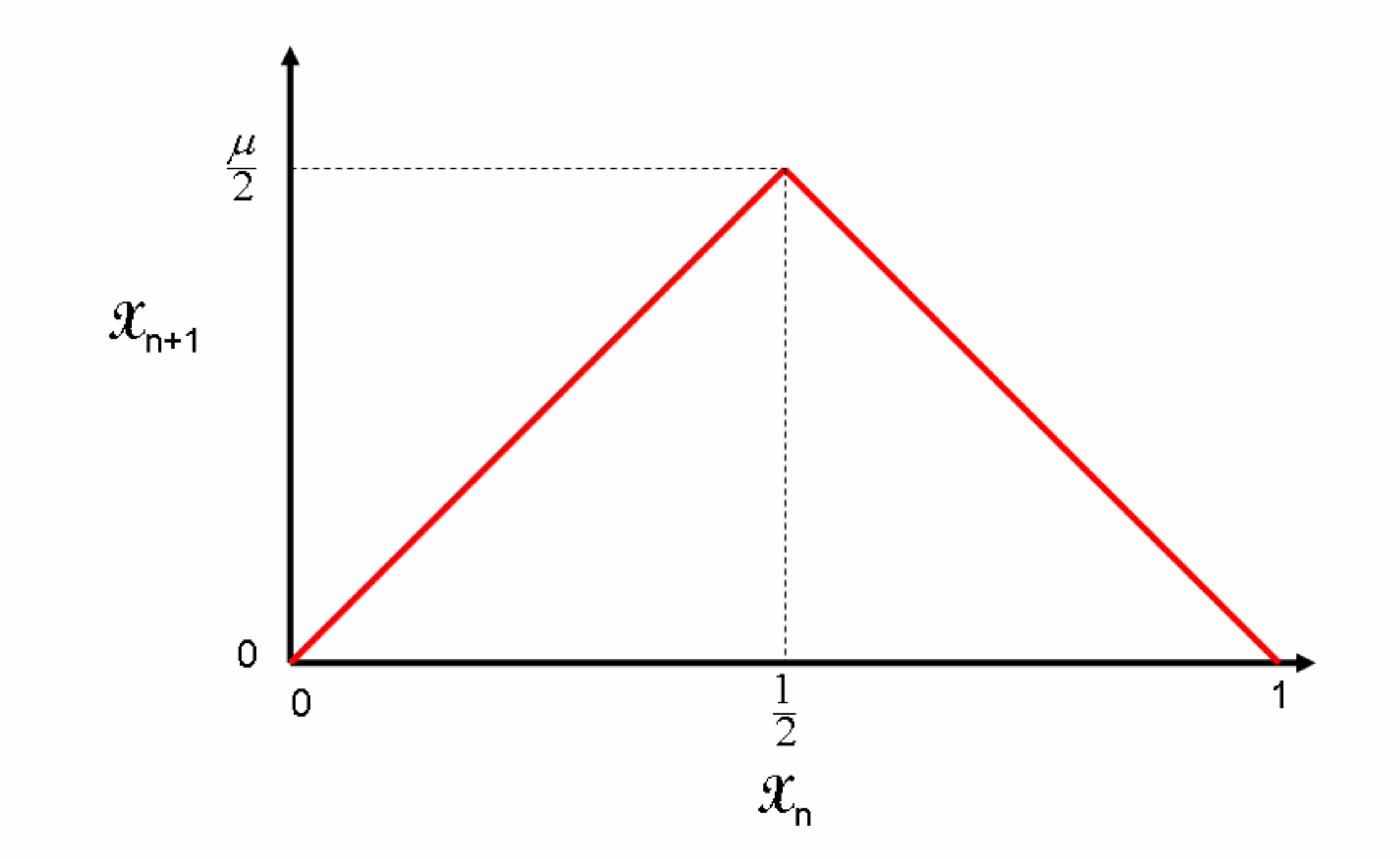}
\caption{The tent map.}
\label{fig:tent}
\end{minipage}
\begin{minipage}[t]{0.44\linewidth}
\centering
\includegraphics[width=\linewidth]{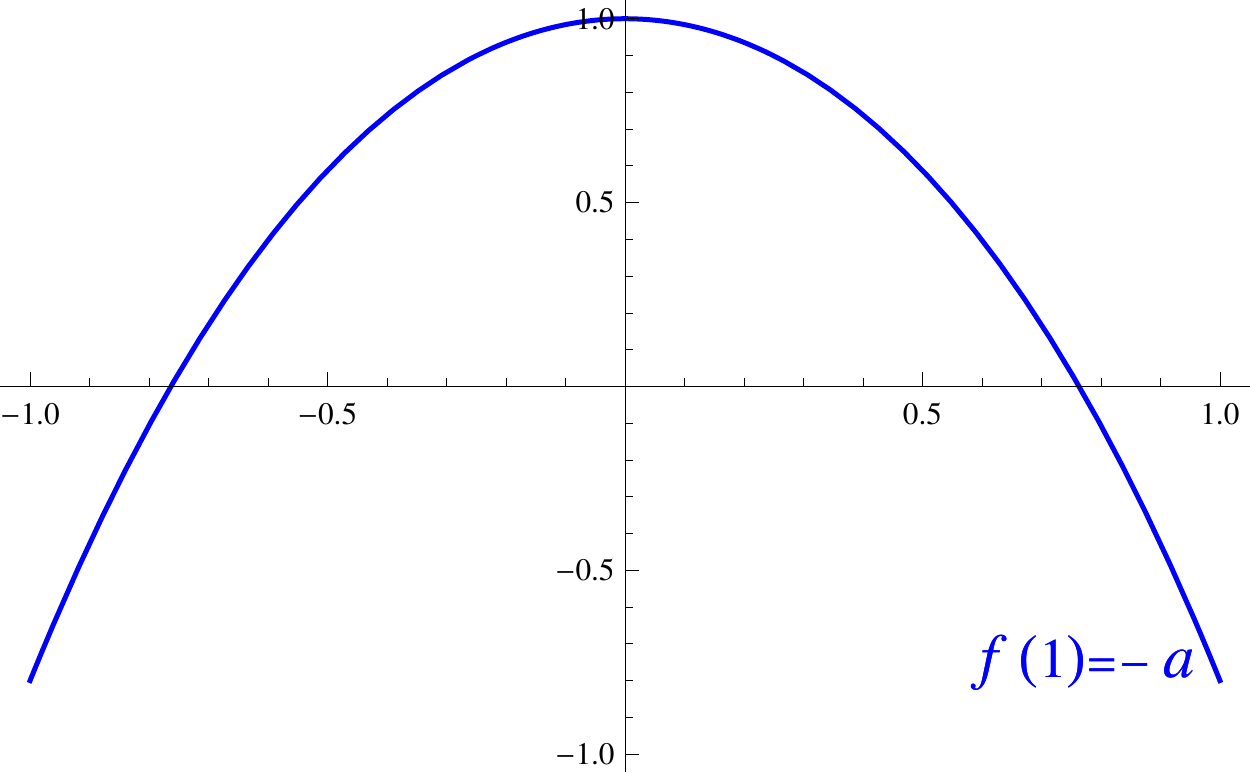}
\caption{Logistic map rescaled.}
\label{fig:rescaledmap}
\end{minipage}
\end{figure}

It turns out that these requisites are indeed necessary. It is reasonable to require some kind of regularity, since we used it in the previous
section to explain period-doubling. Let us consider the ``tent'' map of figure Figure \ref{fig:tent}, defined by
\be
f_\mu(x)=\begin{cases} \mu\,x & x \leq 1/2 \\
\mu(1-x) & x>1/2 \end{cases}\,,\quad\quad 0\leq\mu\leq2\,.
\ee
It is an unimodal map, but it is not differentiable at $x=1/2$. In fact, since it is piecewise linear, its Lyapunov exponent is simply
$\gamma=\log\mu$ so that the transition to chaos happens abruptly at $\mu=1$ with no period-doubling bifurcations.
 As for unimodality, it is also quite clear that a very general $f_\mu(x)$ with many maxima and minima may have a richer dynamics than
 the simpler examples we considered. As for the third requirement, it is hard to justify it \textsl{a priori}, and we will 
 take it as an ``experimental'' evidence. What turns out is that maps that satisfy all requisites but have a maximum of higher
 order show the same period-doubling cascade, but with \emph{different universal constants} in place of $\delta$ and $\alpha$. For 
 instance, if we were to consider a family $f_\mu$ with \emph{quartic} maximum, we would find $\al\approx-1.69$ and
 $\de\approx7.28$.\footnote{Accurate values of such constants can be found e.g. in \cite{1991MaCom..57..435B}.}

In the next section we will see how all these features can be explained in a ``renormalization group'' framework, which also yields 
quantitative predictions for $\delta$ and $\alpha$.

\subsubsection{Renormalization group for dynamical systems}
\label{sec:feigenbaum}
In this section we will discuss how the universal behavior of period-doubling bifurcations can be explained in terms of renormalization 
group (RG) techniques. In this context, both words ``renormalization'' and ``group'' make little sense (to be fair, this name is a bit misleading
in practically any context). They just indicate a rather general set of ideas of broad application, from quantum field theory to the physics of phase transitions.

\bigskip\noindent
\textbf{Heuristic}\\
We are concerned with several one-parameter families of maps $f_\mu$ of the interval $I$ into itself that exhibit similar period-doubling
cascade of bifurcations. It makes sense to consider the space $\cU$ of all maps with ``good properties'',
\be
\cU=\{f:\ I\to I\,,\quad f\ {\rm regular,\ unimodal\ and\ with\ quadratic\ tip}\}\,.
\ee
Observe that the families $\{f_\mu\}\subset\cU$ are curves in $\cU$.
Clearly $\cU$ is a subset of some space of functions, and to proceed rigorously further formalization (e.g. on the metric of this space)
would be needed; here we will be qualitative. 

Let us chose explicitly $I$. Earlier we picked $I=[0,1]$, but to better keep track of the maximum of $f_\mu(x)$ we will set $I=[-1,1]$ in 
such a way that the maximum is in $x=0$ and takes value $f_\mu(0)=1$; we further restrict to even maps to simplify the figures and the discussion. 
A typical map is shown in Figure \ref{fig:rescaledmap}.

To single out the fundamental characteristics of the universal behavior, it is easier to think in terms of superstable maps: we have that, 
for any $n=1,2,\dots$ at $\mu=\tilde{\mu}_n$ there exists a superstable map of period $2^n$, and of characteristic size $d_n$. The 
convergence of $\tilde{\mu}_n\to\mu_\infty$ has universal rate $\delta$, whereas the one of $d_n\to0$ has rate $\alpha$. 

Therefore, universality should emerge as a property of $\cU$ under the action of some ``renormalization'' operator $\Ren:\cU\to\cU$. This operator
\begin{itemize}
\item should relate maps with a $2^n$-cycle to maps with a $2^{n-1}$ cycle,
\item should relate superstable maps to superstable maps, up to a rescaling of $\alpha$.
\end{itemize}
The first property suggests that it must be $\Ren(f)\sim f\circ f$, which however is not an operator on $\cU$. In fact, it is easy to see 
that if $f(x)$ is unimodal, $f\circ f(x)$ is not. It is clear that some kind of rescaling is needed.

\begin{figure}[t!]
  \begin{center}
  \includegraphics[width=\textwidth]{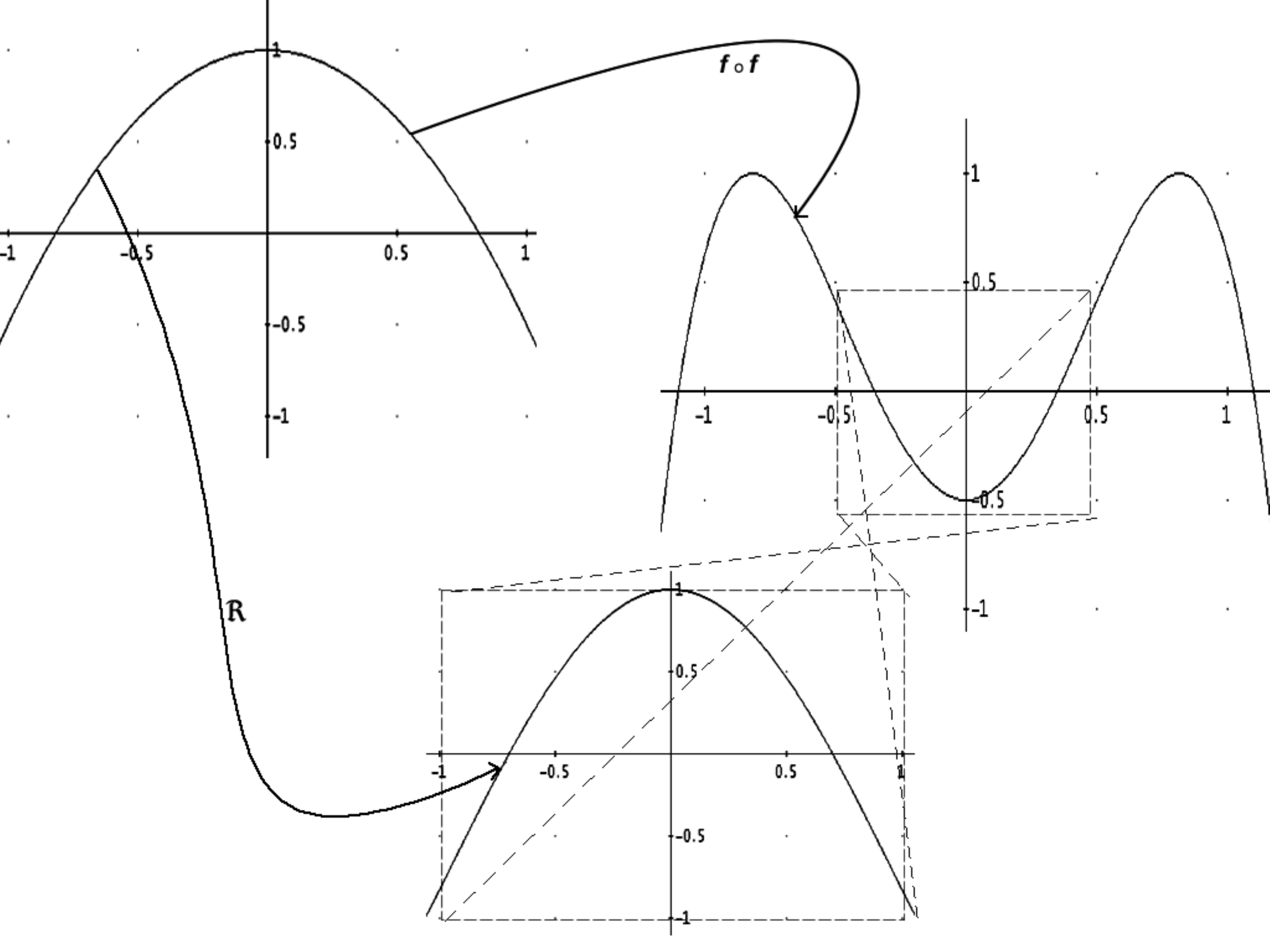}
  \end{center}
  \caption{The renormalization operator $\mathfrak{R}$.}
  \label{fig:R}
\end{figure}

A generic map $f\in\cU$ has its minimum on $I$ exactly at the boundaries of the interval. Looking at Figure \ref{fig:rescaledmap}, define 
\be
a=-f(1)\,,\quad\quad b=f(a)\,.
\ee
With a drawing, one can convince oneself that the following inclusions hold
\be
f([-1,1])=[-a,1]\,,\quad f([-a,a])=[b,1]\,,\quad f([b,1])=[-a,f(b)]\subset[-a,a]\,,
\ee
 provided that one has
 \be
 \label{eq:renormalizability}
 0<a<b<1\,,\quad\quad f(b)<a\,.
 \ee
In this case therefore one has that
\be
f\circ f:\ [-a,a] \to [-a,a]\,.
\ee
Then, there is no problem to act with $\Ren$ on $f$, if we rescale all variables in an appropriate way (see Figure \ref{fig:R}):
\be
\label{eq:ren}
\Ren (f)\,(x)=-\frac{1}{a}f\circ f(-a\,x)\,,
\ee
provided that (\ref{eq:renormalizability}) holds. In other words,  (\ref{eq:renormalizability}) are conditions of $f\in\cU$ to be in the
domain of $\Ren$ or, as sometimes it is said, for $f$ to be renormalizable. 
 
Even if in general it is not immediate to see whether also $\Ren(f)$ is in the domain of $\Ren$, for superstable maps things are easier.
In fact, superstable maps are renormalizable, and if $f$ is a period-$2^n$ superstable map, then $\Ren(f)$ is a period-$2^{n-1}$ superstable map, 
as it is easy to show.
The values $\mu=\mu_\infty$, where each family of maps gets to the onset of chaos, identify maps that have $2^\infty$-period, and are therefore 
infinitely renormalizable.

It is now the time to change point of view, and start studying a new dynamical system, where $\cM=\cU$ and $\Phi=\Ren$; here we want to understand 
the orbits of the ``points'' $f\in\cU$ (that are, in fact, functions in an infinite-dimensional space)  under action of $\Ren$. This analysis
may be more mathematically complicated and subtle than the previous ones, but we will reason by analogy with what we have seen until now.

Unfortunately, it is very hard to establish the properties of this infinite dimensional dynamical system, but a lot of progress can be made if
we accept some conjectures, originally put forward by Feigenbaum \cite{Feigenbaum:1977ys,Feigenbaum1979375,feigenbaum1980}, see also \cite{Feigenbaum198316,cvitanovic1989universality}.
\begin{enumerate}
\item There is a fixed point $\phi^*\in\cU$, i.e. $\Ren(\phi^*)=\phi^*$.
\item The fixed point is hyperbolic, meaning that the derivative of the renormalization operator $\ren=d\,\Ren$ at $\phi^*$ has no eigenvalue of modulus one.
\item Only one of its eigenvalues has modulus larger than one; we will call it $\de$.
\end{enumerate}

Under these assumptions, it is reasonable to assume that there exists an \emph{unstable manifold} $\Wu$ of dimension one which generalizes the
eigenspace relative to $\de$, and a \emph{stable manifold} $\Ws$ of codimension one, that generalizes the eigenspace of stable eigenvectors. 
Let us make an additional assumption.
\begin{enumerate}
\item[4.] Let $\Sigma_n$ be the manifolds of period-$2^n$ superstable maps in $\cU$. Then $\Wu$ intersects $\Sigma_1$ transversally at $\phi^*_0\in\Wu\cap\Sigma_1$.
\end{enumerate}

We have already remarked that $\Ren$ sends superstable maps into superstable maps. Therefore we have the inclusions
\be
\Ren^n (\Sigma_{n+1})\subset\Sigma_1\,,
\ee
and it is not hard to imagine that all the $\Sigma_n$'s will intersect transversally $\Wu$, at points $\phi^*_{n-1}$. It is also clear that the
sequence $\phi^*_n$ converges geometrically to the fixed point $\phi^*$, with rate $\de$. One can imagine that not only the points
$\phi^*_n\to \phi^*$, but also the manifolds $\Sigma_n$ accumulate toward the unstable manifold, and that their distance decreases geometrically with rate $\de$. 

\begin{figure}
\centering
\includegraphics[width=10cm]{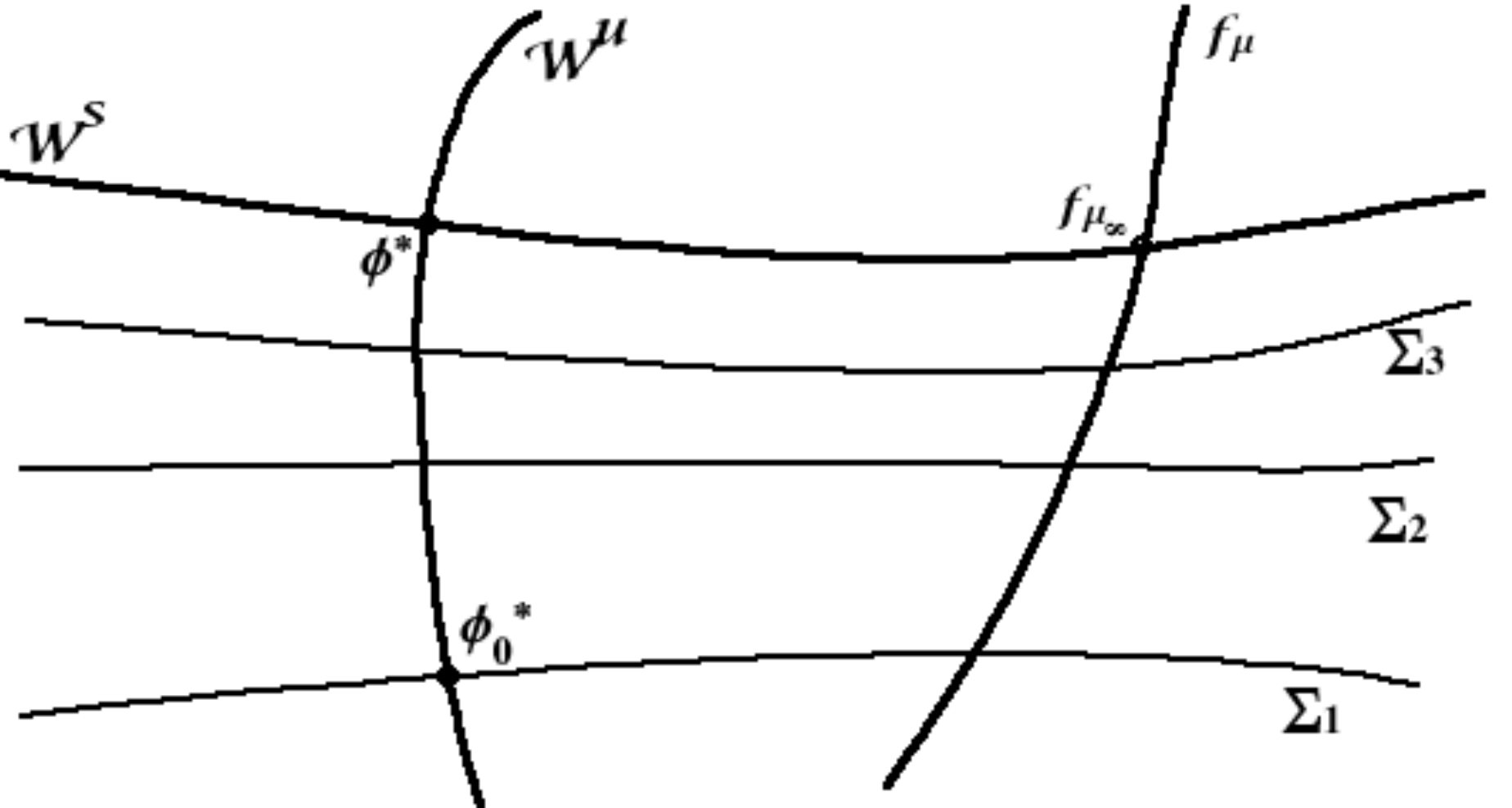}
\caption{The stable and unstable manifold for $\Ren$.}
\label{fig:asymptoticmanifold}
\end{figure}

The whole picture is summarized in Figure \ref{fig:asymptoticmanifold}. This also suggest how to explain the period-doubling cascade in a generic 
family of maps $\{f_\mu\}\subset\cU$. In fact, the sequence of bifurcations occurring at $\mu_n$, or equivalently the sequence of superstable 
maps at $\tilde{\mu}_n$ can be described in terms of the behavior of the sequence of manifolds $\Sigma_n$. We will return on this later, in order
to make the relation with $\de$ more quantitative.

As remarked, this whole discussion has been very qualitative. A rigorous treatment would bring us too far from the points of our discussions; 
the interested reader is invited to consult e.g. \cite{collet2009iterated}. Here it is worth mentioning that, once Feigenbaum's conjectures are accepted, 
it is not hard to prove that the scenario we described happens. What is much harder is to establish the existence of the hyperbolic fixed point. 
Remarkably, all this could be done rigorously \cite{CoulletTresser,Collet:1980iw, lanford,davie1996, ETS:4659272}. Finally, let us stress that had we relaxed 
the condition that our maps are even, we would still have found a single hyperbolic fixed point $\phi^*$, which turns out to be even.

\bigskip\noindent
\textbf{Predictions from renormalization group}\\
Now that we have formulated the RG ideas for maps on the interval, let us try to obtain some quantitative predictions out of them.

The first step is to find some information of the RG fixed point $\phi^*$. We immediately encounter a difficulty: we must understand the role of
$a$ which appears in the definition (\ref{eq:ren}). It has to do with the rescaling of the $x$-direction, and we know that for the fixed point
this amounts to shrinking by (negative) $\alpha$. This leads to the identification, at the fixed point
\be
a=-1/\al\,.
\ee

This still leaves $\al$ undetermined. However, it is easy to see how this is fixed by the normalization of the maximum to $\phi^*(0)=1$. Let us
consider the ansatz for a symmetric $\phi^*(x)$
\be
\phi^*(x)=1+\sum_{n=1}^N c_n\,x^{2n}+O(x^{2N+2})\,.
\ee
Plugging this into the fixed point equation
\be
\phi^*(x)=\Ren(\phi^*)(x)=\al\,\phi^*(\phi^*(x/\al))\,,
\ee
we find e.g. for $N=3$ the solution
\be
\al\approx-2.479\,,\quad
c_1\approx-1.522\,,\quad
c_2\approx0.073\,,\quad
c_3\approx0.046\,.
\ee
Going to $N=6$ yields $\al\approx-2.502897$, an estimate which turns out to be correct up to order $\sim10^{-6}$. Incidentally, 
the above procedure illustrates the importance of the order of the maximum, which is a crucial ingredient in our ansatz: a different 
choice would have lead to a different result for $\phi^*(x)$.

Let us now try to find a more quantitative relation between $\de$ and the sequence $\mu_n\to\mu_\infty$ for a family of maps $\{f_\mu\}\subset\cU$. 
It is convenient to introduce the short-hand notations
\be
F(x)\equiv f_{\mu_\infty}(x)\,,\quad\quad\quad \varphi(x)\equiv\left.\frac{\partial\,f_\mu}{\partial\,\mu}\right|_{\mu_\infty}\!.
\ee
Then a function $f_\mu$ can be written, when $\mu$ is close to $\mu_\infty$, as
\be
f_\mu(x)\approx F(x)+(\mu-\mu_\infty)\,\varphi(x)\,,
\ee
and similarly we can expand
\be
\Ren(f_\mu)(x)\approx \Ren(F)(x)+(\mu-\mu_\infty)\,\ren_F\cdot\varphi(x)\,.
\ee
Recall that $F$ lies exactly in the intersection $\Ws\cap\{f_\mu\}$. Therefore, $\Ren(F)\in\Ws$ is closer 
to $\phi^*$ than $F$, and indeed due to the geometric convergence on the stable manifold, $\Ren^n(F)\approx \phi^*$ 
after a few iterations. We can thus write
\be
\Ren^n(f_\mu)(x)\approx \phi^*(x)+(\mu-\mu_\infty)\,\ren_{\phi^*}^n\cdot\varphi(x)\,.
\ee
Let us expand $\varphi$ on a basis of eigenfunctions of $\ren_{\phi^*}$,
\be
\varphi(x)=c_\delta \,\varphi_\delta(x)+\sum_j c_j\,\varphi_j(x)\,,
\ee
where we have distinguished the eigenfunction pertaining to $\de$. Since all eigenvalues except $\de$ have modulus smaller
than one, their eigenvectors $\varphi_j$, $j\neq\de$ are sent to zero by $\ren_{\phi^*}^n$. Only the eigenvector
$\varphi_\delta$ of $\de$ plays a role, so that we can eventually write
\be
\Ren^n(f_\mu)(x)\approx \phi^*(x)+(\mu-\mu_\infty)\,c_\delta\,\delta^n\,\varphi_\delta(x)\,.
\ee

Let us now specialize the above expression to the case $\mu=\tilde{\mu}_n$, i.e. the case where the map is superstable of
period $2^n$, and evaluate it at $x=0$. On the one hand, we have
\be
\Ren^n(f_{\tilde{\mu}_n})(0)=(f_{\tilde{\mu}_n})^{2^n}(0)=0\,,
\ee
due to the presence of the $2^n$-cycle and the fact that $x=0$ is a point of the cycle. On the other hand we have
\be
 \phi^*(0)+(\tilde{\mu}_n-\mu_\infty)\,c_\delta\,\delta^n\,\varphi_\delta(0)=\frac{1}{\al}+(\tilde{\mu}_n-\mu_\infty)\,\delta^n\ \,c_\delta\,\kappa_\delta\,,
\ee
where we emphasized that $\kappa_\delta=\varphi_\delta(0)$ does not depend on $n$. Therefore, at least up to higher order
terms in $\tilde{\mu}_n-\mu_\infty$ it must be
\be
(\tilde{\mu}_n-\mu_\infty)\,\delta^n\approx -\frac{1}{\al\, c_\de\, \kappa_\de}={\rm const.}\,,
\ee
\emph{for any $\de$}, which means exactly that the rate of convergence $\tilde{\mu}_n\to\infty$ is $\delta$.

The only thing that remains to be done is to compute $\de$, using the explicit expression for the differential $\ren$
\be
\label{eq:frechetder}
\ren_\phi\cdot\varphi(x)=-\frac{1}{a}\varphi\big(\phi(-a\,x)\big)-\frac{1}{a}\varphi(-a\,x)\,\phi'\big(\phi(-a\,x)\big)\,,
\ee
and inserting the approximate result for $\phi^*(x)$ found by means of the power series expansion into the eigenvalue equation, we get
\be
\al\,\varphi_\de\big(\phi^{*}(x/\al)\big)+\al\,\varphi_\de(x/\al)\,\phi^{*}{}'\big(\phi(x/
\al)\big)=\de\,\varphi_\de(x)\,,
\ee
and can be solved approximately by using an ansatz for $\varphi_\de(x)$ too. The result, taking $N=6$ in the ans\"atze, is
\be
\delta\approx4.66914\,,
\ee
with an accuracy of order $\sim10^{-5}$ with respect to the known result \cite{1991MaCom..57..435B}.

We have considered systems that are described by iterated 
maps on the interval (that is among the simplest dynamics one can imagine) where the experimenter is able to tune one parameter
$\mu$; some class of these systems exhibit similar properties as the parameter approaches a critical value $\mu_\infty$. We have 
explained this by using the properties of the renormalization of operator $\Ren$.

What is the physical interpretation of $\Ren$? When we are looking at the dynamics, $\Ren$ operates a rescaling of the time-scale
($f\mapsto f\circ f$) together with a rescaling of the $x$-scale. In this sense it is similar to Kadanoff's coarse-graining 
transformation \cite{Kadanoff:1966wm}: acting with the renormalization operators corresponds to changing the description
of the problem, ``zooming out'' in such a way as to preserve the interesting (universal) properties of the dynamics.

In the language of statistical physics, we would say that $\mu$ plays the role of some adjustable ``knob'' (temperature,
external magnetic field, etc.), and that at $\mu_\infty$ a phase transition occurs. The divergence of the correlation 
length in our case is mimicked by the appearance of an infinite-period cycle. All the systems at the phase transitions 
are points on the stable manifold $\Ws$, and due to this they are very similar. In fact, under the action of $\Ren$, all 
these points get to the fixed point $\phi^*$, so that the properties invariant under $\Ren$ (the large scale properties, 
in a statistical system) are common to all of them. We did not investigate at all what happens to our maps at $\mu_\infty$
(to avoid the complications of chaotic systems), but it is indeed possible to single out several universal properties.

The role played by $\de$ is that of a \emph{relevant} eigenvalue, and $\varphi_\delta(x)$ is a relevant direction at the
fixed point. This means that if we perturb $\phi^*(x)$ by something proportional to $\varphi_\delta(x)$, acting with $\Ren$
will drive us away from the fixed point. Notice that in Figure \ref{fig:asymptoticmanifold} the stable manifold separates $\cU$ 
in two regions. Maps on either side of the $\Ws$ will have different behavior under $\Ren$ and therefore different properties.
We have seen that all the maps ``below'' the stable manifold ($\mu<\mu_\infty$) are periodic, with negative Lyapunov exponent, 
whereas indeed maps ``above'' $\Ws$ will yield chaotic behavior. Again this is in analogy with what happens in statistical systems, 
where the position in parameter space of a theory determines its large-scale properties.

\subsection{Series convergence and Borel summability}\label{subsec:Borel}

Before going to the core of our review we will take a moment to discuss the status of the perturbative expansion in QFT. 
This expansion is obtained by performing a Taylor expansion of the interaction, and then commute the 
sum with the Gaussian integral to obtain a series indexed by combinatorial maps:
\bea
 \langle 1 \rangle &=& \int d\mu_C  \;\;  e^{-\frac{\lambda}{4!} \int d^dx \phi^4(x)  }\crcr 
 & = & \sum_{n=0}^{\infty} \frac{1}{n!} \frac{(-\lambda)^n}{(4!)^n}  \sum_{\text{labelled combinatorial maps with n vertices}} A(M) \; .
\eea 
The problem of the perturbative series is that it is badly divergent: as there are $(4n-1)!!$ labelled combinatorial maps with $n$ vertices,
the series behaves like
\bea
 \sum_{n=0}^{\infty} \frac{1}{n!} \frac{(-\lambda)^n}{(4!)^n} (4n-1)!! =  \sum_{n=0}^{\infty} \frac{1}{n!} \frac{(-\lambda)^n}{(4!)^n} \frac{(4n)!}{2^{2n}(2n)!} 
\eea 
which has zero radius of convergence. 

The root of the problem is that we performed an expansion in $\lambda$ around $\lambda=0$. However the original integral is convergent for
$\lambda>0$ but it is divergent for $\lambda<0$. The partition function $\langle 1 \rangle $ is analytic 
in some domain in the complex plane outside the negative real axis. Hence $\lambda=0$ belongs to the boundary of the analyticity
domain of $\langle 1 \rangle $. A Taylor expansion around a point belonging to the boundary of analyticity domain of some function is 
not absolutely convergent. 

A legitimate question is therefore how much of the information we extract from the perturbative expansion is of any relevance to  
QFT? Does a divergent Taylor series encode any relevant information about the original function? The answer for a smooth 
function of a real variable in general is no: an asymptotic
series does not encode any relevant information (the typical example is the function $f(\lambda) =0$ for $\lambda \le 0$ 
and $f(\lambda) = e^{-\frac{1}{\lambda}}$ for $\lambda >0$ whose
asymptotic series at $\lambda=0$ is $0$).

Only under very special circumstances does a divergent Taylor series encode some information about the function it is coming from: 
in fact under certain assumptions such a series can uniquely fix the function. Fortunately for us, this is the case in QFT: in 
many cases, while divergent, the perturbative series is {\it Borel summable}. 

\begin{theorem}[Nevanlinna-Sokal, \cite{Sokal:1980ey}]
 A function $f(\lambda,N)$ with $\lambda\in \mathbb{C}$ and $N \in \mathbb{R}_+$ is said to be Borel summable in $\lambda$ uniformly in $N$ if
 \begin{itemize}
  \item $f(\lambda,N)$ is analytic in a disk $\Re{(\lambda^{-1})}>R^{-1}$ with $R\in \mathbb{R}_+$ independent of $N$.
  \item $f(\lambda,N)$ admits a Taylor expansion at the origin
        \bea
        f(\lambda,N) = \sum_{ k =0}^{r-1} f_{N,k} \lambda^k + R_{N,r}(\lambda) \; , \qquad |R_{N,r}(\lambda)| \le K \sigma^r r! |\lambda|^r \;,
        \eea
       for some constants $K$ and $\sigma$ independent of $N$.
 \end{itemize}

If $f(\lambda,N)$ is Borel summable in $\lambda$ uniformly in $N$ then 
\[
 B(t,N) =\sum_{k=0}^{\infty} \frac{1}{k!} f_{N,k} t^k 
\]
is an analytic function for $|t|<\sigma^{-1}$ which admits an analytic continuation in the strip
 $\{ z | \; | \Im z | < \sigma^{-1} \} $ such that $|B(t,N)| \le B e^{t/R}$ for some constant $B$ independent of $N$ and $f(\lambda,N)$ is
represented by the absolutely convergent integral
 \bea
   f(\lambda,N ) = \frac{1}{\lambda} \int_0^{\infty} dt \; B(t,N) e^{-\frac{t}{\lambda}} \; .
 \eea
\end{theorem}

\begin{figure}[htb]
   \begin{center}
 \includegraphics[width=4cm]{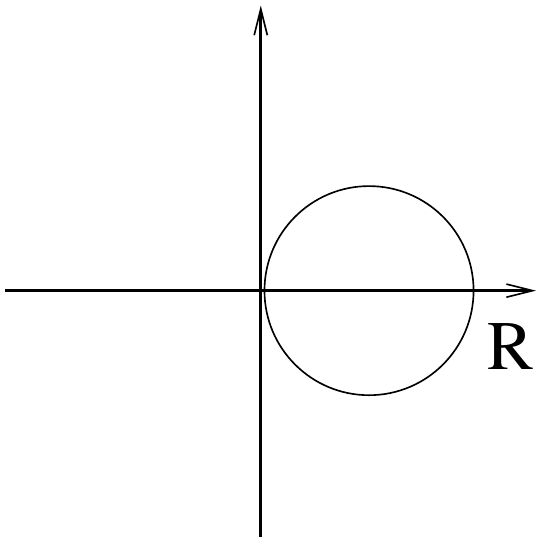}
 \caption{A Borel disk.}
  \label{fig:borel}
   \end{center}
\end{figure}

That is the Taylor expansion of $f(\lambda,N)$ at the origin is Borel summable, and $f(\lambda,N)$ is its Borel sum.
The set $\{\lambda | \Re{(\lambda^{-1})}>R^{-1}, R\in \mathbb{R}_+ \} $ is a disk (called a Borel disk) in the complex
plane with center at $\frac{R}{2}$ and of radius $\frac{R}{2}$(see Figure \ref{fig:borel}) 
as, denoting $\lambda = \frac{R}{2}+ae^{\imath \gamma}$,
\bea
\Re{(\lambda^{-1})}>R^{-1} \Leftrightarrow \frac{R^2}{4} > a \; . 
\eea

Borel summability provides a {\it uniqueness} criterion: if a series 
is the Taylor expansion of a Borel summable function $f(\lambda,N)$ at $\lambda=0$, then $f(\lambda,N)$ 
is the {\it unique} Borel summable function whose Taylor series is the original series.
This is the case of the $\phi^4_d$ model and of QFT in general. 

In fact Borel summability allows to recover $e^{\frac{1}{\lambda}}$ effects. Although such effects cannot be captured by 
the perturbative series, they are recovered by analytic continuation. In fact one expects a $e^{\frac{1}{\lambda}}$ behaviour
for $\lambda\in \mathbb{R}, \lambda<0$: this is the typical contribution of ``instantons'' (that is non trivial solutions of 
the classical equations of motion) which exist for $\lambda<0$. The instanton contributions are recovered by first Borel summing
the perturbative series (which yields a convergent expression for $\Re\lambda>0$) and then analytically continuing to the $\Re\lambda<0$ 
half complex plane. The negative real axis is a cut and the partition function $\langle 1 \rangle$ (or more precisely its logarithm, the free energy)
can be continued on a Riemann surface. The discontinuity of this analytic continuation at the cut captures the instanton contributions.

\newpage

\section{Renormalization essentials}
\label{sec:renorm-essetials}

We refer loosely to \cite{Rivasseau:2011ri} and references therein for this section.
Quantum transition probabilities are represented by a ``sum over histories'' of particles, i.e. a functional integral. 
Expanding this functional integral in term of the number of collisions or interactions of the particles 
leads to {\emph{perturbative}} quantum field theory, indexed by combinatorial maps. In these maps, 
the particle collisions or interactions are represented by vertices. The propagation of particles  is 
represented by the edges joining these vertices. Any correlation is computed by the weighted sum
over infinitely many maps. The maps to be summed and the weights (a.k.a. Feynman amplitudes) 
are determined from the classical action of the theory by the  Feynman rules.

The simplest functional integrals are the one describing free theories, that amount to considering Gaussian measures.

\subsection{Gaussian measures}

A finite dimensional centered normalized Gaussian measure $d\mu_C$
is defined through its covariance. 
Consider a finite dimensional space ${\mathbb R}^N$. The field $\phi$ is a function $\phi:\{1,2,\dots N\} \to \mathbb{R}$.
Let $C$ be a symmetric, positive definite, $N$ by $N$ matrix .
The naive definition of the centered normalized Gaussian measure of covariance $C$ is
\bea  \label{gauss1}
d\mu_C = \frac{1}{(2\pi)^{N/2} \sqrt{\det C} }    \;\;  e^{- \frac{1}{2} \sum_{i,j} \phi_i ( C^{-1} )_{ij} \phi_j } d^N \phi ,  
\eea
so that $\int  d\mu_C = 1$. Taking eq. \eqref{gauss1} as the definition of Gaussian measures has two drawbacks:
\begin{itemize}
 \item eq. \eqref{gauss1} is well defined only if $C^{-1}$ exists. 
 \item more importantly, if one tries to generalize this definition for $\phi$ a genuine field $\phi: \mathbb{R}^d\to \mathbb{R}$, one runs into troubles: 
the functional measure involves a factor $\prod_{x\in \mathbb{R}^d} d\phi(x)$, which is an ill defined infinite product of 
Lebesgue measures.
\end{itemize}

This is clearly not a solid starting point for the study of QFT. 
The correct starting point is the following. Any function can be approximated by polynomials, hence probability measures are characterized by 
the expectations of polynomials in the random variables (the {\it moments} of the measure). For Gaussian measures, this can be rendered 
rigorous by using Minlos' Theorem \cite{Glimm:1987ng} and the general theory of Gaussian processes. For any operator 
$C : L^2(\mathbb{R}^d) \to L^2(\mathbb{R}^d)$, with kernel $C(x,y)$, we define 
\begin{definition} \label{def:gauss} 
The centered, normalized Gaussian measure of covariance $C$, $\mu_C$ is defined by its moments 
\bea
\int d\mu_C(\phi) \; \prod_{i=1}^{n}\phi(x_i)=
\begin{cases}
1   &\text{if }n=0 \\
0   &\text{if }n \text{ is odd} \\
\sum_{G}\prod_{\ell\in G } C(x_i,x_j) \;  & \text{if }n(\neq 0) \text{ is even} 
\end{cases} \; ,
\eea
where $G$ denotes all the pairings of $n$ elements into pairs $\ell=(i,j)$ (that is distinct partitions of the set $1\dots n$ 
into subsets of two elements). 
\end{definition}
 
The moments of a Gaussian measure admit a simple graphical representation in Feynman diagrams. 
We represent the insertions $\phi(x_1), \dots \phi(x_n)$
as vertices, and the pairs $\ell$ as edges connecting the vertices. A pairing $G$ is thus a Feynman graph (a combinatorial map).  
 
The Euclidean quantum field theory of a free scalar field amounts to studying the (normalized) functional 
measure 
\bea
 d\mu_{C}(\phi) \; ,\;\; C(x,y)  = \Bigl( \frac{1}{-\Delta + m^2} \Bigr) (x,y) = \int_0^{\infty} \frac{d\alpha}{(4\pi\alpha)^{ d/2 }} 
  \;\; e^{-\frac{|x-y|^2}{4\alpha} - \alpha m^2 }\; .
\eea 
The QFT predictions are the moments of this measure
\begin{equation}
\label{eq:correlfunct}
\Big{\langle} \phi(x_1)\cdots\phi(x_k) \Big{\rangle}=\int d\mu_C    \;   \phi(x_1)\cdots\phi(x_k) \;,
\end{equation}
which follow from the definition \ref{def:gauss} and make perfect sense for $\phi(x)$ a function on $\mathbb{R}^d$. 
In particular the apparent problem that $\prod_{x\in \mathbb{R}^d} d\phi(x)$ is an ill defined infinite product of 
Lebesgue measures simply does not exist.

The moments can be obtained as the functional derivatives of the generating functional $Z[J]$
\bea \label{logcorr}
&& Z[J]\equiv \int d\mu_C\; e^{\int d^dx\,J(x)\phi(x)} \; , \crcr
&& \Big{\langle} \phi(x_1)\cdots\phi(x_k) \Bigr{\rangle}
  = \frac{\delta}{\delta J(x_k)}\cdots\frac{\delta}{\delta J(x_1)}  Z[J] \Big{|}_{J=0} \; .
\eea

The generating functional can be evaluated exactly in this case, by performing the Gaussian integral
\bea\label{explicitfreeZ}
&&Z(J) = \sum_{n\ge 0} \frac{1}{n!} \int d^dx_1 \dots d^dx_n \; J(x_1) \dots J(x_n) \int d\mu_C \; \phi(x_1) \dots \phi(x_n) \crcr
&& = \sum_{n\ge 0} \frac{1}{n!} \int d^dx_1 \dots d^dx_n \; J(x_1) \dots J(x_n) \sum_G \prod_{\ell\in G} C(x_{a(\ell)}, x_{b(\ell)} ) \crcr
&& = \sum_{p\ge 0} \frac{1}{(2p)!} (2p-1)!! \Bigl( \int d^dx d^dy  J(x) C(x,y) J(y) \Bigr) \crcr
&& =  e^{\frac{1}{2} \int d^dxd^dy J(x) C(x,y) J(y) } \; ,
\eea
where we used the fact that there are $(2p-1)!! = \frac{(2p)!}{2^pp!}$ pairings of $2p$ labels.
The two-point Schwinger function is just the free Euclidean propagator\footnote{Of course, if we wanted to consider the Minkowskian 
theory, we would have to worry about a prescription to obtain the correct Green's function by analytic continuation.}. 
Similarly, we can find the explicit expression for all correlations (\ref{eq:correlfunct}), which can be expressed by the familiar 
Feynman diagrams. For a Gaussian measure, it is immediate to see that the connected moments and their generating functional are:
\bea
\label{eq:freeconnected}
&& e^{ W[J]}\equiv Z[J] \; ,\crcr
&& \Big{\langle} \phi(x_1)\cdots\phi(x_k) \Bigr{\rangle}_c
  = \frac{\delta}{\delta J(x_k)}\cdots\frac{\delta}{\delta J(x_1)}   W(J)   \Big{|}_{J=0} \; .
\eea 
Here the connected moments are particularly simple: the only non-vanishing one is the two-point connected moment
which equals the covariance itself. For general (non Gaussian) measures, the covariance is defined as the connected two 
point function, and as we will prove in the next section the generator of connected moments is also given by~\eqref{eq:freeconnected}. 

For an arbitrary function $F(\phi)$,
\bea
&&\int d\mu_C(\phi) F(\phi)=\int d\mu_C(\phi) 
\Bigg[ F \bigl( \frac{\delta}{\delta J} \bigr) e^{ \int_x J(x) \phi(x)} \Bigg]  \Bigg{|}_{J=0} \crcr
&& = \Bigg[ F \bigl( \frac{\delta}{\delta J} \bigr)  \int d\mu_C(\phi)   e^{  \int_x J(x) \phi(x)} \Bigg]  \Bigg{|}_{J=0}  
= \Bigg[ F \bigl ( \frac{\delta}{\delta J}\bigr) e^{  \frac{1}{2}\int_{xy} J(x) C(x,y) J(y)   } \Bigg] \Bigg{|}_{J=0} \crcr
&& = \Bigg[  e^{  \frac{1}{2}\int_{xy} \frac{\delta}{ \delta \phi(x)} C(x,y) \frac{\delta}{\delta \phi(y) }  }  F(\phi) \Bigg] \Bigg{|}_{\phi=0} 
\; .
\eea 

This expression of the normalized Gaussian measure is by far the most useful for explicit computations. As a direct application 
we have the following lemma.
\begin{lemma}[Split of a normalized Gaussian Measure]
Consider a normalized Gaussian measure $\mu_C(\phi)$. For any decomposition of the covariance $C=C_1+C_2$ there exists an associated 
decomposition of the field $\phi(x)=\phi_1(x)+\phi_2(x)$, with the fields $\phi_1$ and $\phi_2$ independent distributed 
on Gaussians with covariances $C_1$ and $C_2$ such that
\bea
d\mu_{C}(\phi)=d\mu_{C_1}(\phi_1)d\mu_{C_2}(\phi_2) \; .
\eea
\end{lemma}

\prf We have 
\bea
&&  \Bigg[  e^{  \frac{1}{2}\int_{xy} \frac{\delta}{ \delta \phi_1(x)} C_1(x,y) \frac{\delta}{\delta \phi_1(y) } +
  \frac{1}{2}\int_{xy} \frac{\delta}{ \delta \phi_2(x)} C_2(x,y) \frac{\delta}{\delta \phi_2(y) }}  F(\phi_1 + \phi_2) \Bigg] \Bigg{|}_{\phi_1=\phi_2=0} \crcr
&& = \Bigg{[}  e^{  \frac{1}{2}\int_{xy} \frac{\delta}{ \delta \phi_1(x)} C_1(x,y) \frac{\delta}{\delta \phi_1(y) } } \Bigg[ 
 e^{ \frac{1}{2}\int_{xy} \frac{\delta}{ \delta \phi (x)} C_2(x,y) \frac{\delta}{\delta \phi (y) }}  F(\phi) \Bigg] \Bigg{|}_{\phi= \phi_1} \Bigg{]}_{\phi_1=0} 
 \crcr
&& = \Bigg{[}  e^{  \frac{1}{2}\int_{xy} \frac{\delta}{ \delta \phi_1(x)} C_1(x,y) \frac{\delta}{\delta \phi_1(y) } } \Bigg[ 
 e^{ \frac{1}{2}\int_{xy} \frac{\delta}{ \delta \phi_1(x)} C_2(x,y) \frac{\delta}{\delta \phi_1 (y) }}  F(\phi_1) \Bigg] \Bigg{]}_{\phi_1=0} \crcr
&& = \Bigg{[}  e^{  \frac{1}{2}\int_{xy} \frac{\delta}{ \delta \phi_1(x)} [ C_1(x,y) + C_2(x,y) ] \frac{\delta}{\delta \phi_1(y) } }   F(\phi_1)  \Bigg{]}_{\phi_1=0} 
\; .
\eea  

\qed

 Another useful lemma is the following
\begin{lemma} The measure 
 \bea
  d\mu_C(\phi) \;   \sqrt{ \det (1- C^{-1}M ) }  \; e^{\frac{1}{2}\int_{x,y}\phi(x) M(x,y) \phi(y) } \; ,
 \eea 
 is also a normalized Gaussian measure and its covariance is
 \bea
  \bar C(x,y) = \Bigg(C\frac{1}{1-MC} \Bigg)(x,y) \; .
 \eea 
\end{lemma}

\prf In self evident shorthand notations, the expectation of an arbitrary function of the field is
\bea
    \sqrt{ \det (1-C^{-1}M ) } \;  \Big[ e^{\frac{1}{2} \frac{\delta}{ \delta \phi } C \frac{\delta}{ \delta \phi} }
     \; e^{ \frac{1}{2} \phi M \phi} F(\phi) \Bigr]_{\phi=0} \; ,
\eea 
which rewrites 
\bea
 \Bigl[ F \bigl(\frac{\delta}{\delta \phi} \bigr) 
  \Bigl( e^{   \frac{1}{2}  \mathrm{Tr}  [\ln(1 - C^{-1}M)]}  \; \; 
  e^{  \frac{1}{2} \frac{\delta}{ \delta \phi } M \frac{\delta}{ \delta \phi }   } e^{\frac{1}{2} \phi C\phi} \Bigr) \Bigr]_{\phi=0} \; .
\eea 
To evaluate
\bea
  e^{\frac{1}{2}\frac{\delta}{ \delta \phi } M \frac{\delta}{ \delta \phi }   } e^{\frac{1}{2} \phi C\phi}  \; ,
\eea 
we note that this expression is a sum over graphs with labelled vertices. The graphs have two valent vertices (which bring a weight $\frac{C}{2}$), 
and edges (with weight M). Its logarithm is then a sum over connected graphs. The connected graphs can be 
\begin{itemize}
 \item Cycles of length $p$. The combinatorial factor of a cycle of length $p$ is $\frac{1}{ p!}\frac{1}{2^p} (p-1)! 2^{p-1} 
     = \frac{1}{2 p}$ and the amplitude is $ \mathrm{Tr}[(MC)^p]$.
 \item Chains of length $p$. The combinatorial factor of a chain of length $p$ is $\frac{1}{ p!}\frac{1}{2^p}  p! 2^{p-1}  = \frac{1}{2}$,
 and its amplitude $ \phi C(MC)^p \phi $.
\end{itemize}
Thus 
\bea
  && e^{\frac{1}{2}\frac{\delta}{ \delta \phi } M \frac{\delta}{ \delta \phi }   } e^{\frac{1}{2} \phi C\phi}  
  = e^{\sum_{p\ge 1} \frac{1}{2p} \mathrm{Tr}[(MC)^p] + \sum_{p\ge 0} \frac{1}{2}\phi C(MC)^p \phi } \crcr
  && = e^{-  \frac{1}{2}  \mathrm{Tr}  [\ln(1 - C^{-1}M)] + \frac{1}{2} \phi C\frac{1}{1-MC} \phi  } \; ,
\eea
and the lemma follows

\qed

\subsection{Generating functionals}
Let us review two generating functionals that are most useful in the study of the measures associated to QFTs.

\subsubsection{The generating functional of connected graphs}

We have already encountered this functional in the case of free theories. Here we prove that even for an interacting theory~\eqref{eq:freeconnected} expresses the generator of connected moments, i.e. of moments given by connected Feynman graphs. Let ${ Z }$ be the formal power series which writes as a sum over graphs $G$ with $n(G)$ labelled vertices 
and with ``amplitudes'' $A(G)$, depending on the graph $G$ but not on the labels:
\bea
{ Z }= \sum_{G} \frac{1}{n(G)!} A(G) \; .
\eea 
We want to see that if the amplitude of $G$ factors over its connected components, then 
\bea
 W \equiv \ln {Z} = \sum_{G_c} \frac{1}{n_c(G_c)!} A(G_c) \;, 
\eea  
where $G_c$ runs over the {\it connected} graphs with labelled vertices.

To see this, we list the connected graphs $G_1$, $G_2$, and so on. 
We denote $n_1$, $n_2$ and so on the number of vertices of $G_1$, $G_2$, etc.
Any graph $G$ has a unique decomposition $G = G_1^{q_1} \cup G_2^{q_2} \dots \cup G_k^{q_k}\dots $.
There are $\frac{n!}{  \prod_i q_i!  (n_i!)^{q_i} }$ 
distinct ways to distribute the $n$ labels of the vertices of $G$ into $q_1$ boxes of size $n_1$, $q_2$ boxes of size $n_2$, op to $q_k$ boxes of size $n_k$,
corresponding to the vertices of $G_1$, $G_2$, and so on. We therefore have 
\begin{eqnarray*}
&& Z = \sum_G \frac{1}{n(G)!}  A(G)  = \sum_{ G_1^{q_1} \cup G_2^{q_2} \dots \cup G_k^{q_k} \dots } \frac{1}{n(G)!} \prod_{c}  \Bigl( A(G_c) \Bigr)^{q_c } \crcr
&& = \sum_{q_1\ge 0,\dots q_k \ge 0\dots } 
  \frac{1}{(\sum_{c\ge 1} n_cq_c) !} \prod_{c} \Bigl( A(G_c) \Bigr)^{q_c }   \;\; \frac{(\sum_{c\ge 1} n_cq_c )!}{  \prod_i q_i!  (n_i!)^{q_i} } \crcr
&& = \sum_{q_1\ge 0,\dots q_k \ge 0\dots } \frac{1}{ \prod_c q_c! }  \prod_c \Bigl(\frac{1}{n_c!} A(G_c) \Bigr)^{q_c} \crcr
&& = \sum_{q \ge 0} \frac{1}{q!} \Bigl( \sum_c \frac{1}{n_c!} A(G_c)  \Bigr)^q = e^{ \sum_c \frac{1}{n_c!} A(G_c)    } \; .
\end{eqnarray*}

\subsubsection{The quantum effective action}

Another generating functional, which will be useful later on is the Legendre transform of $W[J]$.
We define the ``effective field'' $\varphi(x)$ as
\begin{equation}
 \varphi(x) =  \frac{\delta W[J]}{\delta J(x)} \; .
\end{equation}
The above equation can be solved for $J$ as a function of $\varphi$, $J_\varphi$.
The Legendre transform of $W[J]$ is then 
\bea\label{legendretransform}
\Gamma[\varphi]&=&\inf_{J} \left( - W[J] + \int d^dx J(x)\varphi(x)\right) \crcr
    &=&
  - W[J_\varphi] + \int d^d x   J_\varphi   (x) \varphi(x)  \; .
\eea
Observe that in the free case, $J_{\varphi}$ can be found explicitly
\begin{equation}\label{effectivefreefield}
 \varphi(x) =  \frac{\delta W[J]}{\delta J(x)}  = \int d^dy \;  C(x,y)J(y) \Rightarrow J_\varphi(x) := (-\Delta+m^2)\varphi(x)    \; .
\end{equation}

The derivative of $\Gamma$ is 
\bea\label{JphiGamma}
&& \frac{\delta\Gamma[\varphi]}{\delta \varphi(x) }= 
- \int d^d y  \; \frac{\delta W}{\delta J(y)}\Big{|}_{ J_\varphi}  \frac{\delta J_\varphi (y) }{\delta \varphi (x) } 
  \crcr
  && \qquad \qquad +\int d^d y \frac{\delta J_\varphi  (y)}{\delta \varphi (x) } \varphi(y)
   + J_\varphi(x)= J_\varphi(x)\,.
\eea

This is a constraint for the effective field: in absence of currents $\varphi(x)$
must solve a variational equation similar to the one for $S[\phi]$ in the classical theory. 
For this reason,  $\Gamma[\varphi]$ is called the \emph{quantum effective action}. 

We have two 
complementary interpretations for $\Gamma$: on the one hand, in a diagrammatic approach, it can be proven 
that $\Gamma$ can be obtained from the one particle irreducible (1PI) Feynman diagrams, 
\footnote{This was actually the way the effective action was originally defined \cite{Goldstone:1962es},
whereas the functional definition (\ref{legendretransform}) was given in \cite{DeWitt:1964oba,JonaLasinio:1964cw,weinberg1995quantum,weinberg1996quantum}; see also \cite{Jackiw:1974cv} for a 
derivation of the $n$-loop expansion of $\Gamma$.}.

We can also think of $\Gamma$ as given by an infinite series in $\varphi^n$, whose coefficients depend on the loop 
integrals\footnote{Actually, on the renormalized loop integrals, as it will be clearer later.}. Such an expansion in effective 
vertices will be\lq\lq semi-local\rq\rq, since we are not just summing tree diagrams:
\begin{equation}\label{1PIexpansion}
\Gamma[\varphi]=\sum_{n}\frac{1}{n!}\int d^4x_1\cdots d^4x_N\,\Gamma^{(n)}[x_1,\dots,x_N]\varphi(x_1)\cdots\varphi(x_N).
\end{equation}
This gives a second interpretation: the full quantum theory generated by the action $S$ is equivalent to the classical (tree-level) 
theory for an action $S_{new}=\Gamma$ built out of the effective vertices. As a result we can write
\begin{equation}
\Gamma[\varphi]=S[\varphi]+\mathrm{quantum\ corrections}.
\end{equation}
For more on this, see \S 16 in \cite{weinberg1995quantum,weinberg1996quantum} and \S 8 in \cite{toms}.

If we go back to the free theory, we can explicitly substitute (\ref{effectivefreefield}) in the definition of $W[J]$ to find that
\begin{equation}
\Gamma[\varphi]=\int d^4x \frac{1}{2}\varphi(x)\left(-\Delta+m^2\right)\varphi(x)\,.
\end{equation}
This shows that a free theory, even after taking into account  quantum corrections, just 
describes the propagation of noninteracting particles, as was already clear from (\ref{explicitfreeZ}).
Furthermore, we see that
\begin{equation}\label{GammaWrel}
\frac{\delta^2\Gamma[\varphi]}{\delta\varphi(x)\,\delta\varphi(y)} = \frac{\delta J_{\varphi}(y)}{\delta \varphi(x)}
   = \left[\frac{\delta^2 W[J]}{\delta J(x)\,\delta J(y)}\right]^{-1}\Bigl{|}_{J = J_\varphi} \; .
\end{equation}
One also obtains from  (\ref{legendretransform}) that\footnote{For a proof of this last
 expression, see \cite{weinberg1995quantum,weinberg1996quantum} \S 18, or \cite{salmhofer} \S 2.5 where a different strategy is used: 
 the above formula is taken as definition, and the familiar properties of $\Gamma$ are derived from it.}
\bea\label{gammafull}
\exp\left(-\Gamma[\varphi]\right)&=\frac{1}{\mathcal{N}}\int \mathcal{D}\rho\,
   \exp\left\{-S[\rho+\varphi]+\int d^dx\frac{\delta\Gamma[\varphi]}{\delta\varphi(x)}\,\rho(x)\right\}
\eea 
which can be taken as an alternative definition of $\Gamma[\varphi]$.

\subsection{Warm-up examples}
Before tackling interacting QFTs on~$\mathbb{R}^d$, let us consider  two simpler examples that illustrate how to work with generating functionals and their expansions.

\subsubsection{The single point universe}
 
Let us start with a point-like universe, hence "field theory in zero dimension". The field is then a random variable 
and we consider the one dimensional Gaussian measure of covariance $1$, which can be written as 
\bee \label{xquat1} d\mu (\phi)= e^{-\phi^2/2 } \frac{1}{\sqrt{2 \pi}} d\phi  .
\ee 
The moments of this measure are 
\be\label{xquat2} 
 \bigl{\langle} \phi^{2p+1}\big{\rangle} = 0 \; , \qquad \bigl{\langle} \phi^{2p}\big{\rangle} = (2p-1)!! \; . 
\ee
 
The Feynman graph representation pictures $(2p-1)!!$ as a sum over graphs. Each variable
$\phi$ is pictured as a vertex with a "half-edge" or "field" hooked to it.
The result of Gaussian integration is then expressed as a sum over all possible pairings
between these fields. Each such pairing  is represented as an edge 
between the corresponding vertices. The set $G$ of all vertices and edges
is called the Feynman graph associated to the pairing. It is still quite trivial, as each connected component
is made of a single edge and its two ends.

\begin{figure}
\centerline{\includegraphics[width=6cm]{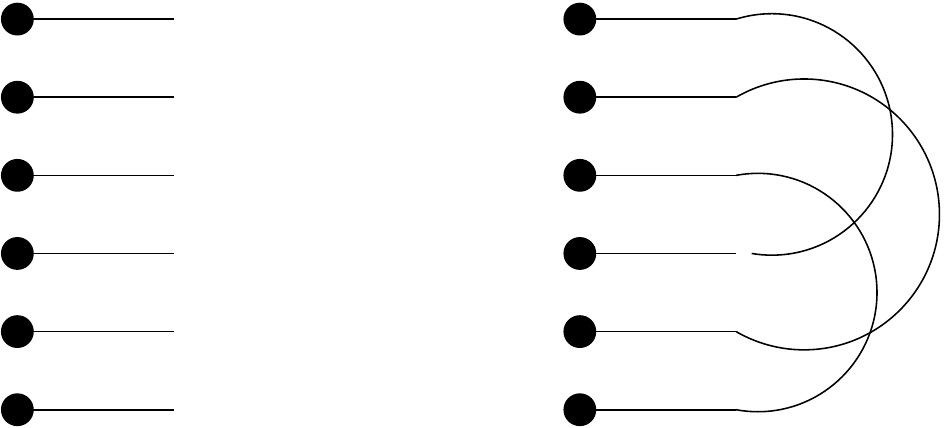}}
\caption{A contraction scheme on 6 fields.}
\label{graph1h}
\end{figure}

Interactions are  then added to the measure to make it non-Gaussian and lead to
interesting Feynman graphs with vertices of degree higher than 1. 
To illustrate our purpose let us consider the following one dimensional integral
\bee Z(\lambda ) = \int_{-\infty}^{+\infty}
e^{-\phi^2/2 - \lambda \phi^4 } d\phi   \; \label{xquat3}
\ee 
which is the partition function of the $\phi^4_0$ model. 
We choose to consider a  $\lambda \phi^4$ perturbation of the Gaussian measure rather than
the lower degree $\lambda \phi^3$ because the function $Z$ is then well defined as a convergent integral for $\lambda $ positive.

Expanding the exponential as a power series in the coupling constant $\lambda$, and commuting sums and integrations without caring
for convergence we get
\bee Z(\lambda) = \sum_{n=0}^{\infty}  \frac{(-\lambda)^n}{n!}  (4n-1)!! \;.  \label{seriesba}
\ee

Instead of considering $4n$ vertices of degree 1, we consider $n$ vertices, each of degree 4, hence with four 
half-edges or `fields" hooked, and the pairings now build up Feynman graphs (combinatorial maps) with 
$n$ vertices, each of degree 4. Such drawings can now be connected or not. 
 
\begin{figure}
\centerline{\includegraphics[width=10cm]{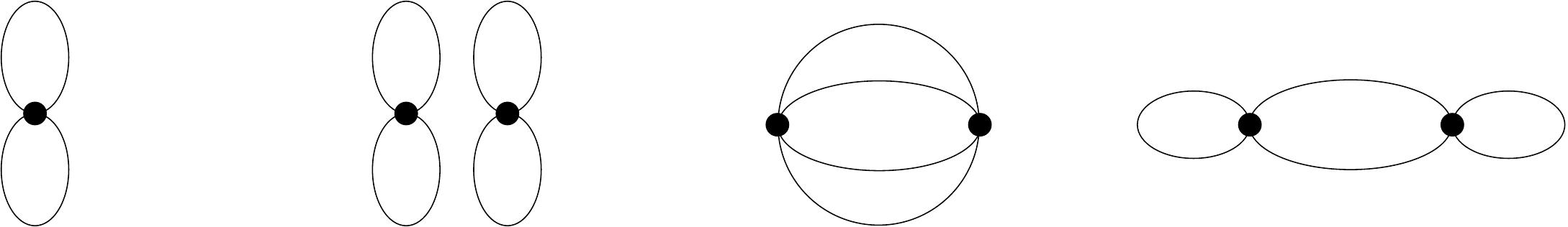}}
\caption{Vacuum Graphs with $n=1, 2$.}
\label{graphvacuumcontr}
\end{figure}

At order $n=1$ we have 4 fields, hence $ 3$ combinatorial maps which all correspond to the same connected graph.
At order $n=2$ we have 8 fields, hence $105$ combinatorial maps, which give three graphs, one made of 
two connected components, of weight $9$, the two others connected, of weights 24 and 72.
 
If we compute the first two orders of $Z$ we have therefore
\[  Z(\lambda) = 1 -3 \lambda + \frac{105}{2} \lambda^2  + O(\lambda^3) \; , \qquad 
   \log Z(\lambda) = 1 -3 \lambda + \frac{96}{2} \lambda^2  + O(\lambda^3) 
\]
and we remark that 96= 24+72 is, as expected, the number connected maps with two four valent vertices.

The higher moments of the interacting measure
\bee  S_N = \frac{1}{Z(\lambda) }\int_{-\infty}^{+\infty}   \phi^N
  e^{-\phi^2/2 - \lambda \phi^4 } d\phi \;,
\ee
are given by sums over all Feynman graphs (combinatorial maps) with $N$ external vertices of degree 1 and any number $n$ of internal 
vertices of degree 4. All these sums are divergent, but Borel summable.

\subsubsection{Quantum field theory on a finite graph}

The single-point universe was too simple an example to illustrate the role of graphs, contractions and combinatorics that are ubiquitous in QFT. To do so without introducing all of the QFT complications, let us now consider a universe with many points, made of a fixed finite graph $\Gamma$, on which particles propagate and interact.
This is obtained by the following generalization of the functional integral \eqref{xquat3}:

\begin{itemize}

\item the field $\phi$ is replaced by a function $\phi : V(\Gamma) \to \R$ on the vertices of the graph 
(which we note $x$, to suggest positions in the $\Gamma$ universe).

\item the Gaussian normalized measure \ref{xquat1} is replaced by
\bee  d\mu_{C_\Gamma}(\phi)  =  \frac{1}{Z_0} e^{-\frac{1}{2} \phi (L_\Gamma + \mu \bbbone ) \phi  
} \prod_{x\in V(\Gamma)} d\phi(x) , \label{eq:mesuremu}
\ee
where $L_{\Gamma}$ is the Laplacian matrix (see section~\ref{sec:LaplacianMat}) of the graph $\Gamma$ and $Z_0$ is some normalization constant.

\item  the interaction measure is 
\bee
d\nu_\Gamma (\phi )  =  {1 \over Z(\Gamma, \lambda)}  e^{- \lambda \sum_{x \in V(\Gamma)} \phi^{4} (x) }  d \mu_{C_\Gamma} (\phi) \;,
\ee
where the normalization is
\bee Z(\Gamma, \lambda)= \int e^{- \lambda \sum_{x \in V(\Gamma)} \phi^{4} (x) }  d \mu_{C_\Gamma} (\phi) = \int d\nu_\Gamma (\phi ) .
\label{mesurenu}\ee

\item the correlations (Schwinger functions) of the $\phi^{4}$ model on the universe $\Gamma$ are
the normalized moments of this measure:
\bee  S_{N} (z_1,...,z_N )  = \int     
\phi(z_1)...\phi(z_N) \; d\nu_\Gamma (\phi ) , 
\ee
where the $z_i$ are external positions hence fixed vertices of $\Gamma$.
\end{itemize}

Let us consider first the free measure \eqref{eq:mesuremu}. The covariance of the Gaussian measure admits a sum-over-paths representation. 
We expand around the local part of the Gaussian measure:
\bee  C(x,y) = \frac{1}{Z_{0}}  \int \phi(x) \phi(y) \; e^{\frac{1}{2}\sum_{v,w} \phi(v) A(v,w)  \phi(w)}  \;  d \mu_{\text{loc}} \;,
\ee
where we recall that $A(v,w)$ is the number of edges with end vertices $v$ and $w$ in $\Gamma$, and 
where $Z_{0} = \int e^{\frac{1}{2}\sum_{v,w} \phi(v) A(v,w)  \phi(w)}   d \mu_{ \text{loc} }$ and 
\bee   d \mu_{\text{loc}}  =   \prod_v \sqrt{\frac{ (d(v) + \mu ) }{2\pi}} e^{- \frac{1}{2}(d(v) +\mu)  \phi^2(v)} d\phi(v) \; ,
\ee
is a normalized \emph{local} Gaussian measure factorized over the vertices of the graph.
We get 
\bea  \label{pathrep1}
C(x,y) &=& \frac{1}{Z_{0}} \biggl( \sum_n  \frac{1}{n!}  \sum_{v_i, w_i} 
\bigl[\prod_i\frac{1}{2}A(v_i, w_i)\bigr]  \int   \phi(x) \phi(y)  \prod_i \phi(v_i) \phi(w_i) d \mu_{\text{loc} } \biggr) \nonumber  \\
&=& \frac{1}{Z_{0}} \biggl( \sum_n  \frac{1}{n!2^n}  \bigl[\prod_i \sum_{v_i, w_i} A(v_i, w_i) \bigr] \prod_v \frac{[n(v)-1]!!}{ (d(v) +\mu )^{n(v)/2}  } \biggr) \; ,
\eea
where $n(v)$ is the total number of fields $\phi$ at the vertex $v$.

This formula can be recast as a sum over paths. Before computing the Gaussian integral with measure $d\mu_{\text{loc}}$, 
at order $n$ we have a set of oriented edges edges of the graph 
\[
A(v_i,w_i)\phi(v_i) \phi(w_i) = \sum_{e,e=(v_i,w_i)} \phi(v_i) \phi(w_i)  \; .
\]
The $1/n!$ factor is canceled by the relabeling of the edges, while $1/2^n$ is canceled by the sum over orientations of the edges. 
We thus obtain a set of unlabeled, unoriented edges. A contraction scheme yields a pairing of the half edges at every vertex, 
that is it designates for any edge its successor when going trough the vertex. A contraction scheme is then a configuration of paths $\omega:\mathbb{N} \to V$,
with $\omega(i+1)$ being one of the neighbors of $\omega(i)$. Note that this induces an orientation of the edges, the path orientation, which has 
nothing to do with the a priory orientation canceled by the $1/2^n$ factor. The paths can either be closed or open, connecting the vertices 
$x$ and $y$. Dividing by $Z_0$ selects only the paths going from $x$ to $y$ and we obtain 
\bea
 C(x,y)= \sum_{\omega:\;  x\to y} \prod_{v\in V(\Gamma)} \frac{1}{ (d(v) +\mu )^{n(\omega, v) } } \; ,
\eea 
where $n(\omega,v)$ is the number of times the path $\omega$ passes trough the vertex $v$. 

\begin{figure}
\centerline{\includegraphics[width=12cm]{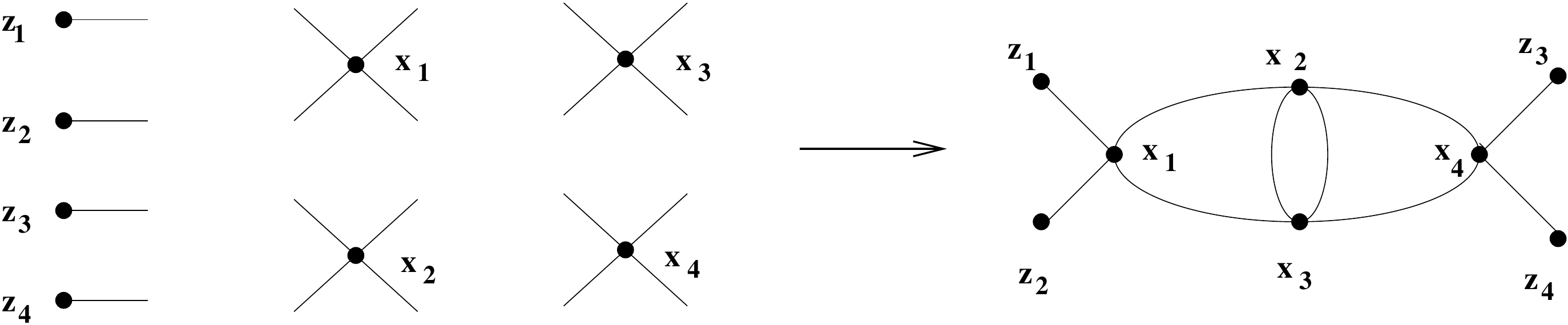}}
\caption{A possible contraction scheme with $n=N=4$.}
\label{fig:feynrulesgraph}
\end{figure}

In order to analyze the interaction measure, we expand the exponential as a power series in the coupling constant $\lambda$, and commuting
again sums and integrations without care for convergence one obtains the formal series expansion for the Schwinger functions in powers of $\lambda$:
\bea \label{schwi}  && S_{N} (z_1,...,z_N )  \\
&& \quad =  {1 \over Z(\Gamma,\lambda)}  \sum_{n=0}^{\infty} \frac{1}{n!} \int
\bigl[ - \lambda \sum_{x \in V(\Gamma)} \phi^{4} (x) \bigr]^{n}  
\phi(z_1)...\phi(z_N) d \mu_{C_\Gamma} (\phi) \;   . \nonumber
\eea
Labeling the $n$ dummy integration variables in (\ref{schwi})
as $x_1,...,x_n$, we draw an edge $\ell$ for each contraction of two fields.
Each position $x_1,...,x_n$ is then associated to a four-legged vertex
and each external source $z_i$ to a one-legged vertex, as shown in 
Figure \ref{fig:feynrulesgraph}.

At order $n$ the Gaussian integral is a sum over $(4n+N-1)!!$
contractions schemes $\cW$,  i.e. ways of pairing together 
$4n+N$ fields into $2n+N/2$ pairs. 
The evaluation of a a given scheme is the factor 
${1 \over n!}({-\lambda })^{n}$  times the product
$\prod_{\ell \in \cW} C_\Gamma(x_\ell, y_\ell ) $, where $(x_\ell, y_\ell ) $ are the ends of the edge $\ell$.
Hence the amplitude of a graph is, up to normalization constants, a sum over the positions of the internal vertices of the product
of covariances (or propagators) for all edges
\bee  A_{G}(z_1,...,z_N) \equiv  \sum_{x_1\in V(\Gamma), \cdots x_n \in V(\Gamma)}
 \prod_{\ell \in E(G)}  C_\Gamma(x_\ell,y_\ell) . \ee

The Schwinger functions are therefore formally given 
by the sum over all combinatorial maps with the right number of external edges of the 
corresponding Feynman amplitudes:
\bee S_{N} =   \frac{1}{Z(\Gamma, \lambda)}\sum_{\phi^{4}{\rm \ graphs \ } G {\rm\ with \ } N(G)=N} 
{(-\lambda)^{n(G)} \over s(G)} A_G \; ,
            \label{series}
\ee
where the factor $s(G)$ takes into account the fact that there are several labelled combinatorial maps associated to an unlabeled graph $G$.
The partition function $Z(\Gamma, \lambda)$ itself is given by the sum of graphs with no external edges:
\bee Z(\Gamma, \lambda)=   \sum_{\phi^{4}{\rm \ graphs \ } G {\rm\ with \ } N(G)=0} 
{(-\lambda)^{n(G)} \over s(G)} A_G. 
\ee
and again $\log Z(\Gamma, \lambda) $ is given by the same sum but restricted to connected graphs.

Returning to \eqref{pathrep1} we have another possible expansion for the Schwinger functions, by expanding the interacting measure around its local part.
This leads to the self-repelling paths representation of Schwinger  functions. We decompose the interacting measure 
$d\nu$ as 
\bee  d \nu_{\Gamma} = \frac{1}{Z(\Gamma,\lambda)} e^{  \frac{1}{2}  \sum_{v,w} \phi (v) A(v,w) \phi (w)}   d\nu_{\text{loc} }
\ee
with
\bea  
&& d\nu_{\text{loc}}  = \frac{1}{Z_{\text{loc} }(\lambda) }\prod_{v \in V(\Gamma)}   e^{-\frac{1}{2} (d(v)+\mu )\phi(v)^2  -\lambda \phi^4}d\phi (v)   \crcr
&& Z_{\text{loc} }(\lambda) = \int  \prod_{v \in V(\Gamma)}   e^{-\frac{1}{2} (d(v)+\mu )\phi(v)^2  -\lambda \phi^4}d\phi (v) \; .
\eea 
For the two point Schwinger function for instance we obtain the representation:
\[S(x,y) = \frac{1}{Z_{\text{loc} }(\lambda)}  \Bigl(\sum_n  \frac{1}{n!} 
\sum_{v_i, w_i} [\prod_i\frac{1}{2}A(v_i, w_i)]  \int   \phi(x) \phi(y)  \prod_i \phi(v_i) \phi(w_i) d \nu_{\text{loc} } \Bigr) \; ,
\]
which rewrites
\bea 
   \sum_n  \frac{1}{n!2^n}  \sum_{v_i, w_i} [\prod_i A(v_i, w_i)]  \prod_v \frac{ [ n(v)-1]!!}{ (d(v) +\mu )^{n(v)/2}  } P(n(v),\lambda) \;,  \label{pathrep2}
\eea
where $P(n(v), \lambda)$ is a correcting factor, non-zero only for $n(v)$ even, and which is 1 at $\lambda =0$
\[ N(n(v), \lambda) = \frac{(d(v) +\mu )^{n(v)/2} }{ [n(v)-1]!! } \frac{ 
   \int \phi^{n(v)} \; \; e^{-\frac{1}{2} (d(v)+\mu) \phi(v)^2  -\lambda \phi^4}d\phi (v) }
   {  \int  e^{-\frac{1}{2} (d(v)+\mu) \phi(v)^2  -\lambda \phi^4}d\phi (v) } .
\]
The factors $\frac{1}{n!2^n} \prod_v \frac{[n(v)-1]!!}{ (d(v) +\mu )^{n(v)}  }$ yield a sum over paths representation of the Schwinger function.
However the correcting factors $P(n(v), \lambda)$ do not factorize between the path from $x$ to $y$
and the closed loops, hence it is not easy to compute the quotient. For a single path or a single loop, however, it is 
possible to get the strong coupling limit $\lambda \to \infty$ and to check that it is represented
by a sum over self-avoiding walks. To divide correctly the normalization factor one can use either a supersymmetric representation
or a replica trick. In this way one can access to a quantum field theoretic representation of self-avoiding random walks or polymers on a graph.

\subsection{Quantum field theory on \texorpdfstring{$\mathbb{R}^d$}{R**d}}

We now consider a genuine field theory, that is the field is a function (actually, a distribution) 
$\phi:\mathbb{R}^d\to \mathbb{R}$. The free field theory has already been
discussed, and it is the theory of the normalized Gaussian measure of covariance
\[
  C(x,y) = \Bigl( \frac{1}{-\Delta +m^2} \Bigr) (x,y) = \int_{0}^{\infty} \frac{d\alpha}{(4\pi\alpha)^{d/2}} \; e^{-\frac{(x-y)^2}{4\alpha} -m^2 \alpha} \; .
\]
Note that for $d\ge 2$, $C(x,x)$ is not well defined, as the integral over $\alpha$ diverges at $\alpha \simeq 0$. This 
divergence is called an {\it ultraviolet divergence} and it is the source of the most interesting phenomena in QFT.
The covariance $C(x,y)$ must be understood as a distribution. In order to make sense of QFT one must consider 
regularized covariance kernels. 

A particularly useful regularization is the \emph{heat kernel} regularization consisting 
in replacing the distribution $C(x,y)$ by the smooth function 
\bee     
C_{\kappa}(x,y) = \int_{\kappa}^{\infty} \frac{d\alpha}{(4\pi\alpha)^{d/2}}  e^{-\alpha m^{2}- (x-y)^{2}/4\alpha} \; .
\ee

Such a regulator $\kappa$ is called
an ultraviolet cutoff, and we have (in the distribution sense)
$\lim_{\kappa \to 0} C_{\kappa}(x,y)= C(x,y)$. Remark that due to the
non zero $m^2$ mass term, the kernel $C_{\kappa}(x,y)$ decays
exponentially at large $\vert x-y\vert$ with rate $m$. 
For some constant $K$ and $d>2$ we have:
\bee   \vert C_{\kappa}  (x,y) \vert \le K 
\kappa ^{1-d/2} e^{- m \vert x-y \vert } .  \label{decay}
\ee

The fundamental feature of the heat kernel regularization is that it is {\it universal}. While other cutoff schemes might work in 
particular instances, this regularization always works. Any positive operator $C$ admits a representation
\[
 C = \frac{1}{H} = \int_0^{\infty} d\alpha \;  e^{-\alpha H} \; ,
\]
and the ultraviolet divergence {\it always} comes from the $\alpha\simeq 0$ region. The UV cutoff can always be implemented by 
cutting the integration interval at $\kappa$. 

For any quantum field theory, the connected Schwinger functions (i.e. the derivatives of $W[J]=\ln Z[J]$) can be computed 
directly from the Schwinger functions (i.e. the derivatives of $Z[J]$) by M\"obius inversion formula
\bee  C_{N} (z_{1},...,z_{N}) = \sum_{P_{1}\cup ... \cup P_k = \{1,...,N\}; 
\, P_{i} \cap P_j =0}    (-1)^{k+1}  \prod_{i=1}^{k}  S_{p_i} 
(z_{j_1},...,z_{j_{p_i}}) , \ee
where the sum is performed over all distinct partitions of $\{1,...,N\}$ 
into $k$ 
subsets $P_1,...,P_k$, $P_i$ being made of $p_i$ elements called 
$j_1,...,j_{p_i}$. For instance, if the odd Schwinger functions vanish (which is the $\phi^4_d$ theory) 
the connected 4-point function is:
\bea \label{trunca1}
&& C_{4} (z_{1},...,z_{4})  =   S_{4} (z_{1},...,z_{4}) - S_{2}(z_{1},z_{2}) S_{2}(z_{3},z_{4}) \crcr
&& \qquad \qquad \qquad \;\; - S_{2}(z_{1},z_{3})S_{2}(z_{2},z_{4}) -S_{2}(z_{1},z_{4})S_{2}(z_{2},z_{3})  \; . 
\eea
 
The full, normalized, interacting measure of the $\phi^4_d$ theory is defined as the multiplication of the Gaussian 
measure $d\mu(\phi)$ by the interaction factor:
\bee \label{mesurenui}
 d\nu  =  \frac{1}{Z(\lambda)} e^{-\frac{\lambda}{4!} \int \phi^{4}(x) d^dx} \;  d\mu_C(\phi) \; ,
\ee
and the Schwinger functions are the normalized moments of this measure:
\bee   
S_{N} (z_1,...,z_N )  = \int  d\nu (\phi)  \; \phi(z_1)...\phi(z_N)  \; . 
\ee
Expanding the exponential as a power series in the coupling constant $\lambda$, and commuting the sum and the integral
one obtains the perturbative expansion for the Schwinger functions:
\bea\label{schwiQFT}
&& S_{N} (z_1,...,z_N )  \\
&& \qquad = \frac{1}{ Z(\lambda)}  \sum_{n=0}^{\infty} \frac{(-\lambda)^{n}}{4!^n n!} \int d\mu_C (\phi)
\Bigl[ \int \phi^{4}(x) dx \Bigr]^{n}  
\phi(z_1)...\phi(z_N)   \; . \nonumber
\eea 

As we have already seen, the Gaussian integral yields a sum over contractions schemes (combinatorial maps), 
which we regroup in Feynman graphs. As in the previous section, the graphs have either four valent 
vertices coming from the $\phi^4(x)$ factors, or univalent external vertices coming from the $\phi(z)$ factors. 
Each contraction is represented as an edge connecting two half edges, as in Figure \ref{fig:feynrulesgraph}.
The amplitude of each scheme is an integral over the positions of the internal vertices of the product
of covariances for all edges
\bee\label{eq:directrrep}
A_{G}(z_1,...,z_N) \equiv  \int d^dx_1 \dots d^dx_n  \prod_{\ell \in E(G)}  C(x_\ell,y_\ell) \; . 
\ee
Such amplitudes are functions (in fact distributions) of the external positions $z_{1},...,z_N$.
They may diverge either because the integrand is typically unbounded due to the ultraviolet 
singularities in the propagator $C$ at coinciding points, or because they are integrals 
over all of ${\R}^{d}$ (no volume cutoff). 

The regrouping of combinatorial maps into corresponding 
unlabeled graphs, and the subtleties related to the symmetry factor are identical to the ones of the previous section.
The {\it unrenormalizaed} Schwinger functions are again
\bea 
&& S_{N} = \frac{1}{Z(\lambda)}  \sum_{\phi^{4}{\rm \ graphs \ } G {\rm\ with \ } N(G)=N} 
 \frac{(-\lambda)^{n(G)} } {s(G)} A_G \; , \crcr
&&  Z(\lambda)=   \sum_{\phi^{4}{\rm \ graphs \ } G {\rm\ with \ } N(G)=0} 
 \frac{(-\lambda)^{n(G)} }{ s(G)} A_G \; . 
\eea

To ensure the convergence of $A_G$, we introduce the ultraviolet cutoff $\kappa$ and the volume cutoff $\Lambda$, and replace the amplitudes $A_G$ by the 
{\it cutoff} amplitudes 
\bea
 A^{\kappa}_{G,\Lambda}(z_1,...,z_N) \equiv  \int_{\Lambda^n} d^dx_1 \dots d^dx_n  \prod_{\ell \in E(G)}  C_{\kappa}(x_\ell,y_\ell) \; .
\eea 
  
From translation invariance, we do not expect $A_{G,\Lambda}^{\kappa}$ to have 
a thermodynamic limit $\Lambda \to \infty$ if $G$ contains subgraphs with no external arguments (a global shift of all the positions for all the 
internal vertices does not change the integrand). However, the  thermodynamic limit $\Lambda\to \infty$ 
can always be taken at fixed external arguments (due to the exponential decay of the propagator) after one divides out the vacuum graphs.
  
To summarize:

\begin{itemize}

\item The great advantage of Feynman graphs is that they form a combinatorial species \cite{bergeron1998combinatorial}
whose logarithm  can be computed as the species of connected graphs. As we already remarked, the computation of
this logarithm is the key physical problem. 

\item However the number of Feynman graphs (combinatorial maps) with $n$ vertices is $a_n \sim (4n-1)!!$ and 
their generating function $\sum \frac{1}{n!}a_n \lambda^n$ has zero radius of convergence in $\lambda$.
At the heart of any constructive strategy \cite{Glimm:1987ng, Rivasseau:2011ri}, lies the replacement of the proliferating species of Feynman graphs by a 
better one, typically  the species of forests. The corresponding connected species is the species of trees. The number of trees 
over $n$ vertices behaves like $a_n \sim n^{n-2}$ and their generating function $\sum_n \frac{1}{n!}a_n \lambda^n$ has a finite, non zero,
radius of convergence. The constructive expansions converge, while ordinary perturbative expansions do not. 

\item The computation factorizes over the connected components
of the graphs. These components may or may not have external arguments. 
In the expansion for the normalized functions the {\it vacuum}  components 
(i.e. those without external arguments) drop out and  only graphs whose
connected components all contain external arguments remain. 

\item If we further
search for elementary bricks of the expansion, we can consider the
{\it connected} Schwinger functions like \eqref{trunca1}, which are evaluated by graphs 
with a single connected component containing all external
arguments.

\end{itemize}

In the reminder of this subsection we will give several formulae for the amplitudes of the Feynman graphs.

\subsubsection{Direct, momentum and parametric representation}

The direct space representation of the Feynman amplitude is given by eq. \eqref{eq:directrrep}
\bee
A_{G}(z_1,...,z_N) \equiv  \int d^dx_1 \dots d^dx_n  \prod_{\ell \in E(G)}  C(x_\ell,y_\ell) \; . 
\ee
The momentum space representation is obtained by substituting the Fourier transform of the covariance
\[
   C(x,y) = \frac{1}{(2\pi)^d} \int d^dp \; e^{\imath p (x-y)} \frac{1}{p^2+m^2} \; ,
\]
and integrating the positions of the internal vertices to obtain
\bee 
  A_{G}  (z_1,...,z_N) =
\int d^dp_1...d^dp_N  \; e^{ \imath \sum p_iz_i \epsilon(v_i, \ell_i) } \;  \tilde A_G (p_1,...,p_N )  \; ,
\ee
where $v_i$ denotes the external univalent vertex with position $z_i$ and
\bee\label{momrep}
   \tilde A_G (p_1,...,p_N ) =  { 1 \over  (2\pi)^{d(n+N)}} \int  \prod_{\ell \ {\rm internal \ edge \ of\ }G} 
{d^d p_\ell \over 
p_\ell^2 + m^2}
\prod_{v \in G} \delta ( \sum_\ell \epsilon (v,\ell) \;  p_\ell  )  \; . \nonumber
\ee 
In (\ref{momrep}) the $\delta$ functions ensure momentum conservation at each internal vertex $v$. Each edge (internal or external) 
is oriented in an arbitrary way. The incidence matrix  $\epsilon(v,\ell)$ captures in a nice way the information
on the internal and external edges\footnote{Strictly speaking this is true only
for graphs without tadpoles.}. Remark also that there is an overall momentum conservation hidden in (\ref{momrep}): one of the delta functions 
is redundant and can be factored out as signed sum of the external momenta. 

The drawback of the momentum representation lies in the necessity for practical 
computations to solve the $\delta$ functions by a ``momentum routing" prescription. Such a prescription is 
linked to the non-canonical choice of a tree. The momenta are associated to the external edges of the tree $\cT$; 
every edge $\ell \in \cT$ has then as momentum flowing through the sum of the momenta entering any of 
the two subtrees obtained by removing that edge in $\cT$.

A more canonical representation is the parametric representation. It is obtained by reinstating the $\alpha$ parameters 
to write (up to an overall factor of $2\pi$) 
\bea
&&   A_G(z_1,\dots z_N) =  
\int_{0}^{\infty} \prod_{\ell} \frac{d\alpha_{\ell}}{   \alpha_{\ell} ^{d/2} } \int \prod_{\text{internal }v }d^dx_v \crcr
&& \qquad \qquad    e^{-\frac{1}{4} \sum_{\ell} \frac{1}{\alpha_\ell} \bigl( \sum_v \epsilon(v,\ell) x_v \bigr)^2   - \sum_{\ell}\alpha_\ell m^2 } \; ,
\eea 
where in the second line  $x_v$ is understood to be replaced by $z_i$ for the external vertices. We use the unified notation $y_v$ for the 
position of a vertex (internal or external) and write the Fourier transform of the amplitude 
\bea
&& \tilde A_G (p_1,...,p_N ) = \int_{0}^{\infty} \prod_{\ell} \frac{d\alpha_{\ell}}{   \alpha_{\ell} ^{d/2} } \int \prod_{v }d^dy_v 
 \crcr
&& \qquad \qquad    e^{-\frac{1}{4} \sum_{\ell} \frac{1}{\alpha_\ell} \bigl( \sum_v \epsilon(v,\ell) y_v \bigr)^2   - \sum_{\ell}\alpha_\ell m^2
- \imath \sum_{\text{external }v} y_v P_v } \; ,
\eea 
where $P_v$ is the $(\mathbb{R}^d)^N$ vector of external momenta, which by convention have all been taken to be entering the external vertices.

By shifting all the variables with $y_1$, the position of one chosen vertex (which we call $1$), 
the integral over $y_1$ can be computed explicitly to yield a global $\delta(\sum_{ \text{external }v } P_v)$ of conservation of momenta.
As the integral over $y_v$ is Gaussian, it can also be explicitly computed. 
The quadratic form in $y_v$ in the exponential can be rewritten as 
\[
 -\frac{1}{4} \sum_{v,v'} y_v ([d_{G}]_{\bar 1\bar 1} )_{vv'} y_{v'} -  \imath \sum_{\text{external }v} y_v P_v  \; ,
 \;\; ([d_{G}]_{\bar 1\bar 1} )_{vv'} = \sum_{\ell} \epsilon(v,\ell) \frac{1}{\alpha_{\ell}} \epsilon(v',\ell) \;, 
\]
that is the variance is just the weighted Laplacian on the graph (including the univalent external vertices) with the vertex $1$ deleted.
The result of the Gaussian integration of $y_v$ is, up to overall constants
\bea
 \frac{1}{\det\Bigl( [d_{G}]_{\bar 1\bar 1} \Bigr) } e^{- P ([d_{G}]_{\bar 1\bar 1} )^{-1} P } \; ,
\eea 
and both the determinant and the inverse of $[d_{G}]_{\bar 1\bar 1}  $ are computed by the weighted matrix tree theorem \ref{wmtt}
\bea
&& \det\Bigl( [d_{G}]_{\bar 1\bar 1} \Bigr) = \sum_{T \text{ tree of G} } \prod_{\ell \in T} \frac{1}{\alpha_{\ell}} , \\
&&  P ([d_{G}]_{\bar 1\bar 1} )^{-1} P = \frac{1}{ \det\Bigl( [d_{G}]_{\bar 1\bar 1} \Bigr) } \sum_{i,j} p_i \cdot p_j 
    \sum_{T_1,T_2}^{T_1 \cup T_2 = \text{ tree of }G/ij }  \prod_{\ell \in T_1 \cup T_2} \frac{1}{\alpha_{\ell}} \; .\nonumber
\eea 
We finally get 
\bea
&&  \tilde A_G (p_1,...,p_N ) = \int_{0}^{\infty} \prod_{\ell} \frac{d\alpha_{\ell}}{[U_G(\alpha)]^{d/2}}
    e^{- \frac{V_G(p,\alpha)}{U_G(\alpha)} -\sum_{\ell} \alpha_{\ell} m^2 } , \crcr
&&  U_G(\alpha) = \sum_{T \text{ tree in G} } \prod_{\ell \notin T} \alpha_{\ell}\; , \\
&& V_G(\alpha) = \sum_{T_1,T_2 \text{ two trees in G} } 
\Bigl( \prod_{\ell \notin T_1\cup T_2} \alpha_{\ell}\Bigr) \bigl(\sum_{i \text{ external vertex of } T_1}  p_i \bigr)^2 \; . \nonumber
\eea 

\subsection{The renormalization group}

As we argued, a quantum field theory is defined by a partition function 
\bea
{  Z }= \int d \mu_C(\phi) e^{-W(\phi)} \; ,
\eea 
which takes the form of a Gaussian measure with an additional ``interaction'' term $U(\phi)$.
The crucial obstruction with evaluating such an expression is that it is plagued by divergences. The appropriate way to evaluate this is by an iterative procedure,
called the ``renormalization group''. The renormalization group starts with a ``scale decomposition'' of the covariance,
\bea
 C = \sum_{i\ge 0}C^i \; .
\eea 
The steps in this iterative procedure are associated to the 
scale parameters $i$. Using the properties of the Gaussian measure, the field decomposes into a sum of fields
associated to the scales $i$. The renormalization group consists in integrating out the ``high scale'' field and 
recasting the effect of this integration into an ``effective action'' $W^i$.%
\footnote{%
This procedure can also formulated in terms of probability theory, see refs.~\cite{JonaLasinio:1974rh,JonaLasinio:2001gc}. 
}

After integrating all the high scale fields up to the scale $i$ the partition function writes as
\bea
{ Z }= \int d\mu_{ C_{\le i}  } (\psi_i) e^{-W^i(\psi_i )} \; , \qquad  C_{\le i} = \sum_{j\le i} C^i \; . 
\eea 
As we will see in the next sections, a subtlety of the renormalization group is that it must be encoded in a step-by-step procedure. That is, after reaching 
the scale $i$, one must prudently integrate just the field in the slice $i$ in order to derive an effective action at scale $i-1$. 
This comes to realizing that the covariance at scale $i$ can be written as 
\bea
 C_{\le i} = \sum_{k\le i} C^k = \sum_{k < i} C^k + C^i \; ,
\eea 
hence the partition function writes 
\bea
\mathbb{ Z }= \int d\mu_{ C_{\le i-1}  }(\psi_{i-1}) \;  d\mu_{C^i}(\phi_i)  e^{-W^i(  \psi_{i-1} + \phi_i   )} \; ,
\eea 
and the effective action at scale $i-1$ is computed as
\bea
 e^{-W^{i-1}( \psi_{i-1}   )} = \int  d\mu_{C^i}(\phi_i)  e^{-W^i(  \psi_{i-1} + \phi_i   )} \; .
\eea 

The basic renormalization group step is therefore made of two main operations:

\begin{itemize}
\item A functional integration 
\item The computation of a logarithm
\end{itemize}
The effect of the integration of the high scales field $\phi_i$ is to change the value of the coupling
constants of the effective action (this includes adding new couplings, whose initial value at scale $i$ can 
be consider to be zero). The change of the coupling 
constants with the scale index $i$ is called the renormalization group flow. 
The flow from the initial bare action $S=S_{\kappa}$ for the full field to
an effective renormalized action $S_0$ for the last ``slowly varying" 
component $\phi_0$ of the field is the flow of a (complicated) discrete-time dynamical system. Its evolution
is decomposed into a sequence of discrete 
steps from $S_i$ to $S_{i-1}$.

The effective action at scale $W^{i-1}(\psi_{i-1})$ is usually very involved. In order to evaluate the change in the 
coupling constants one needs to identify in $W^{i-1}(\psi_{i-1})$  the operators present in the  original action (and separate some 
rest terms). This is further complicated by the fact that the effective action might contain quadratic terms in $\psi_{i-1}$ which can 
change the Gaussian measure.

This renormalization group strategy can be best understood on the system of Feynman graphs
which represent the perturbative expansion of the theory.
 The first step,
functional integration over fluctuation fields, means that we have to consider 
subgraphs with all their internal edges in higher slices than any of their external edges (this will be explained at length later on).
The second step, taking the logarithm, means that we have to consider only
\emph{connected} such subgraphs. We call such connected subgraphs \emph{quasi-local}.
Renormalizability is then a non-trivial result that combines
locality and power counting for these quasi-local subgraphs. 

\subsubsection{Renormalization and anomalous scaling}

Let us study in more depth what are the physical consequences of the renormalization group flow. We use as an input in the path integral 
the bare action $S_{\kappa}$ which in the $\phi^4_d$ model case is
\begin{eqnarray}
\label{eq:bareact1}
&& S_{\kappa}= \frac{1}{2} \mathcal{Z}_{\kappa} \int d^dx \; \phi(x) (-\Delta)\phi(x) + \frac{1}{2} m^2_{\kappa}\mathcal{Z}_{\kappa} 
 \int d^dx \;  \phi(x)^2 \crcr
 && \qquad \qquad\qquad\qquad +\frac{1}{4!}\lambda_{\kappa}\,\mathcal{Z}_{\kappa}^2  \int d^dx \;  \phi(x)^4\,,
\end{eqnarray}
where we have included the Gaussian measure as part of the action and we introduced a \lq\lq wave-function \rq\rq 
renormalization $\mathcal{Z}_{\kappa}$.
 
The main lesson of renormalization is that the physical quantities, the $n$-points Schwinger functions, {\it change} 
with the experimental external scale $\mu$\footnote{For the purposes of 
this subsection we will assume that $\mu$ varies continuously, and not in discrete steps.}: the quantum effective action at scale $\mu$ is
\bea
&& \Gamma_{\mu}[\varphi] = \sum_{n} \frac{1}{n!} \int d^dx_1\dots d^dx_n \;  [ \mathcal{Z}_\mu^{1/2} \varphi(x_1) ] \dots [ \mathcal{Z}_\mu^{1/2} \varphi(x_n) ] \crcr
&&  \qquad \qquad \qquad \times  \mathcal{Z_\mu}^{-n/2}\Gamma^{(n)}(x_1,\dots,x_n;\mu)    \; ,
\eea
where the wave function renormalization at scale $\mu$, $\mathcal{Z}_\mu $ will be fixed below.
When changing $\mu$ this quantum effective action flows in a {\it theory space}, (the analog of the space 
$\cU$ for maps on the interval), the space of all functionals of the fields compatible with the symmetries of the theory. 
For instance, for the theory of a single real scalar field with a $\phi\leftrightarrow-\phi$ symmetry an adequate theory space is
\begin{eqnarray} 
&& \mathcal{T}=\mathrm{span}\left\{ 1, \phi^2, \phi(-\Delta)\phi,\phi^4 ,\phi^3 (-\Delta) \phi, \phi(-\Delta)^2\phi\dots \right\} \\
&& \qquad =  \mathrm{span}\left\{ \varphi^{n_1}(-\Delta)^{p_1}\varphi^{n_2}(-\Delta)^{p_2} \dots \varphi^{n_q}\;, n:= n_1+\dots n_q = \mathrm{even} \right\} \; .
\nonumber
\end{eqnarray}

The quantum effective action can then be expanded in terms of flowing coupling constants  associated to a basis of operators in this space:
\begin{equation}\label{eq:gammaexp}
\Gamma_{\mu}[\varphi]=\sum [\mathcal{Z_\mu}]^{n/2} \frac{1}{n!}\lambda^{(n,p,\sigma)}_\mu \mathcal{O}^{(n,p,\sigma)}(\varphi)\;.
\end{equation} 
To simplify our discussion we ignore for now all the derivatives operators for $n\ge 4$ and choose the basis of operators:
\bea
 && \Gamma_{\mu}[\varphi]=  \mathcal{Z_\mu}  \frac{1}{2}\lambda^{(2,0)}_\mu \int d^dx \; \varphi(x)^2 +  
 \mathcal{Z_\mu}  \frac{1}{2}\lambda^{(2,1)}_\mu \int d^dx \; \varphi(x) (-\Delta) \varphi(x) \crcr
 && \qquad  \qquad  \qquad + \sum_{n \text{ even }, n \ge 4 }  [\mathcal{Z_\mu}]^{n/2}  \frac{1}{n!} \lambda^{(n,0)}_\mu \int d^dx \; \varphi(x)^{n} \; .
\eea 
The wave function renormalization $ \mathcal{Z_\mu}$ is fixed by the renormalization condition $ \lambda^{(2,1)}:=1$.
Denoting $\eta \equiv \mu \frac{d \mathcal{Z(\mu)} }{d\mu}$, the evolution under the renormalization group flow of the quantum effective action,
\bea
&&  \mu \frac{d  \Gamma_{\mu}[\varphi]}{d\mu} = \frac{1}{2}  \mathcal{Z_\mu} \Bigl(\eta + \mu \frac{ d \lambda^{(2,0)}_\mu  }{d\mu} \Bigr) 
 \int_{\mathbb{R}^d}\; \varphi^2  +  \frac{1}{2}  \mathcal{Z_\mu} \; \eta  
 \int_{\mathbb{R}^d} \; \varphi (-\Delta) \varphi \crcr 
&& \qquad + \sum_{n \text{ even }, n \ge 4 } \frac{1}{n!}  [\mathcal{Z_\mu}]^{n/2}  \Bigl(\frac{n}{2}\eta + \mu \frac{ d \lambda^{(n,0)}_\mu  }{d\mu} \Bigr) 
    \int_{\mathbb{R}^d}\; \varphi^n  \; ,
 \eea 
is captured by the {\it beta functions} of the couplings
\bea\label{eq:betadimless}
  \beta_{\lambda^{(n,0)}} \equiv \mu \frac{ d \lambda^{(n,0)}_\mu  }{d\mu} \; .
\eea 
Note that the coupling constants are dimensionfull quantities. Indeed, the field has dimensions $[\varphi] = [\mu]^{\frac{d-2}{2}}$ hence
$[ \lambda^{(n,0)}_{\mu} ] = [\mu]^{d- \frac{d-2}{2}n}$. Rescaling the coupling constants by the appropriate power of $\mu$ one can rewrite the 
beta functions for dimensionless couplings
\bea
&& \tilde \lambda^{(n,0)}_{\mu} = \frac{\lambda^{(n,0)}_{\mu}  }{ \mu^{d- \frac{d-2}{2}n} } ,\crcr 
&&   \beta_{\tilde\lambda^{(n,0)}} \equiv \mu \frac{ d \tilde \lambda^{(n,0)}_\mu  }{d\mu}
 = - \Bigl(d - \frac{d-2}{2}n  \Bigr) \tilde \lambda^{(n,0)}_\mu + \frac{1}{ \mu^{d- \frac{d-2}{2}n}  } \beta_{\lambda^{(n,0)}} \;.
\eea 
The renormalization group flow translates into a change of the scaling dimension of the coupling constants of the theory.
The first term above is just the flow due to the classical dimension of the coupling constant, while the second term 
encodes the quantum corrections.

\subsubsection{Renormalizable and non-renormalizable theories}\label{sec:powercounting}

We can think of the quantum effective action as a sum over Feynman graphs, whose amplitude is generally divergent. Consider the graphs of the $\phi^n$ model in $d$ dimensions. Using the momentum space representation, we see that we must perform an 
integral over $\mathbb{R}^d$ for every internal edge of a product of a $\frac{1}{p^2+m^2}$ factor for every internal edge times 
a $\delta$ function for every internal vertex. As the $\delta$ functions fix the momenta of the edges of a tree, the {\it superficial 
degree of divergence} of a connected graph with $V$ vertices and $N$ external edges (hence $2E = nV-N $) is 
\bea\label{primitivedivergence}
 \delta = d(E - V+1) -2E = d - \frac{d-2}{2}N - V \Bigl( d - \frac{d-2}{2} n \Bigr) \; .
\eea 

It can happen that a graph converges better than what its \emph{superficial} degree of divergence would suggest\footnote{This 
frequently happens in presence of symmetries, an example being the light-by-light scattering in Quantum Electrodynamics.}. However, 
by a theorem of Weinberg, if $\de<0$ the graph is convergent (we will prove this theorem in section~\ref{sec:uniformbounds}). Therefore we can give the following  perturbative and superficial 
classification of quantum field theories:
\begin{enumerate}
\item  Non-renormalizable, when $d<\frac{d-2}{2}n$. In this case $\delta$ increases with $V$.
\item Renormalizable, when  $d=\frac{d-2}{2}n$, so that $\delta$ does not depend on $V$; this is the familiar case of $\phi^4$ in four dimensions, as $d=4$ 
and $n=4$.
\item Super-renormalizable if $d>\frac{d-2}{2}n$; in this case the UV behavior improves with $V$ and there are only a finite number of divergent graphs.
\item Finite, if there are no divergent graphs at all.
\end{enumerate}

What is special about renormalizable theories? Let us rephrase (\ref{primitivedivergence}) in terms of the
scaling dimension of the couplings. The dimensionful coupling $\lambda^{(n,0)}$ of the $\phi^n$ term is given by graphs with $n$ external edges and therefore 
has scaling dimension 
\begin{equation}
 \Delta_{(n,0)} =  d-\frac{d-2}{2}n  \;.
\end{equation}
 Using the superficial degree of divergence, the coupling at scale $\mu$ scales with the UV cutoff $\kappa$ as
\bea
 \lambda^{(n,0)}_{\mu} \sim \kappa^{\Delta_{(n,0)}} \Bigl( \frac{\lambda_\kappa}{ \kappa^{\Delta_{(n,0)}} }\Bigr)^V \Rightarrow
 \tilde \lambda^{(n,0)}_{\mu} \sim \Bigl( \frac{\kappa}{\mu}\Bigr)^{\Delta_{(n,0)}} (\tilde\lambda^{(n,0)}_{\kappa})^V \; ,
\eea 
which is suppressed in the limit $\kappa\to \infty$ for non renormalizable interactions as $\Delta_{(n,0)}<0$. 

Therefore a first reason for the importance of renormalizable theories is that at sufficiently 
low energy they capture the leading physical effects to a good accuracy. Furthermore, they can be defined by measuring a small number of physical 
parameters (coupling constants), and hence are very predictive. Irrelevant operators can be taken into account as higher order effects, which
however require fixing additional parameters, see e.g.~\cite{Burgess:2007pt}.

This does not mean that non-renormalizable theories are useless. In fact, when the number of renormalizable couplings is reduced by symmetries, 
(even to zero) it can be important to consider the lowest order non-renormalizable ones. This happens for instance in chiral perturbation 
theory, which has a remarkable experimental success describing quark bound states (e.g. pions) at low energies \cite{leutwyler}.
However, such effective theories 
break down at some threshold energy. From then on any possible coupling has to be considered and the theory is no longer predictive. 
It means that at energy scales higher than the threshold energy such QFTs must be replaced with more fundamental theories. For the effective theory of pions,
the fundamental theory is Quantum 
Chromodynamics, but in principle nothing forbids that the fundamental theory cannot even be expressed as a Quantum Field Theory.

\subsubsection{Behavior at fixed points}
\label{sec:betafunctions}

The renormalization techniques in QFT have similar features to the dynamical systems we described earlier. 
In fact, equation \eqref{eq:betadimless} looks a lot like the subdivision of eigenvectors of a hyperbolic fixed point into 
relevant and irrelevant, where $\Delta_{(n,0)}=d- \frac{d-2}{2}n$ plays the role of critical exponents, specifying how the corresponding operator 
behave as the energy scale $\mu$ is varied. The flow of the marginal couplings with $ \Delta_{(n,0)} = 0$ is entirely given by the quantum
correction $\beta_{\lambda^{(n,0)}}$. One is thus naturally lead to classify the fixed points of the renormalization group flow
\[
 \mu \frac{d\Gamma_{\mu}[\varphi]}{d\mu} =0 \; .
\]
At vanishing coupling constants all the $\beta$-functions are zero and we obtain a fixed point. This fixed point is the free theory 
which does not get any quantum corrections and is called the Gaussian fixed point.

\begin{figure}[!t]
  \begin{center}
    \subfigure[$\beta_g>0$ and increasing.]{\label{fig:betalandau}\includegraphics[width=4cm, angle=-90]{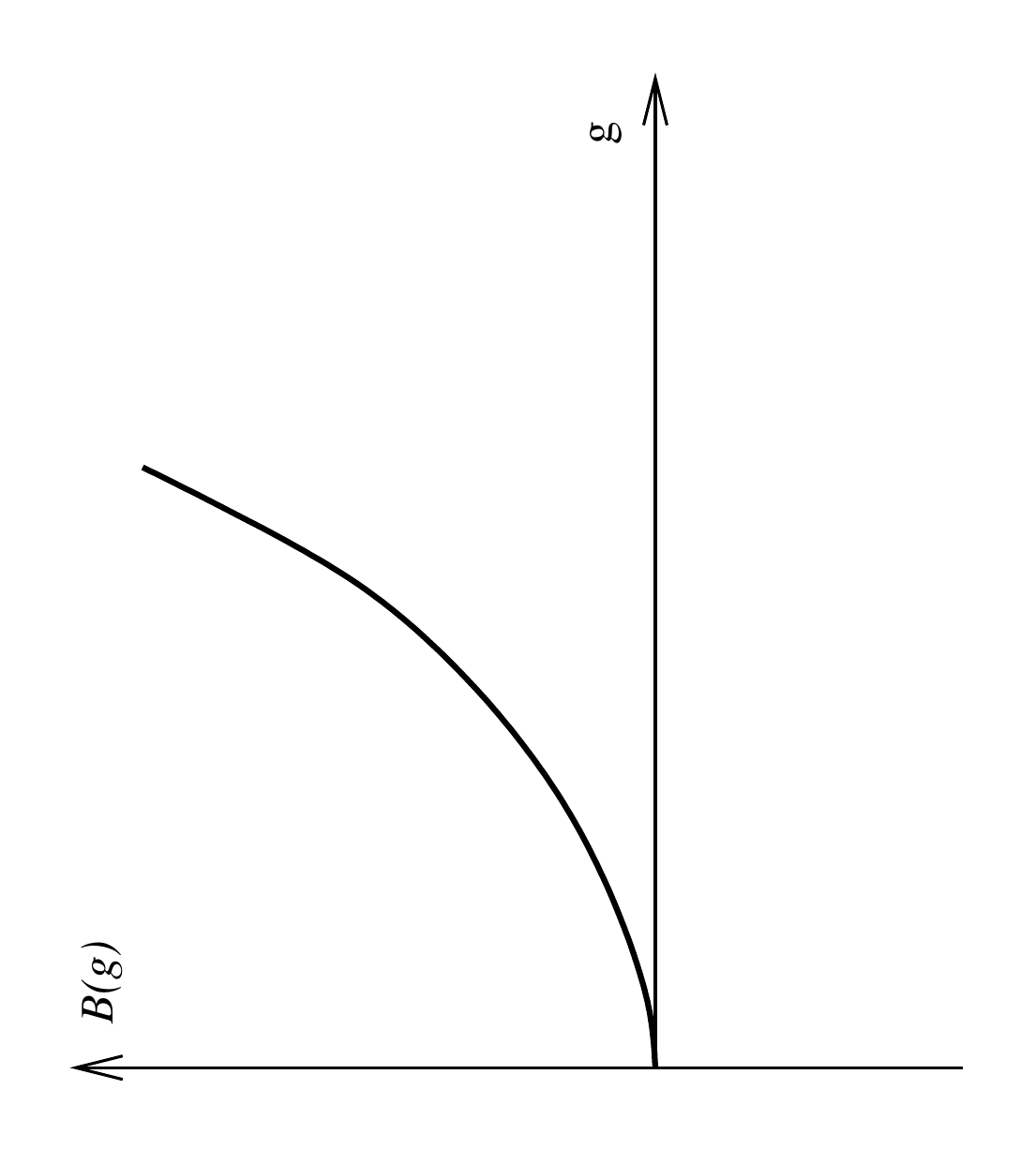}}
    \subfigure[$\beta_g<0$ and decreasing.]{\label{fig:betafree}\includegraphics[width=4cm, angle=-90]{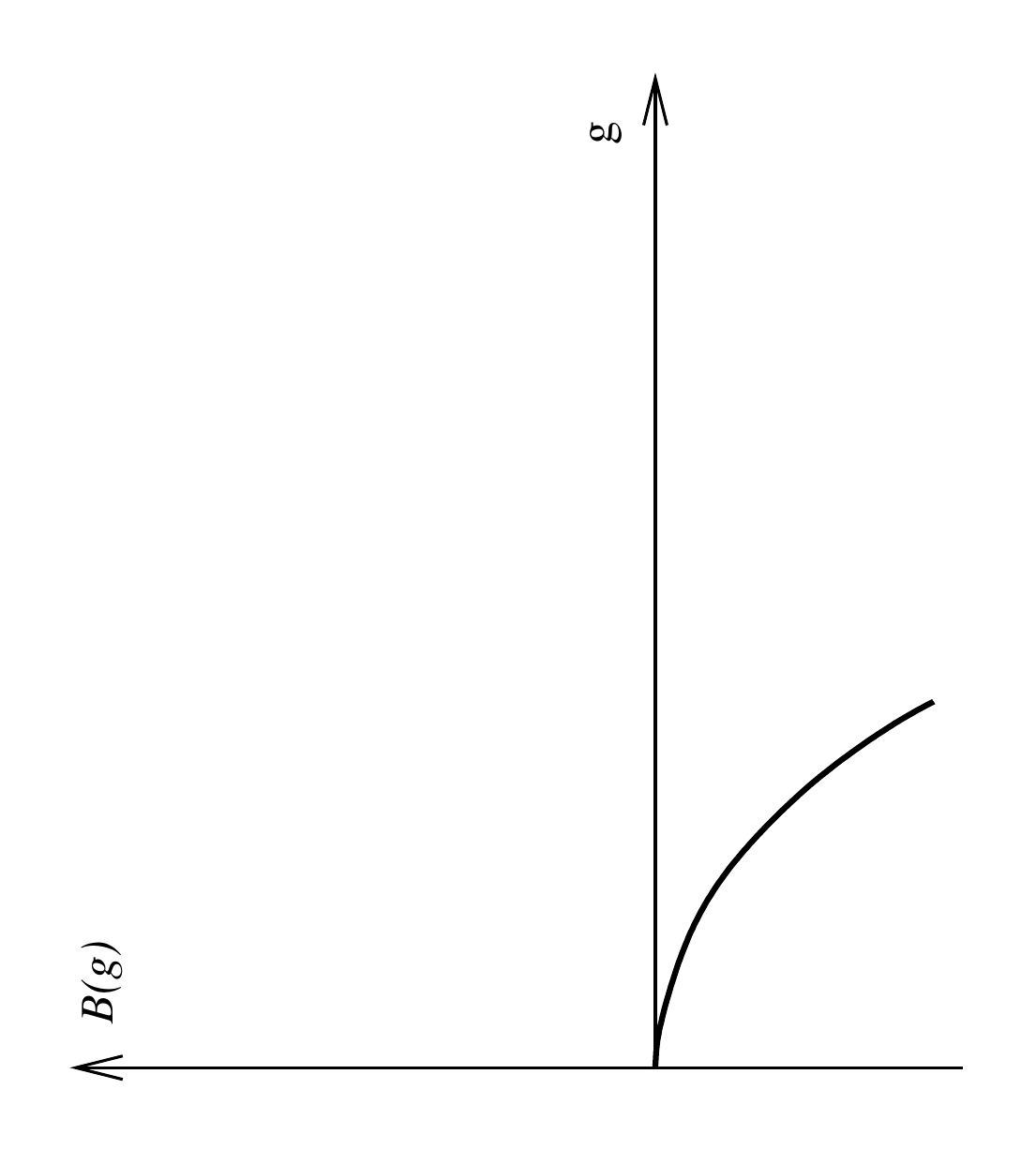}}
     \subfigure[Non Gaussian fixed point.]{\label{fig:betasafe}\includegraphics[width=4cm, angle=-90]{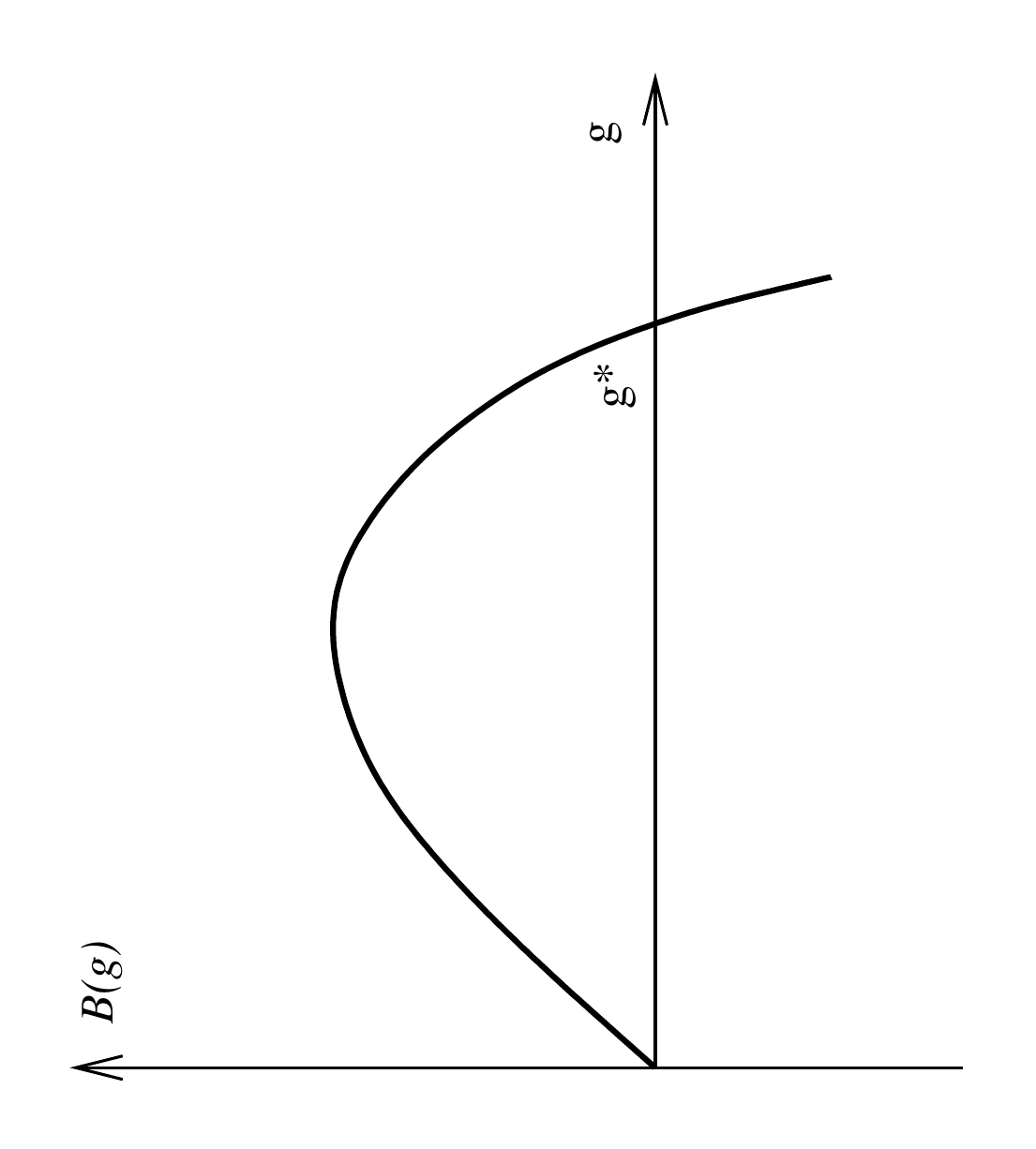}}

    \end{center}
  \caption{Possible qualitative behaviors of $\beta(g)$.}
  \label{fig:betaqualitative}
\end{figure}

The stability properties of any fixed point depend on the sign of the $\beta$-function in its neighborhood.
The most interesting couplings to study perturbatively are the marginal ones (like $\lambda^{(4,0)}$ for the $\phi^4_4$ model)
for which quantum corrections are crucial. Consider $g$ a marginal coupling and say $g\ge 0$ is its physical regime. Qualitatively, we have
\begin{enumerate}
\item Figure \ref{fig:betalandau}, $\beta_g>0$. In this case, $g(t)$ increases towards high energies. Even 
if we measure the coupling to be small at low energies (as happens for the fine structure constant $\alpha_{\rm em}$), we have 
to expect a UV regime where perturbation theory breaks down. On the other hand, in the IR the theory just sinks in the Gaussian fixed point.
Remark that if $\beta_g$ grows with $g$, it is possible that the solution of the differential equation
does not exists for any $t$, i.e. the coupling blows up to infinity in a finite time (i.e. at finite scale).

\item Figure \ref{fig:betafree}, $\beta_g<0$. Since it is the same scenario up to one sign, 
the preceding discussion is valid up to swapping the IR and the UV limits. In this case, the coupling 
becomes smaller and smaller at high energies, and the theory is \emph{asymptotically free}, which is the case of e.g. QCD.

\item Figure \ref{fig:betasafe}. $\beta_g=0$, for some non trivial coupling constant $g^*\neq0$. We have two possibilities, depending on the sign of $\beta_g$. 
Let us take it to be positive as in the figure. In this case, taking an initial condition $g_0<g^*$ the coupling would grow at high energies,
approaching $g^*$ and decreases in the IR approaching the Gaussian fixed point $g=0$. The theory is well behaved both in the UV and the IR, 
even if, depending on how large $g^*$ is, it may not be possible to describe it just perturbatively.
The same picture holds when $\beta_g<0$, up swapping IR and UV.  We call these \emph{asymptotically safe theories}.

\end{enumerate}

A generic dynamical system can have a much richer behavior. If we consider many coupled $\beta$-functions,
limit cycles or even chaotic behavior may arise. For certain classes of QFTs it is possible to constrain these behaviors
\cite{Zamolodchikov:1986gt,Komargodski:2011vj}, at least to some extent \cite{Curtright:2011qg}; this is an active topic of research \cite{PhysRevLett.89.230401,Glazek:2004zz,Fortin:2012ic}.

\newpage

\section{Multi-scale analysis in perturbative renormalization}
\label{sec:multiscale}

In this section we will quantitatively see how the renormalization group procedure introduced in the previous section can be realized in a diagrammatic expansion. The key ingredient to this end is a multi-scale representation.

\subsection{The multi-scale representation}

The renormalization group consists in a combination of three elements: a scale analysis, a locality principle and 
a power counting in the iterated computation of the effective action.

The first ingredient, scales, is essential. Physics could almost be \emph{defined} as mathematics plus 
scales. Scales are related to observations, and to so-called cutoffs, hence to the deep physical question of separating
the observer from the observed physical phenomenon. It is therefore not surprising that scales play
an increasing role from classical physics to quantum physics, in which this separation includes some
new puzzling aspects, and from quantum physics to quantum field theory, in which the infinite number of
degrees of freedom and particle production and annihilation adds further subtleties. 

There are several technical ways to introduce scales and cutoffs in QFT, depending on the specific model under study. 
In  statistical mechanics or for lattice models the technique of block spin transformations is quite natural.
It consists in writing, in a sequence of scaled lattices, each field variable as an averaged field in some 
cube of the next scale, plus a fluctuation field. Another mathematical technique consists 
in decomposing the field on an orthonormal wavelets basis.
Each method has some advantages and drawbacks.

In QFT models which naturally decompose as Gaussian measures perturbed by non-Gaussian interactions,
slicing the propagator is probably the optimal way to introduce scales and organize the theory and its renormalization. 
Usually slicing the propagator is done through its parametric heat-kernel representation.
This is compatible both with perturbative and constructive purposes, and 
extends nicely to essentially all extended types of quantum field theories: those used in condensed matter,
which have no Lorentz invariance, those formuated 
on curved spaces, or even more exotic ones introduced for quantizing space time itself,
such as non-commutative field theories or group field theories.

\subsubsection{Scales and slicing}
We start with the following (physical rather than mathematical) definition:

\begin{definition}
A \emph{physical scale} for a quantum field theory is a gently cut slice of eigenvalues 
of the propagator, according to a geometric progression.
\end{definition}
This definition contains still some arbitrariness.
First to define the geometric progression we need a fixed number $M>1$; the scale $i$ is then composed
of the eigenvalues of the propagator roughly between $M^{-2i}$ and $M^{-2(i+1)}$.
Possible choices for $M$ are $M=10$ (anthropomorphic choice), $M=2$ (computer scientists choice)
or $M=e$ (mathematicians choice); in the last case the scale is also called an e-fold.
Then the slicing itself is done with some arbitrary partitioning of unity which defines the scale cutoffs. 
Gentle smooth cutoffs lead to good decay in various representations.
In practice the sharp parametric slicing is an excellent choice in all concrete cases met up to now.

\begin{definition}[Sharp Parametric Slicing]
Let $C= 1/H$ be the propagator of the theory. The sharp parametric slicing is
\bea  C &=& \int_0^\infty  e^{- \alpha H}  d \alpha =  \sum_{i=0}^{\infty}  C^{i}   ,  \crcr
C^{i} &=&  \int_{M^{-2i}}^{M^{-2(i-1)}} e^{- \alpha H}  d \alpha \; ,
\; \;C^{0} =  \int_{1}^{\infty} e^{- \alpha H}  d \alpha \; . \label{decoab1}
\eea
\end{definition}
The natural ultraviolet cutoff on the theory is then the maximal value $\rho$ we allow the index $i$ to take.
The covariance with UV cutoff is 
\bee  C_{\rho} =  \sum_{i=0}^{\rho}  C^{i}   \label{decocut1} 
\ee
for finite and large integer $\rho$. In the case of the Laplacian plus mass on $\R^d$ we get the following slices
\bea   
C^{i}(x,y) &=&  \int_{M^{-2i}}^{M^{-2(i-1)}} e^{-m^{2}\alpha -  
{\vert x-y \vert^{2} \over 4\alpha }}  {d\alpha \over \alpha^{d/2}}  \crcr
 C^{0}(x,y) &=&  \int_{1}^{\infty} e^{-m^{2}\alpha -  
{\vert x-y \vert^{2} \over 4\alpha }}  {d\alpha \over \alpha^{d/2}} \; .  
\eea
Each propagator  $C^{i}$ corresponds to a theory with
both an ultraviolet and an infrared cutoff. They differ by the fixed 
multiplicative constant $M$, the slice ``thickness". 

The decomposition \eqref{decoab1} is the multislice representation. Associated to the multislice representation 
we have a splitting of the Gaussian measures $d\mu_{\rho}$ of covariance $C_{\rho}$ 
into a product of independent Gaussian measures $d\mu^{i}$ with covariance $C^i$
\bee 
\phi_{\rho} = \sum_{i=0}^{\rho} \phi^i ;\quad d\mu_{\rho}(\phi_{\rho})
= \otimes_{i=0}^{\rho} d\mu^{i}(\phi^{i})    \;,
\ee
where the fields $\phi^{i}$ are independent. It is an easy exercise to derive the following bound:
\begin{lemma}
There exist positive constants
\footnote{By convention we use 
$\delta,\delta',\epsilon,\zeta,...$ as 
generic names for small constants and $K,K_1,c,...$ for large ones.} 
$K>1$ and $\delta<1$ such that:
\bea  C^{i}(x,y)   &\le&   K  M^{(d-2)i}   e^{-  \delta M^{i} \vert x-y \vert}  , \label{bou1} \\
\partial_{\mu_1} \dots  \partial_{\mu_k} C^{i}(x,y)   &\le&   K_k  M^{[(d-2)+k]i}   e^{-  \delta M^{i} \vert x-y \vert}  . \label{bou2}
\eea
\end{lemma}

This bound captures the significant aspects of the ultraviolet and infrared cutoffs. The 
overall factor $M^{(d-2)i}$ signals that the singularity of $C$ at 
coinciding points has been smoothed by the ultraviolet cutoff at a certain 
scale. The scaled spatial decrease $e^{-  \delta M^{i} \vert x-y \vert}$ signals the infrared cutoff.
(In the case $i=0$, $\delta$ can be taken as any number less than $m$, the mass 
appearing in $C$). This is natural from the point of view of Fourier analysis; 
better ``spatial resolution" costs a worse overall power counting factor. The second bound expresses
that arbitrary derivatives of $C^i$ also scale as expected.

Using the slice decomposition we rewrite the bare amplitude for a Feynman 
graph as:
\bea   A_{G} &=&  \sum_{\mu \in {\N}^{\;l(G)}} A_{G,\mu}\; \label{II.1.8} \; ,\\
A_{G,\mu } & =& \int  \prod_{v} dx_{v} \prod_{l {\rm \ internal \ edge \ of 
\ } G} C^{i_{l}} (x_{l},y_{l}) \prod_{l {\rm \ external \ edge \ of 
\ } G} C_{\rho} (x_{l},y_{l}) \; . \nonumber
\eea
where $\mu$ is called a ``scale assignment" (or simply ``assignment").
It is a list of integers, one for each internal edge of $G$, which 
provides for each internal edge $l$ of $G$ the scale $\mu(l)$ (also noted $i_l$) of that edge. 
$A_{G,\mu}$ is the amplitude associated to the pair $(G,\mu)$, and 
\eqref{II.1.8} is called the multi-scale representation of Feynman amplitudes.

To understand the basic aspects of the ultraviolet limit it is simpler not to decompose external 
edges into slices and to use the convention that they have a fictitious scale index -1, lower than any internal scale. 
Decomposition of external edges into slices may be useful for detailed results, e.g. on the asymptotic behavior as some set of 
external momenta are scaled, but is not necessary at this stage.

The bare amplitude in \eqref{II.1.8} appears as an integral over the position of the vertices 
and over scale assignments, i.e. a sum over a $d+1$-dimensional space with $d$ continuous spatial dimensions 
and one discrete dimension, which we call the scale-space\footnote{This space is traditionally called the ``phase space"  in constructive field
theory but this name is slightly misleading; it is obviously not the 
2$d$-dimensional cotangent bundle associated to the $d$-dimensional space.}. 
This discrete scale-space is only a half-space, because we are  interested here in the ultraviolet limit. 

\begin{figure}[ht]
\centerline{   \includegraphics[width=5cm]{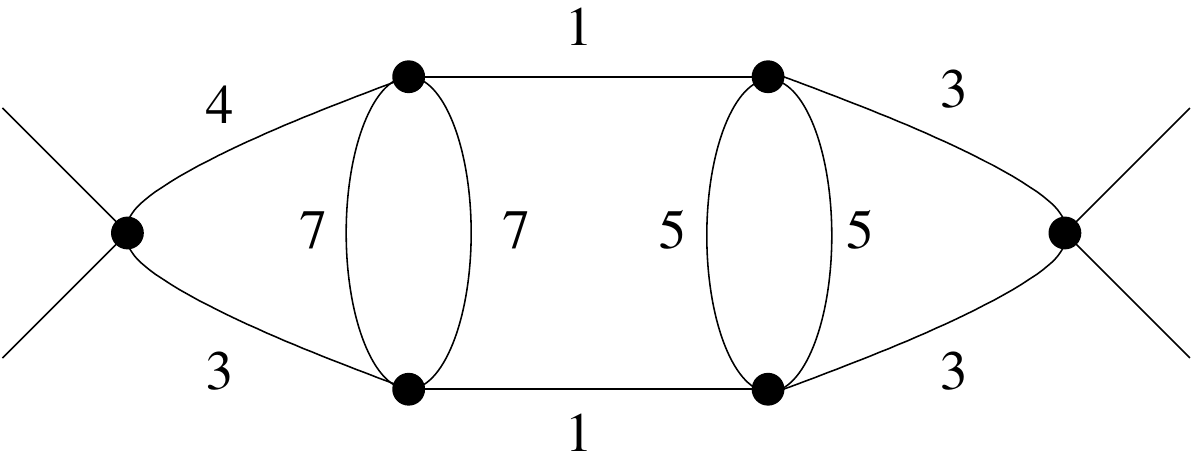}   }
\caption{A graph with a scale assignment}
\label{scales1}
\end{figure}

\begin{figure}[ht]
\centerline{   \includegraphics[width=8cm]{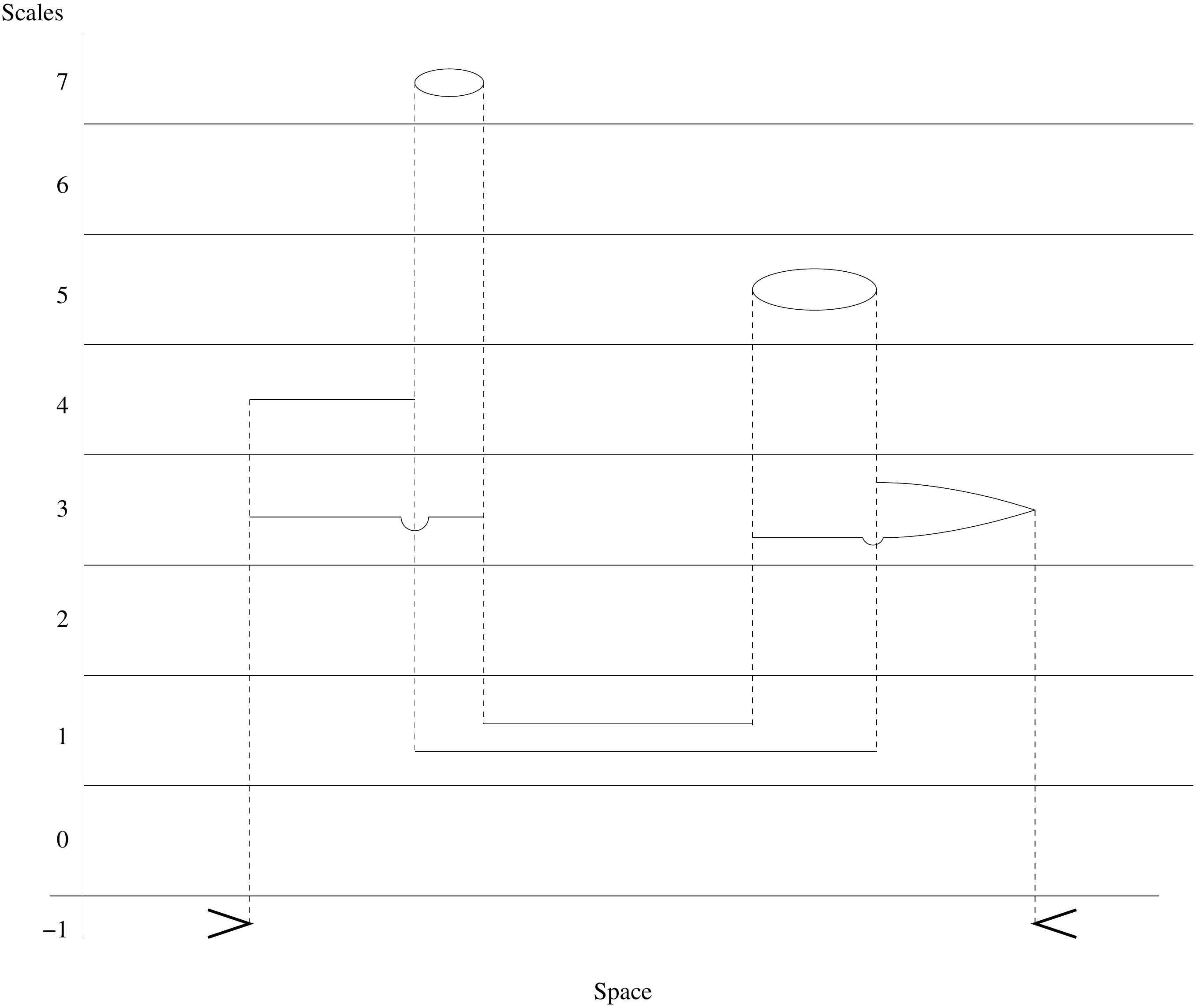}   }
\caption{The corresponding scale-space representation}
\label{scales2}
\end{figure}

In this scale-space edges and 
vertices play a dual r\^ole. To support this 
intuition we draw a two-dimensional picture, 
using the horizontal direction to represent  the ordinary $d$ 
dimensions of the space, and the 
vertical one to picture the discrete scales, the highest ones at 
the top. Then a propagator belongs to the scale of its index and appears as a 
horizontal edge joining two vertices. The internal vertices of the graph sit at a particular 
point in space and join half-edges which may be located in different scales. 
Hence it is convenient to picture them as vertical lines connecting the
horizontal half-edges hooked to them. These lines are dotted to distinguish 
them from the first ones, which correspond to true edges of the graph. Finally the external (half)-edges are pictured in the
fictitious ``-1" scale, hence at the bottom of the picture, see Figures 
\ref{scales1} and \ref{scales2} for an explicit example.


\subsubsection{High subgraphs}

We consider \eqref{II.1.8} and perform first the integration over the positions of 
the vertices in ${\R}^d$. 
To integrate over the positions of internal vertices (save one, $v_{0}$) 
requires at least the decay of a spanning tree, which is a minimal set of edges
connecting all the vertices. In order to obtain the best possible bound according to 
the scale assignment, one should use as much as possible the decay of the 
edges with highest possible index. 

The optimal choice of $T$ for a given $\mu$ 
is made by starting from the highest scale $\rho$ and descending towards lower scales. When reaching the scale $i$ 
we consider the connected components of 
$G^{i}$, the subgraph of $G$ made of all edges with index $j \ge i$. Let us 
call these connected components $G_{k}^{i}$, $k=1,...,k(G^i)$ (where $k(G^i)$ denotes the number of connected components of the graph $G^i$).

There is a systematic way to know whether a given connected
subgraph $g \subseteq G$ is a $G_k^i$
for some $i$ and $k$ or not. We define the internal and external index for $g$ in 
the assignment $\mu$ as:
\bee   i_g(\mu)  = \inf\limits_{l {\rm \ internal \ edge \ of \ } g}  \mu(l)  \label{II.1.10}
\ee
\bee   e_g(\mu)  = \sup\limits_{l {\rm \ external \ edge \ of \ } g}  
\mu(l)  \label{II.1.11}
\ee
(with the $\mu$ dependence sometimes omitted for shortness). 
\begin{definition}[High Subgraphs]
Connected subgraphs verifying the condition 
\bee   e_g(\mu)  < i_g(\mu)  \quad  \quad {\rm (high \ condition)}  
\label{II.1.12}
\ee
are called \emph{high}. This definition depends on the assignment $\mu$. For
a high subgraph $g$ and any value of $i$
such that $e_g(\mu ) <i\le i_g (\mu )$ there exists exactly one value of
$k$ such that $g$ is equal to a $G_k^i$. 
\end{definition}

\begin{figure}[ht]
\centerline{\includegraphics[width=8cm]{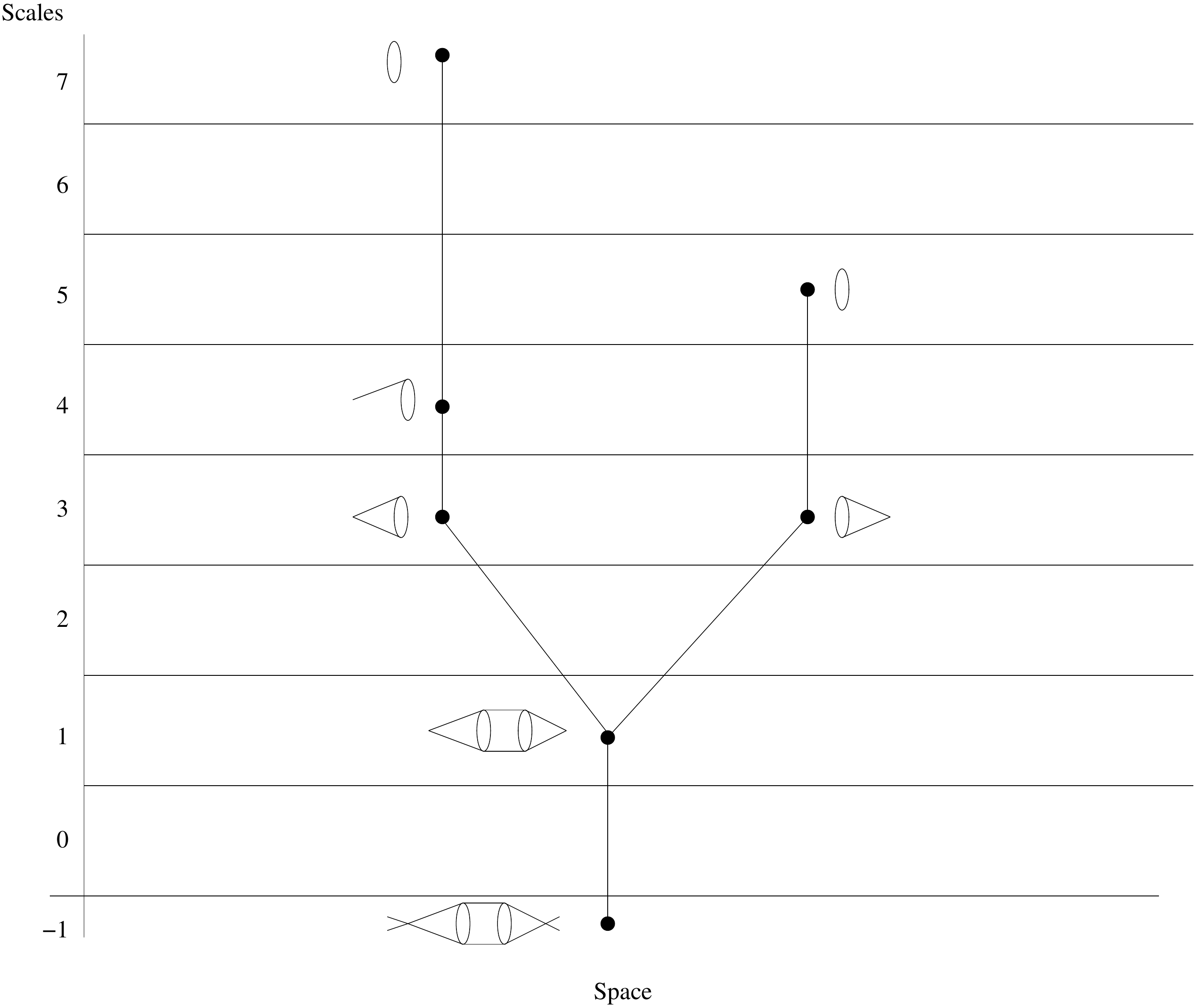}   }
\caption{The i-tree of high subgraphs for the previous graph and scale assignment}
\label{scales3}
\end{figure}

High subgraphs are partially ordered by inclusion. An essential result is that they form an i-forest in the sense of Definition \ref{iforest}
(see Figure \ref{scales3}).
\begin{lemma}
Let $(G,\mu)$ be a fixed graph and scale assignment.
The set of high subgraphs is an i-forest. If $G$ is connected it is an i-tree.
\end{lemma}
\prf
Suppose $(G,\mu)$ is not an i-forest. There exist then $S_1$ and $S_2$ with a non trivial intersection. 
In this case $S_1$ would have an external edge which belongs to $S_2$ and conversely. 
But the scale of any of these two edges should be both strictly larger
and strictly smaller than the other, which is impossible.

\qed

The i-tree of the graph and scale assignment of figure \ref{scales1} is presented in figure \ref{scales3}.
Duplicating the high subgraphs over every scale for which they are high we obtain a
more redundant tree which has a node for every $G^i_k$. 

\subsubsection{Convergence degree, convergent graphs}

Returning to the choice of $T$ to optimize the bound of an amplitude and assignment,
we have to pick a tree $T$ compatible with the i-tree of high subgraphs. 
Such a choice is always possible through Lemma \ref{compat}. 
We recall briefly how this tree is chosen. We start from the highest scale $\rho$ and descend. At slice index $i$, we 
complete the set of edges already chosen to a tree in every subgraph $G^i_k$ (i.e. to a spanning forest of $G^i$).
Note that the tree $T$ thus obtained is not unique~\cite{Gallavotti:1985yj,Gallavotti:1985qc}.

\begin{figure}[ht]
\centerline{\includegraphics[width=5cm]{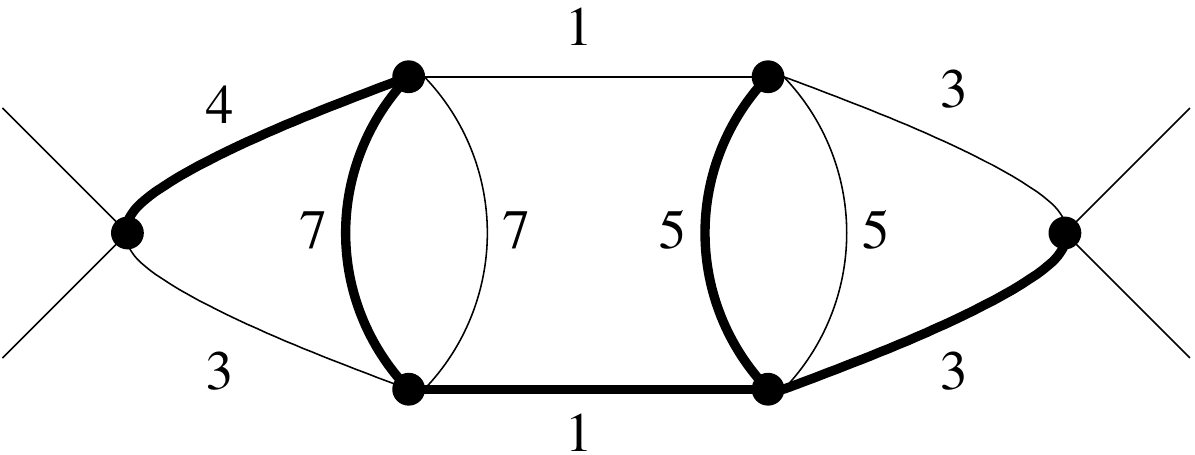}   }
\caption{A tree $T$ compatible with the i-tree of high subgraphs is shown in bold}
\label{scales4}
\end{figure}

A particular tree compatible with the scale assignment for the graph and scale assignment in figure \ref{scales1} is pictured in Figure \ref{scales4}.

The two trees of figures \ref{scales3} and \ref{scales4}  have very different 
meaning. The one in figure \ref{scales3} is an abstract picture of the inclusion 
relations derived from the assignment $\mu$, hence it is an i-tree. The tree $T$ of figure \ref{scales4} is a 
concrete set of edges of $G$, itself with a scale assignment. The essential point is that the tree of figure \ref{scales4}
can always be chosen to be compatible with the one of figure \ref{scales3}.

If we use the decay of the propagators in $T$ to integrate over the positions of 
the vertices  we obtain  an upper bound on the integral over the internal positions of all but one of the vertices of $G$
\bee   \int  \prod_{v \ne v_{0}} d^dx_v  \prod_{l \in G} e^{-\delta 
 M^{i_{l}+1} \vert x_l - y_l \vert }  \le  \int  \prod_{v \ne v_{0}} d^dx_v 
 \prod_{l \in T} e^{-\delta  M^{i_{l}+1} \vert x_l - y_l \vert } 
\ee
\bee \quad \quad \quad =  \prod_{\ell \in T} K M^{-d(i_l+1)} = \prod_{v \ne v_{0}} K M^{-d(i_{v}+1)} 
\label{II.1.13}
\ee
where $K$ is some constant and $i_v$ is the index in $\mu$ of the first edge in the unique path in $T$
starting at $v$ and going to the root $v_0$ (i.e. the edge hooked to $v$ in this path).
  
The choice of $T$ to be compatible with the i-tree of the high subgraphs ensures that  the 
$i_{v}$'s are as large as possible, hence that the bound \eqref{II.1.13} is optimal. 
Remark that for any $G_{k}^{i}$, every vertex $v$ save one has $i_{v} \ge i$ (as $T \cap G_{k}^{i}$ is a tree). 

Taking into account the prefactors $M^{(d-2)i}$ we obtain the bound on the amplitude of $G$ at scale attribution $\mu$
\bea
 |A_{G,\mu}| \le \Bigl( \prod_{l\in G} M^{(d-2)i_l} \Bigr) \Bigl( \prod_{v \ne v_{0}} K M^{-d i_{v} }  \Bigr) \; ,
\eea 
where $K$ denotes some constant. The two products above can be reorganized in terms of a product over the $G^i_k$
as
\bea  \prod_{l\in G} M^{(d-2)i_l} &=&  \prod_{l\in G} \Bigl( \prod_{1\le j \le i_l}    M^{d-2} \Bigr) 
= \prod_{i\ge 1} \prod_{k=1}^{k(G^i)} \prod_{l\in G^i_k} M^{d-2} \crcr
 & = & \prod_{i\ge 1} \prod_{k=1}^{k(G^i)} M^{(d-2)E(G^i_k)} , \crcr
 \prod_{v \ne v_{0}} M^{-di_v}   & = & \prod_{v \ne v_{0}} \Bigl(  \prod_{1\le j \le i_v}    M^{-d} \Bigr) = 
\prod_{i\ge 1} \prod_{k=1}^{k(G^i)} M^{-d[V(G^i_k)-1]} \; ,
\eea 
where $E(G^i_k)$ and $V(G^i_k)$ denotes the number of edges and vertices of $G^i_k$. We then obtain the bound
\bee
|A_{G,\mu}| \le K^{V(G)} \prod_{i \ge 0} \prod_{k=1}^{k(G^i)}   M^{-\omega(G_{k}^{i})}  \;, \label{II.1.14}
\ee
where $K$ is some constant and
\bee  \omega (S)  = -(d-2) E(S)  + d( V(S) -1) \; ,
\ee
is the (superficial) \emph{degree of convergence} of $S$ (that is minus its degree of divergence we introduced before).
Using the topological relation $nV = 2E +N$ for a graph with $n$ valent internal vertices and $N$ external edges we have 
\bee 
\omega (G) = \frac{d-2}{2} N(G) -d - V(G) \Bigl( \frac{d-2}{2} n  -d \Bigr) \; .
\ee
If $d=4,n=4$ then $\omega(G) =N(G)-4 $.

\begin{definition} A completely convergent (connected) graph is a connected graph $G$ for which $\omega (g) > 0 $ $\forall g \subseteq G$. 
\end{definition}

\subsection{Uniform bounds for convergent graphs}
\label{sec:uniformbounds}

\begin{figure}[ht]
\centerline{\includegraphics[width=6cm]{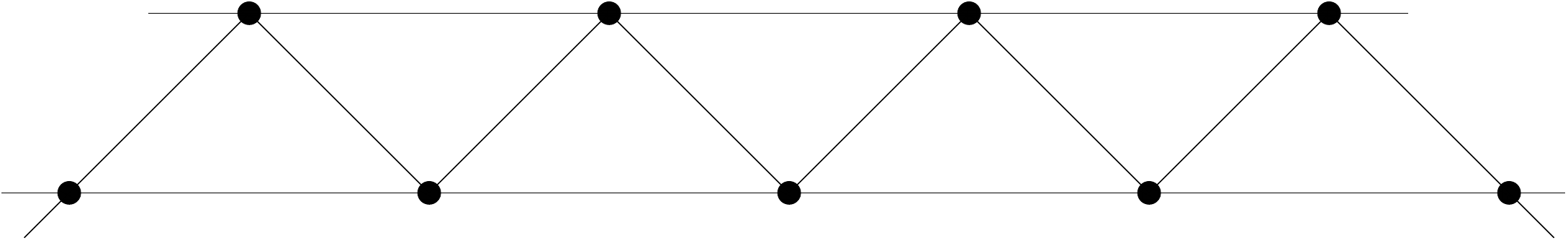}}
\caption{A family of convergent graphs}
\label{graph9}
\end{figure}

We concentrate now on the $\phi^4_4$ model, that is $d=4,n=4$ and show how to
bound a completely convergent graph $G$. A family of such graphs
is shown in Figure \ref{graph9}.

Of course there are no vacuum graphs of this type, since $G$ must have at least six external edges itself to have $\omega (G) >0$. 
Hence we need to precise the treatment of external edges.

\begin{theorem}[Weinberg convergence theorem]\label{convbound}
There exists a constant $K$ such that for any completely convergent 
$\phi^{4}_{4}$ graph the Feynman amplitude of $G$ is  
absolutely convergent and bounded by 
\bee   \vert A_{G} \vert  \le  K^{V(G)} \cdot Ext   \label{II.1.9}
\ee
where the function $Ext$ depends on the way external arguments are treated. 
\end{theorem}

The point to emphasize in \eqref{II.1.9} is the uniform exponential 
character of the bound at large order expressed by the factor
$K^{V(G)}$; the particular form of $Ext$ is not essential here and several
possible forms are admitted
\begin{itemize} 
\item{H1} {Each external vertex is integrated over a standard unit cube of 
${\R}^4$, }
\item{H2} {Each external vertex $v$ is integrated against a test function 
$f_{v}$ on ${\R}^4$,}
\item{H3} {Each external vertex has a fixed external momentum $k_{v}$ entering 
it.}
\end{itemize}
In these three cases a possible choice for $Ext$ would be:
\begin{itemize} 
\item{H1}  $Ext = \sup\limits_{x_{v},v \in V_{E}} \
\prod\limits_{l \in G} e^{-m(1-\zeta) 
\vert x_l - y_l \vert } $,
\item{H2} $Ext = \inf\limits_{v \in V_{E}} \Vert f_{v} \Vert_{L^1} 
\ \prod\limits_{w\ne v \in 
V_{E}} \Vert f_{w} \Vert_{L^{\infty}}$,
\item{H3} $ Ext = \delta(\sum\limits_{v \in V_{E}} k_v ) $.
\end{itemize} 

To underline the essential part of the argument, let us prove 
the theorem in the slightly simpler case of an amplitude for which 
external propagators have been amputated, and exactly one internal
vertex $v_{0}$ is fixed to the origin (not integrated over ${\R}^4$). This 
is in essence the case $H3$, because each external propagator 
$(p^{2}+m^{2})^{-1}$ is bounded by $m^{-2}$, and fixing $v_{0}$ at the origin comes to factoring the 
the overall $\delta$ function of $Ext$.
\medskip

{\noindent \bf Proof of Weinberg's theorem.} We define 
\bee e_v(\mu)  =  \sup_{l {\rm \ hooked \ to \ }v} \mu(l)   \; ,
\qquad  i_v(\mu)  =  \inf_{l {\rm \ hooked \ to \ }v} \mu(l)  \label{II.1.17}
\ee
where the $\inf$ in \eqref{II.1.17} is over $every$  edge hooked to $v$, including 
external edges, which by convention have index $-1$.
Since $G$ is completely convergent (and $N(g)$ is even for all $g\subseteq G$) we have:
\bee 
\omega ( G_{k}^{i}) = N(G_{k}^{i}) -4 \ge { N(G_{k}^{i}) \over 3}   \; .
\label{II.1.15} 
\ee
We remark also that for any $i$, a given vertex $v$ belongs to exactly 
one $G_{k}^{i}$ for $i \le e_v(\mu)$ and to none otherwise. Furthermore 
some external edges of this $G_{k}^{i}$ are hooked precisely at $v$
if and only if $i_v(\mu) < i \le 
e_v(\mu)$. Hence, using \eqref{II.1.15}:
\bee  \prod_{i \ge 0} \prod_{k=1}^{k(G^i)} M^{-\omega(G_{k}^{i})} 
\le \prod_{i \ge 0} \prod_{k=1}^{k(G^i)}   M^{- \frac{1}{3} N(G_{k}^{i})} 
\le  \prod_v   M^{-{1 \over 3}
\vert e_v(\mu) -i_v(\mu) \vert } \; . \label{II.1.18}
\ee
This bound means that for a completely convergent graph, after spatial integration, the vertices 
pictured as dotted lines in Fig. \ref{scales2} acquire an exponential decay in their
length $\vert e_v(\mu) -i_v(\mu) \vert $ in the vertical direction.  It is now intuitively obvious that this decay should make the sum 
over momentum assignments easy, the external edges with their index $-1$ 
breaking vertical translation invariance, and playing 
a dual r\^ole to the one of the fixed vertex $v_0$.

Let us make this intuition more precise, with no efforts to find optimal constants. Using the 
fact that there are at most 4 half-edges, hence at most 6
pairs of half-edges hooked to a given vertex, and that for such a pair 
obviously $\vert e_v(\mu) -i_v(\mu) \vert \ge \vert i_l -i_{l'} \vert$,
we can convert the decay in vertical length of \eqref{II.1.18} into a decay 
associated to each pair of half-edges hooked to the same vertex:
\bee  \prod_{v}  M^{-{1 \over 3}
\vert e_v(\mu) -i_v(\mu) \vert }
\le \prod_v  \Bigl( \; \prod_{(l,l') {\rm \ hooked \ to \ } v} M^{-{1 \over 18}
\vert i_l -i_{l'} \vert }\Bigr) \; ,  \label{II.1.19}
\ee
where the factor 18 is not optimal (it can be improved to 12 with negligible
effort).  

The analog of picking a tree to perform the spatial integration is 
to pick a total ordering of the internal edges of $G$ as $l_1,...,l_{E(G)}$ 
such that $l_1$ is hooked to $v_0$ and such that each subset
$\{ l_1,...,l_m\}$, $m\le E(G)$ is connected, which is clearly possible. For an edge $l_j$, we chose 
$l_{p(j)}$ with $p(j)<j$ one of the edges which share a vertex with $l_j$ and has a lower index. 
Using only a fraction of the decay in \eqref{II.1.19} we have:
\bee \prod_v  \Bigl(  \prod_{(l,l') {\rm \ hooked \ to \ } v} M^{-{1 \over 18}
\vert i_l -i_{l'} \vert } \Bigr)  \le \prod_{j=1}^{E(G)}
M^{-{1 \over 18} \vert i_{l_j} -i_{l_{p(j)}} \vert } \; .\label{II.1.20}
\ee
where by convention $i_{l_{p(1)}}=-1$, hence the total amplitude is bounded by 
\bee 
| A_G | \le   \sum_{\mu=\{i_1,...,i_{l(G)}\}} \prod_{j=1}^{E(G)} M^{-{1 \over 18} \vert i_{l_j} -i_{l_{p(j)}} \vert } 
=  \sum_{i_1,\dots i_{E(G)} }   M^{-{1 \over 18} \vert i_{l_j} -i_{l_{p(j)}} \vert } \; .
\label{II.1.21}
\ee
Starting from the leafs of the tree, we bound
\bea
 \sum_{i_{l_j}\ge 0}  M^{-{1 \over 18} | i_{l_{p(j)}} - i_{l_j} |} 
  \le 2 \sum_{i_{l_j}\ge i_{l_{p(j)}  }  }  M^{-{1 \over 18} | i_{l_{p(j)}} - i_{l_j} | } \le \frac{2}{1-M^{-1/18}} \; ,
\eea 
and iterating we conclude. 

\qed

\subsection{Renormalization}
At this point we move to the renormalization procedure, out of which we will obtain the renormalized series and the asymptotic expansion. We will then comment on the difference between the two, and in particular on the issue of renormalons.

\subsubsection{Locality, power counting}

Consider now the case of a graph which has divergent subgraphs.
These subgraphs create difficulties in summing over scale assignments only when they are high; they look almost local
also only when they are high. This is the basic reason why renormalization works.

\begin{figure}
\centerline{\includegraphics[width=8cm]{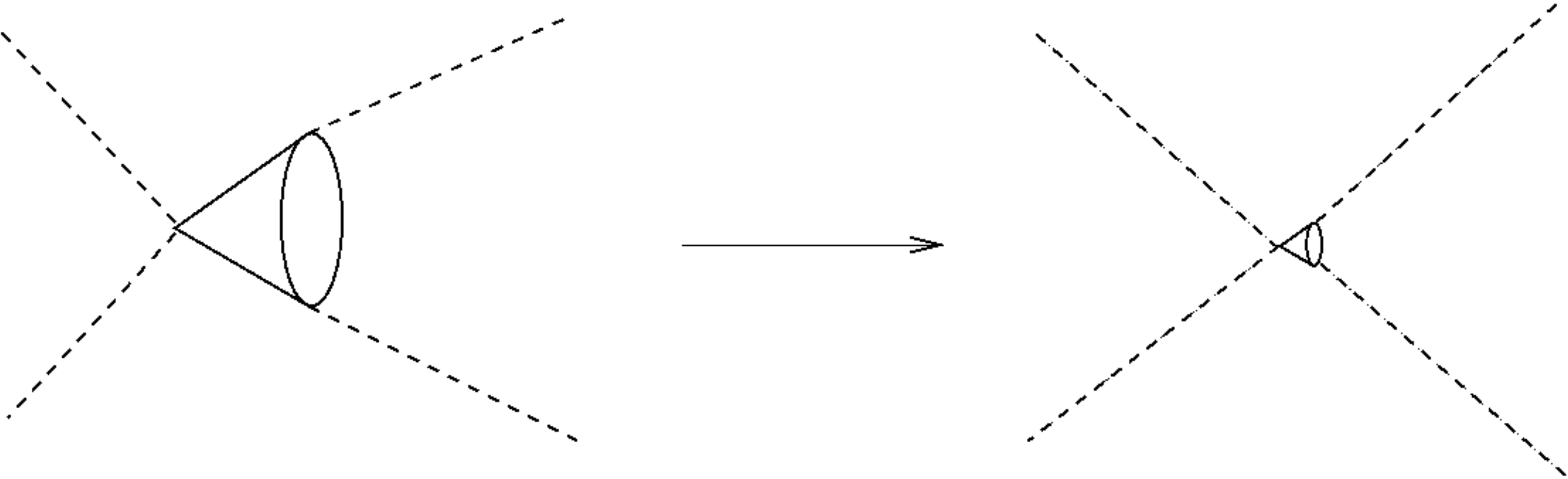}}
\caption{A high subgraph $S$ seen from lower and lower scales looks quasi-local.}
\label{graph2}
\end{figure}

Locality  simply means that \emph{high} subgraphs $S$
look \emph{almost local} when seen through their external edges. Indeed let $g = G^i_k$
for some $i$; since all internal edges have scale $\ge i$, and since $g$ is connected,
all the internal vertices are roughly at distance $M^{-i}$ or less.
But the external edges have scales $\le i-1$, hence only distinguish details larger than $M^{-(i-1)}$.
Therefore they  cannot distinguish the internal vertices of $S$ one from the other. 
Hence high subgraphs look like contracted
``fat dots", when seen through their external edges, see figure \ref{graph2}. 
Obviously this locality principle is completely independent both of the dimension 
and of the type of high subgraph considered. At the combinatorial level it 
corresponds to the contraction of subgraphs within graphs.%
\footnote{The combinatoric aspects of the renormalization group analysis can be understood in terms of Hopf algebras~\cite{Connes:1999yr}.}

Power counting is a rough estimate of the weight of a fat dot such as $S$ in figure \ref{graph2} with $N(S)$ external edges.
By \eqref{II.1.14}, this weight is $  M^{-\omega (S)}$. In contrast with locality, power counting depends on the dimension. For the 
$\phi^4_d$ model there are many possibilities.

\begin{itemize}
\item  If $d=2$, $  \omega (S) = -2 +2V(S)  $, hence primitively divergent
graphs have $V(S)=1$. The only divergence is logarithmic and due to the ``tadpole'' two point graph.
It can be removed by Wick-ordering.

\item  If $d=3$, we find $ \omega (S) = \frac{1}{2} N(S) -3 + V(S)  $, hence primitively divergent
graphs have $ N(S) =0, V(S)\le 3 $, $ N(S)=2, V(S) \le 2$ and the model is superrenormalizable.

\item  If $d=4$, $\omega (S) = N(S)-4$ and the model is just renormalizable.
Every two point graph is quadratically divergent
and every four point graph is logarithmically divergent.
This is in agreement with the superficial convergence degree of 
these graphs being respectively -2 and 0. 
The couplings that do not decay with $i$ all correspond to terms that 
were already present in the Lagrangian, namely
$\int \phi^4$, $\int \phi^2$ and $\int   \phi  (- \Delta)  \phi  $\footnote{Because the graphs 
with $N(S)=2$ are quadratically divergent we must Taylor expand the quasi local fat dots until we get 
convergent remainders. Using parity and rotational symmetry, this generates only a
logarithmically divergent $\int  \phi (-\Delta) \phi $ term beyond the 
quadratically divergent $\int \phi^2$, hence a Laplacian renormalization, also traditionally called ``wave-function renormalization". 
This wave-function renormalization starts only at $V(S)=2$
because the tadpole graph at $N(S)=2$, $V(S)=1$ is \emph{exactly} local.}. 

\item  Finally for $d >4$ the model is non renormalizable. 

\end{itemize}
 
Renormalizability means that we can define an expansion with finite ultraviolet limit (at least order by order).
To check that this is indeed the case in dimension 4, and to see how it works concretely
requires to introduce \emph{localization operators} that compute the 
local parts of high subgraphs and the corresponding remainders.

\subsubsection{Localization of high divergent subgraphs}

For fixed $(G, \mu)$
the i-forest of high subgraphs whose convergence degree is negative or zero is obviously
a subforest of the forest of high subgraphs. It is called the \emph{divergent} i-forest for $\mu$, and noted
$\D(G,\mu)$.

Consider a graph $G$ (which can contain divergent subgraphs). We first consider the assignments $\mu$ such that $\D (G, \mu)$ is empty.
In this case the proof of theorem \ref{convbound} goes through with no modification  because equation \eqref{II.1.18}
holds for all such $\mu$.

When $\D (G, \mu)$ is not empty it is natural, for any $g \in \D (G, \mu)$, to decompose the amplitude of the graph according to 
its local part (or quasi-local part with derivatives) plus a remainder. This is done using a 
localization operator $\tau^\star_g$ which pushes the expansion until the remainder converges.

In dimension 4 connected divergent subgraphs have 2 or 4 external edges, and are called bipeds and quadrupeds in \cite{Rivasseau:2011ri}.
Let us first for simplicity consider the case where the divergent forest   $\D (G, \mu)$ contains only 
quadrupeds. We use the shorthand notation $ {\bf Q} \equiv \D (G, \mu)$. For $g \in {\bf Q}$ we can define 
the localization operator $\tau_g^\star$ which contracts $g$ to a single vertex $v_e(g,\bf{Q})$ and attaches all the
external edges of $g$ to that reduced vertex\footnote{This is the $x$-space equivalent of taking external momenta 
to 0 in the momentum subtraction scheme of Zimmermann. It has the advantage of making the role of locality more transparent.}.

We can define the set of reduced vertices in an inductive way so that product of localization operators 
$ \prod_{g \in {\bf Q}} \tau_g^{*}(v_e (g,{\bf Q}))$, acting on the $x$-space integrand commutes. It is enough to 
chose for all the subgraphs $g$ a preferred or ``fixed" border-vertex $v_e(g,{\bf Q})$ (by border vertex we mean
an internal vertex hooked to an external edge). The following rule is a correct one but it is not unique.

Choose a border vertex arbitrarily for any of the maximal subgraphs of
${\bf Q}$ (the i-trees), but one which, if possible, is also a border vertex for 
$G$ itself. Then choose inductively the other border vertices
according to the natural rule: if $g'$ is the immediate ancestor of $g$ in the 
forest, which we note $g' = B_{{\bf Q}}(g)$ 
and $v_e(g',{\bf Q})$ is a border 
vertex of $g$ also, choose $v_e(g,{\bf Q})=v_e(g',{\bf Q})$. If  $v_e(g',{\bf Q})$ is not a border vertex of $g$ 
but there are some border vertices of $g$ which are also border vertices of $g'$, 
choose $v_e(g,{\bf Q})$ among them. Otherwise choose $v_e(g,{\bf Q})$ arbitrarily among the border vertices of $g$.

With this simple rule we may picture in a graphic way the
action of the product $\prod\limits_{g \in {\bf Q}} \tau_g^{*}
(v_e (g,{\bf Q}))$ on 
the integrand \bee  I_{G;\mu} =
\prod_{l}C^{\mu(l)}(x,y) \; .  \label{integ}
\ee
 Each $\tau_g^{*}$ operation simply replaces for every external edge of $g$ the propagator $C(x,z)$ by $C(x_{v_e(g,{\bf Q})},z)$,
 that is it moves each external edge of $g$ and attaches it to the single border vertex $v_e(g,{\bf Q})$. 

These operations are consistent and commute because 
whenever $g \subseteq g'$ our rule ensures that an external edge common to $g$ 
and $g'$ is never moved by the $\tau_g^{*}$ operator to an internal vertex of $g'$. 
The product  $\prod\limits_{g \in {\bf Q}} \tau_g^{*}(v_e (g,{\bf Q}))$ in fact 
is a commuting product and results in a single
well defined  set of ``moves" for the edges of $G$.
An important point to note is that, as ${\bf Q}$ is an inclusion forest associated to a scale assignment,
by construction if two graph $g_1,g_2\in {\bf Q}$ are not included one into the other, then $g_1$ and $g_2$ not only 
have no common edge, but also they have no common vertex. Indeed, if both $g_1$ and $g_2$ are high subgraphs and share a vertex 
$v$ then $i_{g_1}(v)>e_{g_1}(v)\ge i_{g_2}(v)> e_{g_2}(v) \ge i_{g_1}(v)$ which is impossible.

Returning to scale space, we decompose each amplitude  into scale
assignments by \eqref{II.1.8} and for any divergent high subgraph 
we systematically insert
\bee  1 = \tau_g^*(v_e(g,{\bf Q}) )  + [1 - \tau_g^*(v_e(g,{\bf Q})) ] .
\ee
The part with $ \tau_g^*(v_e(g,{\bf Q}) $ factorizes the amplitude into a local 
vertex corresponding to the contraction of $g$ to a point, times a coefficient.
It will be absorbed into an effective constant for lower scales. 
The part with $ 1- \tau_g^*(v_e(g,{\bf Q}) ) $ is a remainder which has good power counting.

The corresponding effectively renormalized amplitude is
\bee A_G^{ER}  = \sum_{\mu} \int \prod_v dx_v {\bf R_\mu}
\prod_{l}
C^{\mu(l)} (x_l,y_l)   \; ,  \label{II.3.1}
\ee
\bee {\bf R_\mu} \equiv \prod\limits_{g \in {\bf Q}}[1-\tau_g^*(v_e(g,{\bf Q}))]  \; .
\label{II.3.2}
\ee 

What we have gained is that these effectively renormalized parts now obey the same bounds as the convergent ones.
Indeed we can express every move of an external edge through a Taylor formula such as
\bea  
 && C^j(x, z) =  C^j( x_{v_e(g,{\bf Q})}  , z) \\
  && \quad  +  \int_0^1  dt \;  (x- x_{v_e(g,{\bf Q})}  ) \cdot \nabla C^j (x_{v_e(g,{\bf Q})}  + t (x-x_{v_e(g,{\bf Q})} ) , z ) \; . \nonumber  
\eea 

Using the bound \eqref{bou2} for the gradient on an external edge results in an additional factor $M^{e_g (\mu)}$ to the previous bounds. 
Bounding the difference $(x- x_{v_e(g,{\bf Q})}  ) $ by using a fraction of the decay of the inner edges of $g$ results in another factor, which is $M^{-i_g (\mu)}$
Hence the net effect of the $ 1- \tau_g^*(v_e(g,{\bf Q}) ) $ operator is to add a factor 
$M^{- (i_g (\mu) - e_g (\mu))}$ which exactly restores the vertical decay in \eqref{II.1.18} when and where it was missing.
Remark that high divergent graphs have $N \le 4$ so that \eqref{II.1.18} still holds, but with a factor 1/4 instead of 1/3.
Hence the bound \eqref{II.1.21} of the previous section holds\footnote{Remark that if we were to use localization operators for non-high
subgraphs, there would be no improvement but a worse factor than for the initial bounds. This is at the origin of the renormalon phenomenon
discussed in the next chapter.}.

We turn now to the local parts. Since they have the same form as the initial bare vertices
of the theory, adding them to the bare couplings we can \emph{and must} absorb them in the definition of effective or running couplings.

\subsubsection{The effective expansion}

The exactly local part of the high divergent subgraphs  generate
effective constants.

Again for simplicity let us consider first the biped-free piece of the perturbative expansion, 
and the flow of the coupling constant. The results can then be extended to 
the general case.
We fix a cutoff index $\rho$ and the bare coupling $g_\rho$. For each vertex $v$
of a graph $G$ it is useful to define
\bee  e_v (\mu)  =  \max \{ \mu(l) \vert l {\rm \ hooked \ to \  }v \} .
\label{II.4.4}
\ee
Recall that by convention the index of external edges is $-1$, so $e_v(\mu) = 
-1$ is possible but only for the four point graph made of a single vertex. 

The bare expansion for a connected Schwinger function with cutoff $\rho$
is defined as a formal power series in
the bare coupling  $g_\rho$:
\bee  S_{N}^{\rho}  =  \sum_n {(-g_{\rho})^{n} \over n! }\sum_{ \genfrac{}{}{0pt}{}{ G, \; V(G) =n,  }{ \mu, \; \mu \le \rho}  } 
A_{G,\mu} \;  . \label{II.4.5}
\ee
The sum is over assignments $\mu \in [0,\rho]^{l(G)}$ 
and over connected graphs at order $n$ with $N(G)=N$.

For simplicity let us define first an effective expansion for $S_{N,bf}^{\rho} $ the \emph{biped-free}
connected $N$ points function with ultraviolet cutoff $\rho$
which is the sum of all connected amplitudes
with $N$ external edges but \emph{without bipeds}.

\begin{theorem}[Existence of the effective expansion] \label{effexp}
There exist $\rho+1$ formal power series in $g_\rho
\equiv g_{\rho}^{\rho}$, called 
$g_{\rho-1}^{\rho}$, $g_{\rho-2}^{\rho}$, ..., $g_0^{\rho}$ and 
$g_{-1}^{\rho}$ (the upper index is to remind the reader that the entire theory 
has ultraviolet cutoff $\rho$) such that the formal power 
series \eqref{II.4.5} is the same as:
\bee   S_{N,bf}^{\rho} = \sum_{G,\;\mu \le \rho} [ \prod_{v \in G} 
(-g_{e_v(\mu)}^{\rho})] A_{G,\mu}^{ER} \;,
\label{II.4.6}
\ee
where the effective renormalized amplitudes $A_{G,\mu}^{ER} $
are defined by \eqref{II.3.1} and the effective constants $g_i^{\rho}$ 
obey the following inductive definition:
\bea && g_i^{\rho} = g_{i+1}^{\rho} - \sum_{ \genfrac{}{}{0pt}{}{  H {\rm \ quadruped,}\;\; i+1 \le  \mu \le \rho   }{ i_H (\mu) = i+1 }    } 
   \;\;\; \prod_{v \in H} (-g_{e_v(\mu)}^{\rho} ) \\
&& \qquad  \qquad  \qquad 
 \times \int  \Big[\prod\limits_{h \in {\bf D} ( H, \mu ), h \ne H} (1- 
\tau_h^{*} ) \Big](\tau^{*}_H) I_{G;\mu}
 \; , \nonumber \label{II.4.9}
\eea 
with $I_{G,\mu}$ defined in \eqref{integ} and $ i_H (\mu) $ is the common scale of all the edges of $H$.
\end{theorem}

\prf
The amplitude $A_{g,\mu}^{ER} $ is taken at zero external momenta, that is
all vertices are integrated save one.
In \eqref{II.4.9}, the summation over quadrupeds does not include the trivial case of 
the graph reduced to a single vertex, which corresponds in fact to the first 
factor $g_{i+1}^{\rho}$ in the right hand side of \eqref{II.4.9}.
The minus sign in \eqref{II.4.9} is explained by the fact that 
the vertex at scale $i$ has really a coupling $-g_i$, so that the equation reads 
$-g_i = -g_{i+1} + \sum_H {\rm counterterm}(H) $. 

The equation \eqref{II.4.9} defines each $g_i^{\rho }$ (by inductive substitution) as a formal 
power series in $g_{\rho}$ of the form $g_{\rho} + \sum_{n \ge 2} \gamma_n^i 
(g_{\rho})^n$. The induction stops at $g_{-1}^{\rho}$ 
which is the last one for which the sum in \eqref{II.4.9} is not empty.
Let us apply the result \eqref{II.4.6} to $N=4$ and 
put to 0 the four external momenta. When $G$ is a non trivial
quadruped, $G$ itself 
always belongs to ${\bf D} (G, \mu)$, and the $(1-\tau_G)$ operator
makes $A_{G,\mu}^{ER}$ vanish at 0 
external momenta. For the trivial graph with a single vertex 
$v$ we remarked that $e_v(\mu)=-1$. Hence the formal power series in $g_{\rho}$
\eqref{II.4.6} reduces exactly to $-g_{-1}$ for the connected four point function at $0$ external momenta (that is integrated over
all the external positions save one in the direct space)\footnote{This means that in the sense of formal power series in 
$g_{\rho}$ we must identify $g_{-1}^{\rho}$ with 
the renormalized coupling $g_r$ of the BPHZ scheme, which by definition 
is precisely minus the connected four point function 
at 0 external momenta. The renormalization condition
of BPHZ hence include a wave function renormalization ${\cal Z}$ factor, but we do not need to discuss yet this subtlety,
since in the biped-free theory there is obviously no wave function 
renormalization.} 
 
The proof of Theorem \ref{effexp} is a simple combinatoric exercise; no analysis is 
involved, since all integrals have cutoffs and therefore are
absolutely convergent. The combinatoric has to be checked at the level of combinatorial maps defined
as contraction schemes, otherwise we would have to take into account symmetry factors. We go from \eqref{II.4.5} to \eqref{II.4.6} by pulling out inductively 
the useful counterterms hidden in $g_{\rho}$, one slice after the other. At 
slice $i$ an intermediate version of Theorem \ref{effexp} is obtained:
\bee    S_{N,bf}^{\rho}  = \sum_{G,\mu \le \rho} 
[ \prod_{v \in G} (-g_{ \sup ( i,
e_v(\mu)}^{\rho})] A_{G,\mu}^{ER,i} 
\label{II.4.10}
\ee
where:
\bee  A_{G,\mu}^{ER,i} \equiv \int \prod_v dx_v  
\prod_{h \in {\bf Q}^{i}} (1-\tau_h^* )  I_{G,\mu} , \label{II.4.11}
\ee
and
\bee  {\bf Q}^i \equiv \{ h \in {\bf Q} \vert i_h > i \}  .
\label{II.4.12}
\ee

The equation \eqref{II.4.10} is obviously nothing but \eqref{II.4.5} if $i= \rho$. Assuming it at scale 
$i+1$, we prove it at scale $i$ by simply adding and subtracting the 
counterterms which change $ A_{G,\mu}^{ER,i+1}$ into $ A_{G,\mu}^{ER,i}$.
These are the counterterms corresponding to the quadrupeds $\{g_1,...,g_k\}
=\{g \in {\bf Q}  \vert \mu(l)=i+1 , \forall l\in g \}$. Hence we rewrite 
$ A_{G,\mu}^{ER,i+1}$ as:
\bee \sum_{\scriptstyle S \subseteq \{g_1,...,g_k\} \atop \scriptstyle S \ne 
\emptyset}  \prod_{g_j \in S} (1-\tau^*_{g_j} + \tau^*_{g_j}) \prod_{h \in {\bf Q}^{i+1}}
(1-\tau_h^{*} ) I_{G,\mu} .  \label{II.4.13}
\ee

The completely subtracted piece changes $ \prod_{h \in {\bf Q}^{i+1}}(1-\tau_h^{*} )$ 
into $ \prod_{g \in {\bf Q}^{i}} (1-\tau_g^{*} )$ in each 
amplitude, hence it changes $ A_{G,\mu}^{ER,i+1}$ into $ A_{G,\mu}^{ER,i}$.
The second one is developed as a sum over $S$, so as to get:
\bee S_{N,bf}^{\rho}  = \sum_{\scriptstyle (G,\mu,S), \mu \le \rho
 \atop \scriptstyle S 
\subseteq 
{\bf Q}^i - {\bf Q}^{i+1} } 
[ \prod_{v \in G} (-g_{ \sup ( i+1,
e_v(\mu)}^{\rho})] A_{G,\mu ,S}^{ER,i }, \label{II.4.14}
\ee
with
\bee A_{G,\mu ,S}^{ER,i }  \equiv \begin{cases} A_{G,\mu }^{ ER,i}  {\rm \ if \ } S= \emptyset ,\\
\int \prod_v dx_v
\prod_{g_j \in S} (\tau_{g_j} ) \prod_{g \in {\bf Q}^{i+1}}
(1-\tau_g^{*} ) I_{G,\mu}  {\rm \ \ otherwise.}
 \end{cases}
\label{II.4.16}
\ee

We can now define, since the elements of $S$ are disjoint, the contraction map ${\rm Contr}_i$
as an operation acting on triplets $(G,\mu,S)$, $S \subseteq 
{\bf Q}^i - {\bf Q}^{i+1}$, and which sends $(G,\mu,S)$ to $(G', 
\mu', \emptyset)$, $G'$ being obtained from $G$ by reducing each $g_j \in S$ to 
a single vertex, and $\mu'$ being the assignment derived from $\mu$ by simple 
restriction to the edges of $G'$. Remark that every vertex of $G'$ 
corresponding to such a reduction must have $e_v(\mu)=e_v(\mu ') \le i$. 
We reorder now \eqref{II.4.14} as:
\bee S_{N,bf}^{\rho}  =\sum_{(G',\mu ')}
\{ \sum_{\scriptstyle (G,\mu,S), \mu  \le \rho \atop \scriptstyle 
{\rm Contr}_i(G,\mu,S)=(G',\mu ', \emptyset) } 
[ \prod_{v \in G} (-g_{ \sup ( i+1,
e_v(\mu)}^{\rho})]  A_{G,\mu ,S}^{ER,i} \}. \label{II.4.17}
\ee
For each $(G',\mu')$ the corresponding sum in \eqref{II.4.17} is an infinite power 
series which in fact replaces exactly, at each vertex $v$ of $G'$ 
satisfying $e_v(\mu')\le i$, the coupling 
$g_{i+1}^{\rho}$ by the right hand side of
\eqref{II.4.9}, hence by $g_i^{\rho}$; 
the sum over $H$ in \eqref{II.4.9} indeed corresponds 
exactly to the sum over all possible insertions of a $g_j$ which is contracted 
by the ${\rm Contr}_i$ operation to the vertex $v$, in the above 
notation. This achieves the proof of \eqref{II.4.10} at scale $i$, hence by induction, the 
proof of Theorem \ref{effexp}.

\qed

This theorem achieves our goal of an effective expansion which is 
ultraviolet finite. In the next section we will see that, contrary to the usual renormalized series,
it has the great advantage of being free of \emph{renormalons} 
and of complicated sums over forests.
The renormalization of bipeds can be added
along the same lines, pushing further the $\tau_g^\star$ Taylor expansion 
around local parts so as to generate at least three derivatives acting on propagators
for the quadratically divergent two-point subgraphs.

In this way one obtains effective expansion 
with three types of effective parameters, the effective coupling constant, 
the effective mass and the effective wave function constant. Introducing generalized graphs $\hat G$ with effective local two-point vertices 
$W^{0}$ and two-point vertices $W^{1}$ with derivative couplings in the form of a Laplacian (to represent mass and wave-function
renormalizations) we have the generalization of Theorem \ref{effexp}:

\begin{thm} There exist $3(\rho+1)$ formal power series in $g_\rho$, called 
$g_{i}$, $\delta m^{2}_{i}$ and $\delta {\cal Z}_i$, $i= \rho-1,...,0, -1$
(they depend on $\rho$, like those of Theorem  \ref{effexp}, but we drop this 
dependence to avoid too heavy notations),
such that the formal power 
series in $g_{\rho}$ for $ S_{N}^{\rho}$ can be rewritten as:
\[  S_{N}^{\rho}  = \sum_{ \hat G,\mu} [ \prod_{v \in V( \hat G)} 
(-g_{e_v(\mu)})]
[ \prod_{w \in W^{0}( \hat G)} (-\delta m^{2}_{e_w(\mu)})]
 [ \prod_{w \in W^{1}( \hat G)} (-\delta {\cal Z}_{e_w(\mu)})]   A_{ \hat G,\mu}^{ER} ,
\]
where the formula for $A_{\hat G ,\mu}^{ER}$ is:
\bee  A_{\hat G,\mu}^{ER} \equiv \int 
\prod_{v \in V \cup W^{0} \cup W^{1}} dx_v  
\Bigl( \prod_{h \in {\bf Q } } (1-
\tau_h^* ) \prod_{w \in W^{1}( \hat G )} (-\Delta) \Bigr)
I_{ \hat G,\mu} \; ,
\label{II.4.21}
\ee
with $I_{ \hat G,\mu} \equiv  \prod_l  C^{\mu(l)}  (x_l,y_l) $ 
and the operator $\Delta \equiv \partial_{\nu}\partial_{\nu}$ acts, 
for each $w \in W^{1}(\hat G)$, on one of the two propagators hooked to $w$.
\end{thm}

The flow of the coupling constants $g_i$ is fixed by a small number of boundary conditions. 
For the massive Euclidean $\phi_4^4$ theory it is customary to state them for the 1 PI functions in momentum space at zero momenta:
\bee \Gamma^{4} (0) = - \lambda_{-1} {\cal Z}_{-1}^2 \; ,\qquad   \Gamma^{2} (0)  =  m^{2}_{-1} {\cal Z}_{-1} \; , 
\qquad ( \partial_{p^2}  \Gamma^{2} ) (0) = {\cal Z}_{-1} \; .      \ee

To summarize this section:
\begin{itemize}
\item the effective expansion is a way to recast the perturbative expansion into a multiseries, in which each term is not only
finite but \emph{uniformly bounded} in term of its size.
Is does not solve the problem of proliferation of graphs, hence is not yet a full constructive solution of the theory.
But it is as close to it as one can get, starting with the ordinary perturbative premise.
\item an essential aspect of the effective series is that the renormalization group equations are 
deterministic but non-Markovian; to compute the change of the effective coupling at scale $i$
one needs not only the effective coupling at scale $i+1$ but also all the previous couplings from 
the bare scale down to scale $i+1$. Trying to remove this non-Markovian aspect unfortunately
would reintroduce renormalons (see below).
\end{itemize}

\subsubsection{The renormalized series}

\begin{theorem}
The effective expansion reexpressed in 
terms of $g_r$, $m_r$ and $a_r$ in the limit $\rho \to \infty$ is the same 
order by order in $g_r$ as the usual BPHZ renormalized series. 
\end{theorem}

The BPHZ~\cite{Bogoliubov,Hepp:1966eg,Zimmermann:1969jj} renormalized series is a groundbreaking achievement of mathematical physics which 
gives precise mathematical meaning to the notion of renormalizability, using the mathematics
of formal power series. This being said, it is in fact ultimately a very bad way to formulate renormalization. Let us 
try to explain these two apparently conflicting statements.

We can pass from the effective expansion of the previous subsection to the renormalized series by developing all the 
effective constants in term of the renormalized constant at scale $-1$. We will denote the renormalized constants by the 
subscript $ren$ rather than $-1$.

\begin{figure}[ht]
\centerline{\includegraphics[width=10cm]{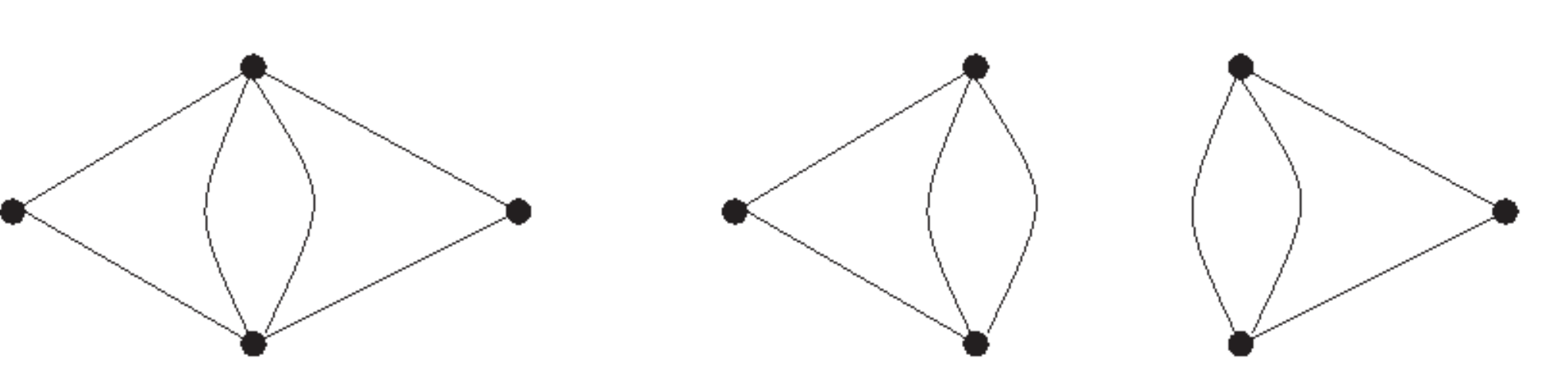}}
\caption{A graph with two overlapping divergent subgraphs}
\label{graph6}
\end{figure}

Using the inversion theorem on formal power series for any \emph{fixed ultraviolet cutoff $\rho$}
it is possible to reexpress any formal power series in $\lambda_{\rho}$ with bare propagators
$1/({\cal Z }_{\rho}p ^2 + {\cal Z }_{\rho} m^2_{\rho} )$
for any Schwinger function as a formal power series 
in $\lambda_{ren}$ with renormalized propagators
$1/({\cal Z }_{ren}p ^2 + {\cal Z }_{ren} m^2_{ren} )$.
The BPHZ theorem then states this power series  
has finite coefficients order by order when the ultraviolet cutoff $\rho$ is lifted. The first proof by Hepp  \cite{Hepp:1966eg}
relied on the inductive Bogoliubov's recursion scheme \cite{Bogolyubov:1980nc}. Then a completely explicit
expression for the coefficients of the renormalized series was written by
Zimmermann and many followers. The coefficients of that renormalized series
can be written as sums of renormalized Feynman amplitudes. They are not the same as the effectively renormalized
amplitudes of the previous section. Renormalized amplitudes involve a sum over localizations operators indexed by \emph{all}
divergent  forests $\cF$, \emph{irrespectively of any  scale assignment}. When the initial graph $G$ contains overlapping divergent graphs,
these divergent forests cannot be described simply as the subforests of a fixed forest. This is particularly obvious when the number
of such divergent forests is not a power of 2 (since the number of subforests in a fixed forest ${\cal F}$ is $2^{\vert {\cal F}\vert}$).
For instance the graph $G$ of Figure \ref{graph6} has two overlapping divergent subgraphs
and 12 divergent forests and 12 is not a power of 2.
Hence the corresponding sum over localizations operators cannot be rewritten as in eg \eqref{II.4.11}, namely not as a single product
of subtraction operators such as $\prod_{g \in \cF}(1- \tau^\star_g) $. 

The solution of this difficult ``overlapping" divergence problem is a kind of tour de force.
In fact to prove finiteness of renormalized amplitudes one \emph{must return} to a scale decomposition (or a continuous analog of it
called Hepp's sectors). In each scale assignment or Hepp's sector there is a \emph{different classification} of forests
into packets so that each packet contains a sum over subforests of a given forest, hence corresponds to a product of subtractions
and leads to a finite integral (see eg \cite{Rivasseau:2011ri}).

\begin{figure}[ht]
\centerline{\includegraphics[width=10cm]{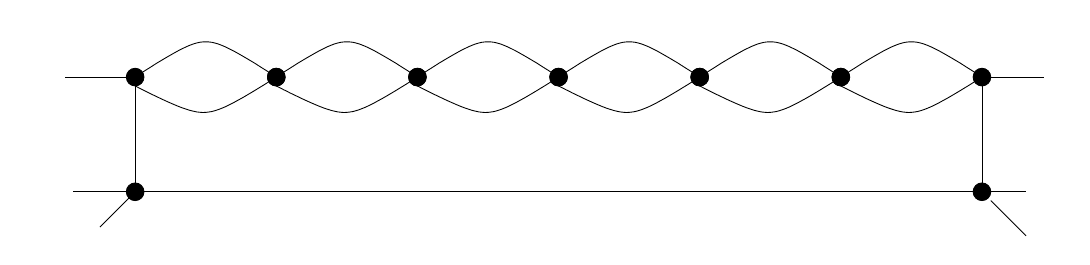}}
\caption{A family of graphs $P_n$ producing a renormalon}
\label{graph8}
\end{figure}

Unfortunately from the physical point of view this tour de force is not only unnecessary but in fact it is also 
misleading. The BPHZ forest formula is unphysical  because it is only for high subgraphs 
that amplitudes should be decomposed into local and renormalized parts. For non-high subgraphs 
introducing this decomposition has a hidden cost. This cost appears
when one considers the \emph{size} of the BPHZ renormalized amplitudes.
Renormalized amplitudes are indeed finite, but they can be enormously large as the size of the graph increases. 
This phenomenon is called \emph{renormalons} and was first detected in \cite{Lautrup:1977hs}.
For instance consider the graphs $P_n$ with 6 external edges and $n+2$ internal 
vertices in Figure \ref{graph8}.
At large momentum $q$ the renormalized amplitude of a bubble subgraph behaves like $\log \vert
q\vert$ hence the total amplitude of $P_n$ behaves as 
\bee  \int 
[\log \vert q \vert]^n  {d^4 q \over [q^2 + m^2 ]^3} \simeq_{n \to\infty} 
c^n n! \; ,
\ee
Therefore after renormalization this family of graphs acquires so large a value that 
it cannot be summed! New infinities are there, created
by the very process which was supposed to remove them. Although
these infinities have been pushed at the non-perturbative level, it is still bad, at least 
from the physical point of view of defining finite numbers out of the theory.

Renormalons are solely due to forests of \emph{not high} subgraphs in a Hepp sector.
The corresponding countyerterms are therefore not necessary to make the amplitude finite, and
they are the ones which create huge contributions to the BPHZ renormalized amplitudes.
In contrast any amplitude of the effective series is not only finite but uniformly
bounded in $c^n$, hence there are no renormalons in the effective series.

We can therefore conclude that subtractions are not correctly organized
in the BPHZ renormalized series. 
 It cannot be the basis of a convergent approximation scheme,
even when the theory is expected to exist non-perturbatively
(eg for the Yang-Mills and QCD theories, the Grosse-Wulkenhaar model \cite{Grosse:2004yu} and so on).

What is wrong from a physical point of view in the BPHZ theorem
is to use the size of the graph rather than the scale decomposition as the relevant parameter to
organize Bogoliubov's induction. 

This leads to the point of view advocated in the previous subsection (and in \cite{Rivasseau:2011ri}): neither the bare
nor the renormalized series are optimal. Perturbation should be organized as a power series in the infinite set of effective couplings. Ultimately this is precisely the \emph{renormalization group} \cite{Wilson:1971bg} point of view.

\subsubsection{The Landau ghost and asymptotic freedom}
\label{Landaughost}

In the case of $\phi^4_4$ only the flow of the coupling constant matters. 

Indeed the ultraviolet limit the flow of $m$ is governed at leading order by the tadpole.
The bare mass $m^2_i$ corresponding to a finite positive physical mass $m^2_{ren}$
is negative and grows as $\lambda M^{2i}$ with the slice index $i$. But
since $p^2$ in the $i$-th slice
is also of order $M^{2i}$ but without the $\lambda$, as long as the coupling $\lambda$ 
remains small it remains much larger than $m^2_i$. Hence the mass term plays no significant role
in the higher slices. Furthermore, as remarked in \cite{Rivasseau:2011ri}, there are no nontrivial overlaps associated to 1PI two point 
subgraphs, hence there is in fact no inconvenience
to use the full renormalized $m_{ren}$ all the way from the bare to the renormalized scales,
with subtractions on 1PI two point subgraphs independent of their scale. In short, mass subtractions
do not create renormalons, so \emph{for them} the BPHZ point of view is acceptable.

The flow of ${\cal Z}$ is also not very important. Indeed it really starts at two loops
because the tadpole is exactly local. So this flow is in fact bounded and does not generate any renormalon. 
In fact (as remarked again in \cite{Rivasseau:2011ri}) for theories of the $\phi^4_4$
type one might as well use the bare value ${\cal Z}_{bare}$ all the way 
from bare to renormalized scales and perform no second Taylor subtraction on any 
1PI two point subgraphs.

\begin{figure}
\centerline{\includegraphics[width=6cm]{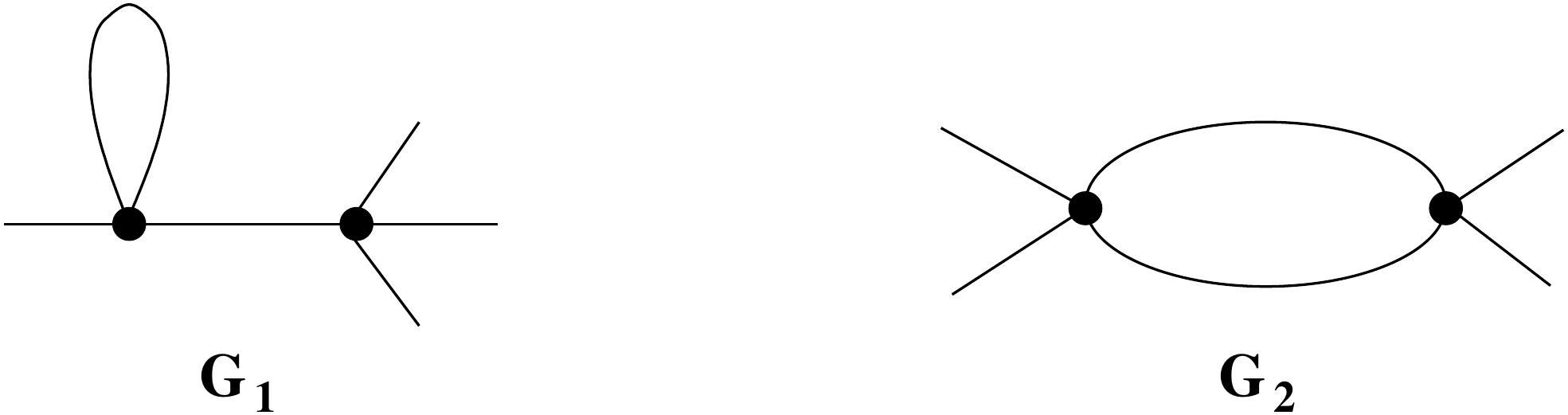}}
\caption{The $\phi^4$ connected graphs with $n=2$, $N=4$.}
\label{oneloop}
\end{figure}

The physics of the $\phi^4_4$ model depends therefore only of the flow of the coupling constant $\lambda$. By a simple second order
computation there are only 2 connected graphs with $n=2$ and $N=4$, $G_1$ and $G_2$, pictured in Figure \ref{oneloop}.

In fact the graph $G_1$ does not contribute to this coupling constant flow because it
is one particle reducible. In ordinary translation-invariant,
(momentum-conserving) theories, such one-particle-reducible graphs
never contribute to the RG flow. Indeed when they are high, hence quasi-local, they become 
also very small when the gap between internal 
and external scales grows. This is because,
by momentum conservation, the momentum of any one-particle-reducible edge has to be the sum
of a finite set of external momenta on one of its sides. But a finite sum of small momenta remains small.
This clashes directly with the intuition that, this edge being internal, its momentum should increase with this gap.
The net result is that in fact the amplitude of such a 1PI high subgraph does not diverge at all when the gap 
tends to infinity. 

At second order the relation between $\lambda_i$ and $\lambda_{i-1}$ is
\bqa 
  - \lambda_{i-1} &\simeq& - \lambda_i + \lambda_{i}^2  \frac{3}{2} \int d^4u [C^i(u)]^2 \Rightarrow 
  \lambda_{i-1} = \lambda_i - \beta \lambda_i^2
  \; , \label{flow1}
\eqa
with $\beta>0$. The theory sinks in the Gaussian fixed point in the IR. Fixing the renormalized coupling 
to some finite value requires a diverging bare coupling, incompatible with perturbation theory. 
This is the Landau ghost problem, which affects both the 
$\phi^4_4$ theory and electrodynamics.

On the contrary, in non-Abelian gauge theories, an extra
minus sign is created by the algebra of the
Lie brackets and makes the theory  asymptotically free. Physically
this means that the interaction is turned off at small distances. This
theory is in agreement with scattering experiments 
which see a collection of almost free particles
(quarks and gluons) inside the hadrons at very high energy.
This was the main initial argument to adopt
quantum chromodynamics, a non-Abelian gauge
theory with $SU(3)$ gauge group, as the theory of strong interactions \cite{Gross:1973id}. 

Remark that in such asymptotically free theories which form the backbone of today's standard model,
the running coupling constants remain bounded between far ultraviolet ``bare" scales 
and the lower energy scale where
renormalized couplings are measured. Ironically the point of view on early 
renormalization theory as a trick to hide the ultraviolet divergences of QFT
into infinite unobservable bare parameters could not turn out to be more wrong
than in the standard model. Indeed the bare coupling constants
tend to 0 with the ultraviolet cutoff, and what can be farther from infinity than 0?

Recently it has been shown to all orders of perturbation theory that there
should be no Landau ghost but an asymptotically safe fixed
point for the similar RG flow of the non-commutative 
Grosse-Wulkenhaar $\phi^{\star 4}_4$ model  \cite{Grosse:2004yu,Disertori:2006nq}. 
Therefore this model is a simple Bosonic renormalizable QFT in which one can presumably fully and rigorously
control at last the phenomenon of ultraviolet renormalization in all its aspects, both
perturbative and constructive\footnote{An equivalent complete ultraviolet control in the Fermionic 
case has been achieved \cite{GN-AK,GN-FMRS} for the Gross-Neveu model in two dimensions \cite{GN-GN}, which is asymptotically free in the ultraviolet limit
\cite{GN-MW}.}.

\newpage

\section{Functional methods and finite coupling aspects}
\label{sec:functional}

Integro-differential renormalization equations can be written that summarize in a compact way the flow 
of the generating functionals of a QFT, hence of all connected or one particle irreducible functions.
We present in the sequel the two most common functional equations, the Polchinski \cite{Polchinski:1983gv}
and Wetterich \cite{Wetterich:1992yh,Wetterich:1989xg} equations. \footnote{More extensive reviews on this 
subject are \cite{Wilson:1973jj,cardy1996scaling, Bagnuls:2000ae,Berges:2000ew,Pawlowski:2005xe, Gies:2006wv,  Delamotte:2007pf,Rosten:2010vm}.}
,
Such equations are sometimes called \lq\lq exact\rq\rq. This is a very unfortunate name, as they are no more exact than
the partition function itself, and suffer from the same ambiguities and pathologies. If they are used for instance to compute 
the flow of the coupling constant as a Markovian equation, that is give the derivative of $g_i$ in terms of a power 
series in $g_i$ itself, that power series suffers from renormalons.
Also using differential rather than difference equations and discrete slices is fine at the perturbative levels but creates additional
difficulties at the constructive level in the few cases we can completely control, such as the Gross-Neveu model 
in two dimensions \cite{GN-AK,GN-FMRS}. See \cite{Disertori:1998qe} for a detailed discussion 
of these difficulties. Roughly speaking this is because iterating a differential equation leads to develop \emph{all} propagators 
of the theory, whereas constructive theory either in the form of cluster expansions or in the form of a loop vertex expansion keeps many 
of them hidden into a better bounded functional integral.

The main advantage of such equations lies instead in the fact that, as they deal with generating functionals, they
allow a very transparent interpretation of the RG flow in the space of coupling constants. 
Each of these two equations has its benefits and drawbacks. Experience teaches us that the Polchinski equation is better adapted 
to mathematical (perturbative) proofs, while the Wetterich equation is better adapted to truncations and numerical computations. The price to pay is that, 
when using the Wetterich equation, one has no control over the rest terms. 

\subsection{The Polchinski equation}

Using the properties of the Gaussian measure, the partition function of a QFT with UV cutoff $\Lambda$ and IR cutoff $\Lambda_0$ writes as 
\bea
 Z = \Bigl[ e^{ \frac{1}{2} \int d^dxd^dy \frac{\delta}{\delta \phi(x)} C^{\Lambda}_{\Lambda_0}(x,y) \frac{\delta}{\delta \phi(y)} }
    e^{-S^{int}(\phi)} \Bigr]_{\phi=0} \; ,
\eea 
where $C^{\Lambda}_{\Lambda_0}$ is the covariance with UV cutoff $\kappa$ and IR cutoff $\kappa_0$, and $S^{int}$ is the non Gaussian 
interaction part.

Introducing an intermediate scale $k$,
the covariance splits into a sum $C^{\Lambda}_{\Lambda_0} = C^{\Lambda}_{k} +  C^{k}_{\Lambda_0}$, and the field is replaced by the 
sum of two independent fields $\phi =\psi_{<k} + \chi_{>k}$ distributed with the covariances $ C^{\Lambda}_{k} $ and $ C^{k}_{\Lambda_0}$ 
respectively. The field $\psi_{<k} $ is a background field consisting of the low energy modes,
while the field $ \chi_{>k}(x)$ is a perturbation consisting of the high energy modes. 

The partition function is then
\bea
&& \Bigl[ e^{ \frac{1}{2} \int d^dxd^dy \frac{\delta}{\delta \psi_{<k}(x)} C^{k}_{\Lambda_0}(x,y) \frac{\delta}{\delta \psi_{<k}(y)} } \\
&& \qquad \times  e^{ \frac{1}{2} \int d^dxd^dy \frac{\delta}{\delta \chi_{>k}(x)} C^{\Lambda}_{k}(x,y) \frac{\delta}{\delta \chi_{>k}(y)} }
 e^{-S^{int}(\psi_{<k}+ \chi_{>k})} \Bigr]_{\psi_{<k}, \chi_{>k}=0} \; , \nonumber
\eea 
and can be rewritten in terms of the {\it effective} action at scale $k$ obtained by integrating the perturbation $\chi_{>k}$,
\bea
 && e^{- S^{int}_k (\psi_{<k}) }  \equiv 
 \Bigl[  e^{ \frac{1}{2} \int d^dxd^dy \frac{\delta}{\delta \chi_{>k}(x)} C^{\Lambda}_{k}(x,y) \frac{\delta}{\delta \chi_{>k}(y)} }
 e^{-S^{int}(\psi_{<k}+ \chi_{>k})} 
 \Bigr]_{ \chi_{>k}=0} \; ,\crcr
 && Z = \Bigl[ e^{ \frac{1}{2}
 \int d^dxd^dy \frac{\delta}{\delta \psi_{<k}(x)} C^{k }_{\Lambda_0}(x,y) \frac{\delta}{\delta \psi_{<k}(y)} } e^{- S^{int}_k (\psi_{<k}) } 
 \Bigr]_{\psi_{<k}=0} .
\eea 
The effective action at scale $k$ respects the equation
\bea
 && - \partial_k S^{int}_k (\psi_{<k})  e^{- S^{int}_k (\psi_{<k}) } = 
 \Bigl[ \frac{1}{2} \int d^dxd^dy \frac{\delta}{\delta \chi_{>k}(x)} \partial_k C^{\Lambda}_{k}(x,y) \frac{\delta}{\delta \chi_{>k}(y)}     \crcr
 && \qquad \qquad  \qquad \times e^{ \frac{1}{2} \int d^dxd^dy \frac{\delta}{\delta \chi_{>k}(x)} C^{\Lambda}_{k}(x,y) \frac{\delta}{\delta \chi_{>k}(y)} }
 e^{-S^{int}(\psi_{<k}+ \chi_{>k})}    \Bigr]_{ \chi_{>k} =0 } \crcr
 && = \frac{1}{2} \int d^dxd^dy \frac{\delta}{\delta \psi_{<k}(x)} \partial_k C^{\Lambda}_{k}(x,y) \frac{\delta}{\delta \psi_{<k}(y)}    \crcr
&& \qquad \qquad \times  \Bigl[ e^{ \frac{1}{2} \int d^dxd^dy \frac{\delta}{\delta \chi_{>k}(x)} C^{\Lambda}_{k}(x,y) \frac{\delta}{\delta \chi_{>k}(y)} }
 e^{-S^{int}(\psi_{<k}+ \chi_{>k})}    \Bigr]_{ \chi_{>k} =0 } \crcr
&& = \frac{1}{2} \int d^dxd^dy \frac{\delta}{\delta \psi_{<k}(x)} \partial_k C^{\Lambda}_{k}(x,y) \frac{\delta}{\delta \psi_{<k}(y)}   e^{- S^{int}_k (\psi_{<k}) }
\; .
\eea 
Computing the functional derivatives on the right hand side, taking into account that the $C^{\Lambda}_{\Lambda_0}$ does not depend on $k$, hence
$\partial_k C^{\Lambda}_{k}=-\partial_k C^{k}_{\Lambda_0} $, and relabeling the field $\phi$ we obtain
\bea\label{eq:pol}
  \partial_k S^{int}_k (\phi) &= & \frac{1}{2}  \int d^dxd^dy \; \; \partial_k C^{k}_{\Lambda_0}(x,y)  \frac{\delta S^{int}(\phi)}{\delta \phi(x) }
  \frac{\delta S^{int}(\phi)}{\delta \phi(y) } \crcr  
& & -   \frac{1}{2}  \int d^dxd^dy \; \; \partial_k C^{k}_{\Lambda_0}(x,y) \frac{\delta^2 S^{int}(\phi)}{\delta \phi(x) \delta\phi(y)  } \; .
\eea 

This equation has a neat graphical interpretation. Indeed, $S^{int}_k$ is the generating functional of connected amputated graphs with 
propagators $C^{\Lambda}_{k}$. The derivative with respect to $k$ can either hit a one particle reducibility edge (the first term on the right 
hand side of \eqref{eq:pol}) or not (the second term). Polchinski stated that it is possible to deduce the BPHZ theorem from 
this renormalization group equation and inductive bounds which does not decompose each order of 
perturbation theory of a Schwinger function into Feynman graphs \cite{Polchinski:1983gv}. This idea was
clarified and applied by C. Kopper and coworkers, see \cite{Keller:1991cj}.

\subsection{The Wetterich equation}

We will now present in more detail Wetterich's equation, which we will be using below in some applications. 
Let us start from the generating functional of the connected moments with UV cut-off 
\begin{equation}
e^{W_\Lambda[J]}=\frac{1}{\mathcal{N}_\Lambda}\int\mathcal{D}_\Lambda \phi\, e^{-S[\phi]+J\cdot\phi}\;,
\end{equation}
where the action is taken now to include the quadratic Gaussian part (which we denote $S_{\rm free}$),
$\mathcal{D}_\Lambda \phi$ is the (ill defined) functional measure with a UV cut-off at 
energy scale $\Lambda$, $\mathcal{N}_\Lambda=\int\mathcal{D}_\Lambda \phi e^{-S_{\rm free}[\phi]}$ and we use the 
short hand notation $J\cdot \phi=\int d^Dx\,J(x)\,\phi(x)$. 

We associate to $W_{\Lambda}[J]$ a one parameter family of generating functionals
\begin{equation}\label{Wk}
e^{W_{k,\Lambda}[J]}=\frac{1}{\mathcal{N}_{k,\Lambda}}\int\mathcal{D}_\Lambda \phi\, e^{-S[\phi]+J\cdot\phi-\Delta S_k[\phi]}\;,
\end{equation}
where now $\mathcal{N}_{k,\Lambda}=\int\mathcal{D}_\Lambda \phi e^{-S_{\rm free}[\phi]-\Delta S_k[\phi]}$ and
$\Delta S_k[\phi]$ is a modification of the action which suppresses low energy modes ($E\lesssim k$) in the path 
integral. Instead of using a sharp cutoff in the functional measure, this can be done by introducing a smooth regulator.

\subsubsection{Regulators}

Let us set
\[
\Delta S_k[\phi]=\!\int\! \frac{d^Dp}{2(2\pi)^D}\,\phi(p)\, \mathcal{R}_k(p)\,\phi(-p)=\!\int\! \frac{d^Dp}{2(2\pi)^D}\,\phi(p)\; k^2\, r(p^2/k^2)\,\phi(-p).
\]
The regulator $\mathcal{R}_k(p)$ is determined by the choice of the shape function $r(z)$. The requirements on $r(z)$ are to be monotonic and to satisfy
\begin{align}
r(0)>0,\ \ \ \ \ &\lim_{p^2/k^2\to0}\mathcal{R}_k(p)>0;\crcr
\lim_{z\to\infty}r(z)=0,\ \ \ \  & \lim_{k^2/p^2\to0}\mathcal{R}_k(p)=0;\crcr
r(z)>0,\ \ 0\leq z\lesssim 1,\ \ \ &  \lim_{k\to\Lambda\to\infty}\mathcal{R}_k(p)=\infty.
\end{align}
The first conditions ensures that $\mathcal{R}_k$ implements an IR cutoff similar to a mass term. The second condition ensures that 
the regulator vanishes (sufficiently fast) when $p^2$ lies in the UV region (no cutoff on the UV modes). Finally, the third one 
ensures that, we send $k\to\Lambda\to\infty$, the path integral is dominated by the quadratic part (which in turn will allow us to set an
initial condition on the flow equation). Some choices of $r(z)$ can be e.g.
\begin{align}
\text{polynomial}\ \ \ \ &r(z)=z^{-\alpha},\ \ \alpha\geq0\;,\crcr
\text{exponential}\ \ \ \ &r(z)=\frac{z}{e^{z^\beta}-1},\ \ \beta\geq1\;,\crcr
\label{litim}
\text{semi-sharp}\ \ \ \ &r(z)=(1-z) H(1-z)\;,
\end{align}
where $H$ is Heaviside's step function, so that the semi-sharp cutoff has a jump discontinuity in its first derivative---hence its name. That last cutoff, due to Litim \cite{Litim:2000ci,Litim:2001up,Litim:2001fd}, is quite convenient for explicit calculations at leading order in a derivative expansion, and for this reason we will be using it extensively. It should be noted however that such a cutoff is not suitable at higher orders in a derivative expansion~\cite{Morris:2005ck}.%
\footnote{%
More specifically, beyond order $\mathcal{O}(\partial^2)$ it yields a more general momentum scale expansion in the sense of ref.~\cite{Morris:1993qb} rather than a derivative one. Such expansions do not generally have good convergence properties~\cite{Morris:1999ba,Morris:2000hm}.
}

\begin{figure}[ht]
  \begin{center}
    \subfigure[The regulator $R_k(p^2)$ for different shape functions. Dotted line: polynomial, $\alpha=1$. Light solid: exponential, $\beta=2$. 
    Dark solid: exponential, $\beta=1$. Dashed: semi-sharp.]{\label{fig:regulatorRG}\includegraphics[width=0.48\textwidth]{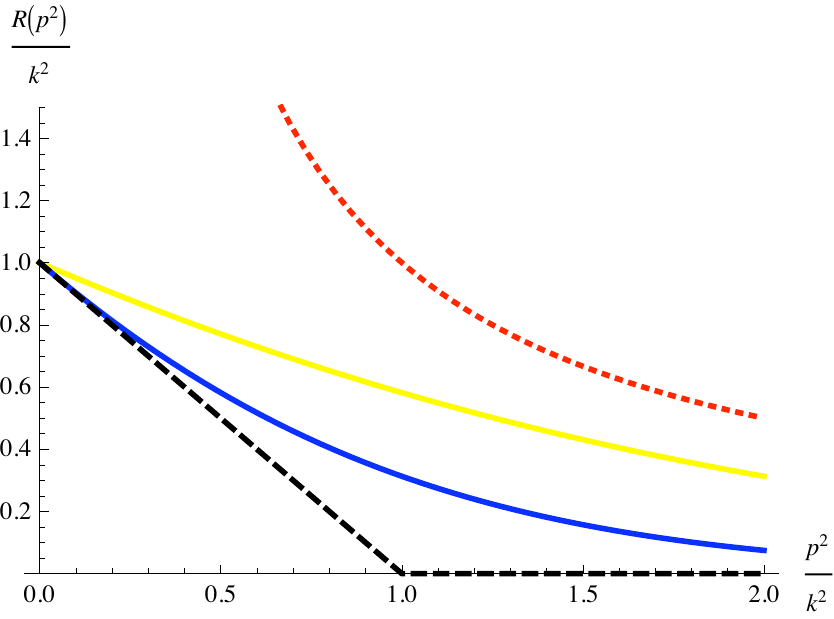}}
    \subfigure[The (normalized) modified propagator $F(p^2)$ for different shape functions, as in the left panel. The gray thin line is 
    the unmodified propagator.]{\label{fig:propagatorRG}\includegraphics[width=0.48\textwidth]{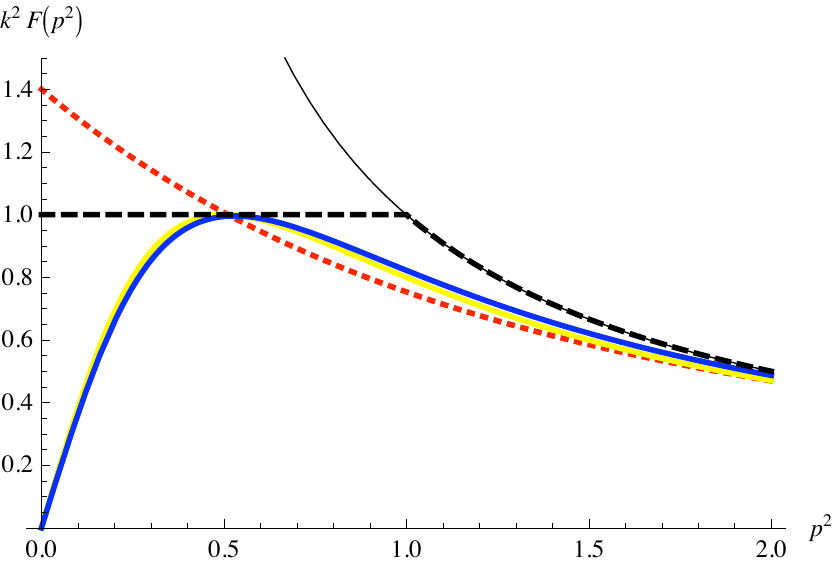}}
  \end{center}
  \caption{Regulator and regularized propagator for different shape functions.}
  \label{fig:regulators}
\end{figure}

These different cutoffs are plotted 
in figure \ref{fig:regulators}, together with the respective shapes of the regularized propagators for $k=1$
\begin{equation}
F(p^2)=\frac{1}{\mathcal{R}_{k=1}(p^2)+p^2}\;.
\end{equation}
The following formal derivation of Wetterich equation will however not depend on details of $r(z)$.

\subsubsection{The quantum effective action at one loop}

To simplify our notation, let us write $W_k\equiv W_{k,\Lambda}$ (and similarly $\mathcal{N}_k$, $\Gamma_k$, etc).
Let us consider the Legendre transform of $W_k[J]$, which is written in term of the $k$-dependent mean field $\varphi(x)=\frac{\delta W_k[J]}{\delta J(x)}$:
\begin{equation}
\tilde{\Gamma}_k[\varphi]=\inf_{J}\left[- W_k[J]+\int d^Dx\; \varphi(x)J(x)\right]\;.
\end{equation}
Notice that the normalization $\mathcal{N}_k$ disappears after integrating out $\phi$, but we have an explicit dependence on the regulator.
To cancel this explicit dependence let us define
\begin{equation}
\Gamma_k[\varphi]=\tilde{\Gamma}_k[\varphi]-\Delta S_k[\varphi]\;.
\label{eq:gammafreek2}
\end{equation}
Formally we retrieve the full quantum effective action when $k\to0$, where $\Delta S_k$ vanishes. Instead, when $k\to\infty$  by a 
saddle-point approximation $\Gamma_\infty[\varphi]\approx S[\varphi]$.

Let us write down the one-loop effective action $\Gamma_k^\mathrm{1loop}[\varphi]$ for a generic interacting theory by a saddle point approximation:
\bea
&& \frac{\delta ( -S - \Delta S_k )}{\delta \phi} [\varphi] + J = 0  \\
&&  \Rightarrow 
 \Gamma_k^\mathrm{1loop}[\varphi]=S[\varphi]+\log\mathcal{N}_k+\frac{1}{2}\mathrm{Tr}\left.\left[\log
\frac{\delta^2 (S[\phi]+\Delta S_k[\phi])}{\delta\phi\,\delta\phi}\right]\right|_{\phi=\varphi}\,, \nonumber
\eea 
where we highlight the presence of the normalization constant. The differential of this one-loop result is inspiring:
\begin{equation}\label{wetterich1loop}
k\frac{d}{d k}\Gamma_k^\mathrm{1loop}[\varphi]=\frac{1}{2}\mathrm{Tr}\left[k\frac{d \mathcal{R}_k}{d k}\,
\left(\left.\frac{\delta^2 (S[\phi]+\Delta S_k[\phi])}{\delta\phi\,\delta\phi}\right|_{\varphi}\right)^{-1}\right]+k\frac{d\log\mathcal{N}_k}{dk}\,.
\end{equation}
Notice the field independent term $k\frac{d\log\mathcal{N}_k}{dk}$. The importance of this term is clear in the free case, where, 
representing the normalization constant as the determinant of a bilinear,  we have that
\begin{equation}
\begin{aligned}
k\frac{d\log\mathcal{N}_k}{dk}&=-\frac{1}{2}k\frac{d}{dk}\mathrm{Tr}\left[\log
\big(-\Delta+m^2+\mathcal{R}_{k}\big)\right]\\
&=
-\frac{1}{2}\mathrm{Tr}\left[
\frac{k\frac{d}{dk}\mathcal{R}_{k}}{-\Delta+m^2+\mathcal{R}_{k}}\right]\,.
\end{aligned}
\end{equation}
Plugging this into~\eqref{wetterich1loop}, we get that the right hand side vanishes, consistently with the fact that free theories do not flow. 
For the purpose of finding the $\beta$-functions in a scalar field theory we see that the running due to the normalization~$\mathcal{N}_k$ plays no role (it represents a vacuum term), 
and in what follows we will ignore it. However, when dealing with gravitational theories such a term may be important and for instance affect the running of the cosmological constant.

\subsubsection{The functional equation for the quantum effective action}

Let us introduce the renormalization time $t=\log(k/\Lambda)$, so that $\frac{d}{dt}=k\frac{d}{dk}$. Differentiating (\ref{Wk}) and dropping the field-independent $\mathcal{N}_k$ term, 
we have the identities
\begin{equation}
\left.\frac{\partial W_k[J]}{\partial t}\right|_{J\ \mathrm{fixed}}=-\left\langle\frac{d \Delta S_k\left[\phi\right]}{dt}\right\rangle_J=-
\frac{1}{2}\mathrm{Tr}\left[\langle \phi\phi\rangle_J \frac{d \mathcal{R}_k}{dt}\right]\;,
\end{equation}
where the trace stands for integration and, were we to consider multiplets of fields, summation on internal indexes. Then, recalling 
that $\varphi=\langle\phi\rangle$ and using the straightforward generalization of (\ref{GammaWrel})
\begin{equation}
\frac{\delta^2\tilde{\Gamma}_k[\varphi]}{\delta\varphi(x)\,\delta\varphi(y)}=\left.\left[\frac{\delta^2 W_k[J]}{\delta J(x)\,\delta J(y)}\right]^{-1}\right|_{J=J_\varphi},
\end{equation}
where $J_\varphi$ is the current used in the Legendre transform, we have
\bea\label{wetterich}
\frac{d \Gamma_k[\varphi]}{dt}&&=-\frac{\partial W_k[J_\varphi]}{\partial t}-\int d^Dx \frac{\delta W[J]}{\delta J(x)}\frac{dJ_\varphi(x)}{dt} \\
&& +\int d^Dx \;  \varphi(x)\frac{dJ_\varphi(x)}{dt}-\frac{d\Delta S_k[\varphi]}{dt}\crcr
\nonumber
&&=-\frac{\partial W_k[J_\varphi]}{\partial t}-\frac{d\Delta S_k[\varphi]}{dt}=-\frac{\partial W_k[J_\varphi]}{\partial t}-\frac{1}{2}\mathrm{Tr}\left[\varphi \frac{d\mathcal{R}_k}{dt}\varphi\right]=\\
\nonumber
&&=\frac{1}{2}\mathrm{Tr}\left[(\langle \phi\phi\rangle_{J_\varphi}-\langle\phi\rangle_{J_\varphi}^2) \frac{d \mathcal{R}_k}{dt}\right] \crcr
&& = \frac{1}{2}\mathrm{Tr}\Bigg{\{}\left[ \Bigl(e^{-W} \frac{\delta^2}{\delta J \delta J} e^W\Bigr)_{J_\varphi} 
- \Bigl( \frac{\delta W}{\delta J} \Bigr)_{J_{\varphi}} \Bigl( \frac{\delta W}{\delta J} \Bigr)_{J_{\varphi}} \right] \frac{d \mathcal{R}_k}{dt} \Bigg{\}} \crcr
&& =\frac{1}{2}\mathrm{Tr}\left[\left.\frac{\delta^2 W[J]}{\delta J\delta J}\right|_{J_\varphi} \frac{d \mathcal{R}_k}{dt}\right]
=\frac{1}{2} \mathrm{Tr}\left[\left(\frac{\delta^2\tilde{\Gamma}[\varphi]}{\delta\varphi\delta\varphi}\right)^{-1} \frac{d \mathcal{R}_k}{dt}\right] \crcr
&&=\frac{1}{2} \mathrm{Tr}\left[\left(\frac{\delta^2\Gamma[\varphi]}{\delta\varphi\delta\varphi}+\mathcal{R}_k\right)^{-1} \frac{d \mathcal{R}_k}{dt}\right]
=\frac{1}{2} \mathrm{Tr}\left[\frac{\frac{d }{dt}\mathcal{R}_k}{\mathcal{R}_k+\Gamma^{(2)}[\varphi]}\right]\; . 
\eea
The above functional renormalization group equation, due to Wetterich, is formally 
a differential equation for the one-parameter family of functionals $\Gamma_k$. Its solution $\Gamma_k$ describes the flow of the effective
action (hence the name \emph{flow equation}) in the \emph{theory space} under changes of the cutoff scale. 
 
The above equation formally accounts for arbitrarily high loop effects. This may seem a bit surprising, because \eqref{wetterich} has a one-loop structure. 
However, the equation is exact \emph{in the full theory space}. Let us illustrate what this means with an example: using the notations of \eqref{eq:gammaexp},
we would have an infinite system of $\beta$-functions 
\begin{equation}\label{manybetafunctions}
 \begin{cases} 
\frac{d}{dt}\lambda^{(2,0)}=\beta_{2}(\lambda^{(2,p,\sigma)},\lambda^{(4,p,\sigma)})\;,\\
\frac{d}{dt}\lambda^{(4,0)}=\beta_{4}(\lambda^{(2,p,\sigma)},\lambda^{(4,p,\sigma)},\lambda^{(6,p,\sigma)})\;,\\
\ \dots\ ,
 \end{cases} 
\end{equation}
so that even if the beta function for the $\varphi^2$ coupling does not explicitly depend on the coupling of $\varphi^6$, it does implicitly via e.g. $\lambda^{(4,\dots)}$. 
Clearly finding a solution to (\ref{manybetafunctions}) is very difficult-- as difficult as solving the full 
QFT, since $\Gamma_k$ contains all the 1PI renormalized diagrams. The advantage is that we can now adopt an approximation scheme that is not a weak-coupling expansion.
 
Note that, formally, the equation \eqref{wetterich} does not depend on the UV cutoff.
Indeed the flow equation does not need it to appear well defined, as the trace converges both at large and small momenta due to the regulator.
However when we integrate the flow equation with respect to $t$, we have no guarantee that the resulting flow does not blow up. If, however, we find that the flow equation can be
integrated up to arbitrary high scales, no problem arises. This is the case of what we have defined in Section \ref{sec:betafunctions} as \emph{asymptotically safe}
theories. In that case, in $\mathcal{T}$ there will be two fixed points, and a certain number of theories that flow from one to another under the RG flow. These span
a critical manifold, whose dimension counts the physical parameters that  must be measured to make a prediction.

In practice it is not possible to address the full equation (\ref{wetterich}), and one has to adopt some approximation scheme. One possibility, which is natural from the point of view of the theory space, is to make an ansatz. This amounts to projecting the flow equation on a linear subspace of~$\mathcal{T}$, which can then taken to be larger and larger. Alternatively, one can work out a recursive relation for the flow of the $n$-point functions, resulting in a systematic expansion scheme. What is the most advantageous method depends both on which observables should be computed and on how complicated the field content of the theory is. Let us briefly describe these approaches.

\subsubsection{Ans\"atze and truncations}
 
The idea here is to expand the flow equations  on a basis of the theory space, for instance taking monomials 
in $\cT$ as basis vector, and retain only some of those in some ansatz. This space \emph{will not be stable under the RG flow}, so that the flow whill have to be projected back onto it (i.e., the right hand side of the equation will have to be truncated). 
While such ans\"atze can lead to satisfactory numerical results, it is usually very difficult to 
control the rest terms.

There are two natural choices for the ansatz:
\begin{enumerate}
\item A derivative expansion. This amounts to retaining terms involving up to a certain number of derivatives. This is advantageous because typically derivative operators
make the explicit computations of $\beta$-functions more involved, and amounts to an expansion in small momentum (with respect to the cutoff scale). Often already the \lq\lq local potential approximation\rq\rq, which amounts to the zeroth 
order of the derivative expansion together with the kinetic term,
\begin{equation}
\Gamma_k[\varphi]=\int d^Dx \left[-\frac{1}{2}\mathcal{Z}_k\varphi \Delta\varphi+ \sum_{n=1}^\infty \mathcal{Z}^n_k\frac{\lambda^{(2n)}_k}{(2n)!}\,\varphi^{2n}\right]\,,
\end{equation}
can be used to make quite accurate physical predictions (the lower script $k$ reminds us that the couplings are scale dependent). Sometimes rather than expanding the potential on a basis, it is convenient to treat it as a whole,
\begin{equation}
\Gamma_k[\varphi]=\int d^Dx \left[-\frac{1}{2}\mathcal{Z}_k\varphi \Delta\varphi-V_k(\varphi^2)\right]\,,
\end{equation}
so that the flow equation results in a differential equation involving the derivatives of $V_k$. This can be advantageous for both analytical and numerical treatment.

\item Expansion according to mass dimension. In this case, one fixes a mass dimension $-\bar{\Delta}\leq0$, and considers all the terms whose couplings have $[\lambda]\geq\bar{\Delta}$. 
For instance, the truncation corresponding to  $\bar{\Delta}=2$ in $D=4$ is
\begin{align}\nonumber
\Gamma_k[\varphi]=&\int d^4x \left[-\mathcal{Z}_k\frac{1}{2}\varphi \Delta\varphi+\mathcal{Z}_k\lambda^{2,2}\frac{1}{2}\varphi \Delta^2\varphi-\mathcal{Z}^2_k\lambda^{(4,1)}_k\frac{1}{3!}\varphi^3 \Delta\varphi+\right.\\
\label{gammadim-2}
&\ \ \ \ \left. +\mathcal{Z}_k^2m_k^2\frac{1}{2}\,\varphi^{2}+\mathcal{Z}_k^2\lambda^{(4)}_k\frac{1}{4!}\,\varphi^{4}+\mathcal{Z}^3_k\lambda^{(6)}_k\frac{1}{6!}\,\varphi^{6}\right]\;.
\end{align}
This can be convenient in the vicinity of the Gaussian fixed point,  where the discarded couplings are irrelevant and should have little effect on the flow.
\end{enumerate}

Any prediction of the flow equation (fixed points, critical exponents, etc.) will depend both on the truncation and on the explicit
form of the regulator $\mathcal{R}_k$, and establishing the regulator dependence of the FRGE in a given truncation (and at finite values of the couplings) is in fact a subtle issue---we refer the reader to refs.~\cite{Rosten:2010vm,Pawlowski:2005xe} for more details on this topic. An empirical  way to make sense of the predictions is to consider larger and larger truncations, and different
regulator schemes, and check which predictions are robust under this changes. In exceptional cases,
such as the Grosse-Wulkenhaar model~\cite{Grosse:2004yu,Sfondrini:2010zm}, one may be able to make statements that
take into account the flow in the whole theory space. Even in such cases it is very hard to provide a rigorous mathematical 
formalization of the flow equation in $\mathcal{T}$. 

Let us also mention that not all couplings are physical, since changes of variable in the path integral do not modify the physical prediction, but lead to redefinitions 
of the couplings. Such couplings are called \emph{redundant}. 

\subsubsection{The Blaizot-Mendez-Wschebor expansion scheme}
As remarked, making an ansatz for the effective action usually offers little control on the rest terms. One advantage of the derivative expansion is that it is systematic, in the sense  that it effectively uses $p/\kappa\ll1$ as expansion parameter, where $p$ are the momenta and $\kappa$ is the smallest mass scale of the theory (either the regulator scale $k$ or the smallest mass~$m$). For several computations, including deriving critical exponents, this can lead to very precise results. However, if we are interested in quantities that depend on non-vanishing external momenta, such a scheme may easily break down.%
\footnote{This is typically the case when considering $k\to0$ in theories at criticality, where therefore~$m=0$.}

A different expansion scheme that does not suffer from this issue was developed by Blaizot, Mendez and Wschebor (BMW) \cite{Blaizot:2005xy,Blaizot:2005wd,Blaizot:2006vr}, see also ref.~\cite{Benitez:2011xx} for a recent, self contained presentation.%
\footnote{
Early attempts in the same direction were also made by Parola, Pini and Reatto~\cite{PhysRevE.48.3321}.
}
The idea is to start from the FRGE written down for the $n$-point functions, which can be formally found by taking appropriate functional derivatives of~\eqref{wetterich}. This results in a tower of equations: it is easy to see that e.g.~$\tfrac{d}{dt}\Gamma^{(2)}_k$ depends on $\Gamma^{(3)}_k$ and $\Gamma^{(4)}_k$, and in fact $\tfrac{d}{dt}\Gamma^{(n)}_k$ depends on all vertex functions up to~$\Gamma^{(n+2)}_k$. An advantage is that one can clearly distinguish between external momenta~$p$ and internal momenta~$q$, where the latter are integrated over in the trace. In this way one can construct an approximation scheme that relies on the internal momenta~$q$ being small, with no condition on~$p$.
This results in an iterative procedure whereby one obtains, at order $2k$, the first $2n$-point functions up to $2n=2k$. Furthermore, this nicely compares with perturbation theory, as the $2n$-point function at order $2k$ accounts for $k-n+1$ loops in the perturbative expansion.

This powerful scheme represents an extension of the derivative expansion, and has been employed to perform numerical and analytical calculations in the case e.g. of the $O(N)$ vector-model (see ref.~\cite{Benitez:2011xx} for an overview) with remarkable success.
For our purpose of illustrating how finite-coupling physics can be extracted out of the FRGE, however, a simple ansatz will suffice.

\subsection{Applications of the FRGE}

We will show how to use the FRGEs to derive $\beta$-functions. 

\subsubsection{Revisiting \texorpdfstring{$\phi^{4}_{4}$}{phi**4(4)} }
Here we will re-derive the perturbative $\beta$-functions of $\lambda\phi^4$ using Wetterich's equation.

Let us consider the minimal ansatz
\begin{equation}
\label{eq:phi4ansatz}
\Gamma_k[\varphi]=\int d^Dx \left[-\frac{1}{2}\mathcal{Z}_k\varphi \Delta\varphi+ \mathcal{Z}_k \frac{m_k^2}{2}\varphi^2 +\mathcal{Z}^2_k\frac{\lambda_k}{4!}\,\varphi^{4}\right]\;,
\end{equation}
and use the one loop equation (\ref{wetterich1loop}), whose r.h.s. depends only on $S^{(2)}$ (and not $\Gamma^{(2)}$). Thus the couplings on the r.h.s. have no explicit 
$k$-dependence. Since we are not interested in higher orders in a derivative expansion it is convenient to use the semi-sharp regulator, which will greatly simplify our computations. In momentum space we have
\begin{align}\label{Rk}
\mathcal{R}_k(p^2)&=\mathcal{Z}_k \left(k^2-p^2 \right)\; H(k^2-p^2)\,.
\end{align}
The flow equation takes the explicit form
\be
\label{lpaphi4}
\frac{d}{dt}\Gamma_k[\varphi]
=\frac{1}{2}\mathrm{Tr}\left[\frac{2k^2+\eta(k^2+ \Delta) }{k^2+m^2_k+\mathcal{Z}_k\frac{\lambda}{2}\varphi^2}\; H(k^2+ \Delta)\right]\,,
\ee
where $\eta=\frac{d}{dt}\log\mathcal{Z}_k$ and $\Delta$ disappears from the denominator due to our choice of $\mathcal{R}_k$. For the moment, let us 
not consider the running of the kinetic term, so that we can put $\eta\approx0$. To recover the field operators of the potential, we can restrict ourselves
to constant fields, $\varphi(x)\equiv\varphi_o$. In this way, the trace can be easily computed as a momentum integral. In fact, if $\mathcal{V}_D$ is a $D$-dimensional 
volume element, we have
 \begin{align}
&\frac{d}{dt}\Gamma_k[\varphi]=\frac{1}{2}\int d^Dx\int\limits_{p^2<k^2} \frac{d^Dp}{(2\pi)^D}\left[\frac{2k^2}{k^2+m^2_k+\mathcal{Z}_k\frac{\lambda}{2}\varphi_o^2}\right]=\\
\nonumber
&=\mathcal{V}_D\frac{1}{(4\pi)^{D/2}\, \Gamma(1+D/2)} \frac{k^{D+2}}{k^2+m^2_k} 
\left[1-\frac{\mathcal{Z}_k\frac{\lambda}{2}\varphi_o^2}{k^2+m^2_k}+\frac{\mathcal{Z}^2_k\frac{\lambda^2}{4}\varphi_o^4}{(k^2+m^2_k)^2}+O(\varphi_o^6)\right]\,,
\end{align}
where in the last line we have expanded the fraction in powers of the fields.\footnote{Accidentally in this case such an expansion equals one in powers of $\lambda$,
but the rationale for the expansion is to be able to match the monomials on the left and right hand side.} Notice how terms of the form $\varphi_o^{2n}$ are generated 
from any $n$, reminding us that we are only in a small sector of $\mathcal{T}$. From the above 
equation we find the beta functions for the dimensionless couplings 
$\tilde{m}^2_k=m^2_k k^{-2}$, $\tilde{\lambda}_k=\lambda_k k^{D-4}$, where the explicit $k$-dependence has washed out.
\begin{align}
\label{mass4}
\frac{d }{dt}\tilde{m}^2_k=&\beta_{\tilde{m}^2}=-2\tilde{m}^2_k-\frac{1}{(4\pi)^{D}\Gamma(1+\frac{D}{2})} \frac{\tilde{\lambda}}{(1+\tilde{m}^2)^2}\,,\\
\label{lambda4}
\frac{d }{dt}\tilde{\lambda}_k=&\beta_{\tilde{\lambda}}=\,(D-4)\tilde{\lambda}+\frac{3!}{(4\pi)^{D/2}\Gamma(1+\frac{D}{2})} \frac{\tilde{\lambda}^2}{(1+\tilde{m}^2)^3}\,,
\end{align}
which in four dimension gives
\be
\frac{d }{dt}\tilde{m}^2_k\approx(-2+\frac{\tilde{\lambda}}{16\pi^2})\tilde{m}^2_k-\frac{\tilde{\lambda}_k}{32\pi^2}\,, \\
\quad\quad
\frac{d }{dt}\tilde{\lambda}_k\approx\frac{3\tilde{\lambda}^2}{16\pi^2} \,.
\ee
This has to be compared with the familiar one-loop result (see e.g. \cite{ramond}, \S 4.7) from perturbation theory with the mass independent renormalization. 
There are two manifest differences: first, we find higher order contributions in $\tilde{m}^2=m^2/k^2$ to both $\beta$-functions. Second, there is a discrepancy 
in the running of the mass term at zeroth order in $\tilde{m}$.
The former difference can be understood as a scheme dependence due to the infrared cutoff, which however does not alter the qualitative behavior of the flow, since $m^2/k^2\ll 1$.
The latter discrepancy amounts to a quadratic divergence, as it can be seen in terms of the dimension full quantities, $k\frac{d}{dk} m^2_k\approx {\tilde{\lambda}_k\;k^2}/{16\pi^2}$,
which does not appear directly in dimensional regularization.

We will now enlarge the truncation to include derivative terms.
We still have to extract the running of the wavefunction. For this purpose it is not enough to limit ourselves to constant fields. We need an $x-$dependent fluctuation term, 
such that $\varphi(x)=\varphi_o+\tilde{\varphi}(x)$. Then we cannot straightforwardly perform the momentum integration, because there are a number of differential operators, 
acting to the right on the $\tilde{\varphi}(x)$. However, we only need to use the commutator
\begin{equation}\label{commrel}
\left[-\Delta,\,\tilde{\varphi}(x)\right]=-\Delta\left(\tilde{\varphi}(x)\right)+2i \partial_\mu \left(\tilde{\varphi}(x)\right)\;i\partial^\mu\,,
\end{equation}
and the cyclic property of the trace to sort them to one side and write formally
\be
\mathrm{Tr}\left[\mathcal{A}(x)\,\mathcal{B}(i\partial)\right]=\sum_{x,p}\langle x| \mathcal{A}(x)|x\rangle\,\langle x|p\rangle\,\langle p|\mathcal{B}(p)|p\rangle\, \langle p|x\rangle\,.
\ee

In the case of a local potential approximation for a scalar field (\ref{lpaphi4}) we can sort all the derivative operators to the right just by the cyclic property of the trace for the 
terms contributing to the flow of $\eta$, i.e. for the ones quadratic in the field (which means that it was actually consistent to set $\eta=0$, which incidentally is a one loop exact result).

The higher loops  effects come into play through the flow of irrelevant couplings. For instance, it is easy to see that including a term such as 
$\int -\varphi^3\Delta\varphi=3\int\varphi^2\,\partial_\mu\varphi\partial^\mu\varphi$ into the ansatz would yield a nonzero $\beta$-function for $\eta$.
Let us sketch the computation. We consider
\[
\Gamma_k[\varphi]=\int d^Dx \left[-\mathcal{Z}_k\frac{1}{2}\varphi \Delta\varphi-\mathcal{Z}^2_k\frac{\lambda^{(4,1)}_k}{3!}
\varphi^3 \Delta\varphi+\mathcal{Z}_k^2\frac{m^2}{2}\,\varphi^{2}+\mathcal{Z}_k^2\frac{\lambda^{(4)}_k}{4!}\,\varphi^{4}\right]\;.
\]
Then
\begin{align}\label{floweta}
\frac{d}{dt}\Gamma_k[\varphi]=\frac{1}{2}\mathrm{Tr}\left[\frac{2k^2+\eta(k^2+ \Delta) }
{k^2+m^2_k-\frac{1}{2}\mathcal{Z}_k\lambda_k^{(4,1)}\Delta\varphi^2+\mathcal{Z}_k\frac{\lambda_k^{(4)}}{2}\varphi^2}\; H(k^2+ \Delta)\right]\;,
\end{align}
where all the derivative operators act to the right and the cyclicity of the trace is understood. To find the $\beta$-functions of the potential
we restrict again to constant configurations $\varphi_o$ (without setting $\eta=0$):
 \begin{align}\nonumber
\frac{d}{dt}\Gamma_k[\varphi]=\frac{1}{2}\int d^Dx& \frac{D\pi^{D/2}}{(2\pi)^D\Gamma(1+\frac{D}{2})}\int_0^k\!dp\,p^{D-1}\times\\
&\times\left[\frac{2k^2+\eta(k^2-p^2)}{k^2+m^2_k+\frac{1}{2}\mathcal{Z}_k\,\lambda_k^{(4,1)}\varphi_o^2p^2+\frac{1}{2}\mathcal{Z}_k \lambda_k^{(4)}\varphi_o^2}\right]+O(\tilde{\varphi})\,,
\end{align}
from which, by the usual expansion, we can extract  
\begin{align} 
\frac{d}{dt}\tilde{m}^2_k=(-2-&\eta)\tilde{m}^2_k-\frac{1}{4(4\pi)^{D/2}\,\Gamma(3+D/2)}\;\times \crcr
&\times \left[(4+D)(2+D-\eta)\frac{\lambda_k^{(4)}}{(1+\tilde{m}^2_k)^2}+D(4+D-2\eta)\frac{\lambda_k^{(4,1)}}{(1+\tilde{m}^2_k)^2}\right]\;,\crcr
\frac{d}{dt}\tilde{\lambda}^{(4)}_k=(D-&4-2\eta)\tilde{\lambda}^{(4)}_k+\frac{3}{4(4\pi)^{D/2}\,\Gamma(4+D/2)} \left[\frac{\scriptstyle(D+4)(D+6)(D+2-\eta)}{(1+\tilde{m}_k^2)^3}
(\tilde\lambda_k^{(4)})^2+\right.
\crcr
& \left.+\frac{\scriptstyle 2D(D+6)(D+4-\eta)}{(1+\tilde{m}_k^2)^3}\tilde\lambda_k^{(4)}\tilde\lambda_k^{(4,1)}+\frac{\scriptstyle D(D+2)(D+6-\eta)}{(1+\tilde{m}_k^2)^3}(\tilde\lambda_k^{(4,1)})^2\right]\;,
\end{align}
and the two remaining $\beta$-functions are obtained by expanding (\ref{floweta}) in terms of $\varphi(x)=\varphi_o+\tilde{\varphi}(x)$ and using (\ref{commrel}):
\begin{align}
\eta=&-\frac{2(D+2-\eta)}{(2\pi)^{D/2}\,\Gamma(2+D/2)}\frac{\tilde{\lambda}_k^{(4,1)}}{(1+\tilde{m}^2)^2}\;,\\
\frac{d}{dt}\tilde{\lambda}^{(4,1)}_k=&(2D-6-2\eta)\tilde{\lambda}_k^{(4,1)}+\frac{3}{8(4\pi)^{D/2}\,\Gamma(3+D/2)}\times \crcr
&\times\left[\frac{(4+D)(2+D-\eta)}{(1+\tilde{m}^2_k)^3}\tilde{\lambda}_k^{(4)}\tilde{\lambda}_k^{(4,1)}+\frac{D(4+D-\eta)}{(1+\tilde{m}^2_k)^3}(\tilde{\lambda}_k^{(4,1)})^2\right]\;, \nonumber 
\end{align}
which yields a nonzero flow for $\eta$.
 
Our results rely on the explicit choice of a regulator. To see its effect, one can repeat the calculation for (\ref{eq:phi4ansatz}) using the slightly more general regulator
 \begin{equation}
R_k(p;\;\alpha)=\alpha\;\mathcal{Z}_k \left(k^2-\frac{1}{\alpha}p^2 \right)\; H(k^2-\frac{1}{\alpha}p^2)\,,\quad\quad\al>0\,.
\end{equation}
It is straightforward to find the one loop results
\begin{align}
\frac{d }{dt}\tilde{m}^2_k=&-2\tilde{m}^2_k-\frac{1}{32\pi^2} \frac{\alpha^3\tilde{\lambda}}{(\alpha+\tilde{m}^2)^2}
\;\approx\;(-2+\frac{\tilde{\lambda}}{16\pi^2})\tilde{m}^2_k-\frac{\tilde{\lambda}_k}{32\alpha\pi^2}\,, \crcr
\frac{d }{dt}\tilde{\lambda}_k=&\,\frac{3}{16\pi^2} \frac{\alpha^3\tilde{\lambda}^2}{(\alpha+\tilde{m}^2)^3}\;\approx\;\frac{3\tilde{\lambda}^2}{16\pi^2}\,. 
\end{align}
We see how the regulator dependence does not modify the universal one-loop coefficient of the $\beta$-functions, but can lead to rather different results at finite values of the couplings.

\subsubsection{The Wilson-Fisher fixed point in \texorpdfstring{$D=3$}{D=3}}\label{sec:wilsonfisher}

The techniques that we have developed up to now will allow us to consider a more complicated statistical system.  It is an experimental fact that several 
three-dimensional magnetic systems exhibit a second order phase transition when their temperature $T$ approaches a critical value $T_\infty$. In that vicinity, 
for a class of them, it is observed that the correlation length $\xi$ diverges as
\be
\label{eq:correldiv}
\xi\approx (T-T_\infty)^{-\nu}\approx\vartheta^{-\nu}\,,\quad\quad \nu\approx0.63\,,
\ee
where we introduced the reduced temperature $\vartheta=(T-T_\infty)/T_\infty$. Such magnets are well described by an \emph{Ising model} which, in the approximation of continuous
spins (and in zero magnetic field), can be described by Euclidean $\lambda\phi^4$ theory in $D=3$. However, the behavior (\ref{eq:correldiv}) is  \emph{universal}, i.e. 
common to many magnetic systems, which may include more general spin interactions.

This behavior can be explained in terms a property of the theory space $\mathcal{T}$ common to all these theories, 
namely the existence of a fixed point with one relevant eigenvalue $\zeta$ (related to $\nu$). Then there will be a codimension-one stable manifold, 
and the phase transition at $T=T_\infty$ will happen as the one-dimensional curve in $\mathcal{T}$ describing the one-parameter family of QFTs under
consideration intersects $\Ws$. This will be confirmed by the RG analysis, which in fact is very reminiscent of the one we did for iterated maps on the interval in section~\ref{sec:feigenbaum}.

In a statistical theory we are interested in the long-wavelength (IR) behavior, i.e. $k\to0$. When we integrate down from a scale $k_1$ to $k_2<k_1$, the free energy 
will transform inhomogeneously, offsetting by a constant term which represents the energy of the modes that have been integrated out. Schematically, recalling that the
free energy is given by $\Gamma^{(0)}$,
\be
\Gamma^{(0)}(k_2;\,\vartheta_2)\approx \Delta \mathcal{F}_{k_1,k_2}+\left({k_2}/{k_1}\right)^{\rm some\ scaling}\Gamma^{(0)}(k_1;\,\vartheta_1)\,,
\ee
where the offset $\Delta \mathcal{F}_{k_1,k_2}$ is a regular function\footnote{This is similar to the offset $\Delta S_k$ that we subtracted from $\tilde{\Gamma}_k$ when deriving the FRGE.}. 
If we restrict to the singular contribution to the free energy we get a homogeneous scaling equation. The scaling exponent is simply given by the dimension, 
since $\Gamma^{(0)}$ has no external edges. As for the relation 
between $\vartheta_1$ and $\vartheta_2$ (similarly to the iterated map of the interval), since we are approaching the stable manifold $\Ws$ 
its scaling is given by the relevant eigenvalue at the fixed point, so that
\be
\Gamma^{(0)}_{\rm sing.}(s\,k;\,\vartheta)\approx s^3\ \Gamma^{(0)}_{\rm sing.}(k;\,s^{\zeta}\,\vartheta)\,,
\ee
where $s=k_2/k_1$, or equivalently
\be
\Gamma^{(0)}_{\rm sing.}((\vartheta_2/\vartheta_1)^{1/\zeta}k;\,\vartheta_2)\approx \left(\vartheta_1/\vartheta_2\right)^{3/\zeta}\Gamma^{(0)}_{\rm sing.}(k;\,\vartheta_1)\,.
\ee
This indicates that, close to the phase transition, the relative scaling of wavenumber and temperature is $k\approx \vartheta^{-1/\zeta}$, so that the correlation length diverges as
\be
\xi\approx \vartheta^{1/\zeta}\quad\quad\Longrightarrow\quad\quad \zeta=-1/\nu\,.
\ee

Now that we know how to relate $\zeta$ to the physics of the problem, it is time to investigate $\mathcal{T}$. As always, there exists a Gaussian fixed point. Clearly it is not 
the one we are after (the Ising model at the phase transition is not a free theory!). The Gaussian fixed point is an UV fixed point. 
If we define a microscopic theory with given values of the corresponding couplings in the UV, these flow away from the perturbative region in the IR,
and the $n$-points interactions are generated as effective vertices. 

To see how this happens we can use the FRGE \cite{Wetterich:1992yh,Adams:1995cv}. The simplest way to proceed is the ansatz (\ref{eq:phi4ansatz}) which at $D=3$ gives
\bea\label{wilsonflow}
&& \eta = 0, \quad  \frac{d }{dt}\tilde{m}^2_k= -2\tilde{m}^2_k- \frac{\tilde{\lambda}_k}{6\pi^2\;(1+\tilde{m}_k^2)^2}, \crcr
&& \frac{d }{dt}\tilde{\lambda}_k=-\tilde{\lambda}_k+ \frac{\tilde{\lambda}_k^2}{\pi^2\,(1+\tilde{m}_k^2)^3}\;.
\eea
Besides the Gaussian fixed point, there is a nontrivial one at
\be
 \tilde{\lambda}^*\approx7.7627,\quad\quad\tilde{m}^{2*}\approx-0.0769\,, 
\ee
and the critical exponents can be estimated from the Jacobian
\begin{equation}
\mathrm{Jac}_{NGFP}=\left(\begin{array}{cc}
-1.6667 & -0.0198\\
-25.229 & 1
\end{array}\right)\cong\left(\begin{array}{cc}
-1.8426 & 0\\
0 & 1.1759
\end{array}\right)\;.
\end{equation}
The numerical estimate for the exponent is
\be
\zeta\approx-1.8426\quad\quad\Longrightarrow\quad\quad
\nu\approx0.5427\,,
\ee
which is not far from the measured value.

\begin{figure}[!t]
  \begin{center}
\includegraphics[width=0.6\textwidth]{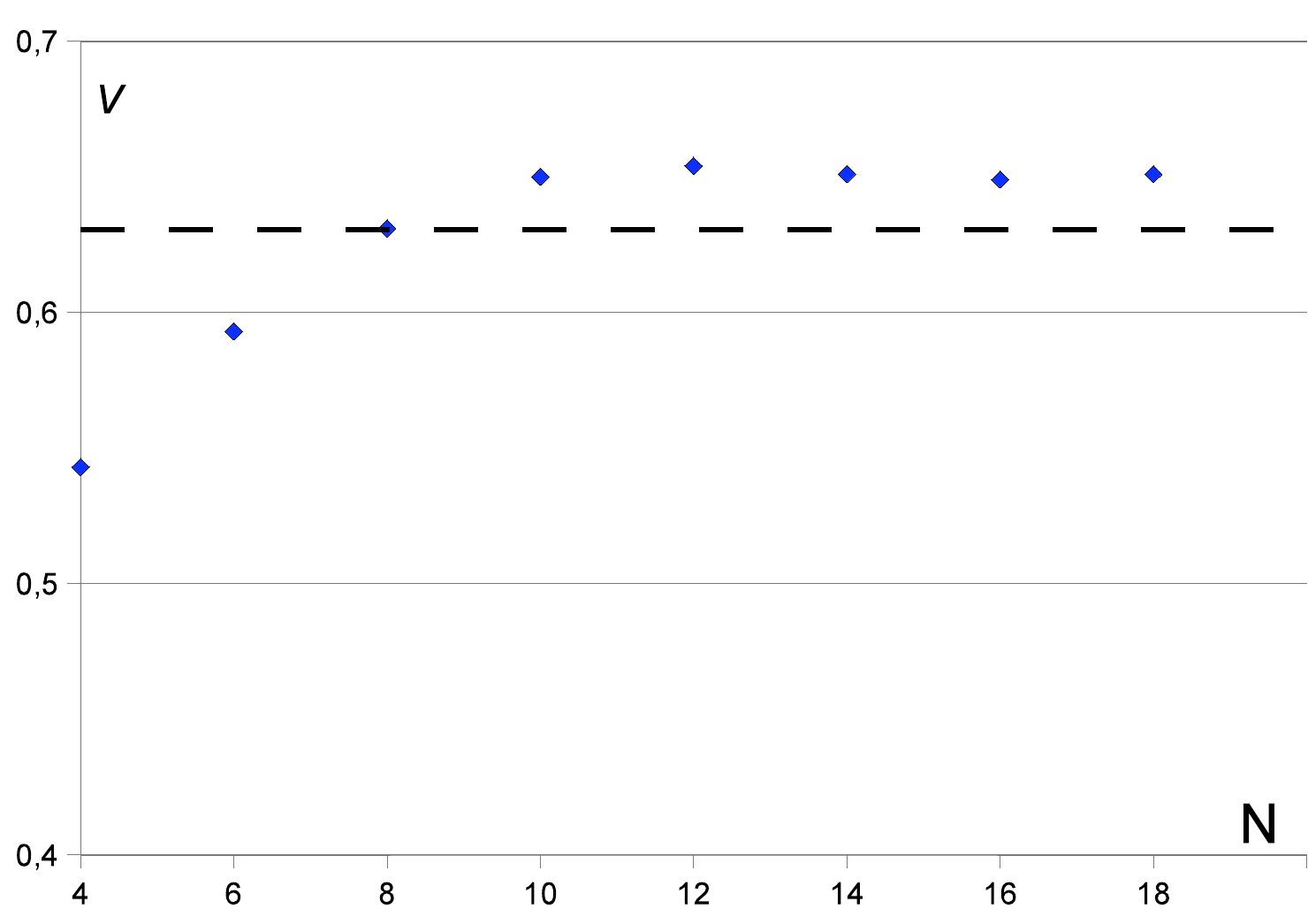}
  \caption{The critical exponent $\nu$ at the Wilson-Fisher fixed point, in various polynomial truncations of degree $N$. The dashed line is the prediction of seven-loops perturbation theory.}
  \label{fig:WF}
  \end{center}
\end{figure}
To be sure that we are dealing with a physical effect, we need to consider more general truncations, and ask ourselves:
\begin{enumerate}
\item Is the Wilson-Fisher fixed point present in all the truncations considered?
\item Is it the only non Gaussian fixed point?
\item Does the dimension of the attractor at the fixed point change?
\item Are critical exponents such as $\nu$ well behaved when considering larger truncations?
\end{enumerate}
Considering more general ans\"atze one sees that, even if some other fixed points may appear, the Wilson-Fisher fixed point is the only one which persists in all 
truncations. Furthermore, it appears that there is only one relevant direction, and the estimate for $\nu$ is quite stable, as shown in Figure \ref{fig:WF}.
Comparing this estimate with the result of resummed seven-loops perturbation theory $\nu=0.6304$, we find a discrepancy of about 3\% \cite{zinnjustin,Pogorelov}.
Remark that we have taken $\eta\equiv0$ whereas in the perturbative scheme $\eta^*=0.0335$. It is also possible to refine these results in the
local potential approximation, i.e. working in terms of a differential equation for the potential~$V_k(\varphi)$, or at higher order in a derivative expansion \cite{Bonanno:2000yp,Berges:2000ew, Canet:2002gs,Canet:2003qd,Canet:2004xe}, as well as in the BMW expansion scheme~\cite{Benitez:2009xg,Benitez:2011xx}. This leads to  spectacular agreement with perturbation theory and as well as Monte Carlo simulations, see e.g. ref.~\cite{Hasenbusch:2011yya}.

We have therefore established a scenario similar to the one of Section \ref{sec:feigenbaum}. This was first done in this context by Wilson and Fisher  \cite{Wilson:1971dc} 
using a different RG based technique\footnote{The interested reader will find nice pedagogical expositions in\cite{cardy1996scaling,weinberg1995quantum,weinberg1996quantum, zinnjustin}.} \cite{Wilson:1971vs,Wilson:1973jj}.
In conclusion, when we consider a one-parameter family of QFTs (i.e. a magnetic system at different temperatures) and approach the phase transition, the critical exponents 
are determined by the RG properties. All the theories on the stable manifold have infinite correlation lengths, and have IR properties similar to the ones of the fixed point theory. 

\begin{figure}[th]
  \begin{center}
\includegraphics[width=0.785\textwidth]{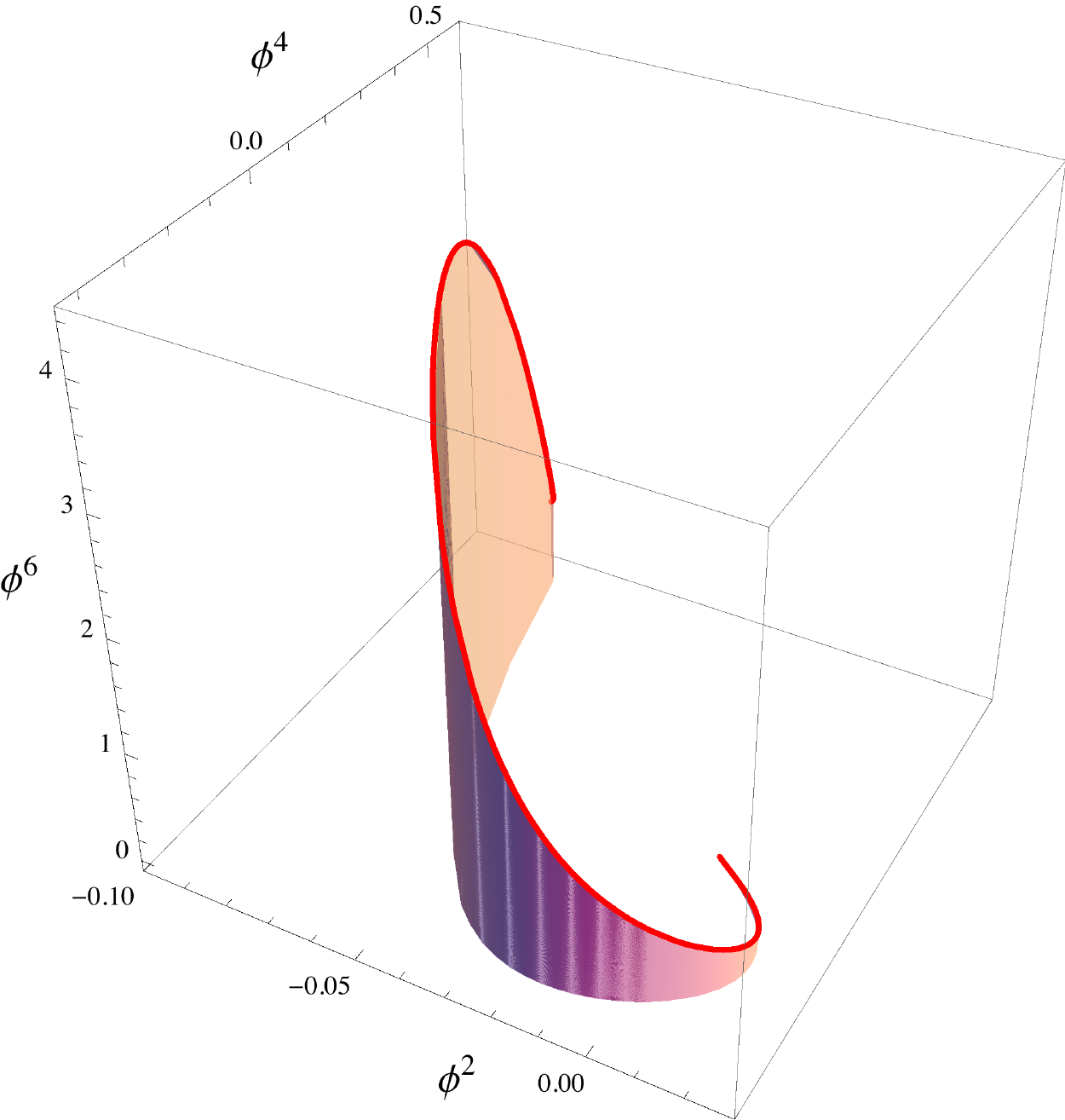}
  \caption{Flow from the Gaussian fixed point to the Wilson-Fisher fixed point. The shaded area represents the strength of the $\phi^6$ coupling.}
  \label{fig:WFflow}
  \end{center}
\end{figure}

Furthermore, the existence of two fixed points means that in this theory space there exists
asymptotically safe theories. In the vicinity of the Gaussian fixed point there exists a two-dimensional
manifold tangent to the plane $\{\tilde{m}^2,\tilde{\lambda}\}$ consisting of theories that in the UV sink 
into the GFP. We can consider theories that in the IR are attracted to the non Gaussian fixed point, so that for them the RG flow
is bounded as we change the scale, justifying \textsl{a posteriori} our crude treatment of the cutoff $\Lambda$. 
Of course in this particular case this would not be an issue, since this theory is not supposed to be fundamental, 
and we could have a natural UV cutoff $\Lambda\approx1/a$ where $a$ is the lattice spacing (nanometers), as well as an 
IR one of order of the meter. Figure \ref{fig:WFflow} depicts the flow from the Gaussian fixed point into the Wilson-Fisher one in a 
truncation including up to $\phi^6$ interactions; notice that the initial condition has no six-points interaction, but the latter is 
is generated as an effective vertex. 

\newpage

\section{Towards non-perturbative renormalization}
\label{sec:nonperturbative}

Constructive field theory \cite{Glimm:1987ng} is a set of techniques to resum perturbative quantum field theory 
and obtain a rigorous definition of quantities such as the Schwinger functions of interacting renormalized models.
It has the reputation of being a difficult technical subject.
In this brief introduction we would like to propose an invitation to the subject by introducing the most
typical formulas and results in increasing order of difficulty: the forest formula, the loop vertex expansion (LVE),
with application to the uniform Borel summability of a quartic combinatorial vector model, and finally a multiscale LVE, 
with application to a combinatorial field theory of the $\phi^4_2$ type \cite{Simon:1974dg,nelson}. We do not treat the most advanced and difficult part of the theory,
namely the so-called multi-scale phase space expansion  which is necessary to construct
just renormalizable asymptotically free models \cite{rivasseau1991perturbative}, as we feel it is both beyond the level of this review and 
has not been yet written into a sufficiently clean and canonical combinatorial form.

\subsection{The Forest Formula}
\label{forformul}

A forest formula expands a quantity defined on $n$ points in terms of forests built on these points. Forest formulas 
are particularly interesting for constructive theory when they have some positivity property. A beautiful such forest formula symmetric 
under action of the permutation group on the $n$ points was discovered in \cite{brydges} and developed with alternative proofs in \cite{Abdesselam:1994ap}\footnote{
Non-symmetric versions appeared earlier in the constructive literature, but won't be treated here (see \cite{Rivasseau:2013tpa} for a recent reference).}.

Consider $n$ points which we identify with the set $V_n$ of vertices of the complete graph $K_n$; the set of pairs of such points has
$n(n-1)/2$ elements $\ell = (i,j)$ for $1\le i < j \le n$ and can be identified with the set $E_n$ of edges $\ell$ of $K_n$.
The forest formula is most often presented as a Taylor expansion for functions $f$
of $n(n-1)/2$ variables $x_\ell$, $\ell \in E_n$ which are smooth e.g. on an open neighborhood
of $[0,1]^{n(n-1)/2}$.  Here we propose a variant formulated in terms of functions defined on positive symmetric matrices. It is closer to 
constructive applications such as the LVE.

Consider the vector space $S_n$ of symmetric $n$ by $n$ matrices $X= \{X_{ij}\}, i,j = 1 ,\cdots n$. It has dimension $n(n+1)/2$.
The set $PS_n$ of positive symmetric matrices whose diagonal coefficients are all equal to 1 and off-diagonal elements are 
between 0 and 1 is compact and convex. Symmetric matrices with diagonal elements equal to one and off-diagonal elements in $[0,1]^{n(n-1)/2}$
do not all belong to $PS_n$, for instance the matrix $\begin{pmatrix} 1 &1&0 \cr 1 &1&1 \cr 0 &1&1 \end{pmatrix}$ is not positive. 
Any matrix $X\in PS_n$ can be parametrized by $n(n-1)/2$ elements $X_\ell$, 
where $\ell$ runs over the edges of the complete graph $K_n$. 

$PS_n$ contains as particularly interesting elements the block matrices $X^\Pi$ for any partition $\Pi$ of $V_n$. The
block matrix $X^\Pi$ has entries $X_{ij}^\Pi =1 $ if $i$ and $j$ belong to the same block of the partition $\Pi$, and 0 otherwise. 
Two extremal cases are the identity matrix $\cId$, corresponding to $X^{sing}$, that is to the maximal partition made of all singletons, and
the matrix $\bbone$ with all entries equal to one, corresponding to $X^{V_n}$, that is to the minimal partition made of a single block.

Let us consider a function $f$ defined and smooth in the interior of $PS_n$ with continuous extensions (together with all their derivatives) to $PS_n$ itself.
The forest formula can be expressed as a multi-variate Taylor formula with integral remainder which expands such a function
between the minimal and maximal block-partition matrices $\bbone$ and $\cId$. 
The important point is that the Taylor remainder integrals stay on the $PS_n$ convex set.
The precise statement is

\begin{theorem}[The Forest Formula]
\bea  f( {\bbone }) =      \sum_{\cF} \int dw_\cF  \;  \partial_\cF  f \, [ X^\cF (w_\cF) ] \label{bkar}
\eea
where
\begin{itemize}

\item The sum over $\cF$ is over forests over $n$ labeled vertices $i = 1, \cdots , n$, including the empty forest with no edge. Such forests
are exactly the acyclic edge-subgraphs of the complete graph $K_n$.

\item  $\int dw_\cF$ means integration from 0 to 1 over one parameter for each forest edge: $\int dw_\cF  \equiv \prod_{\ell\in \cF}  \int_0^1 dw_\ell  $.
There is no integration for the empty forest since by convention an empty product is 1. A generic integration point $w_\cF$
is therefore made of $\vert \cF \vert$ parameters $w_\ell \in [0,1]$, one for each $\ell \in \cF$.

\item  $ \partial_\cF = \prod_{\ell\in \cF} \partial_\ell  $ means a product of first order partial derivatives with respect to 
the variables $X_{\ell}$ corresponding to the edges of $\cF$. Again there is no such derivatives for the empty forest since by convention an empty product is 1.

\item $X^\cF (w_\cF)$ is defined by $X^\cF_{ii} (w_\cF )= 1$ $\forall i$, and for $i \ne j$,
$X^\cF_{ij} (w_\cF)$ is the infimum of the $w_\ell$ parameters for $\ell$
in the unique path $P^\cF_{i \to j}$ from $i$ to $j$ in $\cF$, when such a path exists. 
If no such path exists, which means that $i$ and $j$ belong to different connected components with respect to the forest 
$\cF$, then by definition $X^\cF_{ij} (w_\cF) = 0$.

\item The symmetric $n$ by $n$ matrix $X^\cF (w_\cF)$ defined in this way is positive, hence
belongs to $PS_n$, for any value of $w_\cF$.

\end{itemize}
\end{theorem}
\noindent
Since $X^\emptyset = \cId$, the empty forest term in \eqref{bkar} is $f(\cId)$, hence \eqref{bkar} indeed
interpolates $f$ between $\bbone$ and $\cId$, staying on $PS_n$ as announced.

\medskip
\prf We would like to describe the proof in an informal inductive manner. We introduce first an interpolation parameter $w_1$ and
perform the interpolation $X(w_1) =  (1-w_1)\cId + w_1\bbone $. The interpolation path, $X(w_1)$, remains in $PS_n$ because $PS_n$
is convex. A first order Taylor expansion in the variable $w_1$ between 0 and 1  gives
\bee   f( {\bbone }) =   f(\cId) + \sum_{\ell_1 \in E_n}  \int_0^1 dw_{\ell_1} \;  \partial_{\ell_1}  f \, [ X(w_{\ell_1})  ] .  \label{bkar1}
\ee
This is because all off-diagonal elements in $X(w_1)$ have value $w_1$, hence 
expanding the total derivative $df/dw_1$ into partial derivatives leads to a finite sum over $\ell_1\in E_n$. Once this finite
sum had been commuted with the $w_1$ integral, we then apply a both trivial and subtle relabeling of the dummy 
integration variable $w_1$ as $w_{\ell_1}$.

The first term in \eqref{bkar1}  corresponds to the empty forest. In the second term we can define a first partition $\Pi_1$ of 
$V_n$ into blocks, each block being made of the connected components of $V_n$ with respect to the edge-subgraph $\{ \ell_1\}$. There are exactly
$n-1$ such connected components, namely the $n-2$ vertices untouched by $\ell_1$ and the block of the two end vertices of $\ell_1$. 

In $X(w_{\ell_1}) $ we interpolate the trans-block entries for $\Pi_1$, which have values $w_1$, by a parameter $w_2 \in [0, w_1]$
but we \emph{do not interpolate} the same-block entries for $\Pi_1$:
\bee
  X(w_{\ell_1},w_2) = (1-w_{\ell_1} ) \cId + (w_{\ell_1} - w_2) X^{\Pi_1} + w_2 \bbone \; .
\ee
Applying first order Taylor formula with integral remainder for $w_2 \in [0, w_1]$, then
expanding again the total derivative $df/dw_2$ gives a sum over $\ell_2$, which cannot create a cycle with $\ell_1$. Commuting this sum 
and the $w_2$ integral and performing the ``trivial-subtle" change $w_2 \to w_{\ell_2}$, we obtain a second step formula
\bea   f( {\bbone }) &=&   f(\cId) + \sum_{\ell_1 \in E_n}  \int_0^1 dw_{\ell_1} \;  \partial_{\ell_1}  f \, [ X^{\{\ell_1\}}(w_{\ell_1})  ] \nonumber \\
&+&\sum_{ \{\ell_1 , \ell_2\}\;  {\rm forest\;  of}\;  K_n }  \int_0^1 dw_{\ell_1}  
\int_0^{w_{\ell_1}} dw_{\ell_2} \;  \partial_{\ell_1}  \partial_{\ell_2}  f \, [ X(w_{\ell_1},w_{\ell_2})  ] .
 \label{bkar2}
\eea
Remark indeed that the $w_2 = 0 $ term puts all trans-block entries for $\Pi_1$ to 0 in $X(w_{\ell_1}) $, hence creates exactly $X^{\cF}(w_{\ell_1}) =  (1-w_{\ell_1} ) \cId  + 
w_{\ell_1} X^{\Pi_1} $  for the single-edged forest $\cF = \{\ell_1\} $. The matrix $X(w_{\ell_1},w_{\ell_2}) $ is a convex combination of 
positive symmetric matrices, hence it is in $PS_n$.

Now we can define a partition $\Pi_2$ of $V_n$ into blocks, each block being made of the connected components of $V_n$ 
corresponding to the ordered forest $\{ \ell_1, \ell_2\}$.
Remark that there are exactly $n-2$ such connected components, no matter whether $\ell_1$ and $\ell_2$ are adjacent or not.
Then we interpolate the trans-block entries for $\Pi_2$ in $X(w_{\ell_1},w_{\ell_2}) $, which have values $w_2$, by a parameter $w_3 \in [0, w_2]$ but not 
the ones which are same-block for $\Pi_2$,
\bee 
X(w_{\ell_1},w_{\ell_2},w_{3})  = (1-w_{\ell_1} ) \cId  + (w_{\ell_1} -w_{\ell_2})X^{\Pi_1}  + (w_{\ell_2} -w_{3})X^{\Pi_2}  + w_{3}\bbone \; ,
\ee
obtaining:
\bea   f( {\bbone }) &=&   f(\cId) + \sum_{\{\ell_1\} \;  {\rm forest\;  of}\;  K_n }  \int_0^1 dw_{\ell_1} \;  \partial_{\ell_1}  f \, [ X^{\{\ell_1\}}(w_{\ell_1})  ] \nonumber \\
&+&\sum_{ \{\ell_1 , \ell_2\}\;  {\rm forest\;  of}\;  K_n }  \int_0^1 dw_{\ell_1}  
\int_0^{w_{\ell_1}} dw_{\ell_2} \;  \partial_{\ell_1}  \partial_{\ell_2}  f \, [ X^{\{\ell_1,\ell_2\}}(w_{\ell_1},w_{\ell_2})  ] 
 \nonumber \\
&+&\sum_{ \{\ell_1 , \ell_2, \ell_3\}\;  {\rm forest\;  of}\;  K_n }
\int\int\int_{0\le w_{\ell_3}  \le w_{\ell_2}\le w_{\ell_1}\le1  }  dw_{\ell_1}   dw_{\ell_2} dw_{\ell_3}  \;  \nonumber \\
&& \quad \quad \quad \quad \quad \quad \quad \quad\quad \quad \quad \quad \partial_{\ell_1}  \partial_{\ell_2}\partial_{\ell_3} 
f \, [ X(w_{\ell_1},w_{\ell_2}, w_{\ell_3})  ] \; .
\label{bkar3}
\eea
We iterate this procedure until it terminates, something which must happen after exactly 
$n-1$ of steps (since at step $k$ the partition $\Pi_k$ has $n-k$ blocks). In this way we obtain an \emph{ordered forest formula}
\bea
f( {\bbone }) =  \sum_{\cF, \sigma} \int_\sigma dw_\cF  \;  \partial_\cF  f \, [ X^\cF (w_\cF) ] \label{bkar4}
\eea
where the sum over $\sigma$, for a forest $\cF$ with $k$ edges, runs over the $k!$ ordering of $\cF$ as $(\ell_1, \cdots \ell_k)$, the sign $\int_\sigma dw_\cF$
means that the $w_\ell$ parameters are ordered as $0 \le w_{\ell_k} \cdots \le w_{\ell_1} \le 1$ and the  $X^\cF (w_\cF)$ matrix is the one defined in \eqref{bkar}.
This last fact can be understood as follows; the value of an entry $(i\ne j)$ in $X^\cF (w_\cF)$ is $w_m$ for $m$ the last integer $1 \le mÊ\le k$ such that this entry
is trans-block for $\Pi_m$. But if $i$ and $j$ are same-block entries for $\cF$ 
this is exactly the smallest value of $w_\ell$ for $\ell$ in the path in $\cF$ connecting $i$ to $j$; otherwise it is 0.

The fact that  $X^\cF (w_\cF)$ is positive for any ordering $\sigma$ now stems from the fact that, at each step, it was defined as a convex combination
of block matrices:
\bee 
 X^\cF(w_\cF) = (1-w_{\ell_1}) \cId  + (w_{\ell_1} -w_{\ell_2}) X^{\Pi_1}  + (w_{\ell_2} -w_{\ell_3})X^{\Pi_2} + \dots + w_{\ell_k}  X^{\Pi_k}  \;.
 \label{blockdecomp}
\ee
Remark that this decomposition of $X^{\cF}$ as a barycentric combinations of block matrices
depends on $\sigma$. Hence $X^{\cF}$ is in $PS_n$
for any $w_\cF$, as announced, but for a different reason in each different sector (ordering) of the parameters $w_\cF$.

Now summing in the ordered forest formula \eqref{bkar4} over all orderings $\sigma$ completes the proof, as for any forest $\cF$ 
the sum over orderings reconstructs exactly the integration domain $\int dw_\cF  = \prod_{\ell\in \cF}  \int_0^1 dw_\ell  $.

\qed

We give now a useful corollary of this theorem which expands Gaussian integrals over replicas.
Consider indeed a Gaussian measure $d\mu_C$ of covariance $C_{pq}$ on a vector variable $\vec \tau$ with $N$
components $\tau_p$. To study approximate factorization properties of the integral of a product of $n$ functions of the variable $\vec \tau$ 
it is useful to first rewrite this integral using a replica trick. It 
means writing the integral over $n$ identical replicas $\vec \tau_i$ for $i=1, \cdots , n$ with components $\tau_{p,i}$, with the 
perfectly well-defined measure with covariance $[C\otimes \bbone]_{p,i ; q,j} = C_{pq} \bbone_{ij} = C_{pq} $:
\bee  \int d\mu_C (\vec \tau)    \prod_{i=1}^n  f_i(\vec \tau)  =   \int d\mu_{C\otimes \bbone} (\vec \tau_i)    \prod_{i=1}^n  f_i(\vec \tau_i) .
\ee
Applying the forest formula we obtain the following corollary

\begin{cor}
\bea I= \int d\mu_C (\vec \tau)    \prod_{i=1}^n  f_i(\vec \tau)  =  \sum_{\cF} \int dw_\cF  \int d\mu_{C\otimes X^\cF  (w_\cF)} (\vec \tau_i)      
\;  \partial^C_\cF  \prod_{i=1}^n  f_i(\vec \tau_i)     \label{bkar5}
\eea
where $\partial^C_\cF$ means $\prod_{\ell =(i,j)\in \cF}  \Bigl( \sum_{p,q}\frac{\partial}{\partial \tau_{p,i}} C_{pq} \frac{\partial}{\partial \tau_{q,j}} \Bigr)$.
\end{cor}

Which follows directly by rewriting the Gaussian integral as
\bee 
\int d\mu_C  f (x) =  e^{  \frac{\partial}{\partial \tau_i} C_{ij}  \frac{\partial}{\partial \tau_j}} f \Big{\vert}_{\tau=0} \; .
\ee

Another corollary of the forest formula defines interesting barycentric tree weights $w(G,T)$ for spanning trees $T$ of a fixed connected graph $G$.
Barycentric weights means 
\bee  \sum_{ T \subset G}   w(G, T) = 1 \; ,
\ee
where the sum runs over all spanning trees of $G$.

Consider indeed a fixed connected graph $G$, possibly with self-loops and multiple edges.
Developing the function $f(X) = \prod_{\ell \in G} X_{i(\ell)j(\ell)}$ by the forest formula we obtain
\bee  1 = f( {\bbone }) =  \sum_{T} \int dw_T   \prod_{\ell \in G-\cT}  X^T_{i(\ell)j(\ell)} (w_T) \;,
\ee
where the sum runs over spanning trees $T \subset G$, because the forests with at least two trees will assign a $0$ value to at least one edge 
variable. We can consider this formula as defining the barycentric tree weights
\bee w(G,T) = \int dw_T   \prod_{\ell \in G-T}  X^T_{i(\ell)j(\ell)} (w_T) .  \label{defgoodweights}
\ee 

These barycentric tree weights can also be computed through Kruskal's greedy algorithm. 
For any \emph{Hepp sector} $\sigma$, hence any complete ordering of the edges of $G$, Kruskal greedy algorithm \cite{kruskal} defines a particular  tree $T( \sigma)$,
which minimizes $\sum_{\ell \in T}  \sigma (\ell)$ over all  trees of $G$, where $\sigma(\ell)$ is the order of $\ell$ in $\sigma$, also called the weight of $\ell$.
We call $T( \sigma)$, the \emph{leading tree} for $\sigma$.
The algorithm simply picks the first edge $\ell_1$ (whose weight is minimum) in $\sigma$ which is not a self-loop.
Then it picks the next edge $\ell_2$
in $\sigma$ that does not add a cycle to the (disconnected) graph with vertex set $V$ and edge set $\ell_1$ such that the 
sum of their weights $\sigma(\ell_1)+\sigma(\ell_2)$ is minimal, and so on.
Another way to look at it is through a deletion-contraction recursion: following the ordering of the sector $\sigma$, every edge is either deleted
if it is a self-loop or contracted if it is not. The set of contracted edges is exactly the leading tree for $T(\sigma)$. Then we have
\begin{cor}
\be w(G,T) =\frac{N(G,T)}{|E|!}  \label{def}
\ee
where  $N(G,T)$ is the number of sectors $\sigma$
such that $T(\sigma)=T$.
\end{cor}

\prf We introduce first parameters $w_\ell$ for all the edges in $G-T$, writing
\be X^T_{ij}(\{w \}) = \int_0^1  dw_\ell \bigl[ \prod_{\ell' \in P^T_{i \to j }} \chi(w_\ell < w_{\ell'} ) \bigr] ,
\ee
where $\chi(\cdots)$ is the characteristic function of the event $\cdots$.
Then we decompose the $w$ integrals according to all possible orderings $\sigma$:
\bea
w(G,T)&=& \int_{0}^1 \prod_{\ell\in G}dw_\ell \prod_{\ell \not\in T} \bigl[ \prod_{\ell' \in P^T_\ell} \chi(w_\ell < w_{\ell'} ) \bigr]
\nonumber \\ &=&
\sum_{\sigma} \chi ( T(\sigma) =T) \int_{0< w_{\sigma(|E|)}  < \cdots < w_{\sigma(1)}  < 1} \prod_{\ell\in G}dw_\ell   ,
\eea
as in the domain of integration defined by $0< w_{\sigma(|E|)}  < \cdots < w_{\sigma(1)}  < 1$
the function  $\prod_{\ell \not\in T} \bigl[ \prod_{\ell' \in P^T_\ell} \chi(w_\ell < w_{\ell'} ) \bigr]$
is 1 or zero depending whether $ T(\sigma) =T$ or not, as this function being 1 is
exactly the condition for Kruskal's algorithm to pick exactly $T$. Strict inequalities are easier to use here: of course
equal values of $w$ factors have zero measure anyway. Hence
\be \int dw_T   \prod_{\ell \in G-T}  X^T_{i(\ell)j(\ell)} (w_T)  =\frac{N(G,T)}{|E|!}  \; .  \label{def3}
\ee

\qed

\subsection{LVE for the N-Vector \texorpdfstring{$\phi^4$}{phi**4} Model}

The loop vertex expansion (LVE) combines an intermediate field functional integral representation 
for QFT quantities with the forest formula and the replica trick of the previous section.
It allows the computation of connected functional QFT integrals  such as the free energy or connected Schwinger functions
as convergent sums indexed by spanning trees of arbitrary size $n$ rather than divergent sums indexed by Feynman graphs. 

Initially introduced to analyze \emph{matrix} models with quartic interactions \cite{Rivasseau:2007fr}, the LVE has been
extended to arbitrary stable interactions \cite{Rivasseau:2010ke} and shown compatible with
direct space decay estimates \cite{Magnen:2007uy}. 
It has also been used to analyze random \emph{tensor} models \cite{Magnen:2009at,Gurau:2013pca}.

The LVE expressed any Schwinger function $S$ as a convergent sum over trees of the \emph{intermediate field representation}:
\be  S = \sum_{T} A_T, \quad A_T = \sum_{G \supset T}  w(G,T)  A_G \; ,
\label{const}
\ee
with
\be  \sum_{T}  \vert A_T\vert < +\infty \;.
\label{const1}
\ee 
The usual (divergent) perturbative expansion of $S$ is obtained by the ill defined commutation of the sums over $T$ and $G$,
\bea \label{ordinar}
&& S = \sum_{T} \Bigl(\sum_{G \supset T}  w(G,T)  A_G  \Bigr)  ``="  \sum_{G }  \sum_{T \subset G}  w(G,T)  A_G  = \sum_{G} A_G \; ,\crcr
&& \sum_{G} \vert A_G \vert = \infty \; .
\eea 

We shall limit ourselves here to introduce the LVE in the 
particularly simple case of the quartic $N$-vector models, for which
the $1/N$ expansion is governed by rooted plane trees.

More precisely, consider a pair of conjugate vector fields $\{\phi_p \}, \{\bar \phi_p \},  p=1 ,\cdots , N $, 
with $  (\bar \phi \cdot \phi )^2$ interaction. The corresponding functional integral
\bee  Z(z, N) =  \int \frac{d \bar \phi d \phi }{(2i\pi )^N}  \;  e^{- (\bar \phi \cdot \phi )  + \frac{z}{2N} (\bar \phi \cdot \phi )^2 }  \; ,
\ee
is convergent for $\Re z<0$. Note the slightly unusual sign convention for the interaction term. We rewrite it, using a scalar intermediate field $\sigma$, as:
\bea \label{spliscalar}  Z(z, N) &=&     \int d \sigma \frac{e^{- \sigma^2 /2 } }{\sqrt{2\pi}} \int \frac{d \bar \phi d \phi }{(2i\pi )^N} 
 e^{- (\bar \phi \cdot \phi ) +  \sqrt {z/N} (\bar \phi \cdot \phi ) \sigma } 
  \nonumber \\
&=& \int  \frac{d \sigma }{\sqrt{2\pi}} e^{- \sigma^2 /2 - N \log (1 -   \sqrt { z/N}   \sigma )} .
\eea
Defining  $\tau = \sigma / \sqrt N $ one gets
\bea  Z(z, N) &=&  \int \frac{\sqrt N d \tau }{\sqrt{2\pi}}e^{- N [\tau^2 /2 + \log (1 - \sqrt {z}   \tau )]} . \label{partitfunc}
\eea
The two point function
\bea  G_2(z, N) &=& \frac{1}{Z(z, N)}   \int  \frac{d \bar \phi d \phi }{(2i\pi )^N} \;\; \frac{1}{N} (\sum_p \bar \phi_p \phi_p) e^{- (\bar \phi \cdot \phi )  
+ \frac{z}{2N} (\bar \phi \cdot \phi )^2 } \; , \eea
can be deduced from the free energy by a Schwinger-Dyson equation
\bea  0 = \frac{1}{Z(z, N)}  \int   \frac{d \bar \phi d \phi }{(2i\pi )^N} \;\; \frac{1}{N}\sum_p  \frac{\partial}{\partial \phi_p}
[ \phi_p   e^{- (\bar \phi \cdot \phi )  +\frac{z}{2N} (\bar \phi \cdot \phi )^2 } ]  \;,
\eea 
which yields 
\bea  G_2(z, N) &=& 1  + 2 z  \frac{d}{d z } \biggl(\frac{1}{N}  \log \int \frac{ \sqrt{N}d \tau }{\sqrt{2\pi}}e^{- N [\tau^2 /2 + \log (1 - \sqrt {z}   \tau )]}  \biggr) . \label{2pointint}\eea

A simple saddle point evaluates the integral \eqref{partitfunc} as
$\frac{K e^{- N f(\tau_c) } }{ \sqrt{  f" (\tau_c)}}$,
where the saddle point of $f(\tau) = \tau^2 /2 + \log (1 - \sqrt {z}   \tau )]$ is at $\tau_c$ with $f'( \tau_c)=0$ hence 
$\tau_c = \frac{1}{{2 \sqrt z}}  [1 - \sqrt{1 - 4 z}]$. Also 
\bea \label{limpart} \lim_{N \to \infty}  \frac{ \log  Z(z, N)}{N} &=&  - f(\tau_c) \;,
\eea
and the two point function in the $N\to \infty$ limit is 
\bea  \lim_{N \to \infty}  G_2(z, N)
&=& 1  + 2 z \Bigl(- \partial_z f(\tau_c) -\partial_{\tau } f(\tau_c) \frac{d\tau_c}{dz}   \Bigr) = 
  1 - 2z \frac{ -\frac{1}{2 \sqrt{z} } \tau_c }{ 1-\sqrt{z}\tau_c }
\crcr
&=&  \frac{1}{2z}   [ 1 -   \sqrt{1 - 4 z} ]    \; , \label{catalansigma}
\eea
which we recognize as the generating function of the Catalan numbers. 

Let us now study Borel summability in  $z$ of these quantities uniformly as $N \to \infty$, using the loop vertex expansion. 
We start from the intermediate field representation of the two-point function \eqref{2pointint} and apply the LVE to get
\bea \label{vectorlve} G_2(z, N) &=& \sum_{\cT} \frac{1}{n!} z^{n} \int dw_\cT \int d\mu_\cT \prod_{c \in C(\cT)} \frac{1}{1- \sqrt {z} \tau_{i(c)}}
\eea
where in \eqref{vectorlve} 
\begin{itemize}

\item the sum over $\cT$ is over rooted plane trees, with one ciliated root vertex labeled $i=0$ plus $n \ge 0$ ordinary vertices labeled $1, \cdots ,n$.

\item $\int dw_\cT$ as in subsection \ref{forformul} means $\big[ \prod_{\ell\in \cT}  \int_0^1 dw_\ell   \big]$

\item $d\mu_\cT $ is the normalized Gaussian measure on the $(n+1)$-dimensional vector field $\vec \tau = (\tau_i)$, $i=0,1, \cdots, n$ 
running over the vertices of $\cT$,
which has covariance $\frac{X_{ij}^\cT (w_\cT)}{N}$ between vertices $i$ and $j$. Recall that $X^\cT (w_\cT)$ is 
defined in subsection \ref{forformul}.

\item the product over $c$ runs over the set $C(\cT)$ of the $2n+1$ corners of the tree, the cilium creating an additional corner 
on the plane tree, and $i(c)$ is the index of the 
vertex to which the corner $c$ belongs.

\end{itemize}

\begin{figure}[ht] 
\centerline{\includegraphics[width=6cm,angle=0]{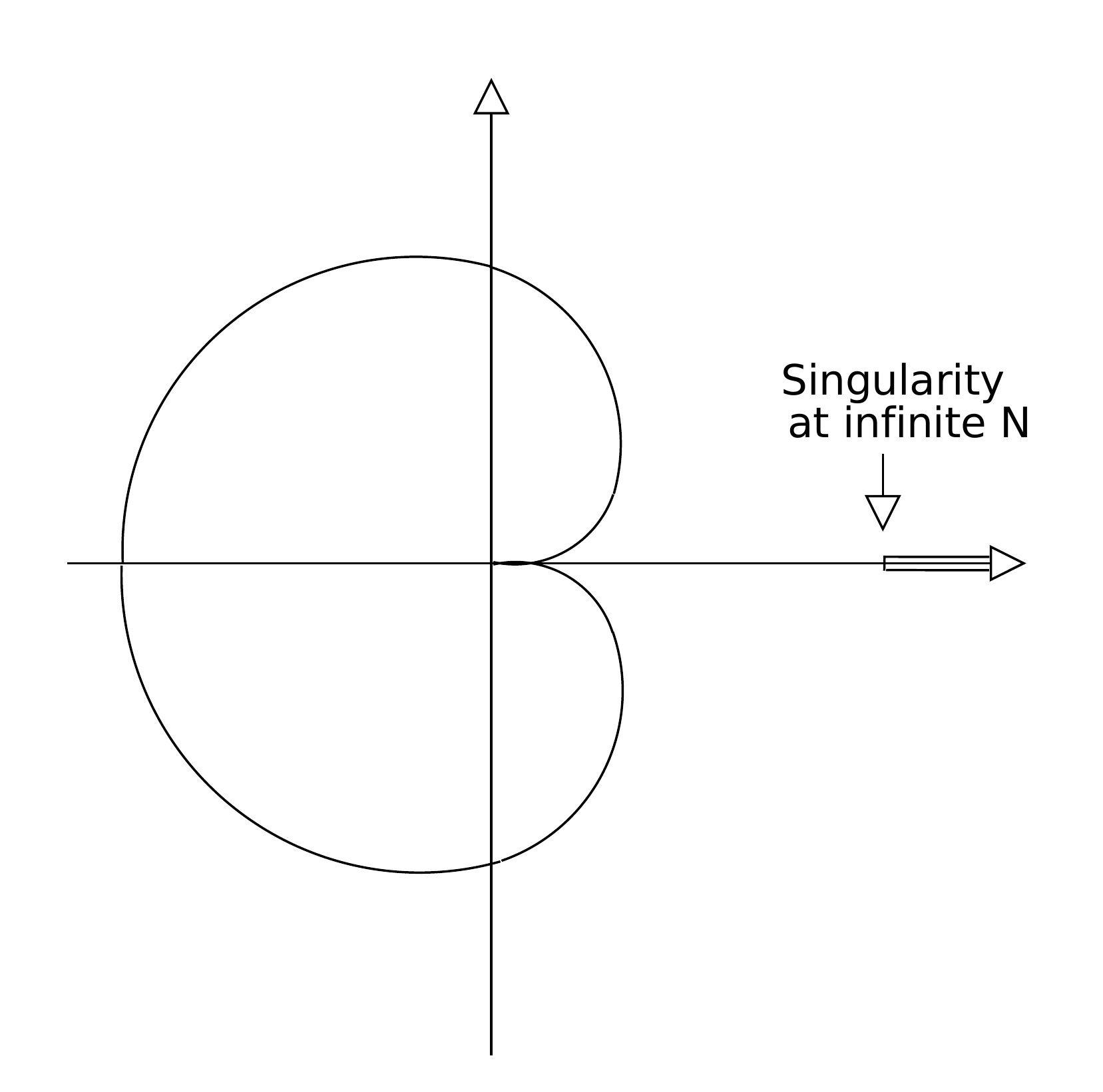}}
\caption{The cardioid domain}\label{cardioidfig}
\end{figure}

It is now obvious why \eqref{catalansigma} is true; since the covariance of the $\tau$ fields vanishes as $N \to \infty$ 
the limit of $G_2(z, N) $ is obtained by putting every $\tau_{i(c)}$ factor to 0 in every corner resolvent, 
in which case we exactly get a weight $z^{n(\cT)}$ for each rooted plane tree, 
hence we recover the Catalan generating function (the $1/n!$ is canceled by the relabellings of the vertices). 

We can now use \eqref{vectorlve} to prove analyticity and Borel summability of the free energy and correlation functions of the model in the variables 
$z$ and $1/N$ in the cardioid domain of Figure \ref{cardioidfig} (see \cite{Frohlich:1982rj} for an early reference
to Borel summability of the $1/N$ expansion of vector models).

Let us set $z = \vert z \vert e^{i \pi + i\phi}$ for $\vert \phi \vert <\pi $. We have 
$\sqrt z = i\sqrt{ \vert z \vert} e^{i \phi/2}$. Each resolvent $\frac{1}{1- \sqrt {z} \tau_{i(c)}}$ is 
bounded in norm by $[\cos (\phi/2)]^{-1}$, hence using the fact that there are $2n+1$ such resolvents,
we obtain analyticity of representation \eqref{vectorlve} for $4 \vert z \vert < [ \cos (\phi/2 )]^{2} $,
the cardioid domain of Figure \ref{cardioidfig}.

\begin{figure}[ht] 
\centerline{\includegraphics[width=6cm,angle=0]{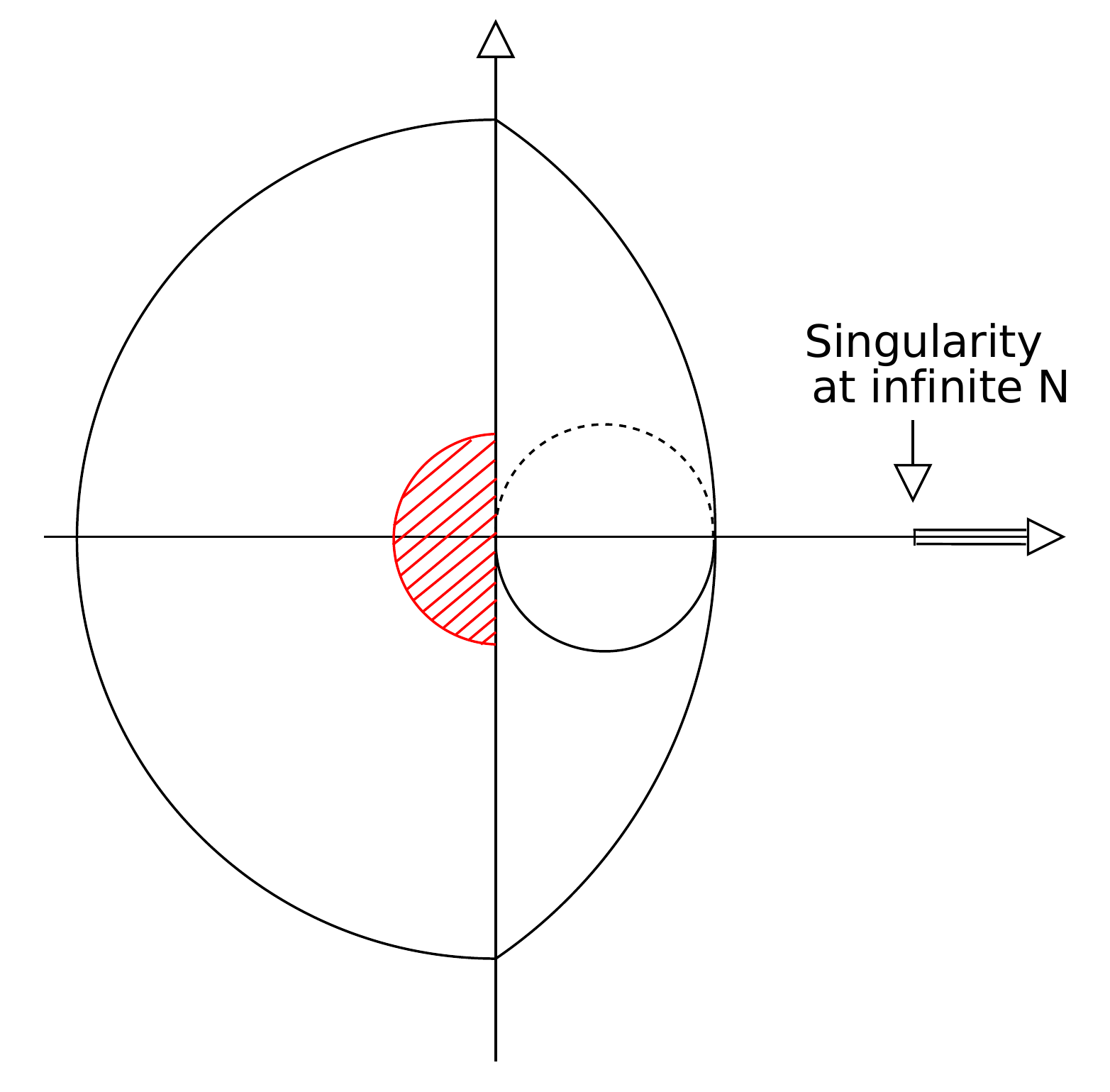}}
\caption{The extended cardioid domain}\label{extendedcardioidfig}
\end{figure}

But in fact we can extend the analyticity domain into the extended cardioid domain of Figure \ref{extendedcardioidfig}, a domain 
introduced for quartic vector models in \cite{Billionnet:1982cd}.
Indeed using the parametric representation of resolvents
\bea \label{resolvents1}  \frac{1}{1- \sqrt {z} \tau}  &=& \int_0^\infty  d \alpha_c   e^{- \alpha_c (1 -  \sqrt {z}  \tau ) }  
\eea
we can explicitly integrate over the measure $d\mu_{\cT}$ and get the integral representation
\bea   G_2(z, N) &=& \sum_{\cT} \frac{1}{n!} z^{n}  \int dw_\cT  \big[ \prod_{c\in \cT} \int_0^\infty  d \alpha_c  e^{- \alpha_c }   \big] \nonumber \\
&&e^{ \frac{z}{2N}  \sum_{ij}(\sum_{c \in i} \alpha_c) X^\cT_{ij}(w_\cT ) (\sum_{c \in j} \alpha_c) } \; .
\eea
The formula above can be further simplified. 
Putting $\beta_i = \sum_{c \in i} \alpha_c $ we have
\bea   G_2(z, N) &=& \sum_{\cT} \frac{1}{n!} z^{n} \big[ \prod_{i =0}^n \int_0^\infty d \beta_i \frac{\beta_i^{d_i-1}}{(d_i-1)!}  e^{- \beta_i}      \big]  \nonumber \\
&& \int dw_\cT   e^{ \frac{z}{2N}  \sum_{ij} \beta_i X^\cT_{ij}(w_\cT )  \beta_{j} } \label{betarep}
\eea
where $d_i$ is the degree of $i$, hence the number of corners of $i$.

Setting  $z = \vert z \vert e^{i \pi + i\phi}$ and $\beta = \vert \beta \vert e^{i \psi}$ we have, for 
$ - \pi/2 \le \phi + 2 \psi  \le \pi/2$ and $- \pi/2  < \psi  < \pi /2$,  both $\cos ( \phi + 2 \psi )  \ge 0 $
and $\cos \psi > 0 $, hence
\bea && \Big{\vert} \big[ \prod_{i =0}^n \int_0^{e^{i\psi}\infty} d \beta_i \frac{\beta_i^{d_i-1}}{(d_i-1)!}   e^{- \beta_i}   \big]  
e^{  \frac{z}{2N}  \sum_{ij}  \beta_i  X^\cT_{ij}(w_\cT ) \beta_j } \Big{\vert} \nonumber \\
&&\le     \prod_{i =0}^n 
\int_0^\infty  d \vert \beta_i \vert   \frac{\vert \beta_i\vert ^{d_i-1}}{(d_i-1)!}  e^{-\vert  \beta_i \vert \cos \psi}    \nonumber \\
&& \le ( \cos \psi )^{-\sum_{i =0}^n  d_i  }  = ( \cos \psi )^{-2n-1} .
\eea
Therefore $G_2(z, N)$ is analytic in the extended cardioid $C = C_+ \cup C_-$, where $C_+$ 
is the union of the quarter-disk $0 \le \phi <  \pi /2$,  $ 4\vert z \vert  < 1 $ and of the domain $\pi/2 \le \phi <  3\pi /2$,  
$ 4\vert z \vert  < [ \cos (\phi/2 -\pi/4 )]^{2} $. $C_-$ is the complex conjugate domain.

To prove that this convergent analytic function is the Borel sum of its perturbative series at any fixed $N$ also requires uniform Taylor estimates
of the type $K^p p!  \vert z\vert^{p}$ for the Taylor remainder at order $p$ in at least a disk tangent to the imaginary axis (Nevanlinna's criterion). 
They follow from Taylor expanding the exponential of the $\beta$ quadratic form with an integral remainder:
\bea &&  e^{  \frac{z}{2N}  \sum_{ij}  \beta_i  X^\cT_{ij}(w_\cT ) \beta_j } =
\sum_{q=0}^{p-1}   \frac{z^q}{q!(2N)^q} [ \sum_{ij}  \beta_i  x^\cT_{ij}(w_\cT ) \beta_j ]^q \\&& +  \int_0^1 dt  \frac{ (1-t)^{p-1}}{(p-1)!} 
\frac{z^{p}}{(2N)^{p} }[ \sum_{ij}  \beta_i  X^\cT_{ij}(w_\cT ) \beta_j ]^{p} e^{ t \frac{z}{2N}  \sum_{ij}  \beta_i  X^\cT_{ij}(w_\cT ) \beta_j } . \nonumber
\eea
The sum over $q$, i.e. the $p$ first terms, are exactly the perturbative expansion up to order $p$ hence support a $K^pp!|z|^p$ bound.
The Taylor remainder term for any tree $\cT$ in the disk
$ - \pi/2  \le \phi  \le  \pi /2$, where we can take $\psi =0$ can be bounded as:
\bea &&   \Big{\vert}  \big[ \prod_{i =0}^n \int_0^{\infty}  \frac{\beta_i^{d_i-1}}{(d_i-1)!} d \beta_i  e^{- \beta_i}   \big]   \int_0^1 dt  \frac{ (1-t)^p}{p!} 
\frac{z^{p}}{(2N)^{p} } \nonumber\\
&&\int dw_\cT  [ \sum_{ij}  \beta_i  X^\cT_{ij}(w_\cT ) \beta_j ]^{p} e^{ t \frac{z}{2N}  \sum_{ij}  \beta_i  X^\cT_{ij}(w_\cT ) \beta_j } \Big{\vert} \nonumber\\
&&\le \frac{\vert z \vert ^{p}}{p!}\big[ \prod_{i =0}^n \int_0^{\infty}  \frac{\beta_i^{d_i-1}}{(d_i-1)!} d \beta_i  e^{- \beta_i}   \big] 
[\sum_{i=0}^n \beta_i ]^{2p} \nonumber\\
&& = \frac{\vert z \vert ^{p}}{p!}  \Bigl( \prod_{i=0}^{n} \frac{1}{(d_i-1)!} \Bigr)
\int_0^{\infty}  d\beta e^{-\beta} \beta^{2p  }
\int_{\beta_1+\dots \beta_n=\beta }  \prod_{i=0}^{n}  \beta_i^{d_i-1}   d\beta_i 
\\
&&= \frac{\vert z \vert ^{p}}{p!}  \Bigl( \prod_{i=0}^{n} \frac{1}{(d_i-1)!} \Bigr) (2p+2n+1)!
\int_{u_1+\dots u_n=1 }  \prod_{i=0}^{n}  u_i^{d_i-1}   du_i \le 4^n K^p p!|z|^p \; . \nonumber
\eea
These Taylor estimates for a single rooted plane tree can be summed over all rooted plane trees (using the $\vert z\vert ^n$ factor in \eqref{betarep})
in the half-disk defined by $16 \vert z \vert <1$ and $ - \pi/2  \le \arg z  \le  \pi /2$ (shown in red on figure \ref{extendedcardioidfig}). 
Hence  in this half-disk (which is uniform in $N$)  we obtain the desired Taylor estimates, which is more than enough to check that
the expansions \eqref{vectorlve} and \eqref{betarep} represent indeed for all $N$ the unique Borel sum of the perturbative series.

Interesting functions are the real and imaginary parts along the real axis $0\le z < 1/8$ which are
\bee G^{mean}_{2} (z, N)= \frac{ G_2(z, N)_+ + G_2(z, N)_-}{2}, G^{cut}_{2} (z, N)= \frac{ G_2(z, N)_+ -  G_2(z, N)_-}{2i},
\ee
where $G_+$ is analytically continued to $\phi = +\pi$ and $G_-$ is analytically continued to $\phi = -\pi$. Taking 
$\psi = -\pi /4$ in the first case and $\psi = + \pi/4$ is the second case, one obtains
explicitly convergent integral representations for these quantities, namely
\bea   \label{niceglouglou} G^{mean}_{2} (z, N) &=&  \sum_{\cT} \frac{1}{n!} z^{n} 
\big[ \prod_{i \in V(\cT)} \int_0^\infty  \frac{\beta_i^{d_i-1}}{(d_i-1)!} d \beta_i   e^{- \frac{\sqrt 2}{2}  \beta_i}  \big] \\
&& \int dw_\cT  \cos \bigl((2n+1) \frac{\pi}{4} + \frac{\sqrt 2}{2} \sum_i \beta_i +  \frac{z}{2N}  \sum_{ij} \beta_i X^\cT_{ij}(w_\cT ) \beta_{j} \bigr)  .\nonumber 
\eea
\bea  \label{nicebeta2} G^{cut}_{2} (z, N) &=&  \sum_{\cT} \frac{1}{n!} z^{n}  \int dw_\cT 
\big[ \prod_{i \in V(\cT)} \int_0^\infty  \frac{\beta_i^{d_i-1}}{(d_i-1)!} d \beta_i   e^{- \frac{\sqrt 2}{2}  \beta_i}  \big] \\
&& \int dw_\cT \sin \bigl((2n+1) \frac{\pi}{4} + \frac{\sqrt 2}{2} \sum_i \beta_i +  \frac{z}{2N}  \sum_{ij} \beta_i X^\cT_{ij}(w_\cT ) \beta_{j} \bigr)  . \nonumber 
\eea
where the factors $(2n+1) \frac{\pi}{4}$ come from the rotation of the $\beta$ integrals, using $\sum d_i = 2n+1$.
These convergent integrals extend half-way to the Catalan singularity $ z_{\text{Catalan}}=1/4 $. Indeed bounding the cosine or sinus function by 1
we obtain convergence, but loosing a factor $(\sqrt 2)^{\sum d_i}  = 2^n \sqrt{2}$.

One can still check easily that the limit for $N \to \infty$ of the mean integral for positive $z$ is the Catalan function.
Indeed the cosine function simplifies in that case. Rotating the $\beta$ integrals back in position we obtain again the factor 1 for each rooted plane tree.

The extended cardioid is an analyticity domain in $z$ which holds for any $N \ge 1$. In other words it is common to all $N$-vector models, including the particular $N=1$ scalar case, 
However as $N \to \infty$ we could hope for larger and larger domains of analyticity which approach the $z = 1/4$ singularity 
when $N\to \infty $; but we do not know, even in this simple vector model case, how to prove this.

In the case of quartically interacting  large $N$ matrix \cite{Rivasseau:2007fr} and large $N$ tensor models \cite{Magnen:2009at,Gurau:2013pca}, the LVE also provides
analyticity in cardioid-like domains.

The constructive treatment of renormalizable models requires a multiscale analysis, hence a multiscale version of the loop vertex expansion (MLVE). 
Following \cite{Gurau:2013oqa}, we sketch now how this expansion works in the case of a superrenormalizable toy model which is a slight modification 
of the vector model above.

\subsection{Multiscale loop vertex expansion}

Consider the same pair of conjugate vector fields $\{\phi_p \}, \{\bar \phi_p \},  p=1 ,\cdots , N $, 
with the same $\frac{\lambda^2 }{2}(\bar \phi \cdot \phi )^2$ bare interaction as in the previous section, 
but with a different Gaussian measure $d\mu (\bar \phi, \phi)$  which breaks the $U(N)$ invariance of the theory. It has
diagonal covariance (or propagator) which decreases as the inverse power of the field index:
\[
 d\eta(\bar \phi, \phi) = \Bigl(  \prod_{p=1}^N p  \frac{ d\bar \phi_p d\phi_p }{2\pi \imath} \Bigr)  e^{-\sum_{p=1}^N  p \;  \bar \phi_p \phi_p } \; ,
\qquad  \int  d\eta (\bar \phi, \phi) \; \bar\phi_p \phi_{q}= \frac{\delta_{pq}}{p} \;.
\]
This propagator renders the perturbative amplitudes of the model finite in the $N \to \infty$ limit, except for a mild 
divergence of self-loops which yields a logarithmically divergent sum
$
L_N = \sum_{p=1}^N \frac{1}{p}   \simeq \log N
$.
These divergences are easily renormalized by using a vector-Wick-ordered $\phi^4$ interaction, 
namely $\frac{1}{2}[\lambda (\bar \phi \cdot \phi -L_N)]^2$. Remark that this interaction 
(contrary to the $\phi^4_2$ case) remains positive for $\lambda$ real at all values of $(\bar\phi, \phi)$.
The renormalized partition function of the model is:
\bea\label{eq:partitionfunction} 
Z(\lambda, N) =  \int  d\eta (\bar \phi, \phi ) \; \;  e^{- \frac{\lambda^2}{2} (\bar \phi \cdot \phi -L_N)^2 }.
\eea
The intermediate field representation decomposes the quartic interaction using an intermediate scalar field $   \sigma  $:
\[
e^{- \frac{\lambda^2}{2} (\bar \phi \cdot \phi -L_N)^2 }    =  \int d\nu  (\sigma) \; 
e^{ \imath \lambda  \sigma   (\bar \phi \cdot \phi -L_N) } \; ,
\]
where $d\nu(\sigma) = \frac{1}{\sqrt{2\pi}} e^{-\frac{\sigma^2}{2} }$ is the standard 
Gaussian measure with covariance 1. Integrating over the initial fields $(\bar \phi_p, \phi_p)$ leads to:
\[
Z(\lambda, N)  =   \int d\nu  (\sigma) \; 
     \prod_{p=1}^N \frac{1}{1- \imath \frac{\lambda \sigma}{p}} e^{-\imath \frac{\lambda \sigma}{p} } 
    =  \int d\nu  (\sigma) \;  e^{- \sum_{p=1}^N \log_2 \bigl(1 - \imath \frac{\lambda  \sigma }{p } \bigr) } \; ,
\]
where $\log_2 (1-x) \equiv x+ \log (1-x) = O(x^2)$.

Applying the ordinary LVE of the previous section to this functional integral would express $\log Z(\lambda, N)$ as a sum over trees, 
but there is no simple way to remove the logarithmic divergence of all leaves of the tree without generating many intermediate fields 
in numerators which, when integrated through the Gaussian measure, would create an apparent divergence of the series. 
The MLVE is designed to solve this problem.

We fix an integer $M>1$ and define the $j$-th \emph{slice}, as made of the indices 
$p \in I_j \equiv [M^{j-1},M^{j} -1]$. The ultraviolet cutoff $N$ is chosen as $N =M^{j_{max}}-1$, with $j_{\max}$ an integer.
We can also fix an infrared cutoff $j_{\text{min}}$. 
Hence there are $j_{\max}-j_{\min}$ 
slices in the theory, and the ultraviolet limit corresponds to the limit $j_{\max} \to \infty$. 
The intermediate field representation writes:
\bea\label{eq:partitionfunctionintfield}
Z(\lambda, N) =  \int d\nu  (\sigma) \; \prod_{j =j_{\min}}^{j_{\max}}     e^{- V_j} \; , \quad 
V_{j} =\sum_{p \in I_{j}} \log_2 \Bigl(1 - \imath \frac{ \lambda \sigma}{p}  \Bigr) \; .
\eea
The factorization of the interaction over the set of slices $\cS = [j_{\min}, \cdots j_{\max}]$ can be encoded into an integral over Grassmann numbers. Indeed,
\[
   a = \int d\bar \chi d\chi \; e^{-\bar \chi a \chi} = \int d\mu (\bar \chi ,\chi ) \; e^{- \bar \chi (a-1) \chi}  
\]
where $d \mu(\bar \chi ,\chi ) = d\bar \chi d\chi \; e^{-\bar \chi \chi}$ is the standard normalized Grassmann Gaussian measure with covariance 1. Hence,
denoting $W_j(\sigma) = e^{-V_{j}}-1$, 
\[
Z(\lambda, N) =  \int d\nu  (\sigma) \; \Bigl( \prod_{j = j_{\min}}^{j_{\max} } d\mu (\bar \chi_j , \chi_j) \Bigr) \; 
    e^{ - \sum_{j = j_{\min}}^{j_{\max}}   \bar \chi_j  W_j   (  \sigma)   \chi_j } .
\]

We introduce slice replicas for the Bosonic fields, that is we rewrite the partition function as:
\bea
 Z(\lambda, N) =  \int d\nu_\cS \; e^{- W} \; ,
\quad d\nu_{\cS}  &=& d\nu_{\bbone_\cS}  ( \{ \sigma_j\}  ) \;  d\mu_{\mathbb{I}_\cS} (\{\bar \chi_j , \chi_j\})  
,\\ 
 \quad W &=& \sum_{j =j_{\min}}^{j_{\max}}   \bar \chi_j  W_j   ( \sigma_j)   \chi_j \; .
\eea
This is the starting point for the MLVE. The first step is to expand to infinity the exponential of the interaction:
\[
Z(\lambda, N) =  \sum_{n=0}^\infty \frac{1}{n!}\int d\nu_{\cS}  \; (-W)^n \; .
\]
The second step is to introduce replica Bosonic fields for all the vertices in $V = \{1, \cdots , n\}$: 
\[
Z(\lambda, N) = \sum_{n=0}^\infty \frac{1}{n!}  \int d\nu_{\cS,V} \;  \prod_{a=1}^n  (-W_a) \; ,
\]
where the $a$-th vertex 
\bee
W_a = \sum_{j =j_{\min}}^{j_{\max}}  W_{a,j}  , \quad  W_{a,j} =    \bar \chi_j W_j   (  \sigma^a_j )  \chi_j \;,
\ee
 has now its own (replicated) Bosonic fields $\sigma_j^a$ and the replica measure is completely degenerate:
\[
d\nu_{\cS,V}  = d\nu_{\bbone_\cS \otimes \bbone_V} (\{  \sigma^a_j\}) \;  d\mu_{\mathbb{I}_\cS  } (\{\bar \chi_j , \chi_j\})   .
\]
No vertex replicas are used yet for the Fermionic fields. 

The obstacle to factorize this integral over vertices lies now in the Bosonic degenerate blocks $\bbone_V$ and 
in the Fermionic fields (which couple the vertices $W_a$). In order to deal with these {\it two} different couplings we will apply {\it two} successive
forest formulas. First, in order to disentangle the block $\bbone_V$ in the measure $d \nu$ 
we introduce the matrix $x_V$ with coupling parameters $x_{ab}=x_{ba}, x_{aa}=1$ between the vertex Bosonic replicas:
\[
 Z(\lambda, N) = \sum_{n=0}^\infty \frac{1}{n!} 
 \int   d\nu_{\bbone_\cS \otimes x_V} (\{  \sigma^a_j\}) \;  d\mu_{\mathbb{I}_\cS  } (\{\bar \chi_j , \chi_j\}) 
  \prod_{a=1}^n \Bigl( -  \sum_{j =j_{\min}}^{j_{\max}}   W_{a,j} \Bigr) 
   \Bigr]_{x_{ab}=1 } \;,
\]
and apply the forest formula. We denote $\cF_B$ a Bosonic forest with $n$ vertices labelled $\{1,\dots n\}$, 
$\ell_{B}$ a generic edge of the forest and $a(\ell_B), b(\ell_B)$ the end vertices of $\ell_B$. The result of the first forest formula is:
\begin{eqnarray*}
&&  Z(\lambda, N) = \sum_{n=0}^\infty \frac{1}{n!}
 \sum_{\cF_{B}}  \int dw_{\cF_B}   \int   d\nu_{\bbone_\cS \otimes X(w_{\ell_B})} 
 (\{  \sigma^a_j\}) \;  d\mu_{\mathbb{I}_\cS  } (\{\bar \chi_j , \chi_j\}) 
\crcr
&& \qquad \qquad \qquad \times   \partial_{\cF_B}
\;
 \prod_{a=1}^n \Bigl( -  \sum_{j =j_{\min}}^{j_{\max}}  W_{a,j}  \Bigr) \Bigg]\; ,
\end{eqnarray*}
where 
\bee \int dw_{\cF_B} = \prod_{\ell_B\in \cF_B }  \int_{0}^1  dw_{\ell_B} , \quad
\partial_{\cF_B}=  \prod_{\ell_B \in \cF_{B}} \Bigl( \sum_{j,k=j_{\min}}^{j_{\max}}  
   \frac{\partial}{\partial \sigma^{a(\ell_B)}_j}\frac{\partial}{\partial \sigma^{b(\ell_B)}_k}  \Bigr)\ee
and $X_{ab}(w_{\ell_B})$ is the infimum over the parameters $w_{\ell_B}$ in the unique path
in the forest $\cF_B$ connecting $a$ and $b$, and the infimum is set to 
$1$ if $a=b$ and to zero if $a$ and $b$ are not connected by the forest.

The forest $\cF_B$ partitions the set of vertices into blocks $\cB$ corresponding to its trees. Remark that the blocks can be singletons 
(corresponding to the trees with no edges in $\cF_B$). We denote $a\in \cB$ if the vertex $a$ belongs to a Bosonic block $\cB$. A vertex
belongs to a unique Bosonic block.
Contracting every Bosonic block to an ``effective vertex'' we obtain a graph which we denote $\{n\}/\cF_B$.
We introduce replica Fermionic fields $\chi^{\cB}_j$ for the blocks of $\cF_B$ (i.e. for the effective vertices 
of $\{n\}/\cF_B$) and replica coupling parameters $y_{\cB\cB'}=y_{\cB'\cB}$. 
Applying (a second time) the forest formula, this time for the $y$'s, leads to a set of Fermionic edges $\cL_F$ forming a forest 
in $\{n\}/\cF_B$ (hence connecting Bosonic blocks). We denote $L_{F} $ a generic Fermionic edge connecting blocks and $\cB(L_F), \cB'(L_F) $ 
the end blocks of the Fermionic edge $L_F$. We obtain: 
\bea\label{eq:intermediate}
&&  Z(\lambda, N) = \sum_{n=0}^\infty \frac{1}{n!} \sum_{\cF_{B}} \sum_{\cL_F}
 \int dw_{\cF_B}      \int dw_{\cL_F} 
\int   d\nu_{\bbone_\cS \otimes X(w_{\ell_B})}  (\{  \sigma^a_j\}) \;   
 \crcr
&&  \times  d\mu_{\mathbb{I}_\cS   \otimes  Y (  w_{L_F})   } (\{\bar \chi^\cB_j , \chi^\cB_j\}) 
\partial_{\cF_B}  \partial_{\cL_F}
  \prod_{\cB}  \prod_{a\in \cB} \Bigl( -  \sum_{j =j_{\min}}^{j_{\max }}   \bar \chi^{\cB }_j W_j   (  \sigma^a_j )  
\chi^{\cB }_j \Bigr) 
\eea 
where 
\begin{eqnarray*}  &&  \int dw_{\cL_F} = \prod_{L_F\in \cL_F } \int_0^1 dw_{L_F},\\
&&\partial_{\cL_F}= \prod_{L_F \in \cL_F} 
\Bigg(\sum_{j=j_{\min} }^{j_{\max}} \Bigl( \frac{\partial}{\partial \bar \chi_j^{\cB(L_F)} } \frac{\partial}{\partial \chi_j^{\cB'(L_F)} }  
+   \frac{\partial}{\partial \bar  \chi_j^{\cB'(L_F)} } \frac{\partial}{\partial  \chi_j^{\cB(L_F)}} \Bigr)\Bigg) \; ,
\end{eqnarray*}
and
$Y_{\cB\cB'}(w_{\ell_F}) $ is the infimum over $w_{\ell_F}$ in the unique path in $\cL_{F}$ connecting $\cB$ and $\cB'$, this infimum being 
set to $1$ if $\cB= \cB'$ and to zero if $\cB$ and $\cB'$ are not connected by $\cL_F$. Note that the Fermionic edges are oriented. 
Expanding the sums over $j$ in the last line of eq. \eqref{eq:intermediate} we obtain a sum over slice assignments $J= \{j_a  \}$ to the vertices $a$, 
where $j_a  \in [ j_{\min} , j_{\max}]$. Taking into account that $\partial_{\sigma^a_j} W(\sigma^{a}_{j_a}) = \delta_{jj_a} \partial_{\sigma^a_{j_a} } W(\sigma^{a}_{j_a}) $
we obtain:
\begin{eqnarray*}
Z(\lambda, N) &=& \sum_{n=0}^\infty \frac{1}{n!} \sum_{\cF_{B}} \sum_{\cL_F} \;\sum_J
 \;  \int dw_{\cF_B}      \int dw_{\cL_F} \crcr
&\times& 
\int   d\nu_{\bbone_\cS \otimes X(w_{\ell_B})}  (\{  \sigma^a_j\}) 
 d\mu_{\mathbb{I}_\cS   \otimes  Y (  w_{L_F})   } (\{\bar \chi^\cB_j , \chi^\cB_j\})   \crcr
&\times&  \partial_{\cF_B}  \partial_{\cL_F} \; \prod_{\cB}  \prod_{a\in \cB} \Bigl( -    \bar \chi^{\cB }_{j_a} W_{j_a}   (  \sigma^a_{j_a} )  
\chi^{\cB }_{j_a} \Bigr) .  
\end{eqnarray*} 
In order to compute the derivatives in $\partial_{\cL_F} $ with respect to the block Fermionic fields $\chi^{\cB}_j$ and $\bar \chi^{\cB}_j$ 
we note that such a derivative acts only on 
$ \prod_{a\in \cB} \Bigl(  \chi^{\cB}_{j_a}   \bar \chi^{ \cB  }_{j_a}  \Bigr)  $ and, furthermore, 
\bea
  \frac{\partial}{\partial \bar \chi_j^{\cB} } 
  \prod_{a\in \cB} \Bigl(  \chi^{\cB}_{j_a}   \bar \chi^{ \cB  }_{j_a}  \Bigr)  
 &=& \Bigl( \sum_{a'\in \cB} \delta_{jj_{a'}}  \frac{\partial}{\partial \bar \chi_{j_{a'}}^{\cB} }  \Bigr)
  \prod_{a\in \cB} \Bigl(  \chi^{\cB}_{j_a}   \bar \chi^{ \cB  }_{j_a}  \Bigr)  
 \crcr
 \frac{\partial}{\partial \chi_j^{\cB} } 
  \prod_{a\in \cB} \Bigl(  \chi^{\cB}_{j_a}   \bar \chi^{ \cB  }_{j_a}  \Bigr)  
  &=& \Bigl( \sum_{a'\in \cB} \delta_{jj_{a'}}  \frac{\partial}{\partial   \chi_{j_{a'}}^{\cB} }  \Bigr)
  \prod_{a\in \cB} \Bigl(  \chi^{\cB}_{j_a}   \bar \chi^{ \cB  }_{j_a}  \Bigr)  \; .
\eea
It follows that the Grassmann Gaussian integral is:
\begin{eqnarray*}
 && \Bigg[  e^{ 
   \sum_{\cB,\cB'} Y_{\cB\cB'}(w_{\ell_F})\sum_{a\in \cB, b\in \cB'} \delta_{j_aj_b}
     \frac{\partial}{\partial \bar \chi_{j_a}^{\cB} } \frac{\partial}{\partial \chi_{j_b}^{\cB'} } } \crcr 
  &&      \prod_{L_F \in \cL_F} 
   \Bigg(\sum_{a\in \cB(L_F),b\in \cB'(L_{F})}  \delta_{j_a j_b}\Big( \frac{\partial}{\partial \bar \chi_{j_a}^{\cB(L_F)} } 
   \frac{\partial}{\partial \chi_{j_b}^{\cB'(L_F)} } +  
   \frac{\partial}{\partial \bar \chi_{j_b}^{\cB'(L_F)} } \frac{\partial}{\partial \chi_{j_a}^{\cB(L_F)} } \Big)
   \Bigg) \crcr
&&
  \qquad \qquad\qquad  \prod_{\cB} \prod_{a\in \cB} \Bigl(  \chi^{\cB}_{j_a}   \bar \chi^{ \cB  }_{j_a}  \Bigr)  \Bigg]_{\chi^{\cB}_j, \bar\chi^{\cB}_j =0 } \; .
\end{eqnarray*}
The sums over $  a\in \cB(\ell_F) $ and  $ b\in \cB'(\ell_F)$ yield a sum over all the possible ways to hook the edge $L_F\in \cL_F$ 
to vertices in its end blocks. Each term represents a detailed Fermionic edge $\ell_F$ in the original graph (having the same $w_{\ell_F}= w_{L_F}$ parameter).
The sum over $\cL_F$ becomes therefore a sum over detailed Fermionic forests $\cF_F$ in the original graph (in which the Bosonic blocks are not contracted)
and we obtain a two-level jungle formula \cite{Abdesselam:1994ap} for the partition function:
\[
Z(\lambda, N) =  \sum_{n=0}^\infty \frac{1}{n!}  \sum_{\cJ} \;\sum_J
\int dw_\cJ   \int d\nu_{ \cJ}  \partial_\cJ  \prod_{\cB} \prod_{a\in \cB}   \Bigl(    W_{j_a}   (  \sigma^a_{j_a} )  
\chi^{ \cB }_{j_a}   \bar \chi^{\cB}_{j_a}  \Bigr)  ,
\]
where
\begin{itemize}
\item  the sum over $J$ means $\sum_{j_1=j_{\min}}^{j_{\max }  } \dots \sum_{j_n=j_{\min}}^{j_{\max} }$,

\item the sum over $\cJ$ runs over all two level jungles, hence over all ordered pairs $\cJ = (\cF_B, \cF_F)$ of two (each possibly empty) 
disjoint forests on $V$, such that 
$\bar \cJ = \cF_B \cup \cF_F $ is still a forest on $V$. The forests $\cF_B$ and $\cF_F$ are the Bosonic and Fermionic components of $\cJ$.
The edges of $\cJ$ are partitioned into Bosonic edges $\ell_B$ and Fermionic edges $\ell_F$.
 
\item  $\int dw_\cJ$ means integration from 0 to 1 over parameters $w_\ell$, one for each edge $\ell \in \bar\cJ$.
$\int dw_\cJ  = \prod_{\ell\in \bar \cJ}  \int_0^1 dw_\ell  $.
A generic integration point $w_\cJ$
is therefore made of $\vert \bar \cJ \vert$ parameters $w_\ell \in [0,1]$, one for each $\ell \in \bar \cJ$.

\item 
\[ \partial_\cJ  = \prod_{\genfrac{}{}{0pt}{}{\ell_B \in \cF_B}{\ell_B=(c,d)}} \Bigl(  \frac{\partial}{\partial \sigma^{c}_{j_{ c}} } 
 \frac{\partial}{\partial \sigma^{d}_{j_{d }  } } \Bigr)
\prod_{\genfrac{}{}{0pt}{}{\ell_F \in \cF_F}{\ell_F=(a,b) } } \delta_{j_{a } j_{b } } \Big(
   \frac{\partial}{\partial \bar \chi^{\cB(a)}_{j_{a}  } }\frac{\partial}{\partial \chi^{\cB(b)}_{j_{b }  } }+ 
    \frac{\partial}{\partial \bar \chi^{ \cB( b) }_{j_{b} } } \frac{\partial}{\partial \chi^{\cB(a) }_{j_{a}  } }
   \Big) \; ,
\]
where $ \cB(a)$ denotes the Bosonic blocks to which $a$ belongs. 

\item the measure $d\nu_{\cJ}$ has covariance $ X (w_{\ell_B}) \otimes \bbone_\cS $ on Bosonic variables and $ Y (w_{\ell_F}) \otimes \mathbb{I}_\cS  $  
on Fermionic variables, hence $\int d\nu_{\cJ} f$ is the value at $\sigma = \bar \chi = \chi =0$ of
\[
 e^{\frac{1}{2} \sum_{a,b=1}^n X_{ab}(w_{\ell_B }) \frac{\partial}{\partial \sigma^a_{j_a} }\frac{\partial}{\partial \sigma^b_{j_b}} 
   +  \sum_{\cB,\cB'} Y_{\cB\cB'}(w_{\ell_F})\sum_{a\in \cB,  b\in \cB' } \delta_{j_aj_b}
   \frac{\partial}{\partial \bar \chi_{j_a}^{\cB} } \frac{\partial}{\partial \chi_{j_b}^{\cB'} } } \; f ,
\] 

\item  $X_{ab} (w_{\ell_B} )$  is the infimum of the $w_{\ell_B}$ parameters for all the Bosonic edges $\ell_B$
in the unique path $P^{\cF_B}_{a \to b}$ from $a$ to $b$ in $\cF_B$. The infimum is set to zero if such a path does not exists and 
to $1$ if $a=b$. 

\item  $Y_{\cB\cB'}(w_{\ell_F})$  is the infimum of the $w_{\ell_F}$ parameters for all the Fermionic
edges $\ell_F$ in any of the paths $P^{\cF_B \cup \cF_F}_{a\to b}$ from some vertex $a\in \cB$ to some vertex $b\in \cB'$. 
The infimum is set to $0$ if there are no such paths, and to $1$ if such paths exist but do not contain any Fermionic edges.

\end{itemize}

Remember that the symmetric $n$ by $n$ matrix $X_{ab}(w_{\ell_B})$ 
is positive for any value of $w_\cJ$, hence the Gaussian measure $d\nu_{\cJ} $ is well-defined. The matrix $Y_{\cB\cB'}(w_{\ell_F})$
is also positive, with all elements between 0 and 1. Since the slice assignments, the fields, the measure and the integrand are now 
factorized over the connected components of $\bar \cJ$, the logarithm of $Z$ is exactly the same sum but restricted 
to the two-levels spanning trees:
\bea \label{treerep}  
\log Z(\lambda, N) =  \sum_{n=1}^\infty \frac{1}{n!}  \sum_{\cJ \,{\rm tree}} \sum_J
 \int dw_\cJ  \int d\nu_{ \cJ}  
 \partial_\cJ  \prod_{\cB} \prod_{a\in \cB}   \bigl[  W_{j_a}   (  \sigma^a_{j_a} )  
 \chi^{ \cB }_{j_a} \bar \chi^{\cB}_{j_a} \bigr] 
\eea
where the sum is the same but conditioned on $\bar \cJ = \cF_B \cup \cF_F$ being a \emph{spanning tree} on $V= [1, \cdots , n]$.
In \cite{Gurau:2013oqa}, it is proven in detail that
\begin{theorem} \label{thetheorem} Fix $j_{\min}\ge 3$ and  $M \ge 10^8$.
The series \eqref{treerep} is absolutely convergent for $\lambda\in [-1,1]$ uniformly in $j_{\max}$.
\end{theorem}

\begin{theorem} \label{thm:theorem2} Fix $j_{\min}\ge 3$ and  $M \ge 10^8$. The series \eqref{treerep} is absolutely convergent for 
$\lambda\in \mathbb{C}$, $\lambda = |\lambda|e^{\imath \gamma}$ in the domain $    |\lambda|^2 <   (\cos2\gamma) $ uniformly in $j_{\max}$. 
\end{theorem}

The conditions $j_{\min}\ge 3$ and  $M \ge 10^8$ are not optimal and were chosen for simplicity of the resulting domain $\lambda\in [-1,1]$. 
We sketch below the proof of theorem \ref{thetheorem}, referring the reader to \cite{Gurau:2013oqa} for details. By Cayley's theorem
the number of two level trees over $n\ge 1$ vertices  is exactly $2^{n-1}n^{n-2}$. 

The Grassmann Gaussian integral evaluates to: 
\bea\label{eq:grassmaint}
&& \Bigl( \prod_{\cB} \prod_{\genfrac{}{}{0pt}{}{a,b\in \cB}{a\neq b}} (1-\delta_{j_aj_b}) \Bigr)
 \Bigl( \prod_{\genfrac{}{}{0pt}{}{\ell_F \in \cF_F}{\ell_F=(a,b) } } \delta_{j_{a } j_{b } } \Bigr) \crcr
&& \times \Bigl( {\bf Y }^{\hat b_1 \dots \hat b_k}_{\hat a_1 \dots \hat a_k}  + 
 {\bf Y }^{\hat a_1 \dots \hat b_k}_{\hat b_1 \dots \hat a_k}+\dots + {\bf Y }_{\hat b_1 \dots \hat b_k}^{\hat a_1 \dots \hat a_k}   \Bigr)  \; ,
\eea 
where the sum runs over the $2^k$ ways to exchange an $a_i$ and a $b_i$. 
Each $ \Big{|}  {\bf Y }^{\hat a_1 \dots \hat b_k}_{\hat b_1 \dots \hat a_k} \Big{|}$
factor is bounded by 1 thanks to Hadamard's inequality, because the matrix $Y$ is positive with diagonal entries equal to 1.

The Bosonic integral is a bit more cumbersome, as one should first evaluate the effect of the Bosonic derivatives on the exponential
vertex kernels $W_j$ through the Fa\`a di Bruno formula, whose combinatoric is easy to control. 
It leads to a sum over similar exponential kernels but multiplied by some polynomials.

To bound the remaining Bosonic functional integral one first separates the exponential kernels from the polynomials by
some Cauchy-Schwarz estimate with respect to the Bosonic Gaussian measure. The exponential terms
being positive, the corresponding piece is bounded by 1. The polynomial piece is then explicitly evaluated.
This generates a dangerous product of local factorials of the number of fields in the Bosonic blocks,
but allows also a good factor $M^{-j}$ from the propagator of scale $j$
for each occupied Bosonic scale $j$.

But here comes the key point. The Grassmann Gaussian integrals ensure that the occupied scales in any Bosonic block
of the first forest formula are all \emph{distinct}. Therefore the good factor collected from the propagator easily beats the local factorials. The worst case
is indeed when the $p$ occupied scales in the block are lowest, in which case $\prod_{j=1}^p  M^{-j}  = M^{-p(p+1)/2}$ which
easily beats $p!$.

For just renormalizable theories it is not so easy to beat the dangerous factors by the decay of the propagators, and the constructive expansion
must proceed even more carefully, essentially expanding the functional integral in each scale in a much more detailed way.

\section*{Acknowledgements}
\addcontentsline{toc}{section}{Acknowledgements}
The authors thank B.~Delamotte and N.~Wschebor for very useful comments on the manuscript. A.S. also thanks G. Torrieri for comments on the manuscript.\\
V.R. thanks J. Magnen for a life-long collaboration on multiscale analysis and renormalization.\\
A.S. acknowledges partial support by the Netherlands Organization for Scientific Research (NWO) under the VICI grant 680-47-602. Furthermore part of the research leading to these results has received funding from the People Programme (Marie Curie Actions) of the European Union's Seventh Framework Programme FP7/2007-2013/ under REA Grant Agreement No 317089. A.S.'s work is also part of the ERC Advanced grant research programme No.
246974, ``Supersymmetry: a window to non-perturbative physics''.

\newpage

\bibliographystyle{nb}
\bibliography{review}

\end{document}